\ifx\pdfoutput\undefined % We're not running pdftex
\documentclass[twoside,titlepage,11pt,a4paper]{report}
\usepackage[all]{xy}                         % xyfig
\usepackage{amssymb,amsmath,amsfonts,amsthm}
\else
\documentclass[pdftex,twoside,titlepage,11pt,a4paper]{report}
\usepackage[all]{xy}                  % xyfig
\usepackage{amssymb,amsmath,amsfonts,amsthm}
\fi

\ifx\pdfoutput\undefined % We're not running pdftex
  \usepackage[dvips]{graphicx}
\else
  \usepackage[pdftex]{graphicx}
\fi

\newlength{\centeroffset}
\newlength{\lengthoffset}
\def\clock{{\count0=\time
           \divide\count0 60
           \ifnum\count0<10 0\fi\the\count0
           \multiply\count0 -60 \advance\count0 \time
           :\ifnum\count0<10 0\fi \the\count0
         }}
\newcommand{\timestamp}{{\small\vbox{\hbox{\tt\jobname.tex}\hbox{\the\day/\the\month/\the\year, \clock}}}}

\newcommand{\Z}{\mathbb{Z}}
\newcommand{\C}{\mathbb{C}}
\newcommand{\R}{\mathbb{R}}

\newcommand{\half}{\mbox{$\frac{1}{2}$}}

\newcommand{\fourth}{\mbox{$\frac{1}{4}$}}

\newcommand{\al}{\alpha} \newcommand{\ga}{\gamma}
\newcommand{\Ga}{\Gamma} \newcommand{\be}{\beta}
\newcommand{\ka}{\kappa} \newcommand{\de}{\delta}
\newcommand{\ep}{\epsilon} \newcommand{\si}{\sigma}
\newcommand{\la}{\lambda} \newcommand{\ta}{\tau}
\newcommand{\om}{\omega} \newcommand{\Om}{\Omega}
 \newcommand{\De}{\Delta}
\newcommand{\La}{\Lambda} \newcommand{\tha}{\theta}

\def\Xint#1{\mathchoice
   {\XXint\displaystyle\textstyle{#1}}%
   {\XXint\textstyle\scriptstyle{#1}}%
   {\XXint\scriptstyle\scriptscriptstyle{#1}}%
   {\XXint\scriptscriptstyle\scriptscriptstyle{#1}}%
   \!\int}
\def\XXint#1#2#3{{\setbox0=\hbox{$#1{#2#3}{\int}$}
     \vcenter{\hbox{$#2#3$}}\kern-.5\wd0}}
\def\dashint{\Xint-} %One dash

\newtheorem*{thmcolemanmandula}{Coleman-Mandula}
\newtheorem*{thmdvsun}{Dijkgraaf-Vafa conjecture; Traceless case}
\newtheorem*{thmdvun}{Dijkgraaf-Vafa conjecture; $\boldsymbol{{\mathrm{U}(N_i)}}$-case}
\newtheorem*{thmdvgen}{Dijkgraaf-Vafa conjecture; General case}

\providecommand{\abs}[1]{\left\lvert#1\right\rvert}
\providecommand{\norm}[1]{\left\lVert#1\right\rVert}

\DeclareMathOperator{\sumlimits}{\sum\limits}

\DeclareMathOperator{\tr}{Tr}

\DeclareMathOperator{\ad}{ad}

\DeclareMathOperator{\im}{Im}

\DeclareMathOperator{\deltafunk}{\de}
\DeclareMathOperator{\deltafunkto}{\de^{(2)}}
\DeclareMathOperator{\deltafunkfire}{\de^{(4)}}

\DeclareMathOperator{\vol}{vol}

\DeclareMathOperator{\supp}{supp}

\newcommand{\defi}{\equiv}

\newcommand{\N}{\mathbb{N}}

\newcommand{\cG}{\mathcal{G}}
\newcommand{\cF}{\mathcal{F}}
\newcommand{\cD}{\mathcal{D}}
\newcommand{\qgen}{\mathcal{Q}}
\newcommand{\qbargen}{\mathcal{\bar{Q}}}
\newcommand{\Nscr}{\mathcal{N}}
\newcommand{\Jgen}{\mathcal{J}}
\newcommand{\Pgen}{\mathcal{P}}
\newcommand{\Agen}{\mathcal{A}}
\newcommand{\Bgen}{\mathcal{B}}
\newcommand{\Cgen}{\mathcal{C}}
\newcommand{\Zgen}{\mathcal{Z}}
\newcommand{\Xgen}{\mathcal{X}}
\newcommand{\lagr}{\mathcal{L}}
\newcommand{\W}{\mathcal{W}}

\newcommand{\cO}{\mathcal{O}}

\newcommand{\sigmabold}{\boldsymbol{\sigma}}
\newcommand{\thetabold}{\boldsymbol{\theta}}
\newcommand{\betabold}{\boldsymbol{\beta}}

\newcommand{\sltoc}{\textrm{SL}(2,\mathbb{C})}
\newcommand{\suto}{\textrm{SU}(2)}
\newcommand{\sun}[1]{\textrm{SU}(#1)}
\newcommand{\son}[1]{\textrm{SO}(#1)}
\newcommand{\on}[1]{\textrm{O}(#1)}
\newcommand{\un}[1]{\textrm{U}(#1)}
\newcommand{\glnc}[1]{\textrm{GL}(#1,\mathbb{C})}
\newcommand{\cun}[1]{\mathcal{U}(#1)}

\newcommand{\idmatr}{\mathbf{1}}

\newcommand{\Qbar}{\bar{Q}}
\newcommand{\Dbar}{\bar{D}}
\newcommand{\sigmabar}{\bar{\sigma}}
\newcommand{\psibar}{\bar{\psi}}
\newcommand{\Psibar}{\bar{\Psi}}
\newcommand{\Phibar}{\bar{\Phi}}
\newcommand{\chibar}{\bar{\chi}}
\newcommand{\thabar}{\bar{\theta}}
\newcommand{\xibar}{\bar{\xi}}

\newcommand{\labar}{\bar{\lambda}}
\newcommand{\Labar}{\bar{\Lambda}}
\newcommand{\phibar}{\bar{\phi}}
\newcommand{\ibar}{\bar{\imath}}
\newcommand{\jbar}{\bar{\jmath}}
\newcommand{\nablabar}{\bar{\nabla}}
\newcommand{\mbar}{\bar{m}}

\newcommand{\aldot}{\dot{\alpha}}
\newcommand{\bedot}{\dot{\beta}}
\newcommand{\gadot}{\dot{\gamma}}

\newcommand{\Mnedop}[2]{M_{#1}^{\phantom{#1}#2}}
\newcommand{\sigmaned}[3]{\left(\sigma_{#1}\right)_{#2\dot{#3}}}
\newcommand{\sigmaop}[3]{\left(\sigma^{#1}\right)_{#2\dot{#3}}}
\newcommand{\sigmabarned}[3]{\left(\bar{\sigma}_{#1}\right)^{\dot{#2}#3}}

\newcommand{\partialop}[1]{\frac{\partial}{\partial\psi^{#1}}}
\newcommand{\partialned}[1]{\frac{\partial}{\partial\psi_{#1}}}
\newcommand{\partialbarop}[1]{\frac{\partial}{\partial\bar{\psi}^{\dot{#1}}}}

\newcommand{\ddthaop}[1]{\frac{\partial}{\partial\theta^{#1}}}

\newcommand{\ddthabarop}[1]{\frac{\partial}{\partial\bar{\theta}^{\dot{#1}}}}

\newcommand{\ph}[1]{\phantom{#1}}

\newcommand{\vep}{\varepsilon}

\newcommand{\inted}{\textrm{d}}
\newcommand{\dx}{\textrm{d}x}
\newcommand{\dtha}{\textrm{d}\theta}
\newcommand{\dthabar}{\textrm{d}\bar{\theta}}
\newcommand{\dtotha}{\textrm{d}^2\theta}
\newcommand{\dtothabar}{\textrm{d}^2\bar{\theta}}
\newcommand{\dfourtha}{\textrm{d}^2\theta^1\textrm{d}^2\theta^2}
\newcommand{\dsuper}{\textrm{d}^4\theta}
\newcommand{\dlor}{\textrm{d}^4x}
\newcommand{\dla}{\textrm{d}\la}
\newcommand{\dtopi}{\textrm{d}^2\pi}
\newcommand{\dmom}{\frac{\textrm{d}^4p}{(2\pi)^4}}
\newcommand{\dmoma}{\frac{\textrm{d}^4p'_a}{(2\pi)^4}}
\newcommand{\dsi}{\textrm{d}s_i}
\newcommand{\dtow}{\textrm{d}^2\mathcal{W'}}
\newcommand{\denw}{\textrm{d}\mathcal{W'}}

\newcommand{\DM}{\!\mathcal{D}M}
\newcommand{\DPhi}{\!\mathcal{D}\Phi}
\newcommand{\DPhibar}{\mathcal{D}\bar{\Phi}}

\newcommand{\kahler}{K\"{a}hler}

\newcommand{\repr}{(\mathbf{r})}
\newcommand{\reprn}{\mathbf{r}}

\newcommand{\wzgauge}{(\textrm{WZ-gauge})}

\newcommand{\cc}{\textrm{c.c.}}

\newcommand{\Ftilde}{\tilde{F}}

\newcommand{\nleft}{\!\left}

\newcommand{\weff}{W_{\mathrm{eff}}}
\newcommand{\leff}{\mathcal{L}_{\mathrm{eff}}}
\newcommand{\wtree}{W_{\mathrm{tree}}}
\newcommand{\wvy}{W_{\mathrm{VY}}}
\newcommand{\weffpert}{W_{\mathrm{eff,pert}}}
\newcommand{\wghost}{W_{\mathrm{ghost}}}

\newcommand{\wpoly}{P_{n+1}}

\newcommand{\Shat}{\hat{S}}
\newcommand{\What}{\hat{\mathcal{W}}}
\newcommand{\lahat}{\hat{\lambda}}
\newcommand{\lahatbar}{\bar{\hat{\lambda}}}
\newcommand{\Phihat}{\hat{\Phi}}
\newcommand{\Vhat}{\hat{V}}
\newcommand{\Dhat}{\hat{D}}
\newcommand{\Fhat}{\hat{F}}

\newcommand{\psihat}{\hat{\psi}}
\newcommand{\psihatbar}{\bar{\hat{\psi}}}
\newcommand{\phihat}{\hat{\phi}}

\newcommand{\Wtilde}{\tilde{\mathcal{W}}}
\newcommand{\Phitilde}{\tilde{\Phi}}
\newcommand{\Dtilde}{\tilde{D}}
\newcommand{\psitilde}{\tilde{\psi}}
\newcommand{\psitildebar}{\bar{\tilde{\psi}}}
\newcommand{\phitilde}{\tilde{\phi}}
\newcommand{\latilde}{\tilde{\lambda}}
\newcommand{\latildebar}{\bar{\tilde{\lambda}}}

\newcommand{\gs}{g_{\mathrm{s}}}
\newcommand{\gm}{g_{\mathrm{m}}}

\newcommand{\Zm}{Z_{\mathrm{matrix}}}
\newcommand{\Zholo}{Z_{\mathrm{holo}}}

\newcommand{\lagrun}{\mathcal{L}_{\mathrm{U}(N)}}

\newcommand{\tauab}{\tau_{\mathrm{abel}}}
\newcommand{\gabel}{g_{\mathrm{abel}}}
\newcommand{\varthetaabel}{\vartheta_{\mathrm{abel}}}

\newcommand{\expect}[1]{\langle #1\rangle}

\newcommand{\vacket}{\lvert\textrm{vac}\rangle}
\newcommand{\vacbra}{\langle\textrm{vac}\rvert}
\newcommand{\ket}[1]{\lvert#1\rangle}

\newcommand{\projchiral}{P_{\mathrm{ch}}}
\newcommand{\projantichiral}{P_{\mathrm{anti-ch}}}

\newcommand{\Phicov}{\Phi^\sharp}
\newcommand{\sqcov}{\square_{\mathrm{cov}}}

\newcommand{\opned}[2]{^{#1}_{\phantom{#1}#2}}

\newcommand{\eucl}{(\mathrm{E})}
\newcommand{\meucl}{m_\mathrm{E}}
\newcommand{\geuclk}{g_{\mathrm{E},k}}

\newcommand{\lavect}{\vec{\la}}
\newcommand{\Wvect}{\vec{\mathcal{W}}}

\newcommand{\nsymm}{n_{\textrm{symm}}}
\numberwithin{equation}{chapter}
\begin{document}

\thispagestyle{empty}

\vspace{4.5cm}
\begin{center}
{\huge

\textbf{On the Dijkgraaf-Vafa Conjecture}}

\vspace{3cm}

{\Large Cand. Scient. Thesis

\vspace{1cm}

by

\vspace{1cm}

Peter Browne R\o nne}

\vspace{0.2cm}

\texttt{roenne@nbi.dk}
%\vspace{3cm}
\vspace{\stretch{1}}

\end{center}

\begin{center}
\textbf{Abstract}
\end{center}
This master's thesis gives a thorough and pedagogical introduction
to the Dijkgraaf-Vafa conjecture which tells us how to calculate
the exact effective glueball superpotential in a wide range of
$\mathcal{N}=\textrm{1}$ supersymmetric gauge theories in four
space-time dimensions using a related matrix model. The
introduction is purely field theoretical and reviews all the
concepts needed to understand the conjecture. Furthermore,
examples of the use of the conjecture are given. Especially, we
find the one-cut solution of the matrix model and use this to
obtain exact superpotentials in the case of unbroken gauge groups.
Also the inclusion of the Veneziano-Yankielowicz superpotential
and problems such as the nilpotency of the glueball superfield are
discussed. Finally, we present the diagrammatic proof of the
conjecture in detail, including the case where we take into
account the abelian part of the supersymmetric gauge field
strength. This master's thesis was handed in May 2004 and
appears here with minor changes.

\vspace{\stretch{1}}

\begin{flushright}

Supervisor: Niels Obers \hspace{\stretch{1}}University of
Copenhagen

The Niels Bohr Institute

Blegdamsvej 17, 2100 Copenhagen \O, Denmark

Handed in May 2004
\end{flushright}

\newpage
\thispagestyle{empty} \ph{phantom}
\newpage
%--------------------------------

\pagenumbering{roman}
%---------------------------------------------------------
\tableofcontents
\newpage
%---------------------------------------------------------
\pagenumbering{arabic}\setcounter{page}{1}
\chapter*{Introduction}
\addcontentsline{toc}{chapter}{Introduction}

One of the main problems in supersymmetric gauge theories is to
understand the low-energy dynamics. This is particularly important
since the $\Nscr=1$ supersymmetric gauge theories in four
space-time dimensions are believed to exhibit some of the same
non-perturbative phenomena as QCD, e.g. confinement, mass gaps and
chiral symmetry breaking. Investigation of the low-energy physics
is, however, difficult due to the strong coupling of the gauge
theory. But due to a remarkable conjecture by R. Dijkgraaf and C.
Vafa it is now possible to systematically obtain the exact
effective glueball superpotential in a wide range of
$\mathcal{N}=\textrm{1}$ supersymmetric gauge theories in four
space-time dimensions. With the glueball superpotential we can
e.g. calculate the values of the gaugino condensates and for the
corresponding chiral symmetry breaking we can find the tension of
the associated domain walls.

The Dijkgraaf-Vafa conjecture tells us that the full glueball
superpotential is a sum of a potential which is perturbative in
the glueball superfield and the Veneziano-Yankielowicz
superpotential~\cite{venezianoyankielowicz} for the supersymmetric
Yang-Mills theory. Furthermore, the perturbative part is related
in a simple way to the free energy of an associated matrix model.
For a $\un{N}$ gauge group and matter in the form of an adjoint
chiral superfield we should take the planar limit for the matrix
model. More generally, for gauge groups and matter representations
allowing a double line notation we should also consider the
contribution to the free energy of the matrix model from diagrams
with Euler characteristic $\chi=1$, that is, with the topology of
the projective plane or the disk. So we have a dramatic
simplification in calculating the glueball superpotential both to
the zero-momentum modes of the matrix model and further to planar
and perhaps $\chi=1$ diagrams. Also the gauge couplings are
related to the free energy of the matrix model.

Originally, the conjecture arose from considerations in string
theory. In a series of articles large $N$ dualities in string
theory were investigated and led to the relation between the
glueball superpotential and the free energy of the matrix model
in~\cite{0206255} and~\cite{0207106}. In~\cite{0208048} this was
summarised and the general conjecture was stated entirely within
gauge theory and further a sketch of a field theoretic proof was
given. Proofs of the conjecture (for the form of the perturbative
part of the effective superpotential) given entirely within gauge
theory followed shortly. For the case of $\Nscr=2$ supersymmetric
Yang-Mills theory broken to $\Nscr=1$ by a tree-level
superpotential a proof was given using factorisation of
Seiberg-Witten curves in~\cite{0210135}. In the slightly more
general case of an $\Nscr=1$ supersymmetric theory with $\un{N}$
gauge group and an adjoint chiral superfield a diagrammatic proof
was given in~\cite{0211017}, and a proof using generalised Konishi
anomalies was given in~\cite{0211170}. Special cases have been
proven using these methods in a large number of articles. A
perturbative proof for the reduction to the zero-momentum modes of
the matrix model for general gauge groups and (massive) matter
representations was given in~\cite{0304271}.

The aim of this thesis will be to give a thorough and pedagogical
introduction to the Dijkgraaf-Vafa conjecture and the concepts
needed to understand the conjecture. We will work entirely within
supersymmetric gauge theories. Our prime example will be an
$\Nscr=1$ supersymmetric gauge theory with a $\un{N}$ gauge group
and matter in the form of an adjoint chiral superfield. But we
will also see how the Dijkgraaf-Vafa conjecture can be used to
obtain exact superpotentials and we will see that the matrix model
also captures the form of the non-perturbative
Veneziano-Yankielowicz superpotential. Furthermore, we will in
detail go through the diagrammatic proof of the conjecture.

The outline of the thesis is as follows. In the first chapter we
will introduce supersymmetry. We will establish the supersymmetry
algebra as the unique extension of the Poincar\'e algebra and
consider its representations, the supermultiplets. The focus will
then be on the $\Nscr=1$ supersymmetric field theories using
chiral and vector superfields, but we will also consider $\Nscr=2$
supersymmetric theories.

The main chapter is the second chapter. We will start by
introducing the Dijkgraaf-Vafa conjecture in our prime case of a
$\un{N}$ gauge group with adjoint chiral matter. Since most of the
concepts in the conjecture needs explaining, the rest of the
chapter will be devoted to this. We will start by introducing the
supersymmetric vacua and the vacuum moduli space and then consider
these in our special case. Especially, we will explain the
breaking of the gauge group and the non-zero masses needed in the
conjecture. We will then consider the quantised theory and
introduce the supergraphs. With these we will prove the
perturbative non-renormalisation theorem. To understand the
effective superpotential we will present the Wilsonian
renormalisation and the integrating out procedure. In
section~\ref{dvsecweff} we will then investigate the Wilsonian
effective superpotential. We will introduce the important concept
of holomorphy and use it together with the symmetries of the
theory to prove the non-renormalisation theorem more generally and
to constrain the form of the non-perturbative corrections. We will
here also need to learn about instantons and chiral anomalies. We
will then go on and introduce the ILS linearity principle and the
concept of ``integrating in'' which we will use to define the
glueball superpotential. After explaining the lore of the low
energy gauge dynamics including phases, confinement, mass gaps and
chiral symmetry breaking, we will use the integrating in procedure
to derive the form of the Veneziano-Yankielowicz superpotential
and then we will consider the glueball superpotential in our
special case.

After introducing the double line notation and the 't Hooft large
$N$ limit in section~\ref{dvsecdouble}, we will consider the
matrix model in section~\ref{dvsecmatrix}. Here we will first
present the diagrammatic approach; also for the broken gauge group
using ghosts. We will then see that the measure of the matrix
model gives a term with the same form as the
Veneziano-Yankielowicz superpotential under the Dijkgraaf-Vafa
conjecture. The exact solution for the planar limit of the matrix
model is then examined and we will obtain algebraic equations for
the one-cut solution. In section~\ref{dvsecexample} we will use
these and the Dijkgraaf-Vafa conjecture to obtain the exact
glueball superpotential for a cubic tree-level superpotential. We
will here see that we get a term in the matrix model which exactly
matches the Veneziano-Yankielowicz superpotential.

At the end of the chapter we will discuss the problem of the
nilpotency of the glueball superfield and how to interpret the
full effective superpotential obtained via the matrix model.
Finally, we will state the Dijkgraaf-Vafa conjecture for classical
gauge groups with adjoint and fundamental matter.

In the third and last chapter we will present the diagrammatic
proof of the conjecture. The main focus we will be on the $\un{N}$
case, but we will also consider general gauge groups and matter
allowing a double line notation. We will give a detailed proof for
the case where we take into account the abelian part of the
supersymmetric gauge field strength. At last we will show the
reduction to zero-modes for general gauge groups and matter
representations.

In appendix~\ref{appnota} we will present our notation. In
appendix~\ref{apppoincare} we show that the Minkowski space can be
seen as cosets in the Poincar\'e group. The spinors and the
notation for these will be introduced in
appendix~\ref{appspinors}. The general Lagrangian for an $\Nscr=2$
supersymmetric Yang-Mills theory is derived in
Appendix~\ref{appn2super}. Finally, in appendix~\ref{appmatrix} we
will show how to calculate some integrals needed in the one-cut
solution of the matrix model.

\thispagestyle{plain}

%---------------------------------------------------------

\chapter{Supersymmetry}\label{chpsusy}
In this chapter we will introduce the concept of supersymmetry.
While being introduced in the early 1970s there has been no
experimental evidence for this theory. But as shown in the next
section it is natural to consider supersymmetric theories since
supersymmetry is the unique extension of the Poincar\'e algebra.
We will, however, give no other motivation than the simple beauty
of the results derived in this thesis. Our introduction to
supersymmetry will follow a theoretic stream of logic rather than
a chronological and it is based
on~\cite{wessandbagger},~\cite{weinberg3},~\cite{0109172},~\cite{argyres},~\cite{9912271}
and~\cite{9701069}.

%---------------------------------------------------------

\section{Supersymmetry Algebras}

In this section we will introduce the supersymmetry algebra.

\subsection{The Coleman-Mandula Theorem}
Given a relativistic quantum field theory we can ask ourselves
which symmetries we can have beyond the manifest Poincar\'e
invariance. Supposing that the symmetry generators form a Lie
algebra in the usual way, the answer was given in 1967 by Coleman
and Mandula (here taken from~\cite{weinberg3}):
\begin{thmcolemanmandula}
    Given that
    \begin{enumerate}
    \item For any mass $M$ there are only a finite number of
    particle types with mass less than $M$.
    \item Any two-particle state undergoes some reaction at all
    energies except perhaps an isolated set.
    \item The amplitudes for elastic two-body scattering are
    analytic functions of the scattering angle at all energies and
    angles except perhaps an isolated set. (Analyticity of the
    S-matrix).
    \end{enumerate}
    then the most general Lie algebra of symmetry generators consists of $\Pgen_\mu$, $\Jgen_{\mu\nu}$
    (i.e. the Poincar\'e algebra) and possibly internal symmetry generators
    commuting with the Poincar\'e generators and being independent
    of spin and momentum.
\end{thmcolemanmandula}
Here symmetry generators are defined as hermitian operators that
commute with the S-matrix, whose commutators are again symmetry
generators, that take single particle states into single particle
states, and that act on multiparticle states as the direct sum of
their action on single particle states. For definition of the
Poincar\'e algebra please see appendix~\ref{apppoincare}. It
should be noted that in the massless case the Poincar\'e algebra
can be extended to the algebra of the conformal group. We will not
go through the rather lengthy proof of the Coleman-Mandula theorem
here, but just remark that it is based on the fact that the
Poincar\'e symmetry of a scattering process only leaves the
scattering angle unknown. Extra symmetry would restrict this angle
to a discrete set. Hence the scattering amplitude would be zero by
analyticity in contradiction with assumption number 2.

\subsection{Superalgebras}
The way to get around the Coleman-Mandula theorem is to introduce
fermionic symmetry generators which do not satisfy commutation
relations like in Lie algebras, but rather anticommutation
relations. Hence instead of searching for the most general Lie
algebra, we search for the most general $\Z_2$-graded algebra also
known as Lie \emph{superalgebra}. This is defined as a
$\Z_2$-graded vector space $V=V_0\oplus V_1$. The elements $\Agen$
of $V_0$ are called \emph{bosonic} and are given a \emph{grading}
$|\Agen|=0$. The elements of $V_1$ are called \emph{fermionic} and
are given grading $|\Agen|=1$. The superalgebra has a bilinear
form $[-,-\}$ fulfilling
\begin{align}\label{susyn1}
    &[-,-\}:V_i\times V_j\rightarrow V_{i+j\
    (\mathrm{mod}\
    2)},\nonumber\\{}
    &[\Agen,\Bgen\}=-(-1)^{|\Agen||\Bgen|}[\Bgen,\Agen\}.
\end{align}
The bilinear form must obey the generalised super-Jacobi identity:
\begin{equation}\label{susyn2}
    (-1)^{|\Agen||\Cgen|}[[\Agen,\Bgen\},\Cgen\}+(-1)^{|\Bgen||\Agen|}[[\Bgen,\Cgen\},\Agen\}+(-1)^{|\Cgen||\Bgen|}[[\Cgen,\Agen\},\Bgen\}=0.
\end{equation}
We note that the bosonic space with the restricted bilinear form
is a ´Lie algebra. If it is possible to take products of operators
we see that the brackets are realisable as:
\begin{equation}\label{susyn3}
    [\Agen,\Bgen\}=\Agen\Bgen-(-1)^{|\Agen||\Bgen|}\Bgen\Agen.
\end{equation}
These brackets reduce to commutators on the bosonic space and
anticommutators on the fermionic space. Thus in general we will
use $[-,-]$ on bosonic operators or a mix of bosonic and fermionic
operators, and we will use $\{-,-\}$ on fermionic operators. Now
according to~\eqref{susyn1} the bracket of a bosonic operator with
a fermionic operator is again a fermionic operator. The
super-Jacobi identity tells us that this actually gives a
representation: The fermionic space furnishes a representation of
the bosonic Lie subalgebra.

\subsection{The Haag-Sohnius-Lopuszanski Theorem}\label{susysechslthm}

Let us now ask ourselves the question: What is the most general
Lie \emph{super}algebra of symmetries under the assumptions of the
Coleman-Mandula theorem? The answer was given by Haag, Sohnius and
Lopuszanski. In four space-time dimensions (which we will use
throughout this thesis) there is one unique solution namely the
(extended) supersymmetry algebra:
\begin{subequations}\label{susyalgebra}
\begin{eqnarray}
  [\Pgen_\mu,\Pgen_\nu] &=& 0, \label{susyn4}\\{}
  [\Pgen_\mu,\qgen^A_\al] &=& [\Pgen_\mu,\qbargen_{B\aldot}]=0, \label{susyn5}\\
  \{\qgen^A_\al,\qbargen_{B\bedot}\} &=& 2\sigmaop{\mu}{\al}{\be}\Pgen_\mu\de^A_B, \label{susyn6}\\
  \{\qgen^A_\al,\qgen^B_\be\} &=& \vep_{\al\be}\Zgen^{AB}, \label{susyn7}\\
  \{\qbargen_{A\aldot},\qbargen_{B\bedot}\} &=&
  \vep_{\aldot\bedot}\Zgen^\dagger_{AB}.\label{susyn8}
\end{eqnarray}
\end{subequations}
Here the fermionic generators are the \emph{supercharges}
$\qgen^A_\al$ and
$\qbargen_{A\aldot}=\left(\qgen^A_\al\right)^\dagger$ where
$A=1,\ldots, \Nscr$. As the indices $\al$ and $\bedot$ indicate,
the supercharges transform under the homogenous Lorentz group in
the $(\half,0)$ and $(0,\half)$ representations respectively
(left/right Weyl spinors). For spinorial representations please
see appendix~\ref{appspinors} where also $\sigma^\mu$ is defined
in~\eqref{appspinorn9}. As shown in the appendix these two
conjugate Weyl spinors can be put together to form a Majorana
spinor. Hence the fermionic generators consist of $\Nscr$ Majorana
spinorial generators. The representation of the supercharges
determines their commutation relations with $\Jgen_{\mu\nu}$ and
hence we have left these out. We have also left out commutation
relations with the internal symmetries as we will not need them.
$\Zgen^{AB}$ and its hermitian conjugate are central charges i.e.
they commute with all other symmetry generators including the
internal symmetries. Please note that they are antisymmetric by
definition of the anticommutator.

The Haag-Sohnius-Lopuszanski theorem has three assumptions besides
those given in the Coleman-Mandula theorem: The fermionic
operators $\qgen$ operate in a Hilbert space with positive
definite metric. If $\qgen$ is a fermionic generator then so is
the hermitian conjugate $\qgen^\dagger$. At last we will assume
physical states with $-\Pgen^2\geq0$ and $\Pgen^0>0$. We will not
prove the whole theorem here, but we will see why we only have
Majorana spinors and how~\eqref{susyn6} comes about.

As described above the bosonic part of the superalgebra constitute
a Lie algebra and by the Coleman-Mandula theorem this Lie algebra
is generated by $\Pgen_\mu$, $\Jgen_{\mu\nu}$ and possibly
internal symmetries. As we also saw the fermionic space is a
representation of this Lie algebra. The Lorentz generators
constitute a subalgebra of the bosonic algebra and hence the
$\qgen$'s furnish a representation of the homogenous Lorentz
group. Referring to appendix~\ref{appspinors} the representations
of the proper orthochronous Lorentz group can be labelled by
$(j_+,j_-)$ corresponding to the product of conjugate spin
$j_+\in\N_0/2$ and spin $j_-\in\N_0/2$ representations. Now
(following~\cite{weinberg3}) we can label the $\qgen$'s according
to this: $\qgen^{j_+j_-}_{m_+m_-}$ with $m_\pm=-j_\pm, \ldots,
j_\pm$. By the spin statistics theorem $j_++j_-$ must be a half
integer since the $\qgen$'s anticommute . But we can do better.
Let us look at
\begin{equation}\label{susyn9}
 \{\qgen^{j_+j_-}_{m_+m_-},\left(\qgen^{j_+j_-}_{m_+m_-}\right)^\dagger\},
\end{equation}
where we used the assumption that the hermitian conjugate of a
fermionic generator is again a fermionic generator. Since it is
the conjugate, $\left(\qgen^{j_+j_-}_{m_+m_-}\right)^\dagger$ must
transform in the $(j_-,j_+)$ representation. Actually
$\left(\qgen^{j_+j_-}_{m_+m_-}\right)^\dagger\sim\qgen^{j_-j_+}_{-m_-,-m_+}$.
By the usual addition of spin~\eqref{susyn9}, which belong to the
bosonic space by~\eqref{susyn1}, can now be expanded in bosonic
operators $\Xgen^{CD}_{cd}$ transforming in the $(C,D)$
representation where $C,D=|j_+-j_-|, \ldots, j_++j_-$. Using the
properties of Clebsch-Gordan coefficients one can see that the top
component fulfils:
\begin{equation*}
    \Xgen^{j_++j_-,j_++j_-}_{j_++j_-,-j_+-j_-}= \{\qgen^{j_+j_-}_{j_+,-j_-},\left(\qgen^{j_+j_-}_{j_+,-j_-}\right)^\dagger\}.
\end{equation*}
The Coleman-Mandula theorem tells us that the right hand side,
being a part of the bosonic Lie algebra, is a linear combination
of internal symmetries, $\Pgen_\mu$'s, and $\Jgen_{\mu\nu}$'s. The
internal symmetries belong to the $(0,0)$ representation since
they are Lorentz invariants, the $\Pgen_\mu$'s belong to the
$(\half,\half)$ representation (appendix~\ref{appspinors}) while
$\Jgen_{\mu\nu}$ transforms in the $(1,0)\oplus(0,1)$
representation.\footnote{The last part is not shown in
appendix~\ref{appspinors}, but it is shown that on spinors the
representation of $\Jgen_{\mu\nu}$ namely $\Sigma_{\mu\nu}$ is a
linear combination of
$\left(\sigma_{\mu\nu}\right)_\al^{\ph{\al}\be}$ and
$\left(\sigmabar_{\mu\nu}\right)^{\aldot}_{\ph{\aldot}\bedot}$. By
lowering the indices these are actually symmetric. Hence by spin
addition $\left(\sigma_{\mu\nu}\right)_{\al\be}$ is the symmetric
part of spin zero plus spin one. Since the antisymmetric tensor
$\vep_{\al\be}$ is invariant (i.e. spin zero)
$\left(\sigma_{\mu\nu}\right)_{\al\be}$ must be spin one or rather
in the $(1,0)$ representation ($0\oplus_s 1=1$). Hence
$\left(\sigma_{\mu\nu}\right)_\al^{\ph{\al}\be}$ and
$\left(\sigmabar_{\mu\nu}\right)^{\aldot}_{\ph{\aldot}\bedot}$
transform respectively in the $(1,0)$ and $(0,1)$
representations.} Looking at the left hand side this leaves us
with the possibilities $j_++j_-=0,\half$. Since $j_++j_-$ should
be half integer we must have $j_++j_-=\half$. For
$j_++j_-\neq\half$ we conclude that $\qgen^{j_+j_-}_{j_+,-j_-}=0$
since the Hilbert space by assumption has a positive definite
metric. By raising and lowering we then have
$\qgen^{j_+j_-}_{m_+m_-}=0$ in the case of $j_++j_-\neq\half$ and
for all $m_\pm$. Consequently we can choose a basis of fermionic
generators as $\qgen^A_\al$ and their hermitian adjoints
$\qbargen_{A\aldot}$ as wanted. Now~\eqref{susyn9} reduces to:
\begin{equation}\label{susyn10}
    \{\qgen^A_\al,\qbargen_{B\bedot}\}=2N^A_B\sigmaop{\mu}{\al}{\be}\Pgen_\mu,
\end{equation}
where we have used the correspondence between 4-vectors and the
$(\half,\half)$ representation from appendix~\ref{appspinors}.
$\sigmaop{\mu}{\al}{\be}\Pgen_\mu$ is (according
to~\cite{weinberg3}) positive definite if (and only if) we use the
assumption that the states fulfil $-P^2\geq0$ and $P^0>0$. Since
the anticommutator is positive definite we see that the $N$-matrix
is positive definite. By~\eqref{susyn10} $N^A_B$ is also hermitian
($\Pgen_\mu$ is hermitian) and hence we can redefine our fermionic
operators such that $N^A_B\mapsto\de^A_B$. We thus end up
with~\eqref{susyn6}. The rest of the proof of the
Haag-Sohnius-Lopuszanski theorem follows using spin addition as
above and the super-Jacobi identity. It should be noted that in
the massless case an extension to a superconformal algebra is
allowed.

If $\Nscr>1$ the algebra~\eqref{susyalgebra} is referred to as the
\emph{extended} supersymmetry algebra. If $\Nscr=1$ it is just
called the supersymmetry algebra. In this case the central charges
are zero by antisymmetry. The $\Nscr=1$ algebra then simplifies
to:
\begin{eqnarray}\label{susyn11}
  [\Pgen_\mu,\qgen_\al] &=& [\Pgen_\mu,\qbargen_{\aldot}]=0. \nonumber\\
  \{\qgen_\al,\qbargen_{\bedot}\} &=& 2\sigmaop{\mu}{\al}{\be}\Pgen_\mu. \nonumber\\
  \{\qgen_\al,\qgen_\be\} &=&
  \{\qbargen_{\aldot},\qbargen_{\bedot}\}=0.
 \end{eqnarray}
This algebra can actually be written in a more compact form using
the Majorana notation for the supercharges \eqref{appspinorn23}:
\begin{equation}\label{susyn12}
\qgen_{\mathrm{M}}\sim\begin{pmatrix}
      \qgen_\al \\
      \qbargen^{\aldot} \\
    \end{pmatrix}.
\end{equation}
Using this, the supersymmetry algebra becomes:
\begin{eqnarray}\label{susyn13}
    [\Pgen_\mu,\qgen_a]=0.\nonumber\\{}
    [\qgen_a,\qgen_b]=-2\ga^\mu_{\ph{\mu}a b}\Pgen_\mu.
\end{eqnarray}
Here we have used Latin indices for the Majorana spinors. The
$\ga$-matrices are defined in~\eqref{appspinorn12.5}. Please note
that the last index on the $\ga$-matrices has been lowered using
the charge conjugation matrix as defined in
appendix~\ref{appspinors}.

\subsection{R-Symmetry}\label{susysecrsymm}

Without central charges the extended supersymmetry
algebra~\eqref{susyn4}--\eqref{susyn8} is invariant under the
unitary transformations:
\begin{equation*}
    \qgen^A_\al\mapsto
    U^{AB}\qgen^B_\al,\qquad\boldsymbol{U}\in \textrm{U}(\Nscr).
\end{equation*}
These automorphisms are called R-symmetries and the group is
denoted $\textrm{U}(\Nscr)_R$. We can split the group in an
abelian and a non-abelian part denoted $\textrm{U}(1)_R$ and
$\textrm{SU}(\Nscr)_R$ respectively. In the case of $\Nscr=1$ only
$\textrm{U}(1)_R$ survives. The R-symmetries are not necessarily
symmetries of the actions we define, but we will generally choose
the actions in such a way that they are symmetric. However,
quantum mechanically $\textrm{U}(1)_R$ will often be broken by
anomaly effects as we will see in
section~\ref{dvsecweffsubnonpert}.

\section{Supermultiplets}\label{susysecsupermultiplets}

In this section we will find the 1-particle finite-dimensional
unitary representations of the supersymmetry algebra.

\subsection{The Wigner Method}\label{susysecwigner}
Since the Poincar\'e algebra is non-compact it is a theoretical
fact that it is not possible to find any finite-dimensional
unitary representations. The way to find unitary representations
is then to use the Wigner method of induced representations. The
idea is simply to fix the non-compact transformations i.e. the
translations and the boosts.

To be more precise let us first remark that\footnote{We use normal
font for the operators since we now look at representations.}
$P^2$ is a Casimir operator of the Poincar\'e algebra and since
$P^\mu$ commutes with the supercharges it is a Casimir for the
whole supersymmetry algebra. Thus $P^2$ is constant on the
irreducible representations by Schur's lemma. Let us now
diagonalise the commuting $P^\mu$'s such that each state is
labelled by $p^\mu$ satisfying the equation of motion $p^2=-M^2$.
Consider now some conventional value of $p^\mu$. The subgroup of
the Poincar\'e group leaving this momentum invariant is called
\emph{the little group}. This group is independent of the choice
of momentum on the mass shell since the invariance groups are
adjointly related by Lorentz transformations and hence isomorphic.
The idea is that the little group is either compact or, if it is
not, we will compactify it. Hence it has unitary
finite-dimensional representations. These induce unitary
representations of the whole Poincar\'e algebra. This is done
since the vector space, $V$, of the representation of the little
group trivially defines a vector bundle on the mass shell. The
states can now be seen as the sections, $\phi(p)$, of this vector
bundle taking values in $V$. These sections are acted upon by the
whole Poincar\'e group in a natural way by simply splitting a
Poincar\'e transformation into the part belonging to the little
group and the part changing $p^\mu$. The first part simply works
under the chosen representation on the vector-valued $\phi$ and
the last part changes the momentum (as defined
in~\eqref{apppoincaren1.5}). The space of sections is of course
infinite-dimensional because of the continuous parameter $p^\mu$,
however, $V$ is finite dimensional.

Since the supercharges commute with $P^\mu$ and hence leave the
momentum parameter invariant they can be included in the algebra
of the little group thus defining the \emph{little supergroup}. It
is the irreducible representations of this group that we are
looking for. Naturally, the representations of the little
\emph{super}group are reducible when we see these as
representations of the little group. Thus different particle
(Poincar\'e-) representations will be related by supersymmetry and
the representations of the little supergroup are hence called
\emph{supermultiplets}.

Since $P^2$ is a Casimir all particles in a supermultiplet have
the same mass. Yet another feature of the supermultiplets is that
they have an equal number of bosonic and fermionic states. This is
readily seen (following~\cite{wessandbagger}) by introducing the
fermion number operator $N_F$ defined such that it commutes with
the bosonic operators and $(-1)^{N_F}$ anticommutes with $Q_\al$.
We thus see:\footnote{The trace is well-defined since we use a
finite-dimensional representation. However, if supersymmetry is
broken states can be mapped out of the Hilbert space as we will
see in section~\ref{dvsecvac} -- hereby dismissing the proof.}
\begin{eqnarray*}
    \tr\nleft((-1)^{N_F}\{Q^A_\al,\Qbar_{B\bedot}\}\right) &=& \tr\nleft((-1)^{N_F}Q^A_\al\Qbar_{B\bedot}+(-1)^{N_F}\Qbar_{B\bedot}Q^A_\al\right)\\
    &=&
    \tr\nleft(-Q^A_\al(-1)^{N_F}\Qbar_{B\bedot}+Q^A_\al(-1)^{N_F}\Qbar_{B\bedot}\right)\\
    &=& 0,
\end{eqnarray*}
where we have used the cyclicity of the trace for the last term in
the second equality. Using~\eqref{susyn6} the left hand side is
equal to
$2\sigmaop{\mu}{\al}{\be}\de^A_B\tr\nleft((-1)^{N_F}P_\mu\right)$.
Choosing the momentum and the indices properly this gives
$\tr\nleft((-1)^{N_F}\right)=0$ proving that we have an equal
number of fermionic and bosonic states. Hence each particle has a
\emph{superpartner} of opposite fermionic parity.

\subsection{Massless Supermultiplets}

The massless supermultiplets are phenomenologically the most
interesting. This is because supersymmetry has not been observed
in nature. Hence supersymmetry must be broken at the energies of
modern accelerators. The masses of the particles are hence
negligible compared to the high energy at which we (might) have
supersymmetry.

In the case of massless multiplets, i.e. $P^2=0$, we boost to the
momentum $p_\mu\sim(-E,0,0,E)$ where $E$ is some conventional
energy. The little group is in this case $\textrm{SO}(2)$ or
rather $\textrm{spin}(2)$ since we represent the spin group rather
than the Lorentz group. However, also the generators $K^1-L^2$ and
$K^2-L^1$ are in the algebra of the little group ($L^i$ and $K^i$
are defined in~\eqref{appspinorn1}). But we remove these by hand
since they render the little group non-compact and the
corresponding continuous parameters are not seen in nature. The
$\textrm{spin}(2)$ group is generated by a single generator which
with our choice of momentum is $L^3$. The representations of this
are one-dimensional and indexed by the eigenvalue of $L^3$, the
helicity $\la$. Because a rotation by $4\pi$ can continuously be
deformed into the identity we have $\la\in\Z/2$.\footnote{As
explained in appendix~\ref{appspinors} the first homotopy group of
the proper orthochronous Lorentz group is $\Z_2$ so we need a
double loop to get the identity: $\exp\nleft(i4\pi L^3\right)=1$}

In the specified frame~\eqref{susyn6} becomes:
\begin{equation}\label{susyn14}
    \{Q^A_\al,\Qbar_{B\bedot}\}=2\begin{pmatrix}
      2E & 0 \\
      0 & 0 \\
    \end{pmatrix}_{\al\bedot}\de^A_B.
\end{equation}
Since we assumed a positive definite inner product,
$\{Q^A_2,\Qbar_{A\dot{2}}\}=0$ (no sum) shows that
$Q^A_2=\Qbar_{A\dot{2}}=0$. Inserting this into~\eqref{susyn7}
and~\eqref{susyn8} further shows that all the central charges
vanish. The only non-zero supercommutator left is:
\begin{equation}\label{susyn14.5}
    \{Q^A_1,\Qbar_{B\dot{1}}\}=4E\de^A_B.
\end{equation}
After rescaling, this is simply the algebra of $\Nscr$ fermionic
creation and corresponding annihilation operators. Using that
$Q^A_\al$ transforms as
(using~\eqref{appspinorn22}):\footnote{Please note the minus sign
on the right hand side. This sign is of course determined by the
super-Jacobi identity. In the same way as we found in
section~\ref{apppoincaresecaction} that $\Pgen^\mu$ transforms in
the vector representation, this sign makes sure that $Q_\al$
transforms in the $(\half,0)$ representation.}
\begin{equation}\label{susyn15}
    [J^{\mu\nu},Q^A_\al]=-i\left(\sigma^{\mu\nu}\right)_\al^{\ph{\al}\be}Q^A_\be,
\end{equation}
and $L^3=J^{12}$ we get:
\begin{equation}\label{susyn16}
    [L^3,Q^A_1]=-\half Q^A_1.
\end{equation}
This shows that $Q^A_1$ lowers the helicity by 1/2 and hence
$\Qbar_{B\dot{1}}$ raises the helicity by 1/2. Now we can
construct the multiplets. Since they should be finite-dimensional
there must be some state $\Omega_{\lambda_{\mathrm{min}}}$ with
lowest helicity $\lambda_{\mathrm{min}}$ defined by:
\begin{equation}\label{susyn17}
    Q^A_1\Omega_{\lambda_{\mathrm{min}}}=0,\qquad
    A=1,\ldots,\Nscr.
\end{equation}
This state must be non-degenerate for the sake of irreducibility
of the supermultiplet. All other states are obtained by raising:
\begin{equation}\label{susyn18}
    \Omega_{\lambda_{\mathrm{min}}+1/2
    n,A_1,\ldots,A_n}=N\Qbar_{A_n\dot{1}}\cdots\Qbar_{A_1\dot{1}}\Omega_{\lambda_{\mathrm{min}}},
\end{equation}
where $N$ is a proper normalisation factor. These states transform
as rank $n$ antisymmetric tensors under the $\textrm{SU}(\Nscr)_R$
symmetries and hence they are $\binom{\Nscr}{n}$-fold degenerate
as helicity eigenstates. We also see that we reach a highest
helicity state with helicity $\lambda_{\mathrm{min}}+\Nscr/2$ by
raising with all the different $\Qbar_{A\dot{1}}$'s. Thus the
total number of states is
$\sum_{n=0}^{\Nscr}\binom{\Nscr}{n}=2^{\Nscr}$. However, CPT
reverses the sign of helicity so if we want a CPT-invariant theory
we must directly sum each multiplet with its CPT-conjugate
antimultiplet having the opposite helicities.\footnote{In the case
of $\Nscr=2$ we have a helicity-symmetric multiplet with 2
helicity zero particles transforming as a $\textrm{SU}(2)_R$
doublet. We could ask if this is not its own antimultiplet. But
this can not be true since the two particles would then be real
and thus could not transform as a $\textrm{SU}(2)_R$ doublet.
However, for $\Nscr=4$ we have a multiplet that is its own
antimultiplet. This is possible since here the 6 helicity zero
particles transform as a rank 2 antisymmetric tensor under
$\textrm{SU}(4)_R$. This is actually the vector representation of
$\textrm{SO}(6)$ which is real.}

Now we can tabulate all massless multiplets. However, we are only
interested in multiplets with helicities $|\la|<\frac{3}{2}$ since
otherwise the only consistent couplings require gravitation -- and
we will only deal with \emph{global} supersymmetries. The result
is given in table~\ref{tablesusyn1} .
\begin{table}
\caption{}\label{tablesusyn1}\centering
\begin{tabular}{|l|c|c|c|c|c|c|}
  \hline
  $\Nscr$ & 1 & 1 & 2 & 2 & 3 & 4 \\\hline
  Name & Gauge & Chiral & Gauge & Hyper & Gauge & Gauge \\\hline
  $\la=1$ & 1 & 0 & 1 & 0 & 1 & 1 \\\hline
  $\la=1/2$ & 1 & 1 & 2 & 1+1 & 3+1 & 4 \\\hline
  $\la=0$ & 0 & 1+1 & 1+1 & 2+2 & 3+3 & 6 \\\hline
  $\la=-1/2$ & 1 & 1 & 2 & 1+1 & 1+3 & 4 \\\hline
  $\la=-1$ & 1 & 0 & 1 & 0 & 1 & 1 \\\hline
  Total number & 4 & 4 & 8 & 8 & 16 & 16 \\\hline
\end{tabular}
\newline\newline  The number of massless states in global
supersymmetry-multiplets for each helicity $\la$. CPT-invariance
of the supermultiplets is assumed. Plus-signs are used to indicate
when states of the same helicity stem from CPT doubling. Based on
table in~\protect{\cite{9912271}}.
\end{table}
We note that the spectra for $\Nscr=3$ and $\Nscr=4$ are the same
-- they are in fact equal. We also observe that the number of
bosonic and fermionic states are the same as expected.

The reason that some of the supermultiplets are called gauge
multiplets is of course that they contain two states with
helicities 1 and $-1$ respectively i.e. making up a massless gauge
boson. These are always followed by the superpartner -- a fermion
made out of helicity $\pm1/2$ states i.e. a Weyl (or Majorana)
fermion called the \emph{gaugino}. Please note that only in
special dimensions it is possible to make a gauge multiplet as
shown in section~\ref{appspinorsecdim}.

The chiral multiplet is a matter multiplet consisting of one
Majorana fermion, say a quark, along with its superpartner a
complex scalar called the \emph{squark}.

\subsection{Massive Supermultiplets}\label{susysecmassive}

In the case of massive supermultiplets we boost to the rest frame
with 4-momentum $p_\mu\sim(-M,0,0,0)$. Thus the little group is
$\mathrm{SO}(3)$ or rather its spin cover $\suto$. Hence we get
states described by spin $j\in\N_0/2$ and $m=-j,\ldots,j$.

We will here assume that the central charges are all zero.
Consequently the only non-zero supercommutator is~\eqref{susyn6}
and it takes the form:
\begin{equation}\label{susyn19}
    \{Q^A_\al,\Qbar_{B\bedot}\}=2M\de_{\al\bedot}\de^A_B.
\end{equation}
This shows that we now have $2\Nscr$ fermionic creation and
corresponding annihilation operators. As before we start from a
``vacuum'', $\Omega$, defined by
\begin{equation}\label{susyn20}
    Q^A_\al\Omega=0,\qquad\al=1,2,\ A=1,\ldots,\Nscr.
\end{equation}
However, this time the state needs not be non-degenerate, but can
be in some spin representation. Again all other states are built
by making all possible raisings ($N$ is a normalisation factor):
\begin{equation}\label{susyn21}
    N\Qbar_{A_n\dot{\al}_n}\cdots\Qbar_{A_1\dot{\al}_1}\Omega.
\end{equation}
This is a rank $n$ tensor under the $\textrm{SU}(\Nscr)_R$
symmetries. But from~\eqref{susyn19} we see that we also have a
$\suto$ symmetry in the spinorial $\al$ index.\footnote{Actually
the full symmetry group of the massive algebra is
$\mathrm{SO}(6).$} The state~\eqref{susyn21} is also a rank $n$
tensor under this symmetry group. However, it is only
antisymmetric if we pairwise change the indices $A_i\al_i$.
Thinking of this paired index as one index taking $2\Nscr$ values,
we see that we have $\binom{2\Nscr}{n}$ states with $n$ raisings.
Thus the total number of states is
$\sum_{n=0}^{2\Nscr}\binom{2\Nscr}{n}=2^{2\Nscr}$ multiplied with
the dimension of the spin representation of the vacuum.

The states~\eqref{susyn21} can be spin summed using Clebsch-Gordan
coefficients to gain spin eigenstates. We can, however, easily
determine the state with maximal spin. To gain this state we must
symmetrise in as many spin-$\half$ indices $\aldot$ as possible
(remembering that a pair of antisymmetric indices is spin zero
since $\vep_{\aldot\bedot}$ is an invariant tensor). However,
since the indices should be pairwise antisymmetric the
$A_i$-indices must simultaneously be antisymmetrised. Hence we can
maximally symmetrise in $\Nscr$ spin indices adding spin
$\half\Nscr$ to the vacuum $\Omega$. Consequently, in order to
avoid states with spin 3/2 or more we can only have $\Nscr=1,2$.

In the case of $\Nscr=1$ we can start from a spin 0 ``vacuum'' and
get the chiral multiplet with a complex scalar and one Majorana
fermion. We can also start from a spin $\half$ vacuum getting a
gauge multiplet with a scalar field, a Dirac fermion and a gauge
boson.

In the case of $\Nscr=2$ we have to start from a spin 0 vacuum and
we will have 5 scalars, 4 Majorana spinors and one gauge boson.

\section{$\Nscr=1$ Supersymmetric Field Theories}\label{susysecN1}

Instead of venturing ahead combining fields to make supersymmetric
Lagrangians it is possible in the case of $\Nscr=1$ supersymmetry
to define the \emph{superspace}. This is the analog of what
Minkowski space is to the Poincar\'e algebra. Supersymmetry can
then be realised as differentials of fields defined on superspace
and it will be easy to make manifestly supersymmetric Lagrangians.

\subsection{Superspace}\label{susysecN1subsuper}

Let us define anticommuting Grassmann numbers $\xi^\al$ and
$\xibar^{\aldot}$. These anticommute with each other and with the
fermionic operators while they commute with the bosonic operators.
Please note that they have indices like Weyl spinors and we will
use the same summing, raising and lowering conventions as for Weyl
spinors. Using $\xi$ and $\xibar$ we can turn the anticommutators
of the $\Nscr=1$ supersymmetry algebra into commutators (here
assuming that the super-bracket is realisable as an
(anti)commutator):
\begin{eqnarray}\label{susyn22}
  [\xi\qgen,\xibar\qbargen] &=& \xi\qgen\xibar\qbargen-\xibar\qbargen\xi\qgen=-\xi^\al\{\qgen_\al,\qbargen^{\bedot}\}\xibar_{\bedot} \nonumber\\
  &=& 2\xi\sigma^\mu\xibar\Pgen_\mu.
\end{eqnarray}
The rest of the commutators with $\xi\qgen$ and $\xibar\qbargen$
are zero.

Since we now only have commutation relations, we can form the
\emph{Lie supergroup} by exponentiating the generators of the
supersymmetry algebra, but with the supercharges having
Grassmannian coefficients. The Baker-Campbell-Hausdorff formula
then applies as usual. We note that the supersymmetry algebra
forms a semi-direct product of the Lorentz generators with the
momentum generators and supercharges.\footnote{That is: The
momentum generators and the supercharges constitute a subalgebra.
Also the Lorentz generators form a subalgebra and the commutator
of a Lorentz generator with a momentum generator or a supercharge
is a sum of momentum generators and supercharges.} This means that
we are in the same situation as in appendix~\ref{apppoincare}
where the Minkowski space is defined as cosets of the Lorentz
group in the Poincar\'e group. Now we define \emph{superspace} as
cosets of the Lorentz group in the \emph{Lie supergroup}. In
analogy with~\eqref{apppoincaren2} the cosets can uniquely be
written as:
\begin{equation}\label{susyn23}
    e^{i\left(-x^\mu\Pgen_\mu+\tha\qgen+\thabar\qbargen\right)}G_{\mathrm{Lorentz}},
\end{equation}
where we denoted the Grassmannian coefficients by $\tha^\al$ and
$\thabar_{\aldot}$. This gives us a one-to-one correspondence
between the cosets modulo the Lorentz group and superspace
coordinates $(x,\tha,\thabar)$.\footnote{$(x,\tha,\thabar)$ should
in the strict mathematical sense be seen as coordinate functions
in a noncommutative geometry.}

In analogy with section~\ref{apppoincaresecaction} the action of
the Lie supergroup on the superspace is determined by left
multiplication of the group on the cosets. This means that
translations work as usual: $x^\mu\mapsto x^\mu+\ta^\mu$.
Multiplication with
$\exp\nleft(i\om^{\mu\nu}\Jgen_{\mu\nu}\right)$ shows that $x^\mu$
transforms as a vector and that $\tha$ and $\thabar$ transform as
left and right Weyl spinors respectively. The
\emph{supertranslations} with $\xi$ and $\xibar$ are easily
obtained using the Baker-Campbell-Hausdorff formula (since only
the first commutator is non-zero):
\begin{eqnarray}\label{susyn24}
    e^{i\left(\xi\qgen+\xibar\qbargen\right)}e^{i\left(-x^\mu\Pgen_\mu+\tha\qgen+\thabar\qbargen\right)}&=&e^{i\left(-x^\mu\Pgen_\mu+\left(\tha+\xi\right)\qgen+\left(\thabar+\xibar\right)\qbargen+i\half[\xi\qgen+\xibar\qbargen,\tha\qgen+\thabar\qbargen]\right)}\nonumber\\
    &=&e^{i\left(-\left(x^\mu+i\tha\si^\mu\xibar-i\xi\si^\mu\thabar\right)^\mu\Pgen_\mu+\left(\tha+\xi\right)\qgen+\left(\thabar+\xibar\right)\qbargen\right)}.
\end{eqnarray}
This corresponds to the transformation
\begin{equation}\label{susyn25}
    (x,\tha,\thabar)\mapsto(x',\tha',\thabar')=(x+i\tha\si^\mu\xibar-i\xi\si^\mu\thabar,\tha+\xi,\thabar+\xibar).
\end{equation}
This representation of the supercharges on superspace we now turn
into a representation, $Q$ and $\Qbar$, on the fields defined on
superspace. However, to get a representation on fields we must
remember that the coordinate should transform in the opposite way
as in~\eqref{susyn25}.\footnote{Suppose the group $G$ acts on $M$
as $g.x$ with $g\in G$ and $x\in M$. This induces an action of $G$
on $\C^M$ namely $g.f(x)=f(g^{-1}.x)$ where $f\in\C^M$.} But
following~\cite{wessandbagger} we do not take this into
consideration so we will get an antirepresentation of the
supersymmetry algebra. But at the same time~\cite{wessandbagger}
at this point tacitly changes the definition of the $Q$'s such
that the infinitesimal change of the field $F$ is (note the
missing $i$):
\begin{align}\label{susyn26}
    \de_{\xi}F=[i\xi\qgen+i\xibar\qbargen,F]\defi\left(\xi
    Q+\xibar\Qbar\right)F,\nonumber\\
    \de_{\xi}F(x,\tha,\thabar)=F(x',\tha',\thabar')-F(x,\tha,\thabar),
\end{align}
where $(x',\tha',\thabar')$ is defined in~\eqref{susyn25}. Thus
the supercharges are realised by the differential operators:
\begin{eqnarray}
  Q_\al &=& \ddthaop{\al}-i\sigmaop{\mu}{\al}{\be}\thabar^{\bedot}\partial_\mu, \label{susyn27}\\
  \Qbar_{\aldot} &=&
  -\ddthabarop{\al}+i\tha^\be\sigmaop{\mu}{\be}{\al}\partial_\mu.
  \label{susyn28}
\end{eqnarray}
These satisfy (note the change in sign compared to~\eqref{susyn11}
and~\eqref{notan1} due to the above mentioned redefinitions):
\begin{equation}\label{susyn29}
    \{Q_\al,\Qbar_{\bedot}\}=2i\sigmaop{\mu}{\al}{\be}\partial_\mu.
\end{equation}
The rest of their mutual anticommutators are zero. However, please
note that according to~\eqref{appspinorn31.2} we have (due to the
missing $i$'s):
\begin{equation}\label{susyn29.1}
    \left(Q_\al\right)^\dagger=-\Qbar_{\aldot}.
\end{equation}
On the other hand according to~\eqref{appspinorn31} we now have
\begin{equation}\label{susyn29.2}
    \left(Q_\al\right)^*=\Qbar_{\aldot}.
\end{equation}

We can also define multiplication from the right on the cosets.
This gives in the same way as above differential operators $D_\al$
and $\Dbar_{\aldot}$ -- the covariant derivatives -- where:
\begin{eqnarray}
  D_\al &=& \ddthaop{\al}+i\sigmaop{\mu}{\al}{\be}\thabar^{\bedot}\partial_\mu, \label{susyn30}\\
  \Dbar_{\aldot} &=&
  -\ddthabarop{\al}-i\tha^\be\sigmaop{\mu}{\be}{\al}\partial_\mu.
  \label{susyn31}
\end{eqnarray}
Since left and right multiplications commute the covariant
derivatives anticommute with the supercharges. They satisfy the
opposite algebra of~\eqref{susyn29}:
\begin{equation}\label{susyn32}
    \{D_\al,\Dbar_{\bedot}\}=-2i\sigmaop{\mu}{\al}{\be}\partial_\mu.
\end{equation}

The fields on superspace transforming as~\eqref{susyn26} are
called \emph{superfields}. Given such a general complex superfield
$F(x,\tha,\thabar)$ we can expand it in component fields by
looking at its power series in $\tha$ and $\thabar$. This power
series must terminate since the $\tha$'s and the $\thabar$'s all
anticommute. Using~\eqref{appspinorn32} and~\eqref{appspinorn38}
we get (the notation of components follow~\cite{9912271}):
\begin{multline}\label{susyn33}
    F(x,\tha,\thabar)=\phi(x)+\tha\psi(x)+\thabar\chibar(x)+\tha\tha
    f(x)+\thabar\thabar g^*(x)+\thabar\sigmabar^\mu\tha A_\mu(x)\\
    +i\tha\tha\thabar\labar(x)
    -i\thabar\thabar\tha\rho(x)+\half\tha\tha\thabar\thabar D(x).
\end{multline}
Supposing $F(x,\tha,\thabar)$ is a bosonic Lorentz scalar we see
that $\phi$, $f$, $g$, and $D$ are complex scalars while $\psi$,
$\chi$, $\la$, and $\rho$ are lefthanded Weyl spinors and $A_\mu$
is a gauge field. Hence we note that the number of bosonic degrees
of freedom equals that of the fermionic. The action of
supercharges on superfields induces an action on the components
simply by defining the transformed components as the components of
the transformed superfield.

The set of superfields are closed to addition and multiplication
since the supercharges are differentials. Thus the superfields
form a linear representation of the supersymmetry algebra.
However, this representation is reducible as can be seen from the
number of components in~\eqref{susyn33} since according to table
\ref{tablesusyn1} there should only be four degrees of freedom in
an irreducible representation after imposing the equation of
motion. Hence we must constrain the superfields.

Let us conclude this subsection by noting that for a set of
superfields $F^i$ any differentiable function of them is again a
superfield. We also note that
using~\eqref{susyn26}--\eqref{susyn28} the supersymmetry variation
of the top component $D(x)$ is a space-time derivative -- hence it
can be used as a manifestly supersymmetric Lagrangian (we assume
that the boundary terms are zero). This makes it easy to build
invariant Lagrangians.

\subsection{Chiral Superfields}\label{susysecchiral}

One way to constrain a (complex) superfield $\Phi$ is to demand
\begin{equation}\label{susyn34}
    \Dbar_{\aldot}\Phi=0,\qquad\aldot=\dot{1},\dot{2}.
\end{equation}
This is a supersymmetric covariant constraint since the
supercharges anticommute with $\Dbar$. Complex superfields
fulfilling~\eqref{susyn34} are called \emph{chiral} superfields.
Correspondingly an anti-chiral superfield is defined by
$D_\al\Phi=0$. We see that the complex conjugate of a chiral field
is anti-chiral and vice-versa hence giving a one-to-one
correspondence between the two sets. We also note that only
constant fields can be both chiral and anti-chiral since then
$\{D_\al,\Dbar_{\bedot}\}\Phi=0$ which by~\eqref{susyn32} can be
used to show that $\partial_\mu\Phi=0$. This is all in analogy
with holomorphic functions.

The components of a chiral field are easily found by noting that
both $x^\mu_+=x^\mu+i\tha\si^\mu\thabar$ and $\tha$ are
annihilated by $\Dbar_{\aldot}$. Expressed in these coordinates
$D_\al=\partial/\partial\tha^\al+2i\sigmaop{\mu}{\al}{\be}\thabar^{\bedot}\partial/\partial
x_+^\mu$ and $\Dbar_{\aldot}=-\partial/\partial\thabar^{\aldot}$.
Hence the most general chiral field can be written as:
\begin{eqnarray}\label{susyn35}
    \Phi(x,\tha,\thabar)&=&\phi(x_+)+\sqrt{2}\tha\psi(x_+)+\tha\tha
    F(x_+)\nonumber\\
    &=&\phi(x)+i\tha\si^\mu\thabar\partial_\mu\phi(x)+\frac{1}{4}\tha\tha\thabar\thabar\square
    \phi(x)+\sqrt{2}\tha\psi(x)\nonumber\\&&-\frac{i}{\sqrt{2}}\tha\tha\partial_\mu\psi(x)\si^\mu\thabar+\tha\tha
    F(x).
\end{eqnarray}
Here we have used~\eqref{appspinorn33} and~\eqref{appspinorn34} to
expand. As usual $\square=\partial_\mu\partial^\mu$ and we have
put in a $\sqrt{2}$ for standard normalisation. $F$ will turn out
to be an auxiliary field and hence after taking the equations of
motion, we see that the components fit that of the chiral
multiplet thus realising this off-shell.

The components can be obtained using the covariant derivatives
(using~\eqref{appspinorn40}):
\begin{equation}\label{susyn36}
    \phi(x)=\Phi|,\qquad\psi_\al(x)=\frac{1}{\sqrt{2}}D_\al\Phi|,\qquad
    F(x)=-\frac{1}{4}DD\Phi|,
\end{equation}
where $|$ means setting $\tha=\thabar=0$.

Since all components higher (i.e. with more $\tha$'s in front)
than $F$ are just derivatives, the supersymmetry variation of $F$
is just a space-time derivative making it usable for constructing
manifestly supersymmetric Lagrangians. Actually $\de_\xi
F=i\sqrt{2}\partial_\mu\nleft(\xibar\sigmabar^\mu\psi\right)$.

Given $N$ chiral multiplets represented by the (Lorentz invariant,
bosonic) chiral fields $\Phi^i$ with $i=1,\ldots,N$ we search for
the most general supersymmetric Lagrangian including these. As we
have seen above there are two ways to create Lagrangians: We can
use the top component, $D/2$, of a general superfield or the $F$
component of a chiral superfield. Such terms are called D-terms
and F-terms respectively.

We can write these terms as ``integrations'' over superspace
parameters. To do this we define Lorentz invariant differentials
by $\dtotha\defi\fourth\textrm{d}\tha_\al\textrm{d}\tha^\al$,
$\dtothabar\defi\fourth\textrm{d}\thabar^{\aldot}\textrm{d}\thabar_{\aldot}$
and $\dsuper\defi\dtotha\dtothabar$. As usual fermionic
integration corresponds to differentiation such that e.g.
$\textrm{d}\tha^\al=\partial/\partial\tha^\al$. By the above
normalisations the $F$ component of a chiral field $\Phi$ is
simply $F=\int\dtotha\Phi$ and the D-term, $D/2$, of a general
superfield $\Psi$ is $D/2=\int\dsuper\Psi$. Here it is assumed
that we put $\tha=\thabar=0$ after differentiating. Since the
space-time integral of a space-time derivative is assumed to be
zero, we see that the covariant derivatives work as differentials
under the space-time integration in the action. Hence:
\begin{eqnarray}\label{susyn37}
    \int\dlor\dtotha\Phi=-\frac{1}{4}\int\dlor DD\Phi|.\nonumber\\
    \int\dlor\dsuper\Psi=\frac{1}{16}\int\dlor DD\Dbar\Dbar\Psi.
\end{eqnarray}
There is no need for a restriction ``$|$'' in the last equation
since all $\tha$-dependence is removed by the covariant
derivatives and the removal of total space-time derivative terms.

This shows us that we have a redundancy in our definition of
F-terms. Actually any D-term can now be rewritten as:
\begin{equation}\label{susyn38}
    \int\dlor\dsuper\Psi=-\frac{1}{4}\int\dlor\dtotha\Dbar\Dbar\Psi,
\end{equation}
where there is also no need to set $\thabar=0$ since these are
again removed by the two $\Dbar$'s along with the removal of total
derivative terms. $\Dbar\Dbar\Psi$ is a chiral field since the
product of three $\Dbar_\al$'s vanish by anti-commutivity. A
superfield that can be written in this way with $\Psi$ being a
local field is called a \emph{chirally exact} superfield. In order
to avoid redundancy we redefine F-terms as the $\tha\tha$-term of
chiral fields that are not chirally exact.\footnote{This
definition is due to~\cite{ferretti}. The material in this
reference can now also be found in~\cite{0311066}.}

The most general D-term obtained from the chiral fields $\Phi^i$
and the corresponding anti-chiral complex conjugates, denoted
$\Phibar^i$, is simply the D-term of a real differentiable
function $K\nleft(\Phi^i,\Phibar^i\right)$ since this is again
superfield. This is called the \emph{\kahler{} potential}. We have
here excluded the possibility of a dependence on space-time
derivatives of $\Phi$ and $\Phibar$. This is because such terms,
when expanded, give rise to more than two space-time derivatives
on bosonic fields and more than one on fermionic fields. Such
terms can be excluded when looking at low energy effective
theories or renormalisable theories. Since the covariant
derivatives anticommute with the supercharges, the covariant
derivative of a superfield is again a superfield. However, we will
also assume that $K$ does not depend on such fields.

The most general F-term can be found by noting that a
differentiable function of chiral fields is again chiral. Hence we
look at the holomorphic function $W\nleft(\Phi^i\right)$. There
can be no dependence on $\Phibar$ because then $W$ would not be
chiral. However, as noted above $\Dbar\Dbar\Phibar^i$ is actually
chiral and could contribute in a given term. But we can move the
$\Dbar$'s to the front of the term since they annihilate all the
chiral fields thus showing that we are really dealing with a
chirally exact superfield not contributing to the F-term. However,
yet another possibility is the space-time derivative of a chiral
field. Since $\partial_\mu$ commutes with the covariant
derivatives this is again a chiral superfield. Even though we
could have such contributions in $W$, we choose to assume that we
do not have such terms. In this case $W\nleft(\Phi^i\right)$ is
called the \emph{superpotential}. Naturally, this contribution to
the Lagrangian is complex so we have to add its complex conjugate
(which is the analogue of the F-term for an anti-chiral field).

With these assumptions the most general (low energy effective)
$\Nscr=1$ supersymmetric Lagrangian density of $N$ chiral fields,
$\Phi^i$, is
\begin{equation}\label{susyn41}
    \lagr=\int\dtotha
    W\nleft(\Phi^i\right)+\int\dtothabar\overline{W\nleft(\Phi^i\right)}+\int\dsuper
    K\nleft(\Phi^i,\Phibar^i\right).
\end{equation}
In the next section we will expand the gauge version of this
Lagrangian in components.

If we further constrain this Lagrangian to be renormalisable, we
must require all coupling constants to have non-negative mass
dimension. Since the lowest component of $\Phi$ is a complex
scalar it must have mass dimension one: $\left[\Phi\right]=1$.
Since the supercharges obey~\eqref{susyn6} they must have
$\left[Q\right]=1/2$. Hence by~\eqref{susyn27} and~\eqref{susyn28}
we have $\left[\tha\right]=[\thabar]=-1/2$. Correspondingly
$\left[\dtha\right]=[\dthabar]=1/2$. Since $\left[\lagr\right]=4$
we must have $\left[K\nleft(\Phi^i,\Phibar^i\right)\right]=2$ and
the only renormalisable possibility is
$K\nleft(\Phi^i,\Phibar^i\right)=K_{ij}\Phi^i\Phibar^j$. Here
$K_{ij}$ is hermitian so we can diagonalise it by a change of the
$\Phi^i$'s. The superpotential must have $\left[
W\nleft(\Phi^i\right)\right]=3$ so it can be at most cubic.
Discarding its constant part we get the most general
renormalisable Lagrangian of chiral fields:
\begin{equation}\label{susyn42}
    \lagr=\int\dsuper
    \Phi^i\Phibar^i+\int\dtotha\left(a_i\Phi^i+\frac{1}{2}m_{ij}\Phi^i\Phi^j+g_{ijk}\Phi^i\Phi^j\Phi^k\right),
\end{equation}
where $m_{ij}$ and $g_{ijk}$ are symmetric in their indices.
Please note that we still have the freedom to perform a unitary
rotation to diagonalise the mass matrix. Also the linear term in
the superpotential can be removed (provided non-zero masses) by
the transformation
\begin{equation}\label{susyn43}
    \Phi^i\mapsto\Phi^i+b^i,
\end{equation}
where $b^i$ is constant (and hence chiral). This Lagrangian
describes (for $N=1$) the Wess-Zumino model.

\subsection{R-Symmetry}\label{susysecchiralsubrsymm}

The Lagrangians can also be restricted by employing the
R-symmetries from section~\ref{susysecrsymm}. Actually, using the
Coleman-Mandula theorem one can prove that for $\Nscr=1$ we can
only have a single generator, $R$, which does not commute with
$Q$. $R$ generates $\textrm{U}(1)_R$ and we normalise it such that
$Q$ has charge $-1$ (denoted by $R(Q)=-1$). Correspondingly the
complex conjugate $\Qbar$ must have charge +1. Since $R$ does not
commute with $Q$, the different components have different charge.
Hence the coordinates $\theta$ have charge. By~\eqref{susyn27}
and~\eqref{susyn28} we see that $R\nleft(\tha\right)=1$ and
$R\nleft(\thabar\right)=-1$. Consequently
$R\nleft(\dtha\right)=-1$ and $R\nleft(\dthabar\right)=1$.
Assuming the Lagrangian is invariant under R-symmetry the overall
charge of the \kahler{} potential must be zero and the charge of
the superpotential must be 2. If the superfields are given
R-charges $R\nleft(\Phi^i\right)$ we see that a renormalisable
\kahler{} potential has charge zero since it is real. The
superpotential is, however, strongly restricted by the R-symmetry.

\subsection{Supersymmetric Gauge Theories}\label{susysecchiralsubgauge}

A second way (and the last necessary) to constrain a superfield,
$V$, is to impose reality:
\begin{equation}\label{susyn44}
    V=V^*.
\end{equation}
Such a superfield is called a \emph{vector superfield}. Please
note that this constraint is also supersymmetric covariant since
by~\eqref{susyn29.2} and~\eqref{appspinorn26} $\xi Q+\xibar\Qbar$
is real, and it is bosonic. When expanding $V$ in components, it
is customary to use the following notation:
\begin{eqnarray}\label{susyn45}
        V(x,\tha,\thabar)&=&v(x)+\tha\chi(x)+\thabar\chibar(x)+\tha\tha
    f(x)+\thabar\thabar f^*(x)+\thabar\sigmabar^\mu\tha A_\mu(x)\nonumber\\
    &&+i\tha\tha\thabar\left(\labar(x)+\half\sigmabar^\mu\partial_\mu\chi(x)\right)
    -i\thabar\thabar\tha\left(\la(x)+\half\si^\mu\partial_\mu\chibar(x)\right)\nonumber\\&&+\half\tha\tha\thabar\thabar\left(D(x)+\half\square  v(x)\right).
\end{eqnarray}
Here $v$, $A_\mu$ and $D$ must be real.  The motivation for
defining the components in this way is that we get nice gauge
transformations. In the abelian case the gauge transformation of
the vector field simply is:
\begin{equation}\label{susyn46}
    V\mapsto V'=V+i\La-i\Labar,
\end{equation}
where $\La$ is a chiral superfield. According to~\eqref{susyn35}
and~\eqref{susyn45} this means that $A_\mu\mapsto
A_\mu+\partial_\mu\nleft(\phi+\phi^*\right)$. Hence the real field
$A_\mu$ transforms exactly as a gauge field which, of course, is
the reason that $V$ is called a vector (or gauge) superfield.
$\la$ and $D$ are gauge invariant in the abelian case which is the
reason for the notation in~\eqref{susyn45}.

To see how the non-abelian gauge transformations work, we must be
a bit more careful. Let us assume that we have a compact gauge
group $G$ with corresponding Lie algebra $\cG$. To obtain a
unitary representation the generators $T_a$ must be hermitian. We
use roman indices $a,b,c$ for the adjoint gauge indices. The gauge
transformations must commute with the supersymmetry
transformations because otherwise the commutator of a gauge
transformation and a supersymmetry transformation would yield a
new type of supersymmetry (since it exchanges bosons and
fermions). However, it would be a local transformation since gauge
transformations are local. But as noted above we will only
consider global supersymmetries thus demanding gauge and
supersymmetry transformations to commute.

In order to find the transformation of the vector superfield we
first have to look at gauge transformations of chiral matter.
Consider a representation $\reprn$ of $G$ not necessarily
irreducible. The representation is furnished by the components of
$\mathrm{dim}(\reprn)$ chiral superfields, $\Phi^i$. Since gauge
transformations commute with supersymmetry transformations each
independent component must transform in the same way:
\begin{equation}\label{susyn47}
    \phi^i\mapsto(e^{-i\La^a(x)T_a^{\repr}})^i_{\ph{i}j}\phi^j,
\end{equation}
and the same for $\psi$ and $F$ with reference to~\eqref{susyn35}.
Here $T_a^{\repr}$ are the generators of the gauge group in the
representation $\reprn$ and $\La^a(x)$ are real functions.
However, according to~\eqref{susyn35} the chiral superfield does
not transform in this way because some of the higher components
involve derivatives. Instead~\eqref{susyn35} shows that we have
the transformation:
\begin{equation}\label{susyn48}
    \Phi\mapsto e^{-i\La^a(x_+)T_a^{\repr}}\Phi.
\end{equation}
Since $x_+$ is not real, this means that $\Phi^\dagger\Phi$ is not
invariant, but:
\begin{equation}\label{susyn49}
    \Phi^\dagger\mapsto\Phi^\dagger e^{i\left(\La^a(x_+)\right)^*T_a^{\repr}}.
\end{equation}
Thus we need a superfield connection to make $\Phi^\dagger\Phi$
invariant. Let us choose this connection hermitian to keep the
product real. We can then write it as $e^{2V}$ where
$V=V^aT^{\repr}_a$ and the $V^a$'s must be vector superfields by
hermiticity. The reason we have $2V$ instead of just $V$ will
become clear later. The gauge transformation of $V$ must be:
\begin{equation}\label{susyn50}
    e^{2V}\mapsto
    e^{2V'}=e^{-i\left(\La^a(x_+)\right)^*T_a^{\repr}}e^{2V}e^{i\La^a(x_+)T_a^{\repr}}.
\end{equation}
By the Baker-Campbell-Hausdorff formula the transformation of $V$
can be written purely with commutators and is hence independent of
the chosen representation. Thus we can think of $V$ as being Lie
algebra valued simply taking the proper representation when
working on the chiral fields. Let us now note that we can preserve
the invariance of $\Phi^\dagger e^{2V}\Phi$ when going to the
larger group of \emph{extended} gauge transformations where we
replace the $\La^a$'s with chiral superfields.
Hence~\eqref{susyn48} becomes:
\begin{equation}\label{susyn51}
    \Phi\mapsto e^{-i\La\nleft(x,\tha,\thabar\right)}\Phi,
\end{equation}
where $\La=\La^aT^{\repr}_a$. This has the advantage that it
preserves the chirality of the fields and hence makes it possible
to build invariant Lagrangians. The transformation of the vector
fields becomes:
\begin{equation}\label{susyn52}
    e^{2V}\mapsto e^{2V'}=e^{-i\La^\dagger}e^{2V}e^{i\La}.
\end{equation}
We note that hermiticity is preserved.

In the abelian case the transformation of the components $v$,
$\chi$ and $f$ in the vector superfield simply is $\de
v=i\phi-i\phi^*$, $\de\chi=i\sqrt{2}\psi$ and $\de f=iF$. This
means that by choosing $\phi$, $\psi$ and $F$ properly these
components can be set to zero. This is called the Wess-Zumino
gauge (WZ-gauge). In the non-abelian case the first two terms in
$V'$ (using the Baker-Campbell-Hausdorff formula) is $V'\sim
V+\half\left(i\La-i\La^\dagger\right)$ similar to the abelian
transformation. This suggests that the Wess-Zumino gauge also is
possible in the non-abelian case. This is in fact true, however,
one has to consider all orders. In this gauge:
\begin{equation}\label{susyn53}
    V(x,\tha,\thabar)=\thabar\sigmabar^\mu\tha A_\mu(x)+i\tha\tha\thabar\labar(x)-i\thabar\thabar\tha\la(x)+\half\tha\tha\thabar\thabar
    D(x)\qquad\wzgauge.
\end{equation}
Here the components are Lie algebra valued.
Using~\eqref{appspinorn33} we see that this gauge has the nice
property that
\begin{equation}\label{susyn54}
    V^2=-\half\tha\tha\thabar\thabar A_\mu A^\mu,\qquad
    V^3=0\qquad \wzgauge.
\end{equation}

The Wess-Zumino gauge does not fix the gauge totally. For
infinitesimal $\La$, and V in Wess-Zumino gauge we get (using the
Baker-Campbell-Hausdorff formula~\eqref{apppoincaren3}):
\begin{equation}\label{susyn55}
    V'=V+\frac{i}{2}(\La-\La^\dagger)+\frac{1}{2}[V,i(\La+\La^\dagger)]+\frac{1}{6}[V,[V,i(\La-\La^\dagger)]]\qquad\wzgauge.
\end{equation}
Using this we see that the infinitesimal gauge transformation that
preserves Wess-Zumino gauge is
\begin{equation}\label{susyn56}
    \La=\omega(x_+)=\omega(x)-i\thabar\sigmabar^\mu\tha\partial_\mu\omega(x)+\frac{1}{4}\tha\tha\thabar\thabar\square
    \omega(x),
\end{equation}
where $\omega$ is Lie algebra valued and hermitian. According
to~\eqref{susyn55} (the last term is zero since it has to many
$\tha$'s) the infinitesimal gauge transformations of the
components in the Wess-Zumino gauge are:
\begin{eqnarray}\label{susyn57}
    \de_\omega
    A_\mu&=&\partial_\mu\omega-i[\omega,A_\mu],\nonumber\\
    \de_\omega\la &=&-i[\omega,\la],\nonumber\\
    \de_\omega D &=&-i[\omega,D].
\end{eqnarray}
This is exactly as usual for a gauge field $A_\mu$, and $\la$ and
$D$ in the adjoint representation supposing that the group
elements are expressed as in~\eqref{susyn47}.

We will see that $D$ is an auxiliary field so we note that in the
Wess-Zumino gauge the field content of the vector superfield,
after imposing the equations of motion, is the same as in the
vector supermultiplet hence realising this off-shell.

Although Wess-Zumino gauge allows a gauged realisation of the
vector multiplet it is breaking supersymmetry (since some
components are zero). This means that supersymmetry should be
realised by adding a gauge transformation going back to the
Wess-Zumino gauge. Thus on the components the anticommutator of
the supersymmetries is no longer just proportional to the momentum
generators. This makes sense since the local gauge transformations
do not commute with the momentum generators, but as assumed in the
beginning they commute with the supersymmetry transformations and
hence also their commutator.

Let us now look for the supersymmetric version of the gauge field
strength. In normal gauge theory this transforms in the adjoint
representation. By~\eqref{susyn57} the lowest component of $V$
that transforms in this way is $\la$. Let us therefore make a
field with $\la$ as its lowest component. The solution
is\footnote{We use calligraphic font to distinguish the
supersymmetric field strength from the superpotential.}
\begin{equation}\label{susyn58}
    \W_\al=-\frac{1}{8}\Dbar\Dbar e^{-2V}D_\al e^{2V}.
\end{equation}
We note that the supersymmetric field strength is a fermionic
spinor superfield which clearly is chiral -- actually chirally
exact.\footnote{$\W_\al$ is actually also constrained by Bianchi
identities.} It is Lie algebra valued since $e^{-2V}D_\al e^{2V}$
can be written as commutators. Actually, in Wess-Zumino gauge we
get (using~\eqref{susyn54}):
\begin{equation}\label{susyn58.5}
    e^{-2V}D_\al e^{2V}=2D_\al V-2[V,D_\al V]\qquad\wzgauge.
\end{equation}
After a bit of calculation we then get that in Wess-Zumino
gauge:\footnote{We note that since $F_{\mu\nu}$ is antisymmetric,
we could have rewritten $\tfrac{1}{2}\si^\mu\sigmabar^\nu
F_{\mu\nu}=\si^{\mu\nu}F_{\mu\nu}$.}
\begin{multline}\label{susyn59}
    \W_\al(x,\tha,\thabar)=-i\la_\al(x_+)+\tha_\al
    D(x_+)-\frac{i}{2}\left(\sigma^\mu\sigmabar^\nu\right)_\al^{\ph{\al}\be}\tha_\be
    F_{\mu\nu}(x_+)+\tha\tha\si^\mu_{\al\bedot}D_\mu\labar^{\bedot}(x_+)\\\wzgauge,
\end{multline}
where the components are Lie algebra valued. Here $F_{\mu\nu}$ is
the usual gauge field strength given by
\begin{equation}\label{susyn60}
    F_{\mu\nu}=\partial_\mu A_\nu-\partial_\nu
    A_\mu+i[A_\mu,A_\nu],
\end{equation}
and $D_\mu$ is the gauge covariant derivative:
\begin{equation}\label{susyn61}
    D_\mu\labar=\partial_\mu\labar+i[A_\mu,\labar].
\end{equation}
Thus $\W_\al$ neatly turns out to contain the usual gauge field
strength thus justifying it to be the supersymmetric gauge field
strength (this is the reason for the involved
definition~\eqref{susyn58}).

The gauge transformation of $\W_\al$ is:
\begin{equation}\label{susyn62}
    \W_\al\mapsto \W_\al'=e^{-i\La}\W_\al e^{i\La}.
\end{equation}
This simply shows that $\W_\al$ transforms in the adjoint
representation. To prove this one uses that $\La$ is chiral and
hence $\La^\dagger$ is anti-chiral along with the
relation
\begin{equation}\label{susyn62.5}
    [\Dbar_{\aldot},\{\Dbar_{\bedot},D_\ga\}]=0,
\end{equation}
which follows immediately from~\eqref{susyn32}.

Let us now look for the most general supersymmetric gauge
invariant Lagrangian. If we only allow two space-time derivatives
on bosonic fields and one on fermionic fields as in
section~\ref{susysecchiral} we can at most have terms of $\W$
squared. To obtain Lorentz invariance we must naturally look at
$\W^\al\W_\al$ which is also chiral. Thus the most general gauge
kinetic and self interaction term is:
\begin{equation}\label{susyn63}
    \lagr_G=\int\dtotha\tau_{ab}\nleft(\Phi^i\right)\W^{\al
    a}\W^b_\al+\cc,
\end{equation}
where we have added the complex conjugate to make the Lagrangian
real. We have here included functions $\tau_{ab}$ holomorphic in
the chiral fields $\Phi^i$ in the spirit of
section~\ref{susysecchiral}. $\tau_{ab}$ must be symmetric and
transform as an invariant tensor in the adjoint representation of
the gauge group hence putting restrictions on the dependence on
the chiral fields. We note that since the gauge field strength
by~\eqref{susyn58} is chirally exact, it can be written as a
D-term using~\eqref{susyn38} (up to total space-time derivative
terms):
\begin{equation}\label{susyn63.5}
    \lagr_G=-\int\dsuper\tau_{ab}\nleft(\Phi^i\right)\frac{1}{16}\left(e^{-2V}D_\al
    e^{2V}\right)^a\left(\Dbar\Dbar e^{-2V}D_\al
    e^{2V}\right)^b+\cc
\end{equation}
However, the integrand of this full superspace integral is not
gauge invariant. Thus it is natural to keep the Lagrangian in the
form~\eqref{susyn63} and in the rest of this thesis we will think
of it as a $\tha\tha$-term.

The \kahler{} term in~\eqref{susyn41} must now have the form
\begin{equation}\label{susyn64}
    \lagr_K=\int\dsuper K\nleft(e^{2V^{\repr}}\Phi,\Phi^\dagger\right),
\end{equation}
where $V^{\repr}$ is the vector field in the appropriate
representation. The \kahler{} potential must be formed such that
it is gauge invariant. The most simple example is $\Phi^\dagger
e^{2V^{\repr}}\Phi$ as we saw above. This is renormalisable
(polynomial) since in Wess-Zumino gauge $e^V=1+V+\half V^2$. By
the assumptions in section~\ref{susysecchiral} the \kahler{}
potential does not include derivatives. This means that it is
enough to require $K$ to be globally (but complex) gauge
invariant.

We can also have a superpotential term $\lagr_W$ as above which
now must be formed such that it is gauge invariant. However, here
it is enough to require the Lagrangian to be globally gauge
invariant since we have neither derivatives nor complex conjugates
of the chiral fields:
\begin{equation}\label{susyn65}
    \lagr_W=\int\dtotha W\nleft(\Phi^i\right)+\cc
\end{equation}

If the gauge group $G$ has an abelian factor we can also include a
\emph{Fayet-Iliopoulos} term. Let $\ka$ be a non-zero element in
the center of the algebra and as usual let $\tr$ denote the
invariant inner product of the Lie algebra (we can think of $\tr$
as trace in the fundamental representation) then:
\begin{equation}\label{susyn66}
    \lagr_{FI}=\int\dsuper\tr\nleft(\ka V\right)=\frac{1}{2}\tr\nleft(\ka D\right),
\end{equation}
where the last part is in Wess-Zumino gauge. This is
supersymmetric since it is the D-term of a superfield and it is
gauge invariant since by~\eqref{susyn57} $D$ transforms in the
adjoint representation.

With the above constraints the most general Lagrangian is:
\begin{equation}\label{susyn67}
    \lagr=\lagr_K+\lagr_G+\lagr_W+\lagr_{FI}.
\end{equation}
One could add a D-term $m^2V^2$ to gain a representation of the
massive gauge supermultiplet, however, this term is not gauge
invariant.

Let us expand these Lagrangians into components
using~\eqref{susyn36} and~\eqref{susyn37}. By simple
differentiation we get:
\begin{equation}\label{susyn68}
    \lagr_W=F^i\frac{\partial W\nleft(\phi^i\right)}{\partial
    \phi^i}-\frac{1}{2}\psi^i\psi^j\frac{\partial^2 W\nleft(\phi^i\right)}{\partial
    \phi^i\partial\phi^j}+\cc
\end{equation}

In order to expand $\lagr_G$ we calculate (in Wess-Zumino gauge):
\begin{multline}\label{susyn69}
    \W^{\al
    a}\W^b_\al=e^{i\tha\si^\mu\thabar\partial_\mu}\left[-\la^a\la^b-i\tha\la^a
    D^b-i\tha\la^b D^a+\half\tha\left(\si^\mu\sigmabar^\nu\right)\left(\la^a F^b_{\mu\nu}+\la^b
    F^a_{\mu\nu}\right)\right.\\
    \left.-\tha\tha\left(i\la^a \si^\mu \left(D_\mu\labar\right)^b+i\la^b \si^\mu \left(D_\mu\labar\right)^a+\half\left(F^{a \mu\nu}F^b_{\mu\nu}+iF^{a \mu\nu}\Ftilde^b_{\mu\nu}\right)-D^a
    D^b\right)\right]\\\wzgauge,
\end{multline}
where $\Ftilde$ is the Poincar\'e dual defined by
\begin{equation}\label{susyn69.5}
    \Ftilde_{\mu\nu}=\frac{1}{2}\vep_{\mu\nu\rho\ka}F^{\rho\ka}.
\end{equation}
In order to get~\eqref{susyn69} we have used~\eqref{appspinorn32},
\eqref{appspinorn37} and~\eqref{appspinorn39}.\footnote{The sign
on the $F\Ftilde$ term is not standard in the literature. The
reason ought to be the also unconventional sign on
$\si^0=-\idmatr$.} The differential operator in the beginning of
the equation simply ensures that we are evaluating in $x_+$. Using
this we get (independent of the choice of gauge):
\begin{eqnarray}\label{susyn70}
    \lagr_G&=&-\la^a\la^b\left(F^i\frac{\partial\tau_{ab}}{\partial\phi^i}-\frac{1}{2}\psi^i\psi^j\frac{\partial^2\tau_{ab}}{\partial\phi^i\partial\phi^j}\right)\nonumber\\
    &&-\frac{1}{2\sqrt{2}}\frac{\partial\tau_{ab}}{\partial\phi^i}\psi^i\left(-i\la^a
    D^b-i\la^b D^a+\frac{1}{2}\left(\si^\mu\sigmabar^\nu\right)\left(\la^a F^b_{\mu\nu}+\la^b
    F^a_{\mu\nu}\right)\right)\nonumber\\
    &&-\tau_{ab}\left(i\la^a \si^\mu \left(D_\mu\labar\right)^b+i\la^b \si^\mu \left(D_\mu\labar\right)^a+\half\left(F^{a \mu\nu}F^b_{\mu\nu}+iF^{a \mu\nu}\Ftilde^b_{\mu\nu}\right)-D^a
    D^b\right)\nonumber\\&&+\cc
\end{eqnarray}
where $\tau_{ab}$ is seen as a function of $\phi^i$. As noted
above one could have used Majorana spinors instead of Weyl spinors
and thereby have written all of this even more compactly. However,
we shall not need this.

The \kahler{} term requires more computation. The result is
(following, but correcting~\cite{9912271}):
\begin{eqnarray}\label{susyn71}
    \lagr_K&=&-g_{i\ibar}D_\mu\phi^i \left(D^\mu\phi\right)^{*\ibar}-ig_{i\ibar}\psibar^{\ibar}\sigmabar^\mu D_\mu\psi^i+\frac{1}{4}R_{i\bar{k}j\bar{l}}\psi^i\psi^j\psibar^{\bar{k}}\psibar^{\bar{l}}\nonumber\\
    &&+g_{i\ibar}\left(F^i-\frac{1}{2}\Gamma^i_{jk}\psi^j\psi^k\right)\left(\bar{F}^{\ibar}-\frac{1}{2}\Gamma^{\ibar}_{\jbar\bar{k}}\psibar^{\jbar}\psibar^{\bar{k}}\right)\nonumber\\
    &&+\left(\frac{1}{2}D^a(T_a^{\repr})^j_{\ph{j}i}\phi^i\frac{\partial
    K}{\partial\phi^j}+i\sqrt{2}g_{i\ibar}(T^{\repr}_a)^i_{\ph{i}k}\phi^k\labar^a\psibar^{\ibar}+\cc\right).
\end{eqnarray}
The added complex conjugate in the last line is only for the last
two terms. Here $g_{i\ibar}$ is the \emph{\kahler{} metric}
defined by
\begin{equation}\label{susyn72}
    g_{i\ibar}=\frac{\partial^2
    K(\phi^i,\phibar^{\ibar})}{\partial\phi^{i}\partial\phibar^{\jbar}}.
\end{equation}
$\Ga$ and $R$ are the corresponding Levi-Civita connection and
Riemann curvature tensor respectively. $D_\mu\phi^i$ is the usual
gauge  covariant derivative while $D_\mu\psi^i$ also contains the
Levi-Civita connection:
\begin{equation}\label{susyn73}
    D_\mu\psi^i=
    \partial_\mu\psi^i+iA_\mu^a(T_a^{\repr})^i_{\ph{i}j}\psi^j+\Ga^i_{jk}\nleft(D_\mu\phi^j\right)\psi^k.
\end{equation}
We have used a notation where the indices of the complex
conjugates are barred to stress that they are treated as
independent coordinates.

We immediately see that the $F^i$'s and $D^a$'s are auxiliary
fields with no derivatives as postulated above. Hence they can be
replaced with their equations of motions. This is also true in the
quantised theory: Because the auxiliary fields appear at most
quadratically they can be integrated out.\footnote{Actually this
is oversimplified. The coefficients of the squares of the
auxiliary fields are functions possibly depending on $x^\mu$. This
means that when we do the functional integration of the auxiliary
fields we encounter the determinants of the coefficient functions.
Let us call such a coefficient function $f(x)$. The corresponding
``matrix'' is $A(x,y)=f(x)\de^4(x-y)$ such that e.g.
$\iint\dlor\mathrm{d}^4y D^a(x)A(x,y)D^a(y)=\int\dlor
D^a(x)f(x)D^a (x)$. The determinant can be rewritten as the
exponential of the trace. Hence we have to add a term to the
action proportional to $\int\dlor \ln A(x,x)=\int\dlor \ln
f(x)\de^4 (0)$. However, (following~\cite{weinberg3}) using
dimensional regularisation of the determinant this term is
eliminated since here $\de^4
(0)=\int\!\frac{\mathrm{d}^dq}{(2\pi)^d}=\frac{\Ga\nleft(1-d/2\right)}{\Ga(0)}\left(\frac{1}{4\pi}\right)^{d/2}=0$
because the $\Ga$-function of zero is infinite and $d$ is the
dimension close to, but not equal 4.} However, the supersymmetry
transformations on these auxiliary fields then give restrictions
which exactly correspond to the equations of motion. Thus on the
remaining fields supersymmetry is only realised on-shell. This
agrees with the supermultiplets from
section~\ref{susysecsupermultiplets} being on-shell.

Let us finish this section by looking at the case where
$\tau_{ab}$ from~\eqref{susyn63} is independent of the chiral
fields. This is the case when we look at renormalisable
Lagrangians because according to~\eqref{susyn59} the mass
dimension of $\W_\al$ is $3/2$ (the lowest component is a spinor)
and hence $\int\dtotha\W^\al\W_\al$ has mass dimension four. We
can now split the gauge group in its abelian and simple parts. The
Lagrangian~\eqref{susyn63} splits into a sum with one term for
each part of the gauge group. There can be no mixed terms due to
gauge invariance.

Let us first look at the abelian part. Here each $\W^a_\al$ is
gauge invariant. This allows us to write the Lagrangian as:
\begin{equation}\label{susyn74}
    \lagr_{G,\mathrm{abelian}}=\int\dtotha\frac{1}{16\pi i}\tau_{ab}\W^{\al
    a}\W^b_\al+\cc,
\end{equation}
where $\tau_{ab}$ is the theta angles and gauge couplings:
\begin{equation}\label{susyn75}
    \tau_{ab}=\frac{\vartheta_{ab}}{2\pi}+i\frac{4\pi}{g^2_{ab}}.
\end{equation}
Plugging into~\eqref{susyn70} yields:
\begin{multline}\label{susyn76}
    \lagr_{G,\mathrm{abelian}}=-\frac{1}{2g^2_{ab}}\left(i\la^a\si^\mu\partial_\mu\labar^b+i\la^b\si^\mu\partial_\mu\labar^a\right)-\frac{1}{4g^2_{ab}}F^a_{\mu\nu}F^{b\mu\nu}-\frac{\vartheta_{ab}}{32\pi^2}F^a_{\mu\nu}\Ftilde^{b\mu\nu}
    +\frac{1}{2g_{ab}^2}D^aD^b.
\end{multline}
Here we have used~\eqref{appspinorn36} and allowed integration by
parts to get the first term.~\eqref{susyn76} is the standard form
of (supersymmetric) Yang-Mills theory and this is the reason for
the chosen normalisation factor in~\eqref{susyn74}.

For a simple factor $\tau_{ab}$ must be proportional to the
Killing form. It can also be shown that we can choose the
generators of the simple factor such that in any irreducible
representation, $\reprn$, we have
\begin{equation}\label{susyn76.5}
    \tr_{\repr}\nleft(T_a^{\repr}T_b^{\repr}\right)=C(\reprn)\de_{ab}.
\end{equation}
Here $C(\reprn)$ is called the quadratic invariant. The gauge
kinetic Lagrangian for the simple factor then becomes:
\begin{eqnarray}\label{susyn77}
    \lagr_{G,\mathrm{simple}}&=&\int\dtotha\frac{\tau}{16\pi iC(\reprn)}\tr_{\repr}\nleft(\W^\al\W_\al\right)+\cc\nonumber\\
    &=&\int\dtotha\frac{\tau}{16\pi i}\W^{a\al
    }\W^a_\al+\cc,
\end{eqnarray}
where the complex $\tau$ contains the theta angle and the gauge
coupling constant:
\begin{equation}\label{susyn78}
    \tau=\frac{\vartheta}{2\pi}+i\frac{4\pi}{g^2}.
\end{equation}
There is one coupling $\tau$ for each simple factor in the gauge
group. Expanding in the same way as in~\eqref{susyn76} gives:
\begin{equation}\label{susyn78.1}
    \lagr_{G,\mathrm{simple}}=-\frac{i}{g^2}\la^a\si^\mu \left(D_\mu\labar\right)^a-\frac{1}{4g^2}F^a_{\mu\nu}F^{a\mu\nu}-\frac{\vartheta}{32\pi^2}F^a_{\mu\nu}\Ftilde^{a\mu\nu}
    +\frac{1}{2g^2}D^aD^a.
\end{equation}

We will see in section~\ref{dvsecweffsubinstchiral} that the theta
terms are total derivative terms which in the non-abelian case can
be non-zero due to instanton effects. The coupling constant $g$,
which here enters by multiplying terms with $1/g^2$, can be put
into the structure constants by rescaling $V\mapsto gV$. This
removes the overall $1/g^2$ (but puts a factor $g^2$ on the
$\vartheta$-term) and the coupling $g$ will then multiply the
structure constants in the definition of $F_{\mu\nu}$ and $D_\mu$
in equations~\eqref{susyn60} and~\eqref{susyn61} -- as we often
see in non-supersymmetric gauge theory.

We also note that our Lagrangians are not invariant under
rescaling of the generators $T_a$. Scaling $T_a\mapsto\al T_a$
will scale the structure constants as
$f_{ab}^{\phantom{ab}c}\mapsto\al f_{ab}^{\phantom{ab}c}$. The
vectors in the Lie algebra should be the same under this change of
basis and hence the components of the vector field and
correspondingly the supersymmetric gauge field strength scale as
$V^a\mapsto\frac{1}{\al}V^a$ and
$\W^a_\be\mapsto\frac{1}{\al}\W^a_\be$ (we have to remember the
scaling of the structure constants in the last scaling). We
conclude that to keep our Lagrangians invariant under the scaling
we must require that the couplings scale as $g\mapsto g/\al$ and
$\vartheta\mapsto\al^2\vartheta$. Accordingly we have to choose
some normalisation of the generators for the theory to make
sense.\footnote{We could also have chosen to obtain an invariant
theory by multiplying with e.g. the quadratic invariant in the
fundamental representation, $C\nleft(\mathrm{fund}\right)$,
in~(\ref{susyn77} since the quadratic invariant scales as
$C\nleft(\reprn\right)\mapsto\al^2 C\nleft(\reprn\right)$.}

\section{$\Nscr=2$ Supersymmetric Yang-Mills Theory}\label{susysecn2subren}

In this section we will briefly discuss the $\Nscr=2$
supersymmetric gauge field theories. These can be obtained from
the results in the last section by noting that $\Nscr=2$
supersymmetric field theories are special cases of $\Nscr=1$
supersymmetric field theories.

\subsection{$\Nscr=2$ Supersymmetric
Lagrangians from $\Nscr=1$ Supermultiplets}

Let us as above denote the supercharges of the $\Nscr=2$
supersymmetry as $\qgen^1$ and $\qgen^2$ where we have suppressed
the spinor indices. In section~\ref{susysecsupermultiplets} we
obtained the supermultiplets as representations of the little
supergroup. The little supergroup was obtained by adding the
supercharges to the algebra of the little group. However, looking
at the supersymmetry algebra~\eqref{susyalgebra} we see that we
also get a group if we only add $\qgen^1$ and its hermitian
conjugate $\qbargen_1$ to the little group. This is exactly the
$\Nscr=1$ little supergroup which accordingly is a subgroup of the
$\Nscr=2$ little supergroup. Thus the $\Nscr=2$ supermultiplets
split into $\Nscr=1$ supermultiplets just as supermultiplets are
multiplets of Poincar\'e representations. However, we could just
as well had looked at the subgroup corresponding to adding
$\qgen^2$ and its hermitian conjugate to the little group thus
giving another splitting in $\Nscr=1$ supermultiplets. This change
corresponds to the R-symmetry:
\begin{equation}\label{susyn79}
    \qgen^1\mapsto\qgen^2,\qquad\qgen^2\mapsto-\qgen^1.
\end{equation}
Now the method to get an $\Nscr=2$ Lagrangian is simply to split
the $\Nscr=2$ supermultiplets into $\Nscr=1$ supermultiplets
corresponding to say $\qgen^1$, then write the most general
$\Nscr=1$ supersymmetric Lagrangian with these supermultiplets,
and finally impose the discrete R-symmetry~\eqref{susyn79}. The
Lagrangian will then be $\Nscr=2$ supersymmetric since it is
invariant under $\qgen^1$ by construction and hence invariant
under $\qgen^2$ by the R-symmetry~\eqref{susyn79}.

\subsection{Renormalisable $\Nscr=2$ Supersymmetric
Lagrangians}

Let us now find the most general renormalisable Lagrangian for the
fields of the $\Nscr=2$ massless gauge supermultiplet. For
simplicity we will assume a simple gauge group $G$. The fields of
this multiplet are according to table~\ref{tablesusyn1} a gauge
field $A_\mu$, two Weyl (or Majorana) fermions $\la$ and $\psi$,
and a complex scalar $\phi$. In figure~\ref{susyfign1} it is shown
how these fields are related by the supercharges $\Qbar_1$ and
$\Qbar_2$ according to equation~\eqref{susyn18}.
\begin{figure}\caption{}\label{susyfign1}

    \begin{displaymath}
        \xymatrix{
            &A_\mu\\
            \ar[ur]^{\Qbar_1}\la\ar@{<-->}^{\mathrm{SU}(2)_R}[rr]&&\ar[ul]_{\Qbar_2}\psi\\
            &\ar[ul]^{\Qbar_2}\phi\ar[ur]_{\Qbar_1}}
    \end{displaymath}
    \begin{center}
        The $\Nscr=2$ gauge supermultiplet with supersymmetry
        transformations. The spinor indices on the supercharges
        have been suppressed.
    \end{center}
\end{figure}
Breaking the supermultiplet into the $\Nscr=1$ supermultiplets of
the supercharge $\qgen^1$ we get a gauge multiplet $(\la,A_\mu)$
and a chiral multiplet $(\phi,\psi)$ corresponding to the
superfields $V$ and $\Phi$ respectively.

$A_\mu$ transforms in the adjoint representation of the gauge
group so the whole $\Nscr=2$ supermultiplet must transform in the
adjoint representation since we argued above that supersymmetry
transformations and gauge transformations must commute. Thus
$\Phi$ is in the adjoint representation and both $V$ and $\Phi$
can be seen as taking values in the gauge Lie
algebra.\footnote{This also means that the $\Nscr=2$
supersymmetric Lagrangian can not fulfil the standard model since
that would require the matter to belong to a complex
representation (the chiral) of
$\mathrm{SU}(3)\times\mathrm{SU}(2)\times\mathrm{U}(1)$, but the
adjoint representation is always real. Even when matter is added
the representation will still be real.} The most general
Lagrangian involving these two fields is given in~\eqref{susyn67}
where we demand renormalisability. This means that the gauge
kinetic and self interaction term takes the form~\eqref{susyn77}
since the gauge group is assumed simple. Using~\eqref{susyn42}
and~\eqref{susyn64} the renormalisable gauge invariant \kahler{}
term takes the form $\int\dsuper \tr\nleft(\Phi^\dagger e^{2\ad\!
V}\Phi\right)$ where $\ad\! V$ is the adjoint defined
in~\eqref{apppoincaren5}. The trace can be taken in any
representation, but we will now have to care about the
normalisation of the terms in the Lagrangian since we have to
impose the symmetry~\eqref{susyn79}. On the components we see from
figure~\ref{susyfign1} that this symmetry takes the form:
\begin{equation}\label{susyn80}
    \psi^a\mapsto\la^a\qquad\la^a\mapsto-\psi^a.
\end{equation}
Thus the kinetic term for $\psi$ and $\la$ must have the same
normalisation. Using~\eqref{susyn68} we see that the
superpotential term for the chiral field must be proportional to
$\Phi$ since we have no mass or interaction terms for the gaugino
field $\la$. A linear superpotential is trivial and hence we will
set it to zero. There can be no Fayet-Iliopoulos term since we
assumed a simple group. The $\Nscr=2$ supersymmetric Yang-Mills
Lagrangian then takes the form (using~\eqref{susyn71}
and~\eqref{susyn78.1}):
\begin{equation}\label{susyn81}
    \lagr_{\Nscr=2}=\frac{\tau}{16\pi iC(\reprn)}\tr_{\repr}\nleft(\int\dtotha\W^\al\W_\al+2\int\dsuper\Phi^\dagger e^{2\ad\!
    V}\Phi\right)+\cc
\end{equation}
Expanded in components this gives (again using~\eqref{susyn71}
and~\eqref{susyn78.1}):
\begin{multline}\label{susyn82}
    \lagr_{\Nscr=2} = \frac{1}{g^2C(\reprn)}\tr_{\repr}\Big(-i\la\si^\mu D_\mu\labar-\frac{1}{4}F_{\mu\nu}F^{\mu\nu}-\frac{g^2\vartheta}{32\pi^2}F_{\mu\nu}\Ftilde^{\mu\nu}
    +\frac{1}{2}DD
    -D_\mu\phi\left(D^\mu\phi\right)^\dagger\\
    -i\psibar\sigmabar^\mu D_\mu\psi
    +F\bar{F}+D[\phi^\dagger,\phi]-i\sqrt{2}[\labar,\psibar]\phi-i\sqrt{2}[\la,\psi]\phi^\dagger\Big),
\end{multline}
where we have used the cyclic property of the trace (properly
signed for the anticommuting fields) to put the commutators in the
above form. The commutator $[\la,\psi]$ simply means
$\la\psi-\psi\la$ thus defining the spinor indices. Here the bars
on the spinor fields mean hermitian conjugates in the gauge
algebra. We immediately see that this Lagrangian is invariant
under the symmetry~\eqref{susyn80}.

As mentioned above we can now eliminate the auxiliary fields using
their equations of motion $F=0$ and $D=-[\phi^\dagger,\phi]$. This
gives the scalar potential:
\begin{equation}\label{susyn83}
    U=\frac{1}{2g^2C(\reprn)}\tr_{\repr}\nleft([\phi^\dagger,\phi]^2\right).
\end{equation}

It is possible to add matter to the model in the form of the
hypermultiplet from table~\ref{tablesusyn1}. Such a hypermultiplet
splits into two chiral multiplets. Adding this to the Lagrangian
allows mixed superpotential terms.

For non-renormalisable Lagrangians one can use an $\Nscr=2$
superspace formulation to get the most general Lagrangian. As is
shown in appendix~\ref{appn2super}, the Lagrangian is then
determined by a holomorphic function known as the prepotential.

%------------------- New Chapter -----------------------------------------

\chapter{The Dijkgraaf-Vafa Conjecture}

In~\cite{0208048} R. Dijkgraaf and C. Vafa formulated a conjecture
telling us how to systematically compute the exact low energy
effective superpotential of a wide range of $\Nscr=1$
supersymmetric gauge theories in four space-time dimensions.

We will start by simply stating the conjecture in the case of a
$\un{N}$ gauge group and adjoint matter. However, most of the
concepts used in the Dijkgraaf-Vafa conjecture needs explaining so
the rest of the chapter will be devoted to understanding the
conjecture. Along the way we will understand some of its
implications and put it into its right context. At the end of the
chapter we will present the Dijkgraaf-Vafa conjecture for general
gauge groups and matter representations.

%-------------New section--

\section[The Dijkgraaf-Vafa Conjecture with $\un{N}$ Gauge Group\ldots]{The Dijkgraaf-Vafa Conjecture with $\boldsymbol{\un{N}}$ Gauge Group and Adjoint
Matter.}\label{dvsecdvun}

Following~\cite{0208048} we will state the Dijkgraaf-Vafa
conjecture in the case of a $\un{N}$ gauge group and adjoint
matter.\footnote{The notation will differ slightly
from~\cite{0208048} and some points are taken
from~\cite{0211170}.}

\subsection{The Traceless Case}\label{dvsecdvunsubdv}

We will study a four-dimensional supersymmetric field theory. The
tree-level Lagrangian of the theory is obtained by first looking
at a renormalisable $\Nscr=2$ supersymmetric Yang-Mills theory
with gauge group $\un{N}$ and corresponding Lie algebra $\cun{N}$.
Assuming no Fayet-Iliopoulos term since this makes it more
difficult to obtain supersymmetric vacua as we will see in
section~\ref{dvsecvac}, the Lagrangian is simply~\eqref{susyn81}
with an abelian part added (in section~\ref{dvsecdvunsublagr} we
will write out the full Lagrangian). Then we add a tree-level
superpotential for the adjoint $\cun{N}$-valued chiral field
$\Phi$ to the Lagrangian:
\begin{equation}\label{dvn1}
    \int\dtotha\wtree\nleft(\Phi\right)=\int\dtotha\tr
    \wpoly\nleft(\Phi\right),
\end{equation}
where $\wpoly$ is a complex polynomial of degree $n+1$ which we in
general will think of as having the form:
\begin{equation}\label{dvn1.5}
    \wpoly\nleft(\Phi\right)=\frac{1}{2}m\Phi^2+\sum_{k=3}^{n+1}\frac{g_k}{k}\Phi^k,
\end{equation}
where $m$ and the couplings $g_k$ are complex.\footnote{The reason
that we have $m$ and not $m^2$ in the tree-level superpotential is
that $\Phi$ has mass dimension one and the whole superpotential
has mass dimension three as we saw in
section~\ref{susysecchiral}.} Assuming $n\geq1$ this breaks
$\Nscr=2$ supersymmetry to $\Nscr=1$ supersymmetry as we saw in
section~\ref{susysecn2subren}. We should think of
$\Phi=\Phi^aT_{a}$ as taking values in the fundamental
representation\footnote{This should not be confused with $\Phi$
\emph{transforming} in the adjoint representation, but means that
$\Phi=\Phi^aT_{a}^{(\mathrm{fund})}$ where
$T_{a}^{(\mathrm{fund})}$ are the generators of the gauge group in
the fundamental representation.} since e.g. in the adjoint
representation the abelian part would vanish. Thus here and in the
following $\tr$ will denote trace in the fundamental
representation.

We will not restrict ourselves to renormalisable superpotentials
and hence allow $\wpoly$ to have a degree higher than three. In
that case we can think of the superpotential as obtained from a
superpotential at a higher energy scale by integrating out other
fields using Wilsonian renormalisation (explained in
section~\ref{dvsecwilson}).

As we will explain in section~\ref{dvsecvac} the supersymmetric
classical vacua are obtained by diagonalising $\Phi$ and demanding
the eigenvalues to be in the set of critical points of the
polynomial $\wpoly$. Since $\wpoly'$ has degree $n$, there must be
$n$ critical points which we denote $a_1,\ldots, a_n$. We will
assume these to be isolated and -- as we will show -- this means
that the vacua are massive. Since $\Phi$ is an $N\times N$-matrix,
a vacua is obtained by choosing a partition:
\begin{equation}\label{dvn2}
    N=N_1+\ldots+N_n,
\end{equation}
corresponding to distributing $N_i$ of the eigenvalues at the
critical point $a_i$. This furthermore breaks the gauge symmetry
group as:
\begin{equation}\label{dvn3}
    \un{N}\mapsto\un{N_1}\times\cdots\times\un{N_n}.
\end{equation}
If $N_i$ is zero we will leave the corresponding factor out.

Looking at the corresponding quantised theory
(section~\ref{dvsecquant}), the lore of the low energy dynamics
(section~\ref{dvsecweffsublore}) tells us that we have confinement
and gaugino condensation in the $\sun{N_i}$ subgroups of the
$\un{N_i}$ factors. The gauge coupling becomes strong at the
(complex) dynamically generated scale $\La$ and a mass gap is
generated. To describe this we introduce the \emph{traceless
glueball superfield}\footnote{This is -- perhaps more properly --
also called the \emph{gaugino condensate chiral superfield}.} for
each factor of $\sun{N_i}$:
\begin{equation}\label{dvn4}
    \Shat_i=-\frac{1}{16\pi^2}\tr\nleft(\What_{(i)}^\al\What_{(i)\al}\right),
\end{equation}
where the supersymmetric gauge field strength for $\sun{N_i}$ is
denoted $\What^\al_{(i)}$. Please note that here and in the
following there is no sum over the factor indices in parenthesis.
As noted above we use the fundamental representation for the Lie
algebra valued fields. Here we use a normalisation such that the
generators for the simple $\sun{N_i}$ part
$T_{(i)a}^{(\mathrm{fund})}$ with $a=1,\ldots, N_i^2-1$ fulfil:
\begin{equation}\label{dvn5}
    \tr\nleft(T_{(i)a}^{(\mathrm{fund})}T_{(i)b}^{(\mathrm{fund})}\right)=\frac{1}{2}\de_{ab},\qquad\tr\nleft(T_{(i)a}^{(\mathrm{fund})}\right)=0.
\end{equation}
Thus from equation~\eqref{susyn69} we see that the lowest
component of the traceless glueball superfield is
$\frac{1}{32\pi^2}\lahat_{(i)}^a\lahat_{(i)}^a$ where, as before,
the hat means restriction to the traceless part of the algebra.
The condensation of the gauginos described by the field
$\lahat^\al_{(i)}$ thus corresponds to $\Shat_i$ getting a
dynamical expectation value. We also see that the chiral half of
the gauge Lagrangian for $\sun{N_i}$ (equation~\eqref{susyn77}) is
given by the $\tha\tha$-component of $\Shat_i$.

The physical quantities of the vacua are determined by the 1PI
effective action. However, as this is too hard to find we focus on
the Wilsonian effective action and we denote the corresponding
Lagrangian $\leff$. It is the \emph{generalised superpotential}
$\weff$ of this effective Lagrangian which we want to determine
(dealt with in detail in section~\ref{dvsecweff}):
\begin{equation}\label{dvn6}
    \leff=\int\dtotha\weff+\cc+\int\dsuper\ldots,
\end{equation}
where the dots denote some local gauge invariant superspace
function which is not in the focus of interest here. Hence the
generalised superpotential consists of the $\tha\tha$-terms that
can not be written as local gauge invariant D-terms. Thus it
includes terms like~\eqref{susyn77} i.e. $\Shat_i$ can contribute
to $\weff$.\footnote{Remember that $\W_\al\W^\al$ is chirally
exact and hence can be written as a D-term as
in~(\ref{susyn63.5}), but not as a gauge invariant D-term!}
Actually, it is the lore that the elementary fields at low energy
exactly are the $\Shat_i$'s. $\weff\big(\Shat_i,g_k\big)$ is thus
called the effective glueball superpotential. The vacuum
expectation value of $\Shat_i$ is then simply determined by:
\begin{equation}\label{dvn6.5}
    \frac{\partial \weff\big(\Shat_k,g_k\big)}{\partial \Shat_i}=0.
\end{equation}

As we will see in section~\ref{dvsecweffsubvy}, even without
matter ($\Phi=0$) we have an effective superpotential called the
Veneziano-Yankielowicz superpotential. This is given by:
\begin{equation}\label{dvn7}
    \wvy=\sum_i
    N_i\Shat_i\bigg(1-\ln\frac{\Shat_i}{\La_i^3}\bigg),
\end{equation}
where we have one term for each gauge group factor and the scales
$\La_i$ are described in section~\ref{dvsecglueball}.

The Dijkgraaf-Vafa conjecture tells us that in order to determine
$\weff$ we have to look at the related bosonic \emph{one matrix
model} (we will entertain ourselves with the details in
section~\ref{dvsecmatrix}) with partition function given by:
\begin{equation}\label{dvn8}
    \Zm=\int\DM e^{-\frac{1}{\gs}\wtree(M)},
\end{equation}
i.e. where the potential is the tree-level superpotential. Here
$M$ are $N'\times N'$ hermitian matrices (spanned by the
generators of $\un{N'}$) and $\gs$ is a simple (dimensionful)
scaling factor -- the ``$\mathrm{s}$'' simply refers to the
stringy origin of the Dijkgraaf-Vafa conjecture, but here it, a
priori, has nothing to do with string theory.

The vacua of the matrix model are, analogously to the gauge theory
case, determined by diagonal matrices with the $N'$ eigenvalues in
the set of critical points, $a_1,\ldots,a_n$, of $\wpoly$. Thus,
in analogy with~\eqref{dvn2} choosing a vacuum (modulo permutation
of eigenvalues) corresponds to a partition:
\begin{equation}\label{dvn9}
    N'=N'_1+\ldots+N'_n.
\end{equation}
To obtain the correspondence with the gauge theory we must here
demand that $N'_i=0$ if $N_i=0$ in order to have the same gauge
symmetry breaking pattern. Otherwise $N'_i$ and $N_i$ are
completely independent. We can now obtain the free energy of the
matrix model by a perturbative expansion around the chosen vacuum.
However, in the case of broken gauge symmetry we must remember to
take into account Faddeev-Popov ghosts in the matrix model. These
actually take the same form as in the gauge theory in accordance
with correspondence between the potential of the matrix model and
the tree-level superpotential.\footnote{This perturbative
(diagrammatic) approach on the matrix model side in the
formulation of the Dijkgraaf-Vafa conjecture can be found
in~\cite{0210238}.}

We should now take the 't Hooft large $N'$ limit (explained in
section~\ref{dvsecdouble}) where $N'_i\gg1$, $\gs\ll1$ while
keeping $\gs N'_i$ fixed and finite. We will see how to
topologically characterise diagrams such that the free
energy\footnote{For now we ignore any contribution from the
measure of the matrix model. Such a contribution could been seen
as giving the Veneziano-Yankielowicz superpotential
(section~\ref{dvsecmatrixsubmeasure}).} can be written as a
topological expansion:
\begin{equation}\label{dvn10}
    \Zm=e^{-\sum_{g\geq0}\gs^{2g-2}\cF_g\nleft(\gs N'_i\right)},
\end{equation}
where $g$ is the genus of the surface i.e. an integer number
greater than or equal zero which corresponds to the number of
handles that has been added to the sphere. The $\cF_g$'s depend
only on $\gs$ and $N'_i$ through the products $\gs N'_i$. We see
that the dominant contribution stems from $g=0$ and we can thus
restrict to $\cF_{g=0}$ which is called the planar limit.  The
connection to the gauge theory is to identify:
\begin{equation}\label{dvn11}
    \Shat_i\defi\gs N'_i.
\end{equation}
We note that this is a formal identification that allows us to
obtain the glueball superpotential from the planar limit of the
free energy in the matrix model. We will see in
section~\ref{dvsecmatrixexact} that when the matrix model is
solved exactly, $\Shat_i$ should be identified with the filling
fractions in the multi-cut solution.

Now we can state the Dijkgraaf-Vafa conjecture in the case where
we ignore the abelian part of the $\W_\al$'s such that the
glueball superfield is traceless -- but we do not disregard the
abelian part of $\Phi$:
\begin{thmdvsun}
\begin{subequations}\label{dvsun}
\begin{eqnarray}
    \weff\big(\Shat_i,g_k\big)&=&\weffpert\big(\Shat_i,g_k\big)+\wvy\big(\Shat_i\big),\label{dvn12}\\
    \weffpert\big(\Shat_i,g_k\big)&=&\sum_i N_i\frac{\partial\cF_{g=0}\big(\Shat_i,g_k\big)}{\partial
    \Shat_i},\qquad\Shat_i\defi\gs N'_i.\label{dvn13}
\end{eqnarray}
\end{subequations}
\end{thmdvsun}
\noindent The dependence on the couplings in the tree-level
superpotential, $g_k$,  has been added for completeness.
$\weffpert$ is perturbative in $\Shat_i$ and the couplings $g_k$.
It is obtained by integrating out the massive chiral field $\Phi$
(and $\Phibar$) while treating $\W_\al$ as a background field. We
can write this as (we are in Minkowski space):
\begin{eqnarray}
  \Zholo &=& \int\DPhi\DPhibar \left.e^{iS_{\mathrm{tree}}}\right|_{\mathrm{holomorphic}}, \label{dvn13.1}\\
  \Zholo &=& e^{i\int\dlor\dtotha\weffpert},\label{dvn13.2}
\end{eqnarray}
where $\Zholo$ is the partition function where we only keep the
contribution to the $\tha\tha$-term. That is, we only include the
holomorphic contributions with no conjugates
(section~\ref{dvsecweffsubholo}). This is what the restriction in
the first line means. $S_{\mathrm{tree}}$ is the tree-level action
including the superpotential~\eqref{dvn1}.\begin{table}
\caption{}\label{dvtable1}
\begin{flushright}
\begin{tabular}{|l|l|}
  \hline
  \multicolumn{2}{|c|}{Dijkgraaf-Vafa conjecture;
  Traceless case}\\\hline
  Gauge theory & Matrix model \\\hline
  $\vphantom{\Big(}\Zholo = \int\DPhi\DPhibar \left.e^{iS_{\mathrm{tree}}}\right|_{\mathrm{holo}}$ &  $\Zm=\int\DM e^{-\frac{1}{\gs}\wtree(M)}$\\
   $\vphantom{\Big(}\Zholo = e^{i\int\dlor\dtotha\weffpert}$ &  $\Zm=e^{-\sum_{g\geq0}\gs^{2g-2}\cF_g\nleft(\gs N'_i\right)}$\\
   $\vphantom{\Big(}\weff = \weffpert+\wvy$ & $\wvy$ as a measure contribution\\
   $\vphantom{\Big(}$Vacua: & Vacua:\\
   $\vphantom{\Big(}N=N_1+\ldots+N_n$ &  $N'=N'_1+\ldots+N'_n$\\
   $\vphantom{\Big(}$No limit &  't Hooft large $N'_i$ limit\\
   $\vphantom{\Big(}\weffpert=\sum_i N_i\frac{\partial\cF_{g=0}(\Shat_i)}{\partial
    \Shat_i}$ &  \\
   \multicolumn{2}{|c|}{$\vphantom{\Big(}\Shat_i=\gs N'_i\quad$}\\\hline
\end{tabular}
\newline
\begin{center}
The Dijkgraaf-Vafa conjecture in the case of a traceless glueball
superfield. Formulae are explained in the text.
\end{center}
\end{flushright}
\end{table}
The Dijkgraaf-Vafa conjecture in this case is summarised in
table~\ref{dvtable1}.\footnote{It appears to be more suggestive to
compare the matrix model with the gauge theory written in
Euclidean space. However, when Wick rotating to Euclidean space
the superpotentials receive a sign change which seems to give the
wrong sign in the comparison. In section~\ref{proofsecdiasubreduc}
we will see that signs cancel in the right way due to the minus
sign in the definition of the glueball superfield~(\ref{dvn4}).}

\subsection[The $\un{N_i}$-Case]{The
$\boldsymbol{\un{N_i}}$-Case}\label{dvsecdvunsubdvun}

Let us now turn to the case where we also consider the abelian
parts of the supersymmetric gauge field strengths. In this case we
define the \emph{glueball superfield} $S_i$ and the field
$w_{i\al}$ as (here we follow~\cite{0211170}, but naturally with
our normalisations):
\begin{eqnarray}\label{dvn14}
    S_i&=&-\frac{1}{16\pi^2}\tr\nleft(\W_{(i)}^\al\W_{(i)\al}\right),\\
    w_{i\al}&=&\frac{\sqrt{2}}{4\pi}\tr\nleft(\W_{(i)\al}\right),\label{dvn14.5}
\end{eqnarray}
where $\W^\al_{(i)}$ is the supersymmetric gauge field strength
for the whole $\un{N_i}$ subgroup. As a standard we will use the
$N_i\times N_i$ identity matrix (with trace $N_i$) to span the
abelian part of the algebra, and the formerly introduced
$T_{(i)a}^{(\mathrm{fund})}$ to span the simple part. Using
equation~\eqref{dvn5} we get the following relation between
$\Shat$, $S$ and $w_{i\al}$:
\begin{equation}\label{dvn15}
    S_i=\Shat_i-\frac{1}{2N_i}w_i^\al w_{i\al}\qquad\textrm{(no sum over i)}.
\end{equation}
Even though $\Shat_i$ are the elementary fields at low energy (the
$w_{i\al}$'s are IR-free) it is easier to write the Dijkgraaf-Vafa
conjecture using $S_i$. The dependence on $w_{i\al}$ will then
simply be quadratic. Now we have to identify $\gs N'_i$ from the
matrix model with $S_i$. The conjecture for the form of
$\weffpert$ is then:
\begin{thmdvun}
\begin{eqnarray}
    \weffpert\nleft(S_i,w_{i\al},g_k\right)&=&\sum_i N_i\frac{\partial\cF_{g=0}\nleft(S_i,g_k\right)}{\partial
    S_i}+\frac{1}{2}\sum_{i,j}\frac{\partial^2\cF_{g=0}\nleft(S_i,g_k\right)}{\partial S_i\partial S_j}w_i^\al w_{j\al},\nonumber\\
     S_i&\defi&\gs N'_i.\label{dvn17}
\end{eqnarray}
\end{thmdvun}
\noindent Here we have again included the dependence on the
couplings $g_k$ from the tree-level superpotential $\wtree$. We
notice that by setting $w_{i\al}=0$ this case reduces to the
traceless case~\eqref{dvsun}.\footnote{In the case of $N_i=1$ one
should be careful since the field $S_i$ should appear as a field
in its own right as above -- contrary to what one might think (see
also section~\ref{dvsecnil}).}

We note that the coefficient function of $w_i^\al w_{j\al}$
in~\eqref{dvn17} is not the abelian complexified gauge coupling
$\tau_{ij}$ as in equation~\eqref{susyn74}. This is because we
have to split the gauge algebra into its simple and abelian part.
Thus $S_i$ should be split in $\Shat_i$ and $w_{i\al}$
using~\eqref{dvn15}. Expanding the functions according to this
yields:
\begin{equation}\label{dvn18}
    \tau_{ij}\propto\frac{\partial^2\cF_{g=0}\big(\Shat_i,g_k\big)}{\partial \Shat_i\partial
    \Shat_j}-\de_{ij}\frac{1}{N_i}\sum_l N_l\frac{\partial^2\cF_{g=0}\big(\Shat_i,g_k\big)}{\partial \Shat_i\partial
    \Shat_l},
\end{equation}
where there is no sum over $i$. However, we note that we also have
(mixed) terms of higher order than quadratic in the $w_{i\al}$'s
when we split $S_i$ into $\Shat_i$ and $w_{i\al}$.
From~\eqref{dvn18} we see that $\sum_j\tau_{ij}N_j=0$ which is a
reflection of the fact that the overall $U(1)$ is decoupled since
$\Phi$ is in the adjoint (see~\eqref{dvn20.5} below where there is
no coupling between the abelian part of the vector field and
$\Phi$).

For the non-perturbative part we should again add the
Veneziano-Yankielowicz superpotential by hand. However, the
literature is unfortunately inconclusive on how to do this.
Naturally, there should be a term like~\eqref{dvn7}, but perhaps
the traceless glueball superfields should here be replaced by the
full glueball superfields $S_i$ with some modification since the
overall $\un{1}$ should be decoupled. We will discuss this briefly
in section~\ref{dvsecexample}.

A couple of remarks are in order. First we note that the $N_i$
dependence in the effective superpotential is extremely simple.
Secondly, we emphasise that the reduction to planar diagrams on
the gauge theory side is exact while we have to take the 't Hooft
large $N'$ limit on the matrix model side. Furthermore, we will
see in section~\ref{dvsecmatrixmm} that for a given diagram in the
matrix model we get one factor of $S_i=\gs N'_i$ for each index
loop indexed by $i$. Thus it is very simple to make an expansion
of $\weffpert$ to a given order in $S_i$. However, this
immediately poses a problem: Looking at the definition of $S_i$
(or $\Shat_i$) it is a sum of products of a finite number of
Grassmannian variables $\W_{(i)}^{a\al}$ where $a$ is the adjoint
index. Hence $S_i$ must be nilpotent. But on the matrix model side
we can go to any order in $\gs N'_i$ i.e. any number of loops. We
will discuss this in section~\ref{dvsecnil}.

\subsection{Proofs}

The Dijkgraaf-Vafa conjecture actually consists of two parts:

First we have the statement that the total Wilsonian effective
action is obtained as a sum of the Veneziano-Yankielowicz
superpotential for the pure super-Yang Mills theory, $\wvy$, and
the effective potential obtained by integrating out the massive
chiral fields, $\weffpert$. This has not been proven, but one can
argue that it is true \cite{0211170} if we assume that the
$\Shat_i$ fields are the elementary fields in the low energy
effective theory as the lore says.

The second part of the conjecture is the exact form of $\weffpert$
using the related matrix model. This conjecture originates in
topological string theory and is proven herein. However, the
conjecture can be proven within supersymmetric gauge theory
itself. This can be done using Feynman diagrams as we will see in
chapter~\ref{chpproof}. One can also use Seiberg-Witten theory and
the ILS linearity principle (section~\ref{dvsecweffsubilsintein})
to derive the effective superpotential~\cite{0210135}. A last and
powerful way to prove the conjecture within supersymmetric gauge
theory is to use the chiral ring and the generalised Konishi
anomalies \cite{0211170}.

The more general form of the Dijkgraaf-Vafa conjecture that we
will present in section~\ref{dvsecgeneral} can also be proven
using the diagrammatic (as we also will see in
chapter~\ref{chpproof}), the Seiberg-Witten, and the generalised
Konishi anomaly method.

\subsection{The Lagrangian}\label{dvsecdvunsublagr}

Let us finish this section by writing out the Lagrangian that we
have assumed in the conjecture -- for completeness, further
reference, and to see how general it is. The fundamental fields
are the vector field $V$ (and hence $\W_\al$) and the chiral field
$\Phi$. $\W_\al$ and $\Phi$ are both $\un{N}$-adjoint fields and
as noted above they should take values in the fundamental
representation. It will be useful for us to split them in their
abelian and simple parts. In accordance with the notation above we
use a hat to designate the projection onto the simple part of the
space and we choose to use a tilde to designate the projection
onto the abelian part. Thus e.g. for $\Phi$ we choose to write:
\begin{eqnarray}\label{dvn19}
    \qquad\qquad\Phi&=&\Phitilde+\Phihat,\nonumber\\
    \Phitilde&=&\Phitilde^0\idmatr_{N\times
    N},\nonumber\\
    \Phihat&=&\Phihat^a T_{a}^{(\mathrm{fund})},\qquad
    a=1,\ldots,N^2-1.
\end{eqnarray}
The generators $T_{a}^{(\mathrm{fund})}$ of the simple part was
introduced above in~\eqref{dvn5}. In the same way we expand
$\W_\al$ and other fields taking values in the fundamental
representation of $\cun{N}$. Please note that we can set:
\begin{equation}\label{dvn19.5}
    T_{0}^{(\mathrm{fund})}\defi\idmatr_{N\times
    N},
\end{equation}
and thus obtain a basis for the fundamental representation of
$\cun{N}$ as $T_{a}^{(\mathrm{fund})}$ with $a=0,\ldots,N^2-1$.
These fulfil:
\begin{equation}\label{dvn19.6}
    \tr\nleft(T_{a}^{(\mathrm{fund})}T_{b}^{(\mathrm{fund})}\right)=c\nleft(a\right)\de_{ab},
\end{equation}
where $c\nleft(0\right)=N$ and $c\nleft(a\right)=1/2$ for
$a=1,\ldots,N^2-1$.

We should find the most general renormalisable $\Nscr=2$
supersymmetric Lagrangian that contains the $\Phi$ and $V$
superfields (making up the $\Nscr=2$ gauge multiplet). Actually,
we also assumed in chapter~\ref{chpsusy} that the Lagrangians
should be used for low energy effective actions as exactly is the
case here. As noted above we also assume no Fayet-Iliopoulos term.
Under these assumptions we found the most general Lagrangian in
section~\ref{susysecn2subren} for a simple gauge group. So we
simply have to consider how to include the abelian part. Let us
first look at the (renormalisable) \kahler{} term which according
to~\eqref{susyn42} and~\eqref{susyn64} has the form:
\begin{equation}\label{dvn20}
    \lagr_K=\int\dsuper \Phi^\dagger e^{2V^{(\mathrm{adj})}}\Phi.
\end{equation}
However, the adjoint representation of $\un{N}$ is not
irreducible. In fact the adjoint representation of the abelian
generator is, naturally, zero. So we can split the \kahler{} term
into two terms corresponding to the simple and the abelian part
respectively. These two parts can have different normalisation.
Now, as we saw in section~\ref{susysecn2subren} the full $\Nscr=2$
Lagrangian is simply the sum of the properly normalised \kahler{}
terms and $\lagr_G$ from~\eqref{susyn63}. We can split $\lagr_G$
into its abelian part~\eqref{susyn74} and its simple
part~\eqref{susyn77}. These two parts can have different $\tau$'s
(actually, we could normalise the abelian part of the gauge field
strength such that the two couplings are equal, but we choose for
generality to keep them different here). To fulfil the
R-symmetry~\eqref{susyn80} (also for the abelian part) we must
normalise the two \kahler{} terms properly. Thus we simply
get~\eqref{susyn81} with a corresponding abelian term added as
promised in section~\ref{dvsecdvunsubdv}. To get the full
Lagrangian used in the conjecture, $\lagrun$, we simply have to
add the superpotential
term~\eqref{dvn1}:\footnote{\label{dvfootsuper}It should here be
noted that in this notation it is not possible to give the gauge
couplings different units. This is because by multiplying the
gauge couplings onto the \kahler{} term we would then give the
$\Phi^a$'s ``gauge-units'' with $\Phi^0$ and $\Phi^a,\, a>0$
having different units. However, the superpotential can not be
made dimensionless because of these different units since the same
superpotential-couplings multiply $\Phi^0$ as well as $\Phi^a,\,
a>0$. In order to fix this problem the superpotential should
really be defined as (using $g$ and $\gabel$ from~(\ref{dvn22})
and~(\ref{dvn23})):
$\tr\wpoly\Big(\frac{1}{\gabel}\Phi^0T_{a}^{(\mathrm{fund})}
+\frac{1}{g}\sum_{a=1}^{N^2-1}\Phi^aT_{a}^{(\mathrm{fund})}\Big)$.
However, we will stick to the notation in~(\ref{dvn20.5}) because
the superpotential looks as in the Dijkgraaf-Vafa conjecture.}
\begin{eqnarray}\label{dvn20.5}
    \lagrun&=&\frac{\tau}{8\pi i}\tr\nleft(\int\dtotha\What^\al\What_\al+2\int\dsuper\Phihat^\dagger e^{2\ad\!
    \Vhat}\Phihat\right)\nonumber\\
    &&+\frac{\tauab}{16\pi i N}\tr\nleft(\int\dtotha\Wtilde^\al\Wtilde_\al+2\int\dsuper\Phitilde^\dagger
    \Phitilde\right)
    + \int\dtotha\tr
    \wpoly\nleft(\Phi\right)\nonumber\\
    &&+   \cc,
\end{eqnarray}
where the traces are in the fundamental representation. We have
put a hat on $V$ in the \kahler{} term for the simple part to
point out that only the simple part contributes. $\tauab$ is the
gauge coupling for the abelian part of the Lagrangian. In the
superpotential we can use~\eqref{dvn19} to split $\Phi$ into
$\Phitilde$ and $\Phihat$ as in the rest of the Lagrangian. The
complex conjugate added in the last line is for the whole
equation.

Now we want to expand this Lagrangian in components. This has
already been done for the simple part of the $\Nscr=2$ Lagrangian
in~\eqref{susyn82} and we can expand the abelian part in the same
way. The superpotential is expanded using~\eqref{susyn68}. We get:
\begin{eqnarray}\label{dvn21}
    \lagrun&=&\frac{2}{g^2}\tr\Big(-i\lahat\si^\mu \Dhat_\mu\lahatbar-\frac{1}{4}\Fhat_{\mu\nu}\Fhat^{\mu\nu}-\frac{g^2\vartheta}{32\pi^2}\Fhat_{\mu\nu}\frac{1}{2}\vep^{\mu\nu\rho\ka}\Fhat_{\rho\ka}
    +\frac{1}{2} \Dhat \Dhat -\Dhat_\mu\phihat\big(\Dhat^\mu\phihat\big)^\dagger \nonumber\\
    &&\qquad -i\psihatbar\sigmabar^\mu \Dhat_\mu\psihat
    +\Fhat\bar{\Fhat}+\Dhat[\phihat^\dagger,\phihat]-i\sqrt{2}[\lahatbar,\psihatbar]\phihat-i\sqrt{2}[\lahat,\psihat]\phihat^\dagger\Big)\nonumber\\
    &&+\frac{1}{\gabel^2 N}\tr\Big(-i\latilde\si^\mu \partial_\mu\latildebar-\frac{1}{4}\Ftilde_{\mu\nu}\Ftilde^{\mu\nu}-\frac{\gabel^2\varthetaabel}{32\pi^2}\Ftilde_{\mu\nu}\frac{1}{2}\vep^{\mu\nu\rho\ka}\Ftilde_{\rho\ka}+\frac{1}{2} \Dtilde \Dtilde\nonumber
    \\&&\qquad-\partial_\mu\phitilde\big(\partial^\mu\phitilde\big)^\dagger
    -i\psitildebar\sigmabar^\mu \partial_\mu\psitilde
    +\Ftilde\bar{\Ftilde}\Big)\nonumber\\
    &&+\Bigg(\tr\nleft(F\wpoly'\nleft(\phi\right)\right)-\frac{1}{2}\tr\Bigg(\psi\sum_{a=0}^{N^2-1}\psi^a\frac{\partial}{\partial\phi^a}\wpoly'\nleft(\phi\right)\Bigg)+\cc\Bigg),
\end{eqnarray}
where we emphasise that the tildes refer to the projection onto
the abelian part of the algebra and not the Poincar\'e dual. Hence
the Poincar\'e duals have been written out using
definition~\eqref{susyn69.5}. We have put hats on the gauge
covariant derivatives for the simple part to emphasise that these
depend only on the simple part of the gauge algebra. $\tau$ and
$\tauab$ have been expanded as in~\eqref{susyn78}
and~\eqref{susyn75}:
\begin{eqnarray}\label{dvn22}
    \tau&=&\frac{\vartheta}{2\pi}+i\frac{4\pi}{g^2},\\
    \label{dvn23}\tauab&=&\frac{\varthetaabel}{2\pi}+i\frac{4\pi}{\gabel^2}.
\end{eqnarray}
In order to obtain the contribution from the superpotential
using~\eqref{susyn68} we had to differentiate $\wtree=\tr\wpoly$
as a function of the fields $\phi^a$ in $\phi=\phi^aT_a$. Since
the trace is linear we can move the differentiation inside the
trace. We can then use
\begin{equation}\label{dvn24}
    \frac{\partial\phi}{\partial\phi^a}=T_{a}^{(\mathrm{fund})},\qquad
    a=0,\ldots,N^2-1,
\end{equation}
to obtain the first term in~\eqref{susyn68} which contains only
one derivative:
\begin{equation}\label{dvn25}
    F^a\frac{\partial W\nleft(\phi^a\right)}{\partial
    \phi^a}=F^a\tr\nleft(\frac{\partial}{\partial
    \phi^a}\wpoly\nleft(\phi\right)\right)=F^a\tr\nleft(T_{a}^{(\mathrm{fund})}\wpoly'\nleft(\phi\right)\right)=\tr\nleft(F\wpoly'\nleft(\phi\right)\right),
\end{equation}
where we used the cyclicity of trace in the second equality to put
the generator $T_{a}^{(\mathrm{fund})}$ to the front. However,
when we make two differentiations as is the case in the last term
in~\eqref{susyn68} we can only fix the placement of one the
generators that are generated. Thus we settle for only carrying
out one of the derivatives explicitly in the last term
of~\eqref{dvn21}.

Since the $F$ and $D$-fields are auxiliary, even in the quantised
theory, we should replace them with their (algebraic) equations of
motion. Using~\eqref{dvn21} we immediately get:
\begin{eqnarray}
  \Dtilde^0 &=& 0. \label{dvn26}\\
  \Dhat^a &=& -[\phihat^\dagger,\phihat]^a,\qquad a=1,\ldots, N^2-1. \label{dvn27}\\
  \Ftilde^0 &=& -\gabel^2\overline{\tr\nleft(T_{0}^{(\mathrm{fund})}\wpoly'\nleft(\phi\right)\right)}=-\gabel^2\overline{\tr\nleft(\wpoly'\nleft(\phi\right)\right)}.\label{dvn28}\\
  \Fhat^a &=& -g^2\overline{\tr\nleft(T_{a}^{(\mathrm{fund})}\wpoly'\nleft(\phi\right)\right)},\qquad a=1,\ldots, N^2-1. \label{dvn29}
\end{eqnarray}
$\bar{F}$ obeys the complex conjugated equations as that of $F$.

Plugging these expectation values back into the Lagrangian gives a
scalar potential:
\begin{multline}\label{dvn30}
    U\nleft(\phi\right)=\frac{1}{g^2}\tr\nleft([\phihat^\dagger,\phihat]^2\right)+g^2\sum_{a=1}^{N^2-1}\abs{\tr\nleft(T_{a}^{(\mathrm{fund})}\wpoly'\nleft(\phi\right)\right)}^2
    +\gabel^2\abs{\tr\nleft(T_{0}^{(\mathrm{fund})}\wpoly'\nleft(\phi\right)\right)}^2.
\end{multline}
The part stemming from the $D$ field in this potential, naturally,
is the same as we found in the $\Nscr=2$ case~\eqref{susyn83}.

Finally we can we get the full Lagrangian in components after
elimination of the auxiliary fields:
\begin{eqnarray}\label{dvn31}
    \lagrun&=&\frac{2}{g^2}\tr\Big(-i\lahat\si^\mu \Dhat_\mu\lahatbar-\frac{1}{4}\Fhat_{\mu\nu}\Fhat^{\mu\nu}-\frac{g^2\vartheta}{32\pi^2}\Fhat_{\mu\nu}\frac{1}{2}\vep^{\mu\nu\rho\ka}\Fhat_{\rho\ka}-\Dhat_\mu\phihat\big(\Dhat^\mu\phihat\big)^\dagger
     \nonumber\\
    &&\qquad -i\psihatbar\sigmabar^\mu \Dhat_\mu\psihat
    -i\sqrt{2}[\lahatbar,\psihatbar]\phihat-i\sqrt{2}[\lahat,\psihat]\phihat^\dagger\Big)\nonumber\\
    &&+\frac{1}{\gabel^2 N}\tr\Big(-i\latilde\si^\mu \partial_\mu\latildebar-\frac{1}{4}\Ftilde_{\mu\nu}\Ftilde^{\mu\nu}-\frac{\gabel^2\varthetaabel}{32\pi^2}\Ftilde_{\mu\nu}\frac{1}{2}\vep^{\mu\nu\rho\ka}\Ftilde_{\rho\ka}\nonumber\\
    &&\qquad-\partial_\mu\phitilde\big(\partial^\mu\phitilde\big)^\dagger
    -i\psitildebar\sigmabar^\mu \partial_\mu\psitilde
    \Big)\nonumber\\
    &&+\Bigg(-\frac{1}{2}\tr\Bigg(\psi\sum_{a=0}^{N^2-1}\psi^a\frac{\partial}{\partial\phi^a}\wpoly'\nleft(\phi\right)\Bigg)+\cc\Bigg)-\frac{1}{g^2}\tr\nleft([\phihat^\dagger,\phihat]^2\right)\nonumber\\
    &&-g^2\sum_{a=1}^{N^2-1}\abs{\tr\nleft(T_{a}^{(\mathrm{fund})}\wpoly'\nleft(\phi\right)\right)}^2-\gabel^2\abs{\tr\nleft(T_{0}^{(\mathrm{fund})}\wpoly'\nleft(\phi\right)\right)}^2.
\end{eqnarray}

Let us end this chapter by discussing how general this Lagrangian
is. We should compare it to the most general $\Nscr=1$ Lagrangian
with $\un{N}$ gauge group and an adjoint chiral matter field for
use in low-energy effective theories, i.e.~\eqref{susyn67}. As
noted above we have left out a Fayet-Iliopoulos term. The gauge
part of the Lagrangian is the most general renormalisable that we
can have, i.e. we have assumed that the $\tau$'s do not depend on
$\Phi$. The \kahler{} term is also assumed renormalisable, but its
normalisation has also been constrained such that we obtained
$\Nscr=2$ supersymmetric invariance when disregarding the
superpotential. However, this normalisation is not important at
all even in the first version of a diagrammatic proof of the
Dijkgraaf-Vafa conjecture in~\cite{0211017}. Thus the only way the
restriction to $\Nscr=2$ supersymmetry plays a role is that the
matter field $\Phi$ is in the adjoint representation. However, the
$\Nscr=2$ supersymmetry was important in the string theory from
which the conjecture emerged. Naturally, it is not at all
important when we introduce the generalised conjecture in
section~\ref{dvsecgeneral}.

At last let us look at the superpotential~\eqref{dvn1}. This we
did not even constrain by renormalisability, however, it is not
the most general superpotential that we a priori could think of.
Rather, the most general superpotential is a multi-trace form:
\begin{equation}\label{dvn31.5}
    \lagr=\int\dtotha \sum_k\sum_{n_1,\ldots, n_k}
    g^{(k)}_{n_1,\ldots,n_k}\tr\big(\Phi^{n_1}\big)\cdots\tr\big(\Phi^{n_k}\big).
\end{equation}
The Dijkgraaf-Vafa conjecture does not apply immediately in this
case. One has to linearise the superpotential to a single trace
form by introducing auxiliary fields. The Dijkgraaf-Vafa
conjecture can then be used to obtain the effective action
including these auxiliary fields. One finally obtains the correct
effective superpotential by integrating out the auxiliary
fields~\cite{0212082},~\cite{0303074}.

One might think that the superpotential~\eqref{dvn1} would include
any superpotential of the form:
\begin{equation}\label{dvn32}
    \lagr=\int\dtotha\sum_{m,n}
    g_{m,n}\big(\Phitilde^0\big)^m\tr\big(\Phihat^n\big),
\end{equation}
i.e. a product of a trace over the abelian part (proportional to
$\Phitilde^0$) and a trace over the non-abelian part of $\un{N}$.
But expanding~\eqref{dvn1} gives (expanding the polynomial as
$\sum_n g_n\Phi^n$):
\begin{multline}\label{dvn33}
    \wtree\nleft(\Phi\right)=\tr\Big(\sum_n g_n\Big(\Phitilde+\Phihat\Big)^n\Big)=\tr\left(\sum_n g_n\sum_{m=0}^n\binom{n}{m}\Phitilde^m\Phihat^{n-m}\right)\\
    =\sum_ng_n\sum_{m=0}^\infty\binom{n}{m}\big(\Phitilde^0\big)^m\tr\big(\Phihat^{n-m}\big)=\sum_{m,n=0}^\infty
    g_{n+m}\binom{n+m}{m}\big(\Phitilde^0\big)^m\tr\big(\Phihat^{n}\big).
\end{multline}
Here we have used the binomial formula which applies since
$\Phitilde$ is abelian. We have also used that $\binom{n}{m}=0$
for integers with $m>n$. Thus we see that our superpotential is a
constrained form of~\eqref{dvn32}.

%-------------New section
\section{Supersymmetric Vacua}\label{dvsecvac}

As we saw in section~\ref{susysecsupermultiplets} the mass is a
Casimir of the supersymmetry algebra and hence both the fermionic
and bosonic particles in a supermultiplet have the same mass. This
is not observed in nature so supersymmetry is broken at our
energies. This makes the spontaneous breaking of supersymmetry to
a very important issue. In spite of this, we will focus our
interest on vacua which do not break supersymmetry spontaneously.
The treatment builds on~\cite{wessandbagger}, \cite{weinberg3},
\cite{0109172}, \cite{argyres}, \cite{9912271},
and~\cite{9701069}.

\subsection{Supersymmetric Vacua}
In section~\ref{susysechslthm} we chose the Hamiltonian of the
field theory realising the supersymmetry algebra to be positive.
That this is true can also be seen directly from~\eqref{susyn6}
using that the Pauli matrices have zero trace:
\begin{equation}\label{dvn34}
    H=\frac{1}{4}\left(\{\qgen^A_1,\qbargen_{A\dot{1}}\}+\{\qgen^A_2,\qbargen_{A\dot{2}}\}\right)\qquad(\textrm{no sum }
    A),
\end{equation}
for any $A$. Since the barred generators are the hermitian
conjugates, the expectation value of $H$ is clearly positive or
zero in any state. Actually, we see that the expectation value in
a vacuum (or any other state) is:
\begin{equation}\label{dvn35}
    \vacbra
    H\vacket=\frac{1}{4}\left(\norm{\qgen^A_1\vacket}^2+\norm{\qbargen_{A\dot{1}}\vacket}^2+\norm{\qgen^A_2\vacket}^2+\norm{\qbargen_{A\dot{2}}\vacket}^2\right).
\end{equation}
Thus we clearly see that \emph{the vacuum $\vacket$ is
supersymmetric if and only if the vacuum energy vanishes}. On the
other hand, supersymmetry is spontaneously broken if and only if
vacuum energy is strictly positive.\footnote{It is not possible to
simply shift the energy -- and thus making this observation
meaningless -- because the Hamiltonian $\Pgen^0$ is a part of the
supersymmetry algebra~\eqref{susyalgebra}.}.

Now let us assume that $\Nscr=1$. Another necessary and sufficient
criterion for spontaneous breaking of supersymmetry is that there
exist a field, $\psi$, with a non-zero supersymmetry variation in
the vacuum:
\begin{equation}\label{dvn36}
    \vacbra \de_{\xi}\psi\vacket\neq0.
\end{equation}
Using that $\de_{\xi}\psi=[i\xi\qgen+i\xibar\qbargen,\psi]$ we
immediately see that this can never be fulfilled if the
supercharges annihilate the vacuum which is the case if the vacuum
energy is zero. As we are dealing with a relativistic quantum
field theory, we assume that the Poincar\'e invariance is
manifest, i.e. the Poincar\'e generators annihilate the vacuum.
Thus only scalar fields that transform trivially under the Lorentz
group can have a non-zero expectation. Consequently,
$\de_{\xi}\psi$ must be a scalar field and hence $\psi$ is
fermionic. This fermion is called the Goldstino since it plays the
same role as the Goldstone boson. The Goldstino is also massless.

Let us look back at the superfield representation of the
supersymmetry algebra. In the case of a chiral superfield the
Goldstino must be the Weyl spinor $\psi$ from the component
expansion~\eqref{susyn35}. The supersymmetry variation of $\psi$
can be found using~\eqref{susyn26}. Using again that only fields
that transform trivially under the Lorentz group can have non-zero
expectation values, one then gets that supersymmetry is
spontaneously broken if the auxiliary field $F$ gets a non-zero
expectation value. This is known as F-term supersymmetry breaking.
Looking at the vector multiplet the Goldstino is the gaugino and
supersymmetry is spontaneously broken if the auxiliary field $D$
gets a non-zero expectation value. This is known as D-term
supersymmetry breaking.

We end this subsection by noting that when the supersymmetry is
spontaneously broken we no longer have equality of bosonic and
fermionic states. This is because the norm of $\qgen\vacket$
becomes infinite (with an infinite space and finite energy
density). Thus the proof in section~\ref{susysecwigner} no longer
works.

\subsection[Breaking and No-Breaking of Supersymmetry\ldots]{Breaking and No-Breaking of Supersymmetry in $\Nscr=1$
Supersymmetric Field Theories}\label{dvsecvacsubbreak}

Let us look at an $\Nscr=1$ supersymmetric field theory with a
compact gauge group and with chiral fields in some representation
$\reprn$. What we want to do here and in the rest of this section
is to consider the tree-level expansion of the theory i.e. the
semi-classical limit. The vacuum expectation values of the
component fields must be translationally invariant since the
vacuum does not break the Poincar\'e invariance. Thus the
expectation value of the scalar field $\phi$ (which we will denote
$\phi_0$) is independent of the space-time coordinate
$x$.\footnote{The space-time independence can also be obtained by
demanding the vacuum energy to vanish. After computing the
Hamiltonian one sees that this implies that the covariant
derivative of $\phi$ is zero; $D_\mu\phi=0$ and that
$F^a_{\mu\nu}=0$. The last equation means that the gauge field
$A_\mu$ is pure gauge and therefore we can set it to zero. Thus
the first equation reduces to $\partial_\mu\phi=0$.} The
expectation values of the rest of the fields, not transforming
trivially under the Poincar\'e group, vanish. Thus the only
contribution to the vacuum energy comes from the scalar potential.
The vacuum expectation value of $\phi$ in the semi-classical
approximation is the value that minimises the scalar potential
constrained to constant fields:
\begin{equation}\label{dvn36.5}
    \left.\frac{\partial U\rvert_{\textrm{constant fields}}}{\partial
    \phi}\right\rvert_{\phi=\phi_0}=0.
\end{equation}
This is, naturally, the semi-classical limit of the quantum
expectation value obtained by minimising the 1-PI effective
potential:
\begin{equation}\label{dvn37}
    \frac{\partial V_{\textrm{eff}}}{\partial
    \phi_{\mathrm{cl}}}=0.
\end{equation}

We will now find the scalar potential for the theory. For
simplicity we will assume that the gauge couplings $\tau_{ab}$
from equation~\eqref{susyn63} is independent of $\Phi$ and
$\tau_{ab}=\fourth\de_{ab}$. Furthermore, we assume that the
\kahler{} metric is invertible. We normalise the Fayet-Iliopoulos
term such that it is equal to $\ka_aD^a$ where $\ka_a$ only has
non-zero values in the abelian directions to ensure gauge
invariance. We have here also assumed that the inner-product on
the Lie algebra is diagonal. Combining all the terms
from~\eqref{susyn67} and expanding into components as in
section~\ref{dvsecdvunsublagr} yields the equations of motion for
the auxiliary fields:
\begin{eqnarray}\label{dvn38}
    F^i&=&-g^{i\ibar}\frac{\partial\overline{W\nleft(\phi\right)}}{\partial\phibar^{\ibar}},\\
    D^a&=&-\ka_a-\left(\frac{1}{2}\frac{\partial
    K\nleft(\phi,\phi^\dagger\right)}{\partial\phi^i}(T_a^{\repr})^i_{\ph{i}j}\phi^j
    +\cc\right).\label{dvn39}
\end{eqnarray}
Here the \kahler{} metric with upper indices is the inverse of the
\kahler{} metric with lower indices defined in~\eqref{susyn72}.
The scalar potential is then:
\begin{multline}\label{dvn40}
    U=\frac{1}{2}D^aD^a+g_{i\ibar}F^i\bar{F}^{\ibar}\\
    =\frac{1}{2}\sum_a\left(\ka_a+\left(\frac{1}{2}\frac{\partial
    K\nleft(\phi,\phi^\dagger\right)}{\partial\phi^i}(T_a^{\repr})^i_{\ph{i}j}\phi^j
    +\cc\right)\right)^2+g^{i\ibar}\frac{\partial
    W\nleft(\phi\right)}{\partial\phi^i}\frac{\partial\overline{W\nleft(\phi\right)}}{\partial\phibar^{\ibar}},
\end{multline}
where the first line is evaluated using~\eqref{dvn38}
and~\eqref{dvn39}. The \kahler{} metric is positive for unitarity
of the theory since it multiplies the kinetic energy
in~\eqref{susyn71}. Thus the potential energy is positive (or
zero) as promised.

We now see that if we can find a solution $\phi_0$ to $U=0$ then
we automatically have a global minimum and hence the vacuum
expectation value of $\phi$. But at the same time it is the
condition that supersymmetry is unbroken in this vacuum. It is
these supersymmetric vacua that we investigate in this thesis.

From~\eqref{dvn40} we immediately see that the condition of
finding vacua with unbroken supersymmetry translates into:
\begin{equation}\label{dvn41}
    F^i=0,\,D^a=0 \textrm{ have solution }\phi_0\textrm{ for all $i$ and $a$}\longleftrightarrow\textrm{Supersymmetric vacuum.}
\end{equation}
This is the same result we found in the last subsection when
giving the F-term and D-term condition for spontaneous
supersymmetry breaking.

If we for any constant field $\phi$ can find a $F^i$ that is
non-zero, we have F-term breaking of supersymmetry. The order
parameter will be $\expect{F}$. In the case where we do have
solutions to $F^i=0$ for all $i$ we talk about F-flatness.
From~\eqref{dvn38} these F-flatness equations can be rewritten as:
\begin{equation}\label{dvn42}
    \frac{\partial W\nleft(\phi\right)}{\partial\phi^i}=0,\qquad
    i=1,\ldots,\dim\nleft(\reprn\right).
\end{equation}
One can also obtain that F-term breaking generically only happens
when an R-symmetry is broken.

When we always are able to find a $D^a\neq0$ we have D-term
breaking. The order parameter is here $\expect{D}$. When we do
have solutions we talk about D-flatness. Using~\eqref{dvn39} the
D-flatness equations can be rewritten as:
\begin{equation}\label{dvn43}
    \ka_a+\frac{1}{2}\frac{\partial
    K\nleft(\phi,\phi^\dagger\right)}{\partial\phi^i}(T_a^{\repr})^i_{\ph{i}j}\phi^j
    +\frac{1}{2}\left(\phi^i\right)^*(T_a^{\repr})^i_{\ph{i}j}\left(\frac{\partial
    K\nleft(\phi,\phi^\dagger\right)}{\partial\phi^j}\right)^*=0,
\end{equation}
where we have assumed a unitary representation such that the
generators are hermitian. In many theories there will be no
solutions to these equations when the Fayet-Iliopoulos term is
non-zero.

On the other hand, we can always find solutions to the D-flatness
equations in the form~\eqref{dvn43} if we assume that $\ka_a=0$
(the proof here is taken from~\cite{9803099}). Simply take an
arbitrary vector $\phi_0$. We can then look at the surface
obtained by performing \emph{complex} gauge transformations on
this vector
$\phi_0\mapsto\phi\nleft(\La\right)=\exp\nleft(-i\La^a(x)T_a^{\repr}\right)\phi_0$
where $\La^a\in\C$. The renormalisable \kahler{} potential,
$\phi^\dagger\phi$, must have a minimum on this surface since it
is real and positive definite when $\phi$ runs through
$\phi(\La)$. Thus the general \kahler{} potential,
$K\nleft(\phi,\phi^\dagger\right)$, must have a local minimum if
we assume that it is the renormalisable \kahler{} potential
perturbed with some extra gauge invariant real terms. The only
thing that could spoil this is if the renormalisable potential has
some flat directions that the full \kahler{} potential does not
share thus making it possible to break the minimum. However, the
flat directions of $\phi^\dagger\phi$ correspond to performing the
standard real gauge transformations under which the full \kahler{}
potential is also invariant. The local minimum must be a
stationary point when varying $\La$:
\begin{equation}\label{dvn44}
    0=\frac{\partial
    K\nleft(\phi,\phi^\dagger\right)}{\partial\phi^{i*}}i\left(T_a^{\repr}\right)^j_{\ph{j}i}\phi^{j*}\de\La^{a*}-\frac{\partial
    K\nleft(\phi,\phi^\dagger\right)}{\partial\phi^{i}}i\left(T_a^{\repr}\right)^i_{\ph{i}j}\phi^{j}\de\La^{a}.
\end{equation}
The terms multiplying the variations $\de\La^a$ and $\de\La^{a*}$
must be zero. Hence the minimum point fulfil the D-flatness
equations~\eqref{dvn43} with $\ka_a=0$ thus finishing the proof.
This is the reason that we do not allow a Fayet-Iliopoulos term in
the Lagrangian of the Dijkgraaf-Vafa conjecture.

As a corollary in the case of $\ka_a=0$, we see that if there
exist a solution to the F-flatness equations~\eqref{dvn42} then
there must exist a solution $\phi_0$ satisfying both
F-flatness~\eqref{dvn42} and D-flatness~\eqref{dvn43} thus showing
the existence of a supersymmetric vacuum. This result simply
follows from the proof of the existence of a solution to the
D-flatness equations using that the superpotential is holomorphic
in $\phi$ and thus invariant under the whole complexified gauge
group. This means that in order to determine if supersymmetry is
unbroken (with no Fayet-Iliopoulos term) the necessary and
sufficient condition is the F-flatness equations~\eqref{dvn42}.

Please note that we have $\dim\nleft(\reprn\right)$ F-flatness
equations and $\dim\nleft(\cG\right)$ D-term equations in
$\dim\nleft(\reprn\right)$ variables $\phi^i$. However, the
F-flatness equations are constrained by gauge invariance of the
superpotential:
\begin{equation}\label{dvn45}
    \frac{\partial
    W\nleft(\phi\right)}{\partial\phi^i}\left(T_a^{\repr}\phi\right)^i=0,
\end{equation}
thus giving $\dim\nleft(\cG\right)$ constraints and hence we have
$\dim\nleft(\reprn\right)$ equations in $\dim\nleft(\reprn\right)$
variables. Generically, we will thus always have solutions,
however, this is not necessarily true when considering specific
theories.

\subsection{Classical Vacuum Moduli Space}\label{dvsecvacsubclmoduli}

The \emph{vacuum moduli space} is defined as the space of
supersymmetric inequivalent vacua. We know from the last section
that in the semi-classical limit this space is parameterised by
the solutions $\phi_0$ to the F- and D-flatness
equations~\eqref{dvn41}, however, we should here keep in mind that
the solutions should be inequivalent. This defines the vacuum
moduli space as a complex manifold and we can endow it with a
metric by pulling back the \kahler{} metric~\eqref{susyn72} from
the target space of $\phi$'s.

Now we must remember that all observables are independent of the
choice of gauge. Thus gauge transformations relate equivalent
vacua. Hence  the parametrisation of the moduli space is simply
obtained by the solutions to the flatness equations modulo gauge
transformations. We noticed in the last subsection that the orbit
of the complexified gauge group, $G_\C$, through any vector $\phi$
contains a solution to the D-flatness equations. Thus the space of
solutions to the D-flatness equations is simply $\{\phi^i\}/G_\C$
which can also be parameterised by holomorphic gauge invariants
modulo algebraic relations. That is there exist a set of
independent holomorphic gauge invariants $X_r\nleft(\phi\right)$
that parameterise the D-flatness space of solutions. The total
moduli space is then simply obtained by restricting these gauge
invariants by the F-flatness equations.

We will comment on the quantised vacuum moduli space in
section~\ref{dvsecweffsubnon}

\subsection[Classical Vacua for $\lagrun$]{Classical Vacua for
$\boldsymbol{\lagrun}$}\label{dvsecvacsublagrun}

Let us now turn to the Lagrangian of our interest namely
$\lagrun$. In~\eqref{dvn30} we found the scalar potential which
clearly is positive and vanishes if and only if the $F$'s and
$D$'s in~\eqref{dvn26}-\eqref{dvn29} are equal to zero. These
flatness equations actually have the same form as~\eqref{dvn42}
and~\eqref{dvn43} and thus the results above also apply in this
case.

Let us first focus on the D-flatness. From~\eqref{dvn26} we see
that there is no restriction on the abelian part, however, as the
abelian part of $\phi$ does not contribute to any commutator we
can reformulate the D-flatness equations obtained
from~\eqref{dvn26} and~\eqref{dvn27} as:
\begin{equation}\label{dvn46}
    [\phi^\dagger,\phi]=0.
\end{equation}
This is true for an arbitrary gauge group with adjoint matter as
can be seen from~\eqref{dvn43} by setting $K=\phi^\dagger\phi$,
$\ka^a=0$, and taking the generators in the adjoint
representation. We immediately see that we get a solution by
requiring $\phi$ to be in a Cartan subalgebra of the gauge group.
For the unitary and symplectic gauge groups this gives all
possible solutions, however, for $\on{N}$ with $N\geq7$ there can
be more general solutions~\cite[Appendix I]{9712028}.

In our case the gauge group is $\un{N}$. Thus $\phi$ is
$\cun{N}$-valued, however, with complex coefficients since the
$\phi^a$'s are complex fields. Let us assume that $\phi$ is a
solution to the D-flatness equation~\eqref{dvn46}. Splitting the
$\phi^a$'s in their real and imaginary part splits the matrix
$\phi$ into its hermitian and anti-hermitian part
$\phi=\phi_1+i\phi_2$ where both $\phi_1$ and $\phi_2$ are
hermitian. Since $\phi_1$ and $\phi_2$ are hermitian they can be
diagonalised by a unitary matrix. Using~\eqref{dvn46} we see that
$\phi_1$ and $\phi_2$ commute and thus can be diagonalised by the
same unitary matrix. Thus $\phi$ is diagonalised by this unitary
matrix. Since the matrix is unitary and $\phi$ transforms
adjointly, this precisely corresponds to a gauge transformation.
Thus, by a suitable gauge transformation we can choose any
solution to~\eqref{dvn46} to be in the Cartan subalgebra of the
diagonal matrices (we could, naturally, have chosen any other
Cartan subalgebra).

The choice of diagonal matrix is in general not unique. Naturally,
adjoint transformations, $\phi\mapsto U\phi U^{-1}$, preserve the
eigenvalue spectrum, but the eigenvalues in the diagonal form can
be permuted. This is the action of the \emph{Weyl group}. For a
general group and a specific choice of a Cartan subalgebra, the
Weyl group is the subgroup of the gauge group that permutes the
generators of the Cartan subalgebra. It thus transforms an element
in the Cartan subalgebra back into the Cartan subalgebra. This
group is finite. In our $\un{N}$-case of diagonal matrices the
Weyl group simply permutes the axes. Since the Weyl
transformations are gauge transformations, they relate physically
equivalent vacua and the parametrisation of the vacuum moduli
space should be independent hereof.

The gauge invariant parametrisation (and thus also Weyl group
invariant) of the vacuum moduli space is easily obtained in this
case. We simply note that the characteristic polynomial
$\det\nleft(\la-\phi\right)$ is invariant under adjoint
transformations. Hence the coefficients must be gauge invariant.
We obtain these by expanding:
\begin{multline}\label{dvn47}
    \det\nleft(\la-\phi\right)=\la^N\det\nleft(\idmatr-\frac{\phi}{\la}\right)=\la^N
    e^{\tr\ln\left(\idmatr-\frac{\phi}{\la}\right)}=\la^N \exp\nleft(-\sum_{n=1}^{\infty}\frac{\tr\nleft(\phi^n\right)}{n\la^n}\right)\\
    =\la^N-\tr{\phi}\la^{N-1}-\frac{1}{2}\left(\tr\nleft(\phi^2\right)-\tr\nleft(\phi\right)\tr\nleft(\phi\right)\right)\la^{N-2}-\ldots
\end{multline}
As we can guess from the expansion (which only holds true for
$\la\gg\phi$), the holomorphic independent gauge invariants are
$\tr{\phi^n}$ with $n=1,\ldots,N$. Please note that there can only
be $N$ parameters since we only have $N$ eigenvalues of $\phi$.
This also holds true for $\sun{N}$ with the exception that
$\tr\nleft(\phi\right)=0$.

Let us now turn to the F-flatness equations.\footnote{In the rest
of this section we will be very thorough since we have not found
any of the following proofs in the literature.} Using
equations~\eqref{dvn28} and~\eqref{dvn29} these can be written as:
\begin{equation}\label{dvn48}
    \tr\nleft(T_{a}^{(\mathrm{fund})}\wpoly'\nleft(\phi\right)\right)=0,\qquad
    a=0,\ldots,
    N^2-1.
\end{equation}
Generally for a Lie algebra, a product of generators can not be
expressed as a sum of generators. Thus, generally, a polynomial of
matrices $\phi$ can not be expanded in
$T_{a}^{(\mathrm{fund})}$'s. However, in our $\un{N}$ case the
complexified span of the $T_{a}^{(\mathrm{fund})}$'s gives the
whole set of complex matrices i.e. the Lie algebra corresponding
to the group of invertible complex matrices, $\glnc{N}$. Thus
$\wpoly'\nleft(\phi\right)$ can be expressed as a complex linear
combination of $T_{a}^{(\mathrm{fund})}$'s. Using~\eqref{dvn19.6}
we then get:\footnote{Another way to get this is to realise that
by~(\ref{dvn48}) the trace of $\wpoly'$ with any matrix vanishes
thus $\wpoly'$ vanishes.}
\begin{equation}\label{dvn49}
    \wpoly'\nleft(\phi\right)=0.
\end{equation}
We note that this would not be true if we had only looked at
$\sun{N}$. In that case we would get
$\wpoly'-\frac{1}{N}\tr\nleft(\wpoly'\right)\idmatr=0$.

If we now assume $\phi$ to be diagonal to solve the D-flatness
equation,~\eqref{dvn49} simply splits into $N$ equations for the
eigenvalues $\phi_{ii}$:
\begin{equation}\label{dvn50}
    \wpoly'\nleft(\phi_{ii}\right)=0,\qquad
    i=1,\ldots,N.
\end{equation}
Thus we obtain the classical vacuum expectation values, $\phi_0$,
by constraining the eigenvalues to be in the set of roots of
$\wpoly'$, i.e. the critical points of $\wpoly$ as promised in
section~\ref{dvsecdvun}. This gives the final constraints to
obtain the vacuum moduli space.

Let us now think of a specific point in the vacuum moduli space.
This is determined by a diagonal matrix $\phi_0$. Each of the
eigenvalues must be equal to one of the critical points $a_i$ of
$\wpoly$. However, the placement of the eigenvalues are
unimportant since the Weyl group relate equivalent vacua. Hence
the vacuum is simply specified by the partition of $N$
in~\eqref{dvn2}. We can simply think of putting the
$a_1$-eigenvalues first in $\phi_0$, then the $a_2$-eigenvalues
and so on. The unbroken gauge group consist of the matrices of
$\un{N}$ that have $\phi_0$ as a fixed point under adjoint
transformations. We note that this is a subgroup of the full gauge
group. The generators of the unbroken group are determined as
corresponding to the infinitesimal $\La^a$'s with
$[\La^aT_{a},\phi_0]=0$. Let us choose a basis for the fundamental
representation of $\un{N}$ as $\mathbf{D}_i$, $\mathbf{A}_{ij}$
and $\mathbf{B}_{ij}$ where $i<j$ and:
\begin{eqnarray}\label{dvn51}
    \left(\mathbf{D}_i\right)_{kl}&=&\de_{ki}\de_{li},\nonumber\\
    \left(\mathbf{A}_{ij}\right)_{kl}&=&\de_{ki}\de_{l
    j}+\de_{kj}\de_{li},\nonumber\\
    \left(\mathbf{B}_{ij}\right)_{kl}&=&-i\de_{ki}\de_{l
    j}+i\de_{kj}\de_{li}.
\end{eqnarray}
The $\mathbf{D}_i$'s span the Cartan subalgebra of the diagonal
matrices and are all in the unbroken subgroup. Using that $\phi_0$
is diagonal we get that
\begin{equation}\label{dvn52}
    \left([a\mathbf{A}_{ij}+b\mathbf{B}_{ij},\phi_0]\right)_{kl}=(a-ib)\left((\phi_0)_{jj}-(\phi_0)_{ii}\right)\de_{ki}\de_{l
    j}+(a+ib)\left((\phi_0)_{ii}-(\phi_0)_{jj}\right)\de_{kj}\de_{li},
\end{equation}
where we have no sums. We thus see that the commutator on the left
hand side only has non-zero indices at $(i,j)$ and $(j,i)$. Thus
the condition $[\La^aT_{a},\phi_0]=0$ splits into
$[a\mathbf{A}_{ij}+b\mathbf{B}_{ij},\phi_0]=0$ for all $i<j$. For
$a$ or $b$ non-zero we see that this is only possible if
$(\phi_0)_{ii}=(\phi_0)_{jj}$ which in turn allow both $a$ and $b$
to be non-zero. This proves that the partition~\eqref{dvn2} has an
unbroken gauge group $\un{N_1}\times\ldots\times\un{N_n}$ as
in~\eqref{dvn3}.

Given a vacuum determined by $\phi_0$ and the rest of the fields
having expectation value zero, we should expand the Lagrangian
around these expectation values:
\begin{equation}\label{dvn52.5}
    \lagr_\si\nleft(\si\right)\defi\lagrun\nleft(\si+\phi_0\right),
\end{equation}
where the rest of the fields are unchanged. We know that
supersymmetry should be unbroken, but that some gauge symmetries,
generally, are broken. Above we found a nice basis $\{T_a\}$ of
generators for the fundamental representation of $\un{N}$ in which
the generators of the unbroken subgroup merely is a subset -- no
need for taking linear combinations. We can easily obtain that in
this basis, the metric defined by $\tr\nleft(T_aT_b\right)$ is
also diagonal and positive definite. Let us work in this basis.
The expansion of the Lagrangian must be done by Taylor expansion
and even in the simple case of getting the mass terms this is
quite elaborate. We will not do the calculation of the masses
here, but refer to~\cite{weinberg3} for such a calculation. The
result is that we will have three types of masses: We  will have
zero masses corresponding to the unbroken gauge multiplets.
Secondly, we will have strictly non-zero masses determined by the
square-root of the non-zero eigenvalues of the matrix with the
$(a,b)$ entry given by:\footnote{As this mass matrix arises from
the term
$-\frac{1}{g^2}\tr\nleft([\phihat^\dagger,\phihat]^2\right)$ we
really should divide by $1/g^2$ to get the masses.}
\begin{equation}\label{dvn53}
    \phi_0^\dagger\{T_a^{(\mathrm{adj})},T_b^{(\mathrm{adj})}\}\phi_0,
\end{equation}
where $T_a^{(\mathrm{adj})}$ is the generator in the adjoint
representation. We note that this entry is zero if
$T_a^{(\mathrm{adj})}$ or $T_b^{(\mathrm{adj})}$ is unbroken and
it is positive when restricted to the broken subspace, i.e. where
$a$ and $b$ are the indices of broken generators. Thus the number
of eigenvectors with positive eigenvalues equals the number of
broken gauge symmetries. These will be the masses of the massive
gauge multiplets that arise by the supersymmetric version of the
Higgs mechanism. At last we will have masses corresponding to the
eigenvalues of the mass matrix:
\begin{equation}\label{dvn54}
    M_{ab}=\left.\frac{\partial^2W\nleft(\phi\right)}{\partial\phi^a\partial\phi^b}\right\rvert_{\phi=\phi_0}.
\end{equation}
These complex masses are the masses of the chiral multiplets in a
representation of the unbroken gauge group. Naturally, the masses
of the components will be real positive. As we see
from~\eqref{dvn30} by expanding around the vacuum, the masses of
the scalars (and thus of the multiplet) are the absolute values of
the eigenvalues of $M_{ab}$ if we disregard the gauge coupling
constants in~\eqref{dvn30} (which would not be there if we had
defined the superpotential as in footnote~\ref{dvfootsuper}).

To be a bit more precise, we have $N_\mathrm{B}$ broken gauge
symmetries with $[T_a,\phi_0]\neq0$. As we saw above these
$[T_a,\phi_0]$ are linearly independent and must span the whole
space of broken symmetries since for any unbroken symmetry
$T_{b,\mathrm{unb}}$ we have:
\begin{equation}\label{dvn55}
    \tr\nleft(T_{b,\mathrm{unb}}[T_a,\phi_0]\right)=\tr\nleft(T_a[\phi_0,T_{b,\mathrm{unb}}]\right)=0,
\end{equation}
since $[T_{b,\mathrm{unb}},\phi_0]=0$. Thus $[T_a,\phi_0]$ is
orthogonal to the space of unbroken symmetries and thus in the
$N_\mathrm{B}$-dimensional space of broken symmetries. In this
space we find $N_\mathrm{B}$ complex scalars of which we find
$N_\mathrm{B}$ real scalars of zero mass. These are the Goldstone
bosons which can be removed by choosing unitary gauge (``eaten'').
The remaining $N_\mathrm{B}$ scalars will have the positive masses
determined by~\eqref{dvn53}. In the same directions we find two
Weyl spinors with the same mass namely the gaugino and the spinor
from the chiral multiplet. With this positive mass we also find a
gauge boson. This makes up the massive gauge multiplet that we
found in section~\ref{susysecmassive}.

As we saw above the orthogonal space to the non-zero
$[T_a,\phi_0]$'s is the unbroken subspace. In these directions we
find massless gauge multiplets and chiral multiplets. The masses
of the chiral multiplets are determined by the mass matrix
in~\eqref{dvn54} restricted to the unbroken subspace. We will end
this section by proving that all the eigenvalues of this matrix,
and hence the masses, are non-zero if the roots of $\wpoly'$ are
different from each other as was claimed above in
section~\ref{dvsecdvunsubdv}. As in~\eqref{dvn25} we get
\begin{equation}\label{dvn56}
    M_{ab}=\left.\frac{\partial}{\partial\phi^b}\tr\nleft(T_a\wpoly'\nleft(\phi\right)\right)\right\rvert_{\phi=\phi_0}=\tr\nleft(T_aT_b\wpoly''\nleft(\phi_0\right)\right),
\end{equation}
where we have used that $T_b$ is unbroken and thus commutes with
$\phi_0$. This made it possible to move $T_b$ to the front. We
note that it here was crucial that we only look at the unbroken
directions because, as noted above, this could not have been done
using the cyclicity of the trace. We immediately see that we must
assume $\wpoly$ to be non-trivial i.e. of degree two or greater.
Now let $v^a$ by an arbitrary vector in the unbroken subspace and
let us assume that:
\begin{equation}\label{dvn57}
    M_{ab}v^b=\tr\nleft(T_a\phi\wpoly''\nleft(\phi_0\right)\right)=0,
\end{equation}
where $\phi=T_bv^b$, and $a$ and $b$ are in the unbroken space.
Since $\phi_0$ is diagonal, $\wpoly''\nleft(\phi_0\right)$ is
diagonal and each diagonal entry is equal to
$\wpoly''\nleft((\phi_0)_{ii}\right)$. This is non-zero by the
assumption that the roots of $\wpoly'$ are all different and that
$(\phi_0)_{ii}$ is one of the roots. This is easily seen by
differentiating and evaluating in one of the roots. What we should
now realise is that since $T_a$ is in the unbroken subspace
\begin{equation}\label{dvn58}
    T_a\wpoly''\nleft(\phi_0\right)=T_a\wpoly''\nleft((\phi_0)_{kk}\right),
\end{equation}
for some suitable $k$ depending on $a$. This we see by checking it
for all the three types of generators given in~\eqref{dvn51}.
Since $\wpoly''\nleft(\phi_0\right)$ is diagonal it can be
expanded in the $\mathbf{D}_i$'s. Thus equation~\eqref{dvn58} is
obviously fulfilled for $T_a=\mathbf{D}_i$ which are all in the
unbroken subspace; here $k=i$. If $T_a=\mathbf{A}_{ij}$ then
by~\eqref{dvn52} $(\phi_0)_{ii}=(\phi_0)_{jj}$ and thus
$\left(\wpoly''\nleft(\phi_0\right)\right)_{ii}=\left(\wpoly''\nleft(\phi_0\right)\right)_{jj}$.
Using this we easily see by calculation that
$\mathbf{A}_{ij}\wpoly''\nleft(\phi_0\right)=\mathbf{A}_{ij}\left(\wpoly''\nleft(\phi_0\right)\right)_{ii}$
thus realising~\eqref{dvn58} with $k=i$. The same holds true for
$\mathbf{B}_{ij}$ thus finishing the proof~\eqref{dvn58}. Using
cyclicity of the trace and that $T_a$ commutes with $\phi_0$, we
can rewrite~\eqref{dvn57} as
\begin{equation}\label{dvn59}
    \tr\nleft(\phi T_a\wpoly''\nleft(\phi_0\right)\right)=\tr\nleft(\phi
    T_a\right)\wpoly''\nleft((\phi_0)_{kk}\right)=0.
\end{equation}
Using that $\wpoly''\nleft((\phi_0)_{kk}\right)\neq0$, we conclude
that $\tr\nleft(\phi T_a\right)=0$ for all generators $T_a$ of the
unbroken subgroup. Using the invertibility of the metric we get
that $\phi=0$ thus concluding that we have no zero-eigenvalue
vector and hence no zero mass. This finishes the proof.

What we should do now is to quantise the theory we have obtained
thus defining a perturbation theory for the fluctuations around
this vacuum. The massive gauge multiplets should be integrated out
since they are not in the focus of our interest. The final theory
will have a very complex structure and thus we will mostly deal
with the case of unbroken gauge symmetry. This happens when all
the eigenvalues in $\phi_0$ are chosen to be the same. If we can
choose all the eigenvalues to be zero the Lagrangian $\lagr_\si$
from~\eqref{dvn52.5} simply is $\lagrun$.

%----------New section

\section{Quantised Theory}\label{dvsecquant}

In the last section we obtained the Lagrangian, which we want to
quantise, by expanding around the expectation values of the chosen
vacuum. We can expand this Lagrangian into components i.e.
scalars, spinors and gauge bosons. These we already know how to
quantise using the path integral technique. However, instead of
quantising the component fields it is possible directly to
quantise the superfields thus making the supersymmetry manifest.
We note, however, that even though our Lagrangian is classically
supersymmetric since we chose a supersymmetric vacuum, we do not
know a priori if supersymmetry is broken in the quantised case. We
will discuss this issue later. We will not go into details in this
section, but just give a short review of the subject.

\subsection{Supergraphs}\label{dvsecquantsubsuper}
Our goal is to develop a perturbation theory for the quantum
fluctuations around the chosen vacuum. As usual, the important
objects to calculate are the $n$-point Green's functions. From a
superspace point of view we want the $n$-point Green's functions
of the superfields. From these one can then obtain the usual
Green's functions by expanding in components as
in~\eqref{susyn33}. The tool to calculate the Green's functions is
as usual the Feynman graphs which in this case are called
\emph{supergraphs}. We will focus on the case where the Lagrangian
only contains chiral fields. Assuming renormalisability, the
Lagrangian is given by~\eqref{susyn42}.

The first thing to do when developing the supergraphs is to find
the propagators. But before finding these we should realise that
we have a problem: The chiral superfield is a constrained
superfield by the chirality condition~\eqref{susyn34} (and the
analogous for the anti-chiral field $\Phibar$). Thus we have a
problem in defining a path integral with integrations over
unconstrained superfields.

Another problem is that when we deal with supersymmetry the points
should be superspace points. But the fermionic integral in the
superpotential term is only over half the superspace. So we have
to do something to put this term in the same form as the \kahler{}
term with a four-dimensional $\tha$-integration.

The propagators that we are looking for are
$\expect{T\Phi\nleft(x,\tha,\thabar\right)\Phibar\nleft(x',\tha',\thabar'\right)}_0$
and the corresponding with $\Phi\Phi$ or $\Phibar\Phibar$. Here
the vacuum expectation values should be found using the
non-interacting Lagrangian, $\lagr_0$, and $T$ is the
time-ordering operator. There are several ways to find these
propagators. Firstly we could simply expand the Lagrangian in
component fields where we know how to do the calculations thus
evading all problems. A second supersymmetric approach used
by~\cite{wessandbagger} imposes the constraint of chirality using
the projection operator onto chiral fields defined by:
\begin{equation}\label{dvn60}
    \projchiral=\frac{1}{16}\frac{\Dbar\Dbar DD}{\square},
\end{equation}
which is well-defined since $\square=\partial_\mu\partial^\mu$
commutes with the covariant derivatives. We immediately see that
$\projchiral\Psi$ is chiral for any superfield $\Psi$ since
$\Dbar_{\aldot}\Dbar_{\bedot}\Dbar_{\gadot}=0$. That it is really
a projection onto the chiral fields then follows since for any
chiral field $\Phi$ we get by simple calculation
using~\eqref{susyn32},~\eqref{susyn62.5}
and~\eqref{appspinorn38.5}:
\begin{equation}\label{dvn61}
    \frac{1}{16}\frac{\Dbar\Dbar DD}{\square}\Phi=\Phi.
\end{equation}
This we could also have stated as the usual projection condition
$\projchiral^2=\projchiral$. Naturally, the corresponding
projection onto anti-chiral fields
\begin{equation}\label{dvn61.5}
    \projantichiral=\frac{1}{16}\frac{DD\Dbar\Dbar}{\square}
\end{equation}
is also needed. These projections also solve the problem that the
superpotential terms only have integrations over half the
superspace. Using~\eqref{susyn38} and~\eqref{dvn61} e.g. the mass
term can be rewritten as:
\begin{equation}\label{dvn62}
    \int\dlor\dtotha\frac{1}{2}m\Phi^2=-\frac{m}{8}\int\dlor\dsuper \Phi \frac{DD}{\square}\Phi.
\end{equation}
At last, the chirality of the fields also has to be imposed when
varying a chiral field. Using that the chiral field $\Phi$ only
can depend on $x_+$ and $\tha$ as in~\eqref{susyn35}, this boils
down to the rule:
\begin{equation}\label{dvn63}
    \frac{\de}{\de\Phi\nleft(x,\tha,\thabar\right)}\Phi\nleft(x',\tha',\thabar'\right)=-\frac{1}{4}\Dbar\Dbar\deltafunkto\nleft(\tha-\tha'\right)\deltafunkto\nleft(\thabar-\thabar'\right)\deltafunkfire\nleft(x-x'\right).
\end{equation}
Now we are able to find the equations of motions and after some
calculations the propagators will be (in Minkowski
space):\footnote{Here taken from~\cite{wessandbagger}. We note
that it is tacitly assumed that the mass $m$ is real.}
\begin{multline}\label{dvn64}
    \expect{T\begin{pmatrix}
      \Phi\nleft(x,\tha,\thabar\right)\Phi\nleft(x',\tha',\thabar'\right) & \Phi\nleft(x,\tha,\thabar\right)\Phibar\nleft(x',\tha',\thabar'\right) \\
      \Phibar\nleft(x,\tha,\thabar\right)\Phi\nleft(x',\tha',\thabar'\right) & \Phibar\nleft(x,\tha,\thabar\right)\Phibar\nleft(x',\tha',\thabar'\right) \\
    \end{pmatrix}}_0=\\
    \frac{i}{\square-m^2}\begin{pmatrix}
      \frac{m}{4}\Dbar\Dbar & \frac{1}{16}\Dbar\Dbar D D \\
      \frac{1}{16}D D \Dbar\Dbar & \frac{m}{4}D D \\
    \end{pmatrix}\deltafunkfire\nleft(x-x'\right)\deltafunkto\nleft(\tha-\tha'\right)\deltafunkto\nleft(\thabar-\thabar'\right).
\end{multline}
A third way (and the last that we will present) to obtain the
propagators is to introduce potential superfields as is done
in~\cite{weinberg3}. This is in analogy with the well-known
problem of the gauge field strength that is constrained by the
homogeneous Maxwell equations forcing us to introduce the
unconstrained gauge potential. In this case we introduce the
unconstrained superfields $\Pi^i$ defined from the chiral fields
$\Phi^i$ as:
\begin{equation}\label{dvn65}
    \Phi^i=\Dbar\Dbar \Pi^i.
\end{equation}
As we see from~\eqref{dvn61} we can always find such a field, but
it need not be a local field due to the $\square^{-1}$. This would
in turn also have meant that any chiral field would be chirally
exact as discussed below equation~\eqref{susyn38} -- this is not
the case. We note that analogous to~\eqref{susyn29.2} $\Phibar^i=D
D \bar{\Pi}^i$. The Lagrangian~\eqref{susyn42} now becomes:
\begin{equation}\label{dvn66}
    \lagr=\int\dsuper\bar{\Pi}^iD D\Dbar\Dbar \Pi^i-4\int\dsuper
    \left(\tilde{W}\nleft(\Pi^i\right)+\cc\right),
\end{equation}
where we have used that we can always integrate $D$ and $\Dbar$ by
parts under the four-dimensional superspace integral since the
superspace integral of a super-covariant derivative is zero by
\eqref{susyn37}. $\tilde{W}$ is defined as $W\nleft(\Dbar\Dbar
\Pi^i\right)$ where one pair of $\Dbar\Dbar$ has been removed in
each term when using~\eqref{susyn38} to change the half superspace
integral to the full superspace integral. But we still have a
problem in defining the propagator since the Lagrangian by the
definition of $\Pi^i$ clearly is invariant under the
transformations:
\begin{equation}\label{dvn67}
    \Pi^i\mapsto \Pi^i+\Dbar F,
\end{equation}
where $F$ is any superfield. This is in analogy with the gauge
transformation of the gauge potential. However, there is no need
for introducing Faddeev-Popov ghosts here to define the path
integral since all Green's functions that we want to determine are
invariant under~\eqref{dvn67}. The solution is simply to project
onto the space orthogonal to the zero-eigenvalue vector
in~\eqref{dvn67} when determining the propagator. This projection
is simply the anti-chiral projection~\eqref{dvn61.5} since
$\projantichiral \Dbar F=0$. Thinking of the mass term as an
interaction term, the propagator for $\Pi^i\bar{\Pi}^j$ is then
by~\eqref{dvn66}:
\begin{equation}\label{dvn68}
   -i D
    D\Dbar\Dbar\Delta_{ij}\nleft(x,\tha,\thabar;x',\tha',\thabar'\right)=\projantichiral\deltafunkfire\nleft(x-x'\right)\deltafunkto\nleft(\tha-\tha'\right)\deltafunkto\nleft(\thabar-\thabar'\right)\de_{ij}.
\end{equation}
Here there can be some unspecified terms of the form $\Dbar F$
that adds to the right hand side. But these are, as explained,
unimportant. Using the definition of the anti-chiral
projection~\eqref{dvn68} immediately solves as
\begin{equation}\label{dvn69}
    \Delta_{ij}\nleft(x,\tha,\thabar;x',\tha',\thabar'\right)=\frac{i}{16\square}\deltafunkfire\nleft(x-x'\right)\deltafunkto\nleft(\tha-\tha'\right)\deltafunkto\nleft(\thabar-\thabar'\right)\de_{ij}.
\end{equation}
Inserting the $\Dbar \Dbar$ and $D D$ from the definition of the
potential superfields we get the propagator for the chiral fields.
The result is the same as~\eqref{dvn64} with $m=0$.

\subsection{Non-Renormalisation}\label{dvsecquantsubnonrenorm}
Using the method of potential superfields the Feynman rules for
the vertices can immediately be obtained since the interactions
from~\eqref{dvn66} simply are given by $\tilde{W}$ and its complex
conjugate. We can use the supergraphs to find the 1PI-effective
action here following~\cite{wessandbagger} and~\cite{weinberg3}.
For a general diagram contributing to the 1-PI effective action we
see from~\eqref{dvn69} that we get a four-dimensional delta
function in $\tha$ from each propagator. These delta functions are
acted on by super-covariant derivatives from the Feynman rules of
the vertices. We also have full superspace integrations for each
vertex which means that we can always integrate the
super-covariant derivatives by parts. For a given delta function
we can then remove the super-covariant derivatives using
integration by parts and then perform the corresponding
four-dimensional $\tha$-integration. Using this method we can
eliminate all $\tha$-integrations but one by the delta functions.
All super-covariant derivatives then act on the external fields
and all delta functions have been removed.\footnote{When
considering the details one needs to use that
$\delta\nleft(\tha\right)=\tha$. Thus $\delta\nleft(0\right)=0$
and $\delta\nleft(\tha\right)\delta\nleft(\tha\right)=0$.} A
simple counting of vertices, internal and external lines shows
that the super-covariant derivatives can be used to change the
external $\Pi^i$'s back to $\Phi$'s except in the case of
tree-level diagrams. Thus the effective action can be written:
\begin{equation}\label{dvn70}
    \int\dsuper\int\dlor_1\ldots\dlor_n
    F_1\nleft(x_1,\tha,\thabar\right)\ldots
    F_n\nleft(x_n,\tha,\thabar\right)G\nleft(x_1,\ldots,x_n\right)+\textrm{tree-level diagrams}.
\end{equation}
Here the $F_i$'s only depend on the external chiral fields and the
super-covariant derivatives, and $G$ is just some translationally
invariant function. We see that contributions to the
superpotential term, which only has integration over half the
superspace, can only come from the tree-level diagrams which in
the 1-PI case are the simple vertices. We have thus reached the
important result that the superpotential is not renormalised
perturbatively and no new terms are introduced. Thus only the
\kahler{} term is renormalised.

There is, however, one flaw in this proof and that is when we
consider the $\square^{-1}$ in the propagator~\eqref{dvn69}. If we
include the mass term in the propagator, the $\square^{-1}$
changes into $1/\left(\square-m_i^2\right)$ where we have assumed
that the mass term is diagonal giving $\Phi^i$ mass $m_i$.
Naturally, with the $\deltafunkfire\nleft(x-x'\right)$ this just
gives the usual Feynman propagator. With the mass we also have
$\Pi\Pi$ and $\bar{\Pi}\bar{\Pi}$ propagators that also have the
Feynman propagator as a factor.\footnote{A remark is in order for
consistency with the second way of obtaining the propagators that
led to~(\ref{dvn64}). When doing diagrams we should not use the
propagators in~(\ref{dvn64}). The reason is that we also have to
use the formula~(\ref{dvn63}) for varying the external chiral
currents one introduces to develop the Feynman graphs. Taking this
into consideration one gets the Grisaru-Ro\v{c}ek-Siegel
propagator which really just is the same as the propagator for the
potential superfields.} Now if we have a zero mass field, we could
have a 1PI-diagram contributing to the first term in~\eqref{dvn70}
of the form:
\begin{equation}\label{dvn71}
    \int\dsuper\frac{DD}{\square}f\nleft(\Phi^i\right)=-4\int\dtotha
    f\nleft(\Phi^i\right),
\end{equation}
where $f$ is a function of the chiral fields $\Phi^i$, and we
used~\eqref{susyn38} and~\eqref{dvn61} as in~\eqref{dvn62}. This
clearly gives a change in the superpotential. The first explicit
example of such a diagram was given in~\cite{westnotnorenorm} as a
two loop diagram. In the article they actually do the calculations
in components, but it is noted that the corresponding supergraph
yields the same result. It is important to realise that such a
contribution arose because we had a massless field and we thus
have an IR divergency in the D-term. The IR divergency stems from
the propagator $\square^{-1}$ which can bring about an integration
$\int_0\mathrm{d}k\frac{1}{k}$ which is logarithmically divergent.
Actually, all loop contributions to the F-term from 1PI diagrams
come from IR divergent D-terms \cite{9704114}.

We will return to the subject of such non-renormalisation theorems
in section~\ref{dvsecweffsubnon}. However, there is another way to
obtain effective actions without the above described problems. We
will introduce this method in the next section.

%----------New section------------

\section{Wilsonian Renormalisation}\label{dvsecwilson}

Suppose we are interested in the dynamics of the quantised theory
at energies below some energy-momentum cut-off $\mu$. All the
physics is then captured by the Wilsonian effective action that we
will introduce in this section. The treatment is based
on~\cite{argyres},~\cite{9912271},~\cite{weinberg1}
and~\cite{peskinandschroeder}.

\subsection{Wilsonian Effective Action}

Our starting point is a quantum field theory regularised by an UV
cut-off $\La$. Naturally, such a sharp momentum cut-off is not
preserved by gauge symmetry, but let us not worry about that here.
Another problem is that the condition $k^\mu k_\mu<\La^2$ in
Minkowski space does not ensure that each 4-momentum coordinate is
bounded -- thus we rotate to Euclidean space. For concreteness let
us think of a scalar field theory with Lagrangian $\lagr_\La$. The
generating functional is then given by (setting the external
current to zero for simplicity):
\begin{equation}\label{dvn72}
    Z=\int\left[\cD\phi\right]_\La
    e^{-\int\dlor\lagr_\La(\phi)},
\end{equation}
where the scalar field only has non-zero momentum modes for
$\norm{k}<\La$, and the functional integration is defined as
$\left[\cD\phi\right]_\La=\prod_{\norm{k}<\La}\textrm{d}\phi\nleft(k\right)$.
To obtain an action that only depends on the energy-momentum below
the cut-off $\mu$, we want to integrate out $\phi\nleft(k\right)$
with $\mu\leq\norm{k}<\La$. To do this we split the scalar field
as $\phi=\phitilde+\phihat$ where
\begin{equation}\label{dvn72.5}
    \phitilde\nleft(k\right)=\left\{ \begin{array}{ll}
    \phi\nleft(k\right) & \textrm{if $\norm{k}<\mu$}\\
    0 & \textrm{if $\mu\leq \norm{k}<\La$} \end{array}\right.
    ,\qquad
    \phihat\nleft(k\right)=\left\{ \begin{array}{ll}
    0 & \textrm{if $\norm{k}<\mu$}\\
    \phi\nleft(k\right) & \textrm{if $\mu\leq \norm{k}<\La$}
    \end{array}\right..
\end{equation}
Now we can split the Lagrangian into the original Lagrangian
evaluated in $\phitilde$ and a mixed term:
\begin{equation}\label{dvn73}
    \lagr_\La\big(\phitilde+\phihat\big)=\lagr_\La\big(\phitilde\big)+\lagr_{\mathrm{mixed}}\big(\phitilde,\phihat\big),
\end{equation}
where there are no terms only depending on $\phitilde$ in the
mixed Lagrangian. Since $\phitilde$ and $\phihat$ are orthogonal
in momentum space, quadratic terms of the form $\phi\phihat$ are
zero. Thus the kinetic term and the mass term in
$\lagr_{\mathrm{mixed}}$ only depend on $\phihat$ making this a
Lagrangian in $\phihat$ with $\phitilde$ as an external field. We
can now integrate the $\phihat$ field out:
\begin{equation}\label{dvn74}
    Z=\int\Big[\cD\phitilde\Big]_\La
    e^{-\int\dlor\lagr_\La\left(\phitilde\right)}\int\left[\cD\phihat\right]_\La
    e^{-\int\dlor\lagr_{\mathrm{mixed}}\left(\phitilde,\phihat\right)}=\int\left[\cD\phitilde\right]_\mu
    e^{-\int\dlor\lagr_\mu\left(\phitilde\right)},
\end{equation}
where the functional integrations only are over the non-zero
modes. $\lagr_\mu$ is the \emph{Wilsonian effective Lagrangian} at
the energy $\mu$. For calculations of processes with energies and
momenta less than $\mu$ it gives precisely the same result as
using the old Lagrangian, however, now it is only necessary to
perform the momentum integrations up to $\mu$.

For small couplings we can make Feynman diagrams representing the
process of integrating out $\phihat$ in~\eqref{dvn74} with
$\phitilde$ as an external field. It is natural to put this
diagrammatic contribution in exponential form. As usual this
corresponds to only taking into account connected diagrams.
$\lagr_\mu$ is then simply the sum of $\lagr_\La$ and these
diagrammatic contributions:
\begin{equation}\label{dvn74.5}
    \lagr_\mu\big(\phitilde\big)=\lagr_\La\big(\phitilde\big)+\textrm{connected
    diagrams}.
\end{equation}
We note that the propagators in the diagrams will only have to be
integrated over the energy-momentum shell $\mu\leq \norm{k}<\La$.
These diagrams thus suffer from neither UV nor IR divergencies
which is an essential feature in the Wilsonian effective action.
Usually the diagrams give an infinite number of non-zero
correction terms -- all those that are allowed by symmetries.
These corrections are new (and renormalisations of the old)
interactions terms that compensate for the degrees of freedom that
has been integrated out. However, we immediately see that
supersymmetry is an exception (actually the only known) since here
there is no renormalisation of the superpotential as we saw in the
last section for a theory with only chiral fields. The exceptions
to the non-renormalisation theorem mentioned in the last section
had their root in IR divergencies which are not a problem here.

\subsection{Renormalisation Group Running of Couplings}\label{dvsecwilsonsubrunning}
We can use the above procedure for a general theory if we
integrate out the high energy-momentum modes for all the particles
simultaneously to obtain $\lagr_\mu$. We can think of $\mu$ as a
continuous parameter. We note that this e.g. can be done by in
steps integrating out infinitesimal energy-momentum shells. The
couplings $g_i\nleft(\mu\right)$ are then continuous functions of
$\mu$. The renormalisation group running in the space of theories
is then given by the Wilson equation:
\begin{equation}\label{dvn75}
    \mu\frac{\partial g_i\nleft(\mu\right)}{\partial\mu}=\be_i\nleft(g(\mu),\mu\right).
\end{equation}
We can constrain the $\mu$-dependence in the $\beta$-function by
introducing dimensionless couplings. If $\De_i$ is the mass
dimension of $g_i$ then we define the dimensionless couplings as
$\cG_i\nleft(\mu\right)=\mu^{-\De_i}g_i\nleft(\mu\right)$.
However, $\cG_i\nleft(\mu\right)$ can only depend on
$\cG\nleft(\La\right)$ and $\mu/\La$ since these are the only
dimensionless parameters. Differentiating and setting $\La=\mu$
then gives:
\begin{equation}\label{dvn76}
    \mu\frac{\partial
    \cG_i\nleft(\mu\right)}{\partial\mu}=\tilde{\be}_i\nleft(\cG\nleft(\mu\right)\right),
\end{equation}
where we note that there is now no explicit dependence on $\mu$ on
the right hand side. Using the dimensionless couplings one can
show a theorem due to J. Polchinski saying (with some assumptions)
that if the initial couplings $\cG\nleft(\La\right)$ lie on
generic $N$-dimensional surface then for $\mu\ll\La$ they will
approach a fixed surface that is independent of the initial
surface and $\La$. This surface is also approximately stable under
further running of the parameter $\mu$. Here $N$ is the number of
renormalisable couplings. Using this we see that the physical
observables are independent of $\La$ as they should
be.\footnote{This can be used to justify the usual renormalisation
scheme where we take $\La$ to infinity and express the couplings
and masses in terms of physical quantities.}

A fixed point is as usual a point where the left hand side
of~\eqref{dvn75} vanishes. We can linearise the $\beta$-function
around the fixed point. The eigenvalues of the
$\mu\frac{\partial}{\partial\mu}$ operator then determines whether
the coupling is damped or grows along the flow. The corresponding
operators are for the former called \emph{irrelevant} and for the
latter \emph{relevant} operators. Zero eigenvalues correspond to
\emph{marginal} operators. The free theory is, naturally, a fixed
point since $\lagr_{\mathrm{mixed}}$ in~\eqref{dvn73} in this case
is independent of $\phitilde$. In the vicinity of this point where
we have weak coupling, we can determine the relevant operators by
dimensional analysis. As above $g_i$ has mass dimension $\De_i$
and thus its natural order of magnitude is $\La^{\De_i}$. Thus a
coupling with positive mass dimension will become increasingly
important at lower $\mu$ and the contrary for a coupling with
negative mass dimension. This, naturally, only holds true as long
as the couplings are not changed too much by the quantum
corrections i.e. in the vicinity of the free fixed point. Thus we
see that the relevant operators here are the super-renormalisable,
the marginal are the renormalisable, and the irrelevant are the
non-renormalisable operators.

One can now ask why we do not drop the irrelevant operators when
letting $\mu$ flow towards low energies. However, when we want to
examine a vacuum we expand the fields around the vacuum
expectation values. In this way irrelevant and relevant terms are
mixed. Thus irrelevant terms can contribute to relevant terms
after the redefinition of the fields. However, by the above
dimensional analysis we see that it is enough to keep two
derivatives on scalar fields and one on the fermionic fields --
just as we did in chapter~\ref{chpsusy} when developing the
general supersymmetric Lagrangians. With these Lagrangians we can
obtain the vacuum expectation values and investigate the relevant
and marginal physics around the vacua. But if we do not want to
let $\mu$ go all the way to zero, we are to some extent making an
approximation here. Also note that we assumed that we were close
to a free fixed point as $\mu\rightarrow0$ so we could
characterise the relevant operators as the renormalisable ones.
This is actually true for a wide range of theories: The
Coleman-Gross theorem tells us that for small couplings a theory
of scalars, spinors and $\un{1}$ gauge bosons has an IR free fixed
point.

In general when wanting to determine the effective action at some
scale $\mu$ one has to guess which degrees of freedom are relevant
(or marginal) at that scale.

\subsection{Integrating Out Massive Fields}\label{dvsecwilsonsubinteout}
We note that both the kinetic terms and the masses are
renormalised as we let $\mu$ float to lower energies. However, we
can as usual renormalise the wavefunctions to keep the kinetic
terms invariant. This will naturally also renormalise the
couplings that are then called canonical couplings. Let us assume
that we have a massive field $\phi$. Normally the canonically
renormalised mass will not decay as the energy-momentum scale
$\mu$ runs to zero so at some point $\mu$ will be less than the
canonically renormalised mass of $\phi$. This means, especially if
$\mu\ll m$, that the mass term dominates the kinetic term which we
can then disregard. Let us here pause the flow of the rest of the
fields and perform the remaining integration of the field $\phi$.
What we should think here is then:\footnote{The literature is
unfortunately a bit vague at this point.} When we want to
integrate out the remaining momentum shell of thickness $\mu$ the
propagator for the field $\phi$ is only given by the inverse mass
(squared). A loop with this propagator then contains a momentum
integration of the thickness $\mu$. The loop must then scale as
$\mu/m$ to some power and can thus be discarded. Thus we only have
tree-level diagrams left i.e. the semi-classical approximation.
This tells us that when we go below the mass of a field we can
completely integrate it out by replacing it with its equation of
motion:
\begin{equation}\label{dvn77}
    \frac{\partial{\lagr_\mu}}{\partial\phi}=0,
\end{equation}
where we used that the dependence in the Lagrangian on
$\partial\phi/\partial x^\mu$ can be removed since terms including
such space-time derivatives give rise to tree-level diagrams that
scales as $\mu/m$ to some power. The integrating out procedure
will be important for us in the next section.

\subsection{Wilsonian vs. 1PI Effective Action}\label{dvsecwilsonsubwilsonvs1pi}

In the limit $\mu\rightarrow0$ we can compare the Wilsonian and
the 1PI effective action. As we saw in the last subsection the
massive fields will then be completely integrated out of the
Wilsonian effective action. However, the massless fields are still
degrees of freedom. This is contrary to the 1PI effective action
where all fields are integrated out. But if there are no massless
fields the 1PI and the Wilsonian action coincide. As discussed
above, when one integrates out the massless fields in the 1PI
effective action one gets IR divergencies. We saw that because
there were no IR divergencies in the Wilsonian effective action,
it fulfilled the non-renormalisation theorem. In the next section
we will discuss that this also ensures the intimately related
concept of holomorphy of coupling constants in the Wilsonian
effective action -- this is not case for the 1PI effective
potential.

It should be mentioned that the Wilsonian effective action
actually suffers from IR divergencies when we are varying in the
space of theories. Simply think of a $\lagr_\mu$ with some field
$\phi$ with mass $m<\mu$. Here the field is integrated out. If we
now vary $m$ to zero we have accidentally integrated out a
massless field and we get an IR divergency. It is a fact that all
singularities in the Wilsonian effective action arise in this way
as fields becoming massless.

Let us end this section by noting that the Lagrangian $\lagrun$ in
the Dijkgraaf-Vafa conjecture can be seen as an effective
Lagrangian obtained by integrating out a very massive field in an
underlying theory. Hence we can understand the inclusion of
non-renormalisable powers in the tree-level superpotential. In the
next section we will see how further renormalisation group running
changes the effective superpotential.

%---------New section-----------
\section{The Wilsonian Effective Superpotential}\label{dvsecweff}

In this section we will study the Wilsonian effective
superpotential at some scale $\mu$. Let us assume that
supersymmetry is unbroken at this scale. Then we can write the
effective Lagrangian in the general supersymmetric
form~\eqref{dvn6}. This allows us to define the Wilsonian
effective generalised superpotential as the $\tha\tha$-term
in~\eqref{dvn6}. As also noted there we have here enlarged the
definition of the superpotential to also include the
supersymmetric gauge field strength hence the name ``generalised
superpotential''.\footnote{We will simple refer to the generalised
superpotential as ``superpotential'' when no confusion should be
possible.} This made sense since the gauge kinetic term, even
though it can be written as a local D-term, can not be written as
a gauge invariant D-term. We will begin our investigation of this
effective superpotential by introducing the concept of holomorphy
which will be crucial to us not only in this section, but also
when we give the diagrammatic proof of the Dijkgraaf-Vafa
conjecture.

\subsection{Holomorphy}\label{dvsecweffsubholo}

Since the mid-eighties it has been known that the effective
superpotential should be holomorphic not only in the chiral fields
as demanded by supersymmetry, but also in the bare coupling
constants. However, it was discovered that it here is essential
that the effective superpotential is the Wilsonian one since the
IR-divergencies from massless particles in the 1PI effective
superpotential can violate this holomorphy - i.e. give an
holomorphic anomaly (this was made clear
in~\cite{shifmanvainshteinholomorphic1}
and~\cite{shifmanvainshteinholomorphic2}, and is reviewed
in~\cite{9704114}).

In~\cite{9309335} N. Seiberg introduced\footnote{Actually, the
trick is also mentioned in~\cite{shifmanvainshteinholomorphic2},
however, it is not used in its full power as in~\cite{9309335}. It
should also be mentioned that~\cite{9309335} is based on an
unpublished argument by N. Seiberg and J. Polchinski.} a trick
well-known from string theory that allows us to prove the
holomorphicity in the (complex!) coupling constants: We think of
the bare couplings $g_k$ in the (not generalised) superpotential
as being the scalar components of background chiral superfields.
E.g. they could be very massive chiral superfields integrated out
at the energy at which we write our bare Lagrangian and with
fine-tuned couplings such that the expectation value of the scalar
components exactly match the bare couplings $g_k$ (and the rest of
the components have zero expectation value). Now, as we flow down
in energy the Wilsonian effective superpotential will still be
holomorphic in the chiral fields due to supersymmetry which we
assume to be unbroken. This now includes the auxiliary background
fields which are just the bare constants. Thus the effective
superpotential is holomorphic in the bare coupling constants, i.e.
independent of $g_k^*$, as long as supersymmetry is
unbroken.\footnote{Naturally, also the term superpotential no
longer makes sense when supersymmetry is broken.} This remains
true also for non-perturbative corrections as it is based only on
the assumption of unbroken supersymmetry. However, we note that it
is important not to use the canonically renormalised couplings
since the \kahler{} term is not holomorphic.

The holomorphicity also applies to the complex gauge coupling
constants $\tau$ from \eqref{susyn75} and~\eqref{susyn78}. And,
naturally, it also applies when we look at the generalised
superpotential.

Using the full power of holomorphy and treating the couplings as
background fields we can prove powerful non-renormalisation
theorems as we will see already in the next section.

\subsection{Non-Renormalisation Theorems}\label{dvsecweffsubnon}

In this section we will use the idea from the last section that
the couplings in the generalised superpotential can be seen as
background chiral superfields. We will prove that the generalised
superpotential is not renormalised perturbatively or to be more
precise that the only renormalisation that takes place is one-loop
renormalisation of the complex gauge coupling $\tau$.

In the article~\cite{9309335} (reviewed in~\cite{9408013}
and~\cite{9509066}) Seiberg introduced the following scheme for
determining the effective superpotentials:
\begin{enumerate}
    \item Holomorphy: As discussed in the previous section
    the dependence in the generalised superpotential on the bare couplings
    should be holomorphic due to supersymmetry.
    \item Symmetries and selection rules: Setting the background
    fields (i.e. the couplings) to zero gives a large global
    symmetry group of the Lagrangian where we also include R-symmetries.
    This symmetry group is spontaneously broken for non-zero expectation
    values of the background fields (non-zero couplings). However, letting the
    background fields transform (assigning transformation rules to the couplings)
    the global symmetry group can be restored for the Lagrangian.
    If non-anomalous, these symmetries must be shared by the
    effective Lagrangian giving powerful restrictions that are
    further sharpened when using the above holomorphy.
    \item Various limits: E.g. in the weak coupling limit it can be possible
    to put restrictions on the
    Lagrangian using perturbation theory. Sometimes one can require
    smoothness in the weak coupling limit or even, when setting some masses to
    zero, use that there must be
    singularities due to massless particles that have been
    integrated out as explained in section~\ref{dvsecwilsonsubwilsonvs1pi}.
\end{enumerate}

This scheme will often determine the effective superpotential
since a holomorphic function is determined by its asymptotic
behaviour and its singularities. We will now use it to prove the
perturbative non-renormalisation theorem. This was formerly done
using the supergraph method as in
section~\ref{dvsecquantsubnonrenorm} which generalises to the case
where vector superfields are involved. Here we will present a
proof based on~\cite{weinberg3} and~\cite{9803099} using the
Seiberg scheme.

Let us assume that our bare Lagrangian at the UV energy $\mu_0$
has the form:
\begin{equation}\label{dvn78}
\lagr_{\mu_0}=\int\dsuper\Phi^\dagger
e^{2V^{\repr}}\Phi+\left(\int\dtotha\frac{\tau}{16\pi
iC(\reprn)}\tr_{\repr}\nleft(\W^\al\W_\al\right)+\int\dtotha
W_{\mu_0}\nleft(\Phi^i\right)+\cc\right).
\end{equation}
Here we have used~\eqref{susyn67} and assumed a renormalisable
\kahler{} term. The gauge group is for simplicity assumed simple
and the gauge kinetic term renormalisable so that we could
use~\eqref{susyn77}. The superpotential could be
non-renormalisable.

Using the background field method we think of the bare complex
gauge coupling $\tau$ as being a background chiral superfield. The
same we do with the couplings $g_k$ in the superpotential
$W_{\mu_0}\nleft(\Phi^i\right)$ which we can write as:
\begin{equation}\label{dvn78.5}
    W_{\mu_0}\nleft(\Phi^i\right)=\sum_k
    g_kO_k\nleft(\Phi^i\right),
\end{equation}
where $O_k\nleft(\Phi^i\right)$ is a gauge invariant multiple of
the chiral fields, i.e. a multiple of the independent holomorphic
gauge invariants, $X_r$, parameterising the vacuum moduli space
introduced in section~\ref{dvsecvacsubclmoduli}.

Now we want to determine the effective Lagrangian $\lagr_\mu$ at
the lower energy $\mu$. We assume that the cut-off respects
supersymmetry and gauge invariance. Thus $\lagr_\mu$ a priori
takes the form of the most general gauge invariant and
supersymmetric Lagrangian i.e.:
\begin{equation}\label{dvn79}
    \lagr_\mu=\int\dsuper
    G_\mu\nleft(\Phi^i,\Phibar^i,V,\tau,\tau^*,g_k,g_k^*,D_\al\ldots\right)+\left(\int\dtotha
    F_\mu\nleft(\Phi^i,\W_\al,\ta,g_k\right)+\cc\right).
\end{equation}
Here $G_\mu$ and $F_\mu$ are very general gauge invariant
functions since we do not know which operators are relevant at the
scale $\mu$. This means that we can not constrain ourselves to the
Lagrangians given in section~\ref{susysecN1} where we assumed
maximally two space-time derivatives on bosonic fields and one on
fermionic fields as explained in
section~\ref{dvsecwilsonsubrunning}. However, we do note that the
generalised superpotential $F_\mu$ is holomorphic in $g_k$ and
$\ta$ as determined by supersymmetry. Also, $F_\mu$ does not
depend on the covariant derivatives $D_\al$ or space-time
derivatives since according to~\cite{weinberg3} such terms can be
reformulated as D-terms as we did with the chirally exact terms.

The next item on the Seiberg scheme is to constrain $\lagr_\mu$ by
extended global symmetries. There are two of those. The first is a
$\textrm{U}(1)_R$ R-symmetry under which the generalised
superpotential must have charge 2 as explained in
section~\ref{susysecchiralsubrsymm} . When all the couplings are
zero we obtain an R-symmetry simply by choosing $\Phi^i$ and $V$
to be R-neutral. In order for this to be an R-symmetry with
non-zero couplings we must assign transformation rules to the
couplings, i.e. the background fields. All couplings $g_k$ must
have charge 2 to give the tree-level superpotential an overall
charge of 2. On the other hand $\W_\al$ given by~\eqref{susyn58}
must have charge 1 since
$R\nleft(\partial/\partial\tha^\al\right)=R\nleft(D_\al\right)=-1$.
We then conclude that $\tau$ has zero charge. We will now assume
that $\lagr_\mu$ is obtained perturbatively. Then the R-symmetry
is non-anomalous as we will see in the next section. Thus
$\lagr_\mu$ must be invariant under the R-symmetry and $F_\mu$
must have charge 2. Since it is holomorphic, it only depends on
couplings and fields having non-negative charges and we conclude
that it takes the form:
\begin{equation}\label{dvn80}
    F_\mu\nleft(\Phi^i,\W_\al,\ta,g_k\right)=W_\mu\nleft(\Phi^i,g_k,\tau\right)+\frac{1}{16\pi i}\tau_{\mu,ab}\nleft(\Phi^i,\tau\right)\W^{\al
    a}\W^b_\al,
\end{equation}
where $W_\mu$ is linear in the $g_k$'s.

The second symmetry we will use is $\tau\mapsto\tau+\xi$ where
$\xi$ is a real number. This is a symmetry of the Lagrangian since
as we will see in the next section the real part of $\tau$ i.e.
$\vartheta$ multiplies a total derivative term that does not
contribute in the perturbative regime. Thus $\tau$ can only appear
as multiplying a $\W^{\al a}\W^b_\al$ term as in the tree-level
Lagrangian. Hence $W_\mu$ is independent of $\tau$ and demanding
gauge invariance gives $\tau_{\mu,
ab}\nleft(\Phi^i,\tau\right)=c_\mu\de_{ab}\tau+d_{\mu,
ab}\nleft(\Phi^i\right)$ since the gauge group is simple.

The last step in the Seiberg scheme is to consider limits. Setting
$g_k$ equal to zero we have a global symmetry of the bare
Lagrangian: $\Phi\mapsto e^{-i\al}\Phi$. This constrains $d_{\mu,
ab}\nleft(\Phi^i\right)$ to be independent of $\Phi^i$ and thus
equal to $\de_{ab}d_\mu$ for the sake of gauge invariance. Thus we
can write $F_\mu$ as:
\begin{equation}\label{dvn81}
    F_\mu\nleft(\Phi^i,\W_\al,\ta,g_k\right)=W_\mu\nleft(\Phi^i,g_k\right)+\frac{c_\mu\tau+d_\mu}{16\pi
iC(\reprn)}\tr_{\repr}\nleft(\W^\al
    \W_\al\right).
\end{equation}

Let us now take the limit where the gauge coupling and the $g_k$'s
are small i.e. weak coupling. Since $W_\mu$ is linear in the
$g_k$'s, the only diagrams that can contribute to $W_\mu$ is the
single vertex diagrams determined from the bare superpotential
$W$. This shows us that the superpotential is not renormalised:
$W_\mu=W$.

To find $c_\mu$ and $d_\mu$ we also use the weak coupling limit.
We note that this for the gauge coupling amounts to taking $\tau$
to infinity. We can think of $\tau$ as being purely imaginary
since, as noted above, $\vartheta$ does not contribute. We can
then develop the supergraph rules for the vector field $V$. The
propagator and self-interaction vertices are derived from the term
$\int\dtotha\frac{\tau}{16\pi
iC(\reprn)}\tr_{\repr}\nleft(\W^\al\W_\al\right)$. This means that
the propagator goes as $1/\tau$ and the self-interaction vertices
go as $\tau$. The interactions with matter come from the \kahler{}
term and are $\tau$ independent. We note that each of these
interaction vertices has two $\Phi$-lines attached since we
assumed a renormalisable \kahler{} term in~\eqref{dvn78}. There
can be no $\Phi$-$\Phi$ interaction since we have no dependence on
the $g_k$'s in $c_\mu$ and $d_\mu$. Now, let there be given a
diagram contributing to the last term of $F_\mu$ in~\eqref{dvn81}.
We will count the $\tau$ dependence in this diagram. Let $V_V$ and
$I_V$ be the number of pure gauge boson vertices and internal
gauge boson lines respectively. The power of $\tau$ in the diagram
then is:
\begin{equation}\label{dvn81.5}
    N_\tau=V_V-I_V.
\end{equation}
Introducing $V_\Phi$ for the number of (not pure gauge)
interaction vertices and $I_\Phi$ for the number of internal
$\Phi$-lines (or $\Phibar$-lines), the number of loops is given by
(assuming a connected momentum space diagram):
\begin{equation}\label{dvn82}
    L=I_V+I_\Phi-V_V-V_\Phi+1.
\end{equation}
However, since there are no external $\Phi$-lines and each
interaction vertex contains two $\Phi$-lines we have
$V_\Phi=I_\Phi$ so $N_\tau=1-L$. Thus only tree-level diagrams
contributes to $c_\mu$ i.e. it takes the same value as in the bare
Lagrangian: $c_\mu=1$. And $d_\mu$ only receives one-loop
renormalisation (we can include the tree-level which is zero). The
renormalised superpotential then takes the form:
\begin{equation}\label{dvn83}
        F_\mu=W\nleft(\Phi^i\right)+\frac{\tau\nleft(\mu\right)}{16\pi
iC(\reprn)}\tr_{\repr}\nleft(\W^\al
    \W_\al\right),
\end{equation}
where $\tau\nleft(\mu\right)$ is the one-loop renormalised complex
gauge coupling. This ends the proof of the perturbative
non-renormalisation theorem. We could also have obtained this
result from a general non-renormalisable bare Lagrangian, however,
the counting of the powers of $\tau$ would have been a bit
harder~\cite{9803099}.

$\tau\nleft(\mu\right)$ can be determined from standard quantum
gauge theory calculations; here taken from~\cite{weinberg3},
\cite{argyres} and~\cite{peskinandschroeder}. The renormalisation
group running~\eqref{dvn75} of the (real) gauge coupling $g$ is
found to be
$\mu\frac{\partial}{\partial\mu}g=-\frac{b}{16\pi^2}g^3$ to
one-loop order. Here $b$ depends on the quadratic invariants,
$C\nleft(\reprn\right)$, defined in~\eqref{susyn76.5} of the
representations of the complex bosons and the Weyl fermions. Using
that we have one adjoint Weyl fermion for each vector superfield,
and we have a Weyl fermion and a complex boson for each chiral
superfield representation $\reprn_n$, one gets:\footnote{This is
often written with the quadratic Casimir
$C_2\nleft(\mathrm{adj}\right)$ instead of
$C\nleft(\mathrm{adj}\right)$ which one can do since the quadratic
Casimir and the quadratic invariant are equal in the adjoint
representation.}
\begin{equation}\label{dvn85}
    b=3C\nleft(\mathrm{adj}\right)-\sum_{n}C\nleft(\reprn_n\right),
\end{equation}
where the sum is over the different representations of the chiral
fields. We note that this is not invariant under scalings of the
gauge group generators since these scale the gauge coupling as
explained at the end of section~\ref{susysecchiralsubgauge}.
Solving the Wilson equation yields:
\begin{equation}\label{dvn86}
    \frac{1}{g^2\nleft(\mu\right)}=-\frac{b}{8\pi^2}\ln\nleft(\frac{\abs{\La}}{\mu}\right),
\end{equation}
where $\abs{\La}$ is the \emph{strong coupling scale} of the
theory. It is simply a constant of integration that by definition
is scale invariant and we can express it as:
\begin{equation}\label{dvn87.5}
    \abs{\La}=\mu e^{-\frac{8\pi^2}{b g^2\nleft(\mu\right)}}.
\end{equation}
The reason for the modulus is that we can now express the running
complex gauge coupling $\tau\nleft(\mu\right)$
using~\eqref{susyn78} as:\footnote{We note that this tells us that
$d_\mu=\frac{b}{2\pi i}\ln\nleft(\frac{\mu_0}{\mu}\right)$. Here
$\mu_0$ is the scale for the tree-level Lagrangian.}
\begin{equation}\label{dvn87}
    \tau\nleft(\mu\right)=\frac{b}{2\pi
    i}\ln\nleft(\frac{\La}{\mu}\right),
\end{equation}
where $\La$ is the (complex) holomorphic scale\footnote{Not to be
mistaken with the UV cut-off in section~\ref{dvsecwilson}.} given
by:
\begin{equation}\label{dvn88}
    \La\defi\abs{\La}e^{i\vartheta/b}=\mu e^{\frac{2\pi i}{b}\tau(\mu)}.
\end{equation}

The strong coupling scale is naturally so called since when $\mu$
approaches $\abs{\La}$ the effective gauge coupling diverges as
seen from~\eqref{dvn86}. We also see that the sign of $b$
determines whether the theory is UV or IR free. For $b$ positive
the theory is UV free and IR strongly coupled i.e. asymptotically
free. From~\eqref{dvn85} we see that this happens for non-abelian
gauge theories with not too much light\footnote{We note that the
fields contributing to the counting in $b$ are only those which
have not been integrated out at the scale $\mu$ i.e. the light
matter. This also means that $\La$ depends on how much matter that
has been integrated out.} charged matter. On the other hand, for
$b$ negative the theory is weakly coupled in the IR and runs to
strong coupling in the UV. This happens e.g. in the abelian case
where $C\nleft(\mathrm{adj}\right)=0$ or for theories with enough
charged matter.

For an asymptotically free theory we must demand that the scale
$\mu$ in the non-renormali\-sation theorem is greater than the
strong coupling scale -- so the theorem does not solve the strong
coupling problem of asymptotically free theories.

We should note that for a general (classical) gauge group we get
one holomorphic scale for each simple factor. We also note that
the 1PI complex gauge coupling does receive higher order loop
contributions. However, the effective theory is not holomorphic in
the 1PI coupling and the relation between the Wilsonian and the
1PI complex gauge couplings is consequently
non-holomorphic~\cite{shifmanvainshteinholomorphic2}.

An important consequence of the non-renormalisation theorem is
that if supersymmetry is unbroken classically, it is unbroken to
any order in perturbation theory. This follows since the
superpotential is not renormalised and thus we still have the
classical solution to the F-flatness equations. But the \kahler{}
term is now taking a general form since we can not constrain this
term with holomorphy. However, the proof in
section~\ref{dvsecvacsubbreak} still works for this general
\kahler{} term and thus shows us that we can always find a
simultaneous solution to both the D- and F-flatness equations.
Consequently, supersymmetry in unbroken perturbatively. However,
this does not mean that the quantum moduli space is the same as
the classical. As noted in section~\ref{dvsecvacsubclmoduli} the
classical moduli space is endowed with the \kahler{} metric pulled
back from the target space. But this metric is, naturally, changed
by the \kahler{} term renormalisation thus changing the moduli
space. It should be mentioned that non-perturbatively it is
possible to break supersymmetry since we here do not have a strict
non-renormalisation theorem in the non-perturbative regime.

The Dijkgraaf-Vafa conjecture provides a systematic way to obtain
the effective generalised superpotential. We have learned in this
section that the non-trivial part of these effective
superpotentials must be non-perturbative contributions. We will
discuss such contributions shortly, but first we will have to take
a look at the $\vartheta$-angle and chiral anomalies.

\subsection[The $\vartheta$-Angle, Instantons and Chiral Anomalies]{The $\boldsymbol{\vartheta}$-Angle, Instantons and Chiral
Anomalies}\label{dvsecweffsubinstchiral}

In this section we will rather briefly study the $F_{
\mu\nu}\Ftilde^{\mu\nu}$-term in the gauge kinetic and
self-interaction Lagrangian~\eqref{susyn70}. The presentation is
based on~\cite{argyres}, \cite{ryder}, \cite{ambjornandlyng}
and~\cite{0004186}. Let us look at a non-abelian gauge group and
concentrate on a simple factor. For simplicity we think of this
subgroup as $\sun{2}$. The results will be true for general
non-abelian gauge groups since they all contain $\sun{2}$
subgroups. From~\eqref{susyn78.1} we see that the term of our
interest takes the form:
\begin{equation}\label{dvn89}
    S_\vartheta=-\int\dlor\frac{\vartheta}{16\pi^2}\tr\nleft(F_{\mu\nu}\Ftilde^{\mu\nu}\right),
\end{equation}
where the trace here and in the rest of this subsection is taken
in the fundamental representation where, as before, the quadratic
invariant from~\eqref{susyn76.5} is chosen to be $1/2$. Using the
definition of the field strength~\eqref{susyn60} we get:
\begin{equation}\label{dvn90}
    S_\vartheta=-\int\dlor\frac{\vartheta}{8\pi^2}\vep^{\mu\nu\rho\si}\partial_\mu\tr\nleft(A_\nu\partial_\rho A_\si+i\frac{2}{3}A_\nu A_\rho
    A_\si\right),
\end{equation}
which, as promised above, is a total space-time derivative term.

For the integral~\eqref{dvn89} to be finite we must require that
$F_{\mu\nu}$ vanishes at infinity.\footnote{Since $F_{\mu\nu}$
transforms adjointly, the condition $F_{\mu\nu}=0$ at infinity is
gauge invariant.} Hence, at infinity the gauge potential is pure
gauge such that $A_\mu=i\left(\partial_\mu g\right)g^{-1}$ where
$g$ is an element in the gauge group.\footnote{We have here used
that the gauge transformation of $A_\mu$ with $g$ is $A'_\mu=g
A_\mu g^{-1}+i\left(\partial_\mu g\right)g^{-1}$ consistent with
the definition of $D_\mu$ in~(\ref{susyn61}).} Plugging
into~\eqref{dvn90} gives:
\begin{equation}\label{dvn91}
    S_\vartheta=\frac{\vartheta}{24\pi^
    2}\int_{S^3}\mathrm{d}^3\xi\,\vep^{ijk}\tr\nleft(g^{-1}(\partial_i g)g^{-1}(\partial_j g)g^{-1}(\partial_k g)\right),
\end{equation}
where we have rewritten the four-dimensional space-time integral
over the total derivative as a surface integral over the 3-sphere
$S^3$ at infinity.\footnote{We have here assumed a regular gauge
with no divergencies.} However, this integral is known from
homotopy theory. It counts (ignoring the $\vartheta$-angle) the
number of times $g\nleft(x\right)$ wraps $S^3$ around $\sun{2}$
which in turn is topologically equivalent to $S^3$. This is an
integer -- the winding number -- and it determines to which
homotopy (\emph{Pontryagin}-) class the gauge potential belongs.
This corresponds to the third homotopy group being
$\pi_3\nleft(\sun{2}\right)=\Z$. The winding number is a
topological invariant and is thus unchanged under continuous
deformations of the fields. We can now write $S_\vartheta$ as:
\begin{equation}\label{dvn92}
    S_\vartheta=\vartheta n,\qquad n\in\Z,
\end{equation}
where $n$ is the winding number. Since the contribution to the
path integral is $e^{iS_\vartheta}$, we see that
\begin{equation}\label{dvn93}
    \vartheta\mapsto\vartheta+2\pi
\end{equation}
is a symmetry of the theory or, more correctly, it is an exact
equivalence of theories. This shows us that we can think of
$\vartheta$ as an angle. This symmetry can be expressed in the
complex gauge coupling as:
\begin{equation}\label{dvn94}
    \tau\mapsto\tau+1.
\end{equation}

To investigate the above field configuration further we switch to
temporal gauge $A_t=0$ and the surface in~\eqref{dvn91} is taken
to be a cylinder parallel to the time axis. Then the only
contribution to our surface integral comes from the caps at
$t=\pm\infty$. The surface integral is thus the difference between
the full space integrals over the pure gauge configurations at
$t=\pm\infty$. The pure gauge configurations at the caps are
determined by the space dependent group elements $g_\infty$ and
$g_{-\infty}$ respectively and the full space integrals are
$\vartheta$ times integers say $n_\infty$ and $n_{-\infty}$. The
gauge field configuration then interpolates between two vacua
configured by $g_\infty$ and $g_{-\infty}$ and the winding number
is the difference $n=n_\infty-n_{-\infty}$. Both vacua can, by
definition, be gauge transformed into the classical vacua with
zero gauge potential, i.e. $g_{\pm\infty}=\idmatr$ and
$n_{\pm\infty}=0$. However, the gauge transformation can not be
continuously deformed into the identity for
$n_\infty,\,n_{-\infty}\neq1$ since the configurations belong to
different homotopy classes. This kind of gauge transformations are
called \emph{large} gauge transformations.

The homotopically different vacua are related by the above field
configurations with non-zero winding number. They are not
physically equivalent as we already see from the fact that the
interpolating gauge configurations have different weights in the
path integral due the $S_\vartheta=\vartheta n$ term. Also, since
$F^{\mu\nu}$ can not vanish identically (this would give zero
winding number) there is an energy barrier between the vacua. The
corresponding quantum mechanical tunnelling amplitude is
$e^{-S_E}$ where $S_E$ is the Euclidean action (focusing on the
gauge part only) with
$S_E=\frac{1}{2g^2}\int\dlor\tr\nleft(F_{\mu\nu}F_{\mu\nu}\right)$.

We thus have an infinity of homotopically inequivalent vacua. This
naturally assumes that the above interpolating gauge
configurations exist and have finite action implying that the
tunnelling amplitudes are not vanishing. Such solutions indeed
exist. In fact, by rotating to Euclidean space we can for each
winding number $n$ find a unique gauge configuration in the
homotopy class which minimises the action, i.e. satisfies the
classical equation of motion $D_\mu F_{\mu\nu}=0$ and has finite
action. These are called instantons and the corresponding winding
number is called the instanton number or charge. It is easy to
show that the field strengths for the instantons are
(anti-)selfdual $F_{\mu\nu}=\pm\Ftilde_{\mu\nu}$.\footnote{It is
important that we are in Euclidean space because here
$\tilde{\tilde{F}}_{\mu\nu}=F_{\mu\nu}$. This is contrary to
Minkowski space where $\tilde{\tilde{F}}_{\mu\nu}=-F_{\mu\nu}$ and
thus (anti-)selfduality only has trivial solutions.} For
instantons the tunnelling amplitude given by the $FF$ term can
then be calculated using~\eqref{dvn92} (without $\vartheta$):
\begin{equation}\label{dvn95}
    e^{-S_E}=\nleft(e^{-8\pi^2/g^2}\right)^{\abs{n}}=\left(\frac{\abs{\La}}{\mu}\right)^{\abs{n}b},
\end{equation}
where we have used the strong coupling scale from~\eqref{dvn87.5}.
Since the strong coupling scale is non-perturbative,\footnote{If
we try to expand the exponential $e^{-1/g^2}$ in a Taylor series
around $g=0$ each coefficient would be zero.} we see that
instantons are non-perturbative effects given by powers of
$\abs{\La}$ and thus become increasingly important in the strongly
coupled regime.

Using the instantons one can calculate non-perturbative
corrections to the effective Lagrangian, but we will not go into
details of this vast field. However, let us just note that the
instanton solutions only exist in non-abelian groups. Thus, as
mentioned before, we do not have such contributions in abelian
subgroups. However, the abelian $\vartheta$-angle should not be
neglected in the IR effective actions since it here couples to
massive sources not described by the same IR physics. So the real
part of the complex gauge couplings~\eqref{dvn18} in the
Dijkgraaf-Vafa conjecture should not be neglected.

The $F_{ \mu\nu}\Ftilde^{\mu\nu}$-term also plays an important
role when we look at anomalies in chiral symmetries. A chiral
symmetry is defined as a symmetry in which the left and right
handed part of the spinors transform differently. As in the rest
of this thesis we here assume that the spinors are Majorana
spinors. Now, if the left-handed Weyl spinor transforms in the
$\reprn_{\mathrm{chiral}}$ representation of some symmetry group
$G_{\mathrm{chiral}}$ then due to the Majorana reality
condition~\eqref{appspinorn14.5} the right-handed Weyl spinor
transforms in the complex conjugate representation
$\bar{\reprn}_{\mathrm{chiral}}$. We thus conclude that for a
chiral symmetry the spinor must transform in a complex
representation. Corresponding to this symmetry we as usual have a
classically conserved current, $j^\mu_a$, for each generator
$T^{\mathrm{(chiral)}}_a$ of $G_{\mathrm{chiral}}$.

\begin{figure}\caption{}\label{dvnfign1}
\begin{center}
    \includegraphics{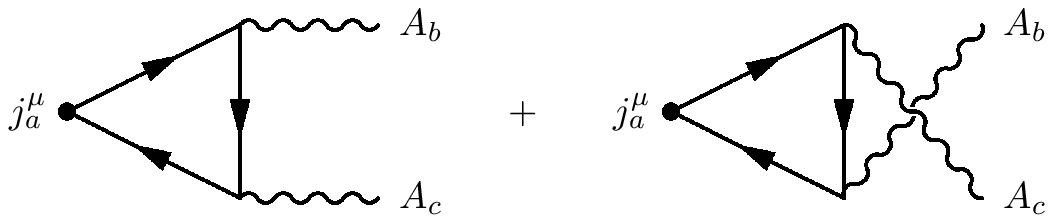}
\end{center}
    \begin{center}
        The diagrams contributing to the chiral anomaly in four space-time
        dimensions. Taken from~\protect{\cite{argyres}}.
    \end{center}
\end{figure}

In four space-time dimensions this symmetry can only be anomalous,
i.e. $\expect{\partial_\mu j^\mu_a}\neq0$, if the fermions are
coupled to gauge fields. We thus assume a gauged theory with gauge
group $G$. The anomaly can be calculated perturbatively using
diagrams and is actually a one-loop effect. The diagrams
contributing to the calculation of $\expect{\partial_\mu j^\mu_a}$
in four dimensions are the triangle diagrams shown in
figure~\ref{dvnfign1}.
The result of the calculation in Minkowski space is (treating the
gauge fields as external fields):\footnote{In this expression all
the generators could be generators of the gauge group. For
consistency we must then demand that we have no gauge anomaly i.e.
$\sum_{\reprn}\tr_{\reprn}\nleft(T_a\{T_b,T_c\}\right)=0$.}
\begin{equation}\label{dvn96}
    \expect{\partial_\mu
    j^\mu_a}\propto\sum_{\reprn}\tr_{\reprn}\nleft(T_a\{T_b,T_c\}\right)F^b_{
    \mu\nu}\Ftilde^{c\mu\nu},
\end{equation}
where the sum is over the different representations of the Weyl
fermions. The right hand side should be seen as external fields.
Setting $T_a$ to be a generator of the chiral symmetry group, and
$T_b$ and $T_c$ to be gauge generators we get the chiral anomaly.
Let us suppose that the Weyl fermions transform in the
representations $(\reprn_i,\reprn_{\mathrm{chiral},i})$ of the
symmetry group $G\times G_{\mathrm{chiral}}$. The anomaly then
depends on
$\sum_i\tr_{\reprn_{\mathrm{chiral}}}\big(T^{\mathrm{(chiral)}}_a\big)\tr_{\reprn_i}\big(T_bT_c\big)$.
We conclude that the anomaly can only be in the abelian factors of
the chiral symmetry group. We thus assume $G_{\mathrm{chiral}}$ to
be a $\un{1}$ symmetry. The different fermion representations have
charge $q_i$, i.e. $\psi_{(i)}\mapsto e^{i\al q_i}\psi_{(i)}$
where $\psi_{(i)}$ is the Weyl fermion in the $i^{\mathrm{th}}$
representation. The exact result of the calculation is
then:\footnote{We note that there here is a factor $\half$
compared to most standard calculations since we look at Majorana
fermions instead of Dirac fermions thus halving the degrees of
freedom. Also, we have changed sign compared to standard texts
(e.g.~\cite{peskinandschroeder}) since the sign depends on
$\tr{\ga^\mu\ga^\nu\ga^\rho\ga^\si\ga^5}$ and our sign here is
unconventional due to the definition of $\si^0$
in~(\ref{appspinorn9}). This was also the reason for the
unconventional sign on the $F\Ftilde$-term (i.e. the
$\vartheta$-term) in~(\ref{susyn69}) as we discussed there.}
\begin{equation}\label{dvn97}
    \expect{\partial_\mu j^\mu}=-\frac{1}{16\pi^2}\sum_{i}q_i\tr_{\reprn_i}\nleft(F_{
    \mu\nu}\Ftilde^{\mu\nu}\right)=-\frac{\sum_i
    q_iC\nleft(\reprn_i\right)}{16\pi^2}F^a_{
    \mu\nu}\Ftilde^{a\mu\nu},
\end{equation}
where we have used the quadratic invariant from~\eqref{susyn76.5}.
We see that the anomaly is proportional to
$\tr\big(F^{\mu\nu}\Ftilde_{\mu\nu}\big)$ and thus, by the
considerations above, the abelian chiral symmetry is anomalous in
the non-perturbative regime with non-abelian gauge
groups.\footnote{Since we saw in~(\ref{dvn90}) that
$\tr\nleft(F_{\mu\nu}\Ftilde^{\mu\nu}\right)=\partial_\mu K^\mu$
for some $K^\mu$, we might think that we could define a
non-anomalous current as $j^\mu-K^\mu$. However, this is not
possible since $K^\mu$ is not gauge invariant under large gauge
transformations as explained above.}

One can also calculate the anomaly in the path integral formalism.
The reason for the anomaly is simply that the measure has a
non-trivial Jacobian under the chiral transformation. To see this
we first promote the chiral symmetry to a local transformation
with $\psi_{(i)}'=e^{i\al(x) q_i}\psi_{(i)}$. One finds that under
this transformation the measure changes as:
\begin{equation}\label{dvn98}
\mathcal{D}\psi\mathcal{D}\psibar=\mathcal{D}\psi'\mathcal{D}\psibar'e^{-i\frac{\sum_i
    q_iC(\reprn_i)}{16\pi^2}\int\!\dlor\,\al(x)F^a_{
    \mu\nu}\Ftilde^{a\mu\nu}}.
\end{equation}
Letting $S[\psi]$ denote the action we then get:
\begin{multline}\label{dvn99}
    \int\mathcal{D}\psi\mathcal{D}\psibar\ldots
    e^{iS[\psi]}=\int\mathcal{D}\psi'\mathcal{D}\psibar'\ldots
    e^{iS[\psi']}\\
    =\int\mathcal{D}\psi\mathcal{D}\psibar\ldots e^{iS[\psi]+i\de S+i\frac{\sum_i
    q_iC(\reprn_i)}{16\pi^2}\int\!\dlor\,\al(x)F^a_{
    \mu\nu}\Ftilde^{a\mu\nu}}\\
    \approx\int\mathcal{D}\psi\mathcal{D}\psibar\ldots
    e^{iS[\psi]}\left(1+i\de S+i\frac{\sum_i
    q_iC(\reprn_i)}{16\pi^2}\int\!\dlor\,\al(x)F^a_{
    \mu\nu}\Ftilde^{a\mu\nu}\right),
\end{multline}
where the last line is to the first order in $\al$. Using that
under the localised chiral transformation the change in the action
is
\begin{equation}\label{dvn100}
    \de S=\int\!\dlor\,\al(x)\partial_\mu j^\mu,
\end{equation}
we immediately get~\eqref{dvn97}. Since the result agrees with the
above one-loop calculation, we conclude that the anomaly is
one-loop exact.

For an $\Nscr=1$ supersymmetric theory with simple gauge group we
see from~\eqref{susyn78.1} and~\eqref{dvn100} that the change in
the Lagrangian under the anomalous chiral symmetry corresponds to
changing the $\vartheta$-angle as
$\vartheta\mapsto\vartheta+2\al\sum_i q_iC(\reprn_i)$. In this way
the anomalous breaking of the chiral symmetry has been transformed
into an explicit breaking by the $\vartheta$-term. The effective
Lagrangian then has the symmetry:
\begin{eqnarray}\label{dvn101}
    \psi_{(i)}&\mapsto&e^{i\al q_i}\psi_{(i)},\nonumber\\
    \vartheta&\mapsto&\vartheta+2\al\sum_i q_iC(\reprn_i).
\end{eqnarray}
Since the anomaly is related to the $\vartheta$-angles, we
conclude that there can only be one independent anomalous chiral
symmetry for each simple factor in the gauge group.

We can use that the $\vartheta$-angle is periodic to see that the
anomalous chiral symmetry group is in fact not completely broken,
but broken to a discrete group (as done in e.g.~\cite{0303160}).
However, in our case we should be a bit careful since the
Lagrangian is not invariant under rescalings of the gauge group
generators as discussed at the end of
section~\ref{susysecchiralsubgauge}. We here appreciate that
in~\eqref{dvn101} $\vartheta$ and $C(\reprn_i)$ transform in the
same way under the rescalings. However, the instanton calculation
that showed $\vartheta$ to be periodic was done for a specific
normalisation chosen such that $C\nleft(\mathrm{fund}\right)=1/2$
for $\sun{N}$. Generalising this to any gauge group the
$\vartheta$-angle is $2\pi$ periodic if we choose a normalisation
(following~\cite{weinberg3}) such that for $T_a$, $T_b$ and $T_c$
in the ``standard'' $\sun{2}$ subalgebra we have the structure
constant:
\begin{equation}\label{dvn101.01}
    f_{abc}=\vep_{abc}.
\end{equation}
Given this normalisation we see that the chiral symmetry is
unbroken for the discrete set of $\al$'s obeying:
\begin{equation}\label{dvn101.2}
    2\al\sum_i q_iC(\reprn_i)\in2\pi\Z.
\end{equation}

We can now use our knowledge of the $F^{
\mu\nu}\Ftilde_{\mu\nu}$-term to constrain the
non-per\-tur\-ba\-ti\-ve corrections to the superpotential.

\subsection{Non-Perturbative Corrections}\label{dvsecweffsubnonpert}
The perturbative non-renormalisation theorem proven in
section~\ref{dvsecweffsubnon} depended on two symmetries: Real
translations of $\tau$ and an R-symmetry. With the knowledge from
the last section we see that both of these symmetries can be
broken non-perturbatively: $\tau$ can only be translated by
integers $\tau\mapsto\tau+n,\, n\in\Z$ (using~\eqref{dvn94}) and
the R-symmetry can be anomalous. However, we will now see that the
Seiberg scheme used in~\ref{dvsecweffsubnon} still constrains the
form of the effective superpotential (based on~\cite{weinberg3},
\cite{argyres}, \cite{9309335} and~\cite{9509066}).

Let us assume the same setup as in section~\ref{dvsecweffsubnon}.
This means that the Lagrangian at the UV energy $\mu_0$ is given
by~\eqref{dvn78} and we want to determine the effective
generalised superpotential at the lower energy $\mu$. The general
form of the Lagrangian at energy $\mu$ is (using holomorphy) again
given by~\eqref{dvn79}.

According to the Seiberg scheme in section~\ref{dvsecweffsubnon}
we shall now use symmetries and limits to constrain the
generalised superpotential. Firstly, the weak coupling limit of
the couplings $g_k$ should be smooth (we do not at this point
integrate out fields so we should also have smoothness in the
limit of masses going to zero). This means that we can rule out
negative powers of $g_k$. But also terms like $e^{-1/g_k^2}$ are
ruled out \cite{ferretti,0311066} since $g_k$ is complex and the
exponential diverges when e.g. $g_k$ goes to zero from the
imaginary direction in the complex plane. This means that we can
simply expand the generalised superpotential in non-negative
powers of the couplings $g_k$.

The dependence on $\tau$ is more complicated. Here we can have a
non-perturbative dependence as in $e^{2\pi i\tau}$ since the
imaginary part of $\tau$ given by~\eqref{susyn78} is $4\pi/g^2$
and thus positive. This means that in the limit of the gauge
coupling $g\rightarrow0$ we have $e^{2\pi i\tau}\rightarrow0$. We
can exchange $e^{2\pi i\tau}$ with $\La^b$ using the holomorphic
scale from~\eqref{dvn88} since the quotient only depends on the
scale. $\La^b$ also has the nice property that it is periodic in
$\vartheta$ and as we saw in~\eqref{dvn95} integer powers of
$\La^b$ can be obtained by instanton corrections. However, we can
have contributions that depend on (not necessarily integer) powers
of $\La$. Below we will actually see an example of an expectation
value with the wrong $\vartheta$-periodicity. The contributions of
$\La^a$ must have $a>0$ since $\La\rightarrow0$ corresponds to the
smooth weak coupling limit $g\rightarrow0$ assuming an
asymptotically free theory with $b>0$. We note that $\La$ depends
on $\tau$ holomorphically. Thus $\La$ should be regarded as a
background chiral superfield and the dependence on it should be
holomorphic. Naturally, we can also have a perturbative dependence
on $\tau$ through $\ln(\La)$. We know how this looks in
perturbation theory from equation~\eqref{dvn83} so let us extract
it from the effective generalised superpotential
(using~\eqref{dvn87}):
\begin{equation}\label{dvn102}
    F_\mu\nleft(\Phi^i,\W_\al,\ta,g_k\right)=-\frac{b}{32\pi^2
C(\reprn)}\ln\nleft(\frac{\La}{\mu}\right)\tr_{\repr}\nleft(\W^\al
    \W_\al\right)+F'_\mu\nleft(\Phi^i,\W_\al,\La,g_k\right).
\end{equation}
Please note that we have not yet proven that the perturbative part
should look like this, but simply extracted it in expectation of
such a term and indeed one can show that it will look like this.
We also note that $F'_\mu$ only depends on $\tau$ through powers
of $\La$. The reason is that there will be chiral anomalies. As we
saw in~\eqref{dvn101} the anomaly is cancelled by the first term
if we translate $\tau$ under the anomalous chiral symmetry and
hence $F'_\mu$ should be invariant under the symmetry. This can
only happen if the dependence of $\tau$ is through powers of $\La$
that under~\eqref{dvn101} has a definite charge.

Let us now turn to the symmetries that can restrict $F'_\mu$.
These are naturally the same as we used in
section~\ref{dvsecweffsubnon}: We have a global symmetry,
$\un{1}_i$, rotating $\Phi^i$, but not the rest of the chiral
fields: $\Phi^j\mapsto e^{i\al\de_{ij}}\Phi^j$. Here $\Phi^i$
transforms in the representation $\reprn_i$ of the gauge group. To
ensure the $\un{1}_i$ symmetry with non-zero couplings we must as
before assign charges, $-N_{k,\Phi^i}$, to the couplings $g_k$
where $N_{k,\Phi^i}$ is the order of $\Phi^i$ in the term with
coupling $g_k$ in the bare superpotential~\eqref{dvn78.5}. Since
this is a chiral symmetry, it can be anomalous and thus we should
also assign a charge to $\La$. The other symmetry that we used in
section~\ref{dvsecweffsubnon} is the R-symmetry, $\un{1}_R$. Here
we found that $R\nleft(\Phi^i\right)=0$ and
$R\nleft(\W_\al\right)=1$. Using~\eqref{susyn35}, \eqref{susyn59}
and that $R\nleft(\tha\right)=1$ this means that the spinor from
the chiral multiplet has charge $-1$ and the gaugino has charge
$+1$. Thus it is a chiral symmetry and it can be anomalous.
Using~\eqref{dvn101} we can find the corresponding charge of
$\La$. The charges for the symmetries are given in
table~\ref{dvtablen2}.
\begin{table}
\caption{}\label{dvtablen2}
\begin{flushright}
\begin{tabular}{|l|cc|}
  \hline
   & $\un{1}_i$ & $\un{1}_R$\\\hline
   $\Phi^j$ & $\de_{ij}$ & $0$\\
   $g_k$ & $-N_{k,\Phi^i}$ & $2$\\
   $\W_\al$ & $0$ & $1$ \\
   $\La^b$ & $2C\nleft(\reprn_i\right)$ & $2C\nleft(\mathrm{adj}\right)-2\sum_iC\nleft(\reprn_i\right)$\\\hline
\end{tabular}
\newline
\begin{center}
Charges for the symmetries constraining the effective generalised
superpotential. $N_{k,\Phi^i}$ is the order of $\Phi^i$ in the
term with coupling $g_k$ in the bare superpotential. Based
on~\protect{\cite{argyres}}.
\end{center}
\end{flushright}
\end{table}

The effective superpotential now heavily depends on the sign of
the charge of $\La^b$ under the anomalous R-symmetry. For a
positive charge there is only one possible power of $\La$ whereas
for a negative charge, $\La$ can be used to compensate the
positive charges of $g_k$ and $\W_\al$ thus allowing arbitrary
powers of these.

Let us focus on the case where the charge is zero and the
R-symmetry is not anomalous. This is exactly the case for the
Lagrangian $\lagrun$ presented in the Dijkgraaf-Vafa conjecture in
section~\ref{dvsecdvun} where we ignore the abelian part of the
gauge group. Here we only have one chiral field in the adjoint
representation exactly compensating the contribution to the
anomaly from the gaugino. Using that the effective generalised
superpotential should have charge two under the R-symmetry we get
two kinds of terms in $F'_\mu$: One term linear in the $g_k$'s and
one term proportional to two $\W_\al$'s. This means that the
effective generalised superpotential takes the form:
\begin{multline}\label{dvn103}
    F_\mu\nleft(\Phi^i,\W_\al,\ta,g_k\right)=-\frac{b}{32\pi^2
C(\reprn)}\ln\nleft(\frac{\La}{\mu}\right)\tr_{\repr}\nleft(\W^\al
    \W_\al\right)+\sum_kg_kO_{\mu,k}\nleft(\Phi^i,\La\right)\\
    +h_{\mu,ab}\nleft(\Phi^i,\La\right)\W^{\al
    a}\W^b_\al.
\end{multline}
We can constrain the dependence on $\Phi^i$ and $\La$ using the
$\un{1}_i$ symmetry. Let us for simplicity assume that we only
have one chiral field $\Phi$ which is in the adjoint
representation. Let $N_\Phi$ and $N_\La$ respectively denote the
powers of $\Phi$ and $\La$ in a given term. Then we see from
table~\ref{dvtablen2} that for $h_{\mu,ab}$ we have
$N_\Phi=-\frac{2C(\mathrm{adj})}{b}N_\La$. Thus the term
independent of $\Phi$ is also independent of $\La$ and will in
fact simply give the one-loop running of $\tau$ that we have
already taken out. The rest of the terms in $h_{\mu,ab}$ have
negative powers of $\Phi$ thus giving new non-perturbative
contributions. For $O_{\mu,k}$ we have
\begin{equation}\label{dvn103.5}
    N_\Phi=N_{k,\Phi}-\frac{2C\nleft(\mathrm{adj}\right)}{b}N_\La.
\end{equation}
Setting $N_\La=0$ we have $N_\Phi=N_{k,\Phi}$ and considering the
weak coupling limit we see that this gives us the bare
superpotential. For $N_\La>0$ we get corrections to terms with
$N_\Phi<N_{k,\Phi}$. Thus a term of order $N_\Phi$ in $\Phi$ only
receives new contributions arising from terms in the bare
superpotential of higher order in $\Phi$.

To be able to perform further analysis of the effective Lagrangian
we will in the next subsection introduce the concept of
integrating in matter.

Let us end this section with an example of the breaking of the
$\un{1}_R$ symmetry to a finite group. We consider the $\Nscr=1$
supersymmetric Yang-Mills theory. In this case we only have one
adjoint fermion, the gaugino, with charge $+1$.
Thus~\eqref{dvn101.2} shows us that the R-symmetry is unbroken if
$2\al C(\mathrm{adj})\in2\pi\Z$. It is here customary to introduce
the dual Coxeter number, $h$. The precise definition of the dual
Coxeter number is~\cite{polchinski2}:
\begin{equation}\label{dvn103.6}
    \tr_{\mathrm{adj}}\nleft(T^{(\mathrm{adj})}_aT^{(\mathrm{adj})}_b\right)=h\psi^2\de_{ab},
\end{equation}
where $\psi^2$ is the length of the highest root. This definition
makes the dual Coxeter number invariant under rescaling of the
generators. The dual Coxeter numbers for the classical and
exceptional groups can be found in table~\ref{dvtablen3} along
with other group theoretical facts.
\begin{table}
\caption{}\label{dvtablen3}
\begin{flushright}
\begin{tabular}{|l|c|c|c|c|c|c|c|c|c|}
  \hline
   & $\sun{N}$ & $\mathrm{SO}(N),\,N\geq4$ & $\mathrm{Sp}(k)$ & $\mathrm{E}_6$& $\mathrm{E}_7$& $\mathrm{E}_8$& $\mathrm{F}_4$& $\mathrm{G}_2$\\\hline
$\mathrm{dim}(G)$&$N^2-1$&$N(N-1)/2$&$2k^2+k$&$78$&$133$&$248$&$52$&$14$\\\hline
$\mathrm{rank}(G)$&$N-1$&$[N/2]$&$k$&6&7&8&4&2\\\hline
$C\nleft(\mathrm{fund}\right)$&1/2&1&1/2&&&&&\\\hline
$h(G)$&$N$&$N-2$&$k+1$&$12$&$18$&$30$&$9$&$4$\\\hline
\end{tabular}
\newline
\begin{center}
Dimensions, $\mathrm{dim}(G)$, ranks, $\mathrm{rank}(G)$, and dual
Coxeter numbers, $h(G)$, for the classical and exceptional groups.
For the classical groups the quadratic invariant in the
fundamental representation, $C\nleft(\mathrm{fund}\right)$, is
also given -- in a normalisation such that the highest root has
length $1$. Based on~\protect{\cite{argyres}} and~\protect{\cite{polchinski2}}.
\end{center}
\end{flushright}
\end{table}
We should now remember that~\eqref{dvn101.2} holds
true for a normalisation such that~\eqref{dvn101.01} is fulfilled
for the standard $\sun{2}$ subalgebra. With this normalisation we
have $\psi^2=1$ such that $C(\mathrm{adj})=h$. Thus
by~\eqref{dvn101.2} the R-symmetry is broken to $\Z_{2h}$ i.e. the
gaugino transforms as:\footnote{\label{dvfootinstbreak}This result
can also be obtained by an instanton calculation
(following~\cite{9701069}). In the case of a $\sun{N}$ gauge group
we have $2N=2h$ zero-modes from the gaugino. The first
non-vanishing correlator must have $2h$ gaugino insertions to soak
the zero modes. Requiring this correlator to be invariant we see
that $\un{1}_R$ breaks to $\Z_{2h}$.}
\begin{equation}\label{dvn104}
    \la_\be\mapsto e^{i\al}\la_\be,\qquad
    \al=\frac{2\pi }{2h}n,\qquad
    n\in\Z_{2h}.
\end{equation}

\subsection{ILS Linearity Principle and Integrating In}\label{dvsecweffsubilsintein}

In the last section we did not consider integrating out massive
fields which, as we explained in
section~\ref{dvsecwilsonsubinteout}, is possible at energy scales
below the masses of the fields. In this section we will do this
and even see that we can integrate the fields back in.

The setup will be the same as in the last subsection and
section~\ref{dvsecweffsubnon} with an $\Nscr=1$ supersymmetric
Lagrangian with simple gauge group. However, we will assume that
the tree level (i.e. the bare) superpotential $\wtree$ that was
given by~\eqref{dvn78.5} now is linear in the basis for
holomorphic gauge invariants $X_k\nleft(\Phi^i\right)$ introduced
in section~\ref{dvsecvacsubclmoduli}:\footnote{The $X_k$ can be
constrained. This can be fixed by including Lagrange multipliers
in the effective superpotential.}
\begin{equation}\label{dvn105}
    \wtree=\sum_k g_k X_k.
\end{equation}

Let us for now ignore the dependence on $\W_\al$ in the effective
superpotential -- we will return to this later. In~\cite{9403198}
K. Intriligator, R. G. Leigh and N. Seiberg (ILS for short)
conjectured that the effective (generalised) superpotential can be
put into the form:\footnote{In writing this we use that gauge
invariance requires the $\Phi^i$ dependence to be expressed
through the basic gauge invariants $X_k$.}
\begin{equation}\label{dvn106}
    \weff\nleft(X_k,\La,g_k\right)=W_{\mathrm{dyn}}\nleft(X_k,\La\right)+\wtree\nleft(X_k,g_k\right).
\end{equation}
This is called the ILS linearity principle since it states that
the dependence on the couplings $g_k$ is linear. However, we note
that it further asserts that the term depending on the couplings
is the tree-level superpotential. This can often be obtained using
the Seiberg scheme from section~\ref{dvsecweffsubnon}, but
actually it is not the form we found in~\eqref{dvn103} for the
case of a single adjoint chiral superfield. Equation
\eqref{dvn103} is nicely linear in the couplings, but the term
depending on the couplings is not the tree-level superpotential.
However, the conjecture in~\cite{9403198} is that one can always
redefine the fields $X_k$ as a function of the couplings $g_k$ to
bring the effective superpotential into the form~\eqref{dvn106}.

Since this is not proven directly let us see that it is true in a
special case using~\eqref{dvn103}. We assume that the gauge group
is $\sun{N}$ and that the tree-level superpotential is given by
$\wtree=\half mX_2+gX_3$ where, as we found in~\eqref{dvn47}, the
basic gauge invariants here are
$X_k=\tr\nleft(\Phi^k\right),\,{k\leq N}$.
Using~\eqref{dvn103},~\eqref{dvn103.5} and assuming only
non-negative integer powers of $\Phi$ we get (dropping constant
terms and using $b=2C\nleft(\mathrm{adj}\right)=2N$ in this case):
\begin{equation}\label{dvn107}
    \weff\nleft(X_k,\La,g_k\right)=\wtree+cg\La X_2,
\end{equation}
where $c$ is a constant and we have used that the only $X_k$ of
order less than three in $\Phi$ is $X_2$ ($X_1=\tr\Phi=0$ in the
non-abelian case). By redefining $X'_3=X_3$ and $X'_2=g X_2$ we
get $\wtree=\frac{1}{2}\frac{m}{g}X'_2+gX'_3$ and
$W_{\mathrm{dyn}}=c\La X'_2$. Here $W_{\mathrm{dyn}}$ is
independent of the couplings $m'=m/g$ and $g'=g$ as wanted.

Let us now continue with the general case following~\cite{9403198}
and~\cite{9407106}. We assume that we are at so low energies that
we can integrate out some massive field say $X_0$
using~\eqref{dvn77} (or rather when working with the elementary
fields $\Phi^i$ we integrate out all $X_k$ that involve $\Phi^i$).
Ignoring the \kahler{}-term at these low energies~\eqref{dvn77}
shows that we integrate $X_0$ out by solving the equation of
motion
\begin{equation}\label{dvn108}
    \frac{\partial\weff}{\partial
    X_0}\nleft(\expect{X_0}\right)=0
\end{equation}
for $\expect{X_0}\nleft(X_1,\ldots,X_n,\La,g_k\right)$ (assuming
$k=0,\ldots, n$) and inserting this back into $\weff$ thus
obtaining the effective action, $W_{\mathrm{eff},\mathrm{L}}$,
with $X_0$ integrated out:
\begin{multline}\label{dvn109}
    W_{\mathrm{eff},\mathrm{L}}\nleft(X_1,X_2,\ldots,X_n,g_k,\La\right)=\weff\nleft(\expect{X_0},X_1,\ldots,X_n,g_k,\La\right)\\
    =W_{\mathrm{dyn}}\nleft(\expect{X_0},X_1,\ldots,X_n,\La\right)+\sum_{k\neq0}g_kX_k+g_0\expect{X_0}.
\end{multline}
The reason for the index ``$\mathrm{L}$'' is that this is nothing
but the Legendre transform of
$W_{\mathrm{dyn}}\nleft(X_k,\La\right)$ in~\eqref{dvn106} as we
see by rewriting~\eqref{dvn108} as $\frac{\partial
W_{\mathrm{dyn}}}{\partial X_0}=-g_0$. Using that the implicit
dependence on $g_0$ through $\expect{X_0}$ is zero by the virtue
of~\eqref{dvn108} we get that:\footnote{We note that if the
tree-level superpotential was given as in~(\ref{dvn78.5}) with the
$O_k$'s being -- not necessarily linear -- functions of the basic
gauge invariants, then the right hand side in~(\ref{dvn110}) would
be $O_k\nleft(\expect{X_0}\right)$.}
\begin{equation}\label{dvn110}
    \frac{\partial W_{\mathrm{eff},\mathrm{L}}}{\partial
    g_0}=\expect{X_0}.
\end{equation}
This is nothing but the inverse Legendre transform. If we know the
$g_0$ dependence in the effective Lagrangian with $X_0$ integrated
out, we can use~\eqref{dvn110} to solve for $g_0$ as a function of
$\expect{X_0}$ and the other fields and couplings, and then
use~\eqref{dvn109} to obtain $W_{\mathrm{dyn}}$. $\weff$ is then
obtained by adding $\wtree$. This is called \emph{integrating in}.
This means that we loose no information in integrating out a field
and we can see $X_k$ and $g_k$ as being dual.

Whereas the dependence on $g_0$ in $W_{\mathrm{eff},\mathrm{L}}$
is complicated, the linearity principle still applies for the rest
of the fields since $\partial W_{\mathrm{eff},\mathrm{L}}/\partial
g_k=X_k$ for $k\neq0$ by~\eqref{dvn108}. This means that we can
continue the integrating out (and in) for the rest of the fields.
Without loss of information we can obtain an effective Lagrangian
only depending on the couplings $g_k$ and $\La$.

The integrating in procedure through equation~\eqref{dvn110}
should not be unfamiliar since this is just the equation we have
for the 1PI effective superpotential. The couplings are here the
external currents. Actually, we could have carried out our
treatment in the 1PI formalism (if well-defined) and if the 1PI
action and the Wilsonian action agrees, this would prove the
linearity principle~\cite{9509066}.

The integrating in method is powerful (following~\cite{9407106})
since it sometimes allows us to determine the effective
superpotential for an ``upstairs'' theory with an extra massive
field $\Phihat$ from a known effective superpotential of a
``downstairs'' theory simply by integrating in the $\Phihat$
field. Two points should, however, be borne in mind. Firstly, the
holomorphic scales depend on the matter representation as noted in
section~\ref{dvsecweffsubnon}. Let $\La_d$ denote the scale for
the downstairs theory and let $b_d$ be the corresponding constant
in the $\beta$-function determined by~\eqref{dvn85}.
Correspondingly for the upstairs theory we have $\La_u$ and $b_u$.
If the gauge group is not simple, we have one such pair for each
simple factor. Assuming simple thresholds (or absorbing factors
into the $\La$'s) we can simply compare the running gauge
couplings $g\nleft(\mu\right)$ (for each factor) at the mass, $m$,
of the field to be integrated in (using~\eqref{dvn86}):
\begin{equation}\label{dvn111}
    \La_d^{b_d}=\La_u^{b_u}m^{C(\reprn)},
\end{equation}
where $C\nleft(\reprn\right)$ is the quadratic invariant for the
representation of $\Phihat$. This follows from~\eqref{dvn85}
yielding $b_u=b_d-C\nleft(\reprn\right)$.

The second point to consider is that when flowing to the upstairs
theory, adding a tree-level superpotential with couplings $g_k$,
and then flowing down again by integrating out $\Phihat$ as
in~\eqref{dvn109} does not give the downstairs effective
superpotential $W_{\mathrm{eff},d}$ back, but rather:
\begin{equation}\label{dvn112}
    W_{\mathrm{eff},\mathrm{L}}=W_{\mathrm{eff},d}\nleft(X_k,\La_d\right)+W_I\nleft(X_k,\La_u,g_k\right),
\end{equation}
where $W_I$ is the renormalisation group irrelevant term that the
downstairs theory does not know about. $W_I$ must vanish for
$g_k=0$ or $m\rightarrow\infty$. Only if such a $W_I$
superpotential can be ruled out, we can use the integrating in
procedure from the downstairs theory to the upstairs.

Let us now consider the dependence on $\W_\al$. We will assume
that the dependence hereon is always\footnote{E.g.
in~(\ref{dvn103}) the dependence on $\W_\al$ is not necessarily
through $\Shat$ so this is an assumption.} through the (traceless
since we study a simple group) glueball superfield, $\Shat$,
defined as in~\eqref{dvn4}, but now, naturally, for any simple
group. \eqref{dvn4} was written for
$C\nleft(\mathrm{fund}\right)=1/2$. Generally we define $\Shat$
as:
\begin{equation}\label{dvn112.5}
    \Shat=-\frac{1}{32\pi^2
C(\reprn)}\tr_{\repr}\nleft(\W^\al\W_{\al}\right).
\end{equation}
This is a massive field so at low energies, which we will assume
to be at, it is integrated out. Looking at the tree-level
Lagrangian with the UV cut-off $\mu_0$ the gauge kinetic
term~\eqref{susyn77} takes the form (using~\eqref{dvn87} as we did
in~\eqref{dvn102}):
\begin{equation}\label{dvn113}
    W_{G}=2\pi i\tau\Shat=\ln\nleft(\frac{\La^b}{\mu^b_0}\right)\Shat.
\end{equation}
Thus we can see $\ln\nleft(\La^b/\mu^b_0\right)$ as the coupling
for $\Shat$. Assuming that the principle of linearity\footnote{As
noted in~\cite{9407106} the assumption of simple thresholds above
is actually a generalisation of the linearity principle for
$\Shat$.} also holds for $\Shat$, we can integrate $\Shat$ back in
analogously to~\eqref{dvn110} by solving:
\begin{equation}\label{dvn114}
    \frac{\partial W_{\mathrm{eff}}}{\partial
    \ln\nleft(\frac{\La^b}{\mu^b_0}\right)}=\expect{\Shat}.
\end{equation}
This gives $\La$ as a function of $\expect{\Shat}$ and $X_r$ (or
$g_k$ if the fields have been integrated out) which we can
substitute back in to obtain $W_{\mathrm{dyn}}$ as a function of
$\Shat$ (in analogy with~\eqref{dvn109}):
\begin{equation}\label{dvn114.5}
    W_{\mathrm{dyn}}\big(X_k,\Shat\big)=W_{\mathrm{eff}}\nleft(X_k,\La\big(X_k,\Shat\big)\right)-\ln\nleft(\frac{\La\big(X_k,\Shat\big)^b}{\mu^b_0}\right)\Shat.
\end{equation}
To obtain the effective superpotential we simply add the
tree-level potential~\eqref{dvn113}. Naturally, the $\mu_0$
dependence then goes out so the low-energy effective
superpotential is independent of this scale. Now we can see
$\ln\nleft(\La^b/\mu^b_0\right)$ and $\Shat$ as a canonical pair
just as $g_k$ and $X_k$ was above. We can also extend this
analysis to the case where we have a semi-simple gauge group with
more than one holomorphic scale and corresponding glueball
superfields. We should mention that we were able to perform the
integrating in procedure since there can be no $W_I$-term as
in~\eqref{dvn112} because the total $\La$ dependence is accounted
for in $\weff$~\cite{0210135}.

We note that the meaning of the superpotential with $\Shat$
integrated in is unclear since it is a massive field that really
should be integrated out at the low energies. However, it is this
potential that we determine in the Dijkgraaf-Vafa conjecture --
with all matter fields integrated out and the glueball superfields
integrated in. This superpotential has the virtue that the
expectation values of the glueball superfields are found by the
equations of motion analogous to~\eqref{dvn108}:
\begin{equation}\label{dvn115}
    \frac{\partial\weff}{\partial
    \Shat}\big(\expect{\Shat}\big)=0.
\end{equation}
Thus we have explained equation~\eqref{dvn6.5}.

We should not think of the Dijkgraaf-Vafa conjecture as depending
on the ILS linearity principle since we in the next chapter will
prove the conjecture (i.e. obtain $\weffpert$) diagrammatically
without using the linearity principle. The linearity principle can
actually quite easily be proven~\cite{0211274} using the
Dijkgraaf-Vafa conjecture and the techniques we develop in the
diagrammatical proof. But one can also prove the Dijkgraaf-Vafa
conjecture using the linearity principle along with Seiberg-Witten
curves~\cite{0210135}. To complete the picture one can quickly
obtain the Konishi anomaly using the linearity principle and the
anomalous $\un{1}_i$ symmetry from table~\ref{dvtablen2} -- and
the other main proof of the Dijkgraaf-Vafa
conjecture~\cite{0211170} actually uses a generalised form of this
Konishi anomaly.

We will use the integrating in procedure to obtain the
Veneziano-Yankielowicz superpotential in
section~\ref{dvsecweffsubvy}, but let us first review some of the
lore of gauge dynamics.

\subsection{The Lore of Gauge Dynamics}\label{dvsecweffsublore}

One of the aims in obtaining the low-energy effective action is to
be able to determine what \emph{phase} the theory is in. The phase
of a theory depends on the parameters of the theory and the choice
of vacuum state. It is characterised by the energy potential,
$V_{\textrm{elec}}\nleft(R\right)$, between two electrical test
charges separated by a large distance, $R$. By electric we here
mean in the abstract gauge group sense. The presentation is based
on~\cite{argyres}, \cite{9704114} and~\cite{9509066} and we are in
four space-time dimensions.

It is here important for us that G. 't Hooft and A. Polyakov
showed the possibility of magnetic monopoles for non-abelian gauge
groups.\footnote{What is meant here is that the UV Lagrangian has
a non-abelian gauge symmetry. Naturally, this gauge group could be
spontaneously broken in the IR.} These 't Hooft-Polyakov monopoles
were obtained as solitonic solutions in the Georgi-Glashow model,
however, it was first with the Seiberg-Witten theory for $\Nscr=2$
supersymmetric theories that exact calculations could be done of
e.g. monopole condensation.
\begin{table}
\caption{}\label{dvtablen4} \begin{displaymath}
        \xymatrix@-1.5pc{
            \textrm{Phase:}&V_{\textrm{elec}}\nleft(R\right)\sim&V_{\textrm{magn}}\nleft(R\right)\sim&\textrm{E-M duality}\\
            \textrm{Coulomb}&\dfrac{1}{R}&\dfrac{1}{R}&\quad\ar@(ur,dr)[]\\
            \textrm{Free
            electric}&\dfrac{1}{R\ln(R\La)}&\dfrac{\ln(R\La)}{R}&\ar@(r,r)@{<->}[d]\\
            \textrm{Free
            magnetic}&\dfrac{\ln(R\La)}{R}&\dfrac{1}{R\ln(R\La)}&\\
            \textrm{Higgs}&\mathrm{constant}&\rho R&\ar@(r,r)@{<->}[d]\\
            \textrm{Confining}&\si R&\mathrm{constant}&
            %\save "1,1"."6,4"*+![F]\frm{} \restore
}
    \end{displaymath}\relax
\begin{center}
The table shows the characterising behaviour of the electric and
magnetic potential for the different phases of gauge theories.
Also shown is the electric-magnetic (E-M) duality.
\end{center}
\end{table}

Table~\ref{dvtablen4} shows the different conjectured phases
characterised by the behaviour of the electric potential,
$V_{\textrm{elec}}\nleft(R\right)$. The behaviour is only
determined up to an additive constant. In the first three phases
(Coulomb, free electric and free magnetic) there are massless
gauge fields and the potentials are of the form $e^2(R)/R$ where
$e(R)$ is the renormalised charge. In the Coulomb phase the charge
is constant. In the free electric phase the massless charged
particles renormalise the charge to zero as $R\rightarrow\infty$
while in the free magnetic phase massless monopoles renormalise
the charge to infinity at large distances. In
table~\ref{dvtablen4} the behaviour of the magnetic potential,
$V_{\textrm{magn}}\nleft(R\right)$, between two magnetic test
charges is also shown. For the first three phases we directly get
the magnetic potential from the electric by the Dirac quantisation
condition $e(R)g(R)\sim1$ where $g(R)$ is the renormalised
magnetic charge.

The Higgs phase is characterised by condensation of an
electrically charged particle. This gives a mass gap to the theory
via the Higgs-mechanism. The potential is then of the Yukawa type
and thus exponentially decays to zero at large $R$. This can be
seen by the charges being screened by the condensate or
equivalently by the gauge bosons acquiring mass. The flux between
two magnetic sources, on the other hand, is confined into a thin
flux-tube with constant tension $\rho$ thus giving the linear
potential. This is in analogy with the Meissner effect from
superconductivity.

The confining phase is the phase of our interest. This phase is
solely for non-abelian groups whereas we can find the above for
both abelian as well as non-abelian groups. Empirically we know
this phase from QCD: In the UV the degrees of freedom are the
gluons and the coloured quarks while in the IR we have the
colourless hadrons - i.e. colour confinement. The qualitative
explanation of confinement can, as suggested by Mandelstam and 't
Hooft, be seen as a dual Meissner effect where the confining phase
is dual to the Higgs phase. The duality here is the
electric-magnetic duality that exchanges the electric and magnetic
charges. As is also indicated in table~\ref{dvtablen4} this
duality exchanges the free electric and free magnetic phase. The
Coulomb phase is self-dual which one finds easily in the abelian
case, but in the non-abelian case one again has to go to
supersymmetric theories where it is part of the Montonen-Olive
duality. Getting back to the confining phase this can then be seen
as the electric-magnetic dual of the Higgs phase. Here it is now
monopoles that form a condensate and the electric flux between the
two electric test charges is confined to a thin tube with constant
string tension $\si$. The corresponding linear potential
$V_{\textrm{elec}}\nleft(R\right)\sim\si R$ shows that it requires
an infinite amount of energy to separate two charged particles.
This explains why we only see the gauge-invariant fields (hadrons
in QCD) at low energy. As in the Higgs phase we also have a mass
gap in the confining phase.

The confining phase has another characteristic: If we only have
adjoint matter, the Wilson loop operator for large loops satisfies
the area law $\expect{\tr\mathcal{P}e^{i\oint\mathrm{d}x^\mu
A_\mu}}\sim e^{-\si\cdot\mathrm{Area}}$ where ``Area'' is the area
of the Wilson loop.\footnote{This is seen by choosing a
rectangular loop with length $T$ in the time direction and length
$R$ in a space direction. The interpretation of the Wilson loop is
then that it measures the Euclidean action of a process where two
heavy charged particles are created and then separated by a
distance $R$ for a time $T$ before they are annihilated. We then
get $\expect{\tr\mathcal{P}e^{i\oint\mathrm{d}x^\mu A_\mu}}=
e^{-TV(R)}$ and inserting $V(R)\sim\si R$ gives the result.} As
opposed to this the Wilson loop operator in the Higgs phase will
rather depend on the perimeter of the loop. However, when we have
particles in the fundamental representation we can not distinguish
the confining and the Higgs phase because virtual pairs can screen
the sources. As a last remark we can also have dyons (particles
with both electric and magnetic charge) if we have a non-zero
$\vartheta$-angle. If the dyons condensate, we get a phase called
oblique confinement.

Let us now turn to supersymmetric gauge theories. Let us
concentrate on supersymmetric Yang-Mills theory with a simple
gauge group i.e. pure superglue. We want to know what happens at
low energy i.e. below the strong coupling scale, $\abs{\La}$.
Based on e.g. lattice simulations it is believed that we have
confinement and a mass gap.

But the theory also shows another phenomenon characteristic for
the strong coupling regime of gauge theories, namely spontaneous
breaking of chiral symmetry. We look at the $\un{1}_R$ symmetry
from table~\ref{dvtablen2}. This is simply measuring the gaugino
number since we have no chiral matter. We have already seen
in~\eqref{dvn104} that the symmetry is anomalous and broken to the
discrete group $\Z_{2h}$ where $h$ is the dual Coxeter number from
table~\ref{dvtablen3}. As explained in
footnote~\ref{dvfootinstbreak} the breaking to $\Z_{2h}$ could
also be seen using that the first possibly non-zero correlator in
an instanton background is $\expect{(\la\la)^h}\propto\La^{3h}$
where $\la^a_\al$ is the gaugino field. The dependence on $\La$
has been found using dimensional analysis and by requiring a
holomorphic dependence on $\tau$ as in
section~\ref{dvsecweffsubholo}. However, at strong coupling the
R-symmetry is spontaneously broken further down to $\Z_2$ by the
gaugino bilinear getting a non-zero dynamical expectation value
i.e. gaugino condensation:
\begin{equation}\label{dvn116}
    \expect{\la\la}\neq 0 \quad\longleftrightarrow\quad
    \Z_{2h}\mapsto\Z_2.
\end{equation}
Here the non-trivial element in $\Z_2$ simply works as a sign
change on $\la^a_\al$ since referring to~\eqref{dvn104} the only
symmetry in $\Z_{2h}$ leaving $\la\la$ invariant is the one with
$n=h$ which simply gives a sign change. We note that the gaugino
condensate can just as well be described by the traceless glueball
superfield getting a non-zero expectation value since its lowest
component is the gaugino bilinear as mentioned in
section~\ref{dvsecdvunsubdv}. The glueball superfield is further
believed to be the relevant field for the low energy theory.

Associated with the breaking of $\Z_{2h}\mapsto\Z_2$ we get $h$
inequivalent vacua. This is because states that are related by the
generators of $\Z_{2h}/\Z_{2}$ are no longer treated as
equivalent. The generators of $\Z_{2h}/\Z_{2}$ are $e^{ik\pi R/h}$
where $k=0,1,\ldots, h-1$ and $R$ is the gaugino number. Working
with these generators on a given vacuum state we get the $h$
inequivalent states $\ket{k}$ with $k$ as above. This also solves
another problem: Using dimensional analysis and holomorphy as
before, we get $\expect{\la\la}=\mathrm{constant}\times\La^3$. In
the proper normalisation we know from~\eqref{dvn93} that
$\vartheta$ should be $2\pi$-periodic. But using~\eqref{dvn85} we
see that $b=3C\nleft(\mathrm{adj}\right)=3h$ and hence
by~\eqref{dvn88} we have $\La^{3}\sim e^{i\vartheta/h}$ which is
only $2\pi h$-periodic. However, using the definition of the $h$
vacua $\ket{k}$ the expectation value of the gaugino condensate
depends on the vacua and is given by:
\begin{equation}\label{dvn117}
    \expect{\la\la}_k=a\La^3e^{2\pi ik/h},
\end{equation}
where $a$ is a constant independent of the chosen vacuum and the
index $k$ refers to in which vacuum the expectation value is
taken. However, using the anomaly~\eqref{dvn101} we see that
effect of $e^{ik\pi R/h}\in\Z_{2h}/\Z_{2}$, which gives a change
in vacuum as $\ket{n}\mapsto\ket{n+k}$ (modulo $h$), is equivalent
to translating $\vartheta\mapsto\vartheta+k2\pi$ which just gives
the exponential in~\eqref{dvn117}. In this way all observables of
the theory are $2\pi$ periodic even though the gaugino condensate
in a given vacuum is not. Each of these $h$ vacua has a mass gap.
We note that this is possible since the breaking of the chiral
symmetry now allows a mass-term for the gaugino.

For any supersymmetric theory we have the gaugino number which now
is not simply $\un{1}_R$ from table~\ref{dvtablen2}, but rather
given by summing the generators of $\un{1}_R$ and the
$\un{1}_i$'s. By gaugino condensation we will again have $h$
vacua.

Let us end this subsection by noting that the existence of the $h$
vacua are strongly suggested by the Witten index being
$\tr(-1)^{N_F}=h$. Here the trace is over the zero energy states
and $N_F$ is the fermion number operator introduced in
section~\ref{susysecwigner}. This corresponds to each of the $h$
vacua contributing with $(-1)^{N_F}=1$ to the Witten index. That
the Witten index for supersymmetric Yang-Mills theory is equal to
the dual Coxeter number is true for both the classical and the
exceptional groups. We note that this also means that
supersymmetry is unbroken in the strongly coupled regime.

\subsection{The Veneziano-Yankielowicz Superpotential}\label{dvsecweffsubvy}

In this section we will consider the low energy effective action
for the $\Nscr=1$ supersymmetric Yang-Mills theory in the case of
a simple gauge group. More precisely, we want to determine the
glueball superpotential with $\Shat$ integrated in as discussed in
the section~\ref{dvsecweffsubilsintein}. The superpotential was
first found, prior to the integrating in method, by G. Veneziano
and S. Yankielowicz in~\cite{venezianoyankielowicz} as the unique
form that fulfils the proper anomaly matching conditions. We will
here derive the superpotential using the integrating procedure (as
done for $\sun{N}$ in~\cite{0211170} and~\cite{argyres}). To do
this we have to know the low energy effective superpotential,
$\weff\nleft(\La\right)$, as a function of the holomorphic scale
$\La$. Assuming that we have chiral symmetry breaking by gaugino
condensation as described in the last subsection, the expectation
value of $\Shat$ must be non-zero as determined
from~\eqref{dvn117}:
\begin{equation}\label{dvn118}
    \expect{\Shat}=a\La^3,
\end{equation}
where $a$ is a constant (different from the one in~\eqref{dvn117})
and we look at a particular vacuum among the inequivalent vacua.
Using~\eqref{dvn115} we can see this as a differential equation
for the glueball superpotential which can then be determined. But
let us take an extra step and first use~\eqref{dvn114} to obtain:
\begin{equation}\label{dvn119}
    \weff\nleft(\La\right)=C\nleft(\mathrm{adj}\right)a\La^3.
\end{equation}
Where we, by~\eqref{dvn85}, have used that
$b=3C\nleft(\mathrm{adj}\right)$. We can then use~\eqref{dvn114.5}
to obtain $W_{\mathrm{dyn}}\big(\Shat\big)$ noting that
by~\eqref{dvn118} $\La\big(\Shat\big)=\big(\Shat/a\big)^{1/3}$.
Adding the tree-level superpotential~\eqref{dvn113} finally gives
the effective superpotential -- the Veneziano-Yankielowicz
superpotential:
\begin{equation}\label{dvn120}
    \wvy\big(\Shat\big)=C\nleft(\mathrm{adj}\right)\Shat\bigg(1-\ln\frac{\Shat}{a\La^3}\bigg).
\end{equation}
The $\Shat\ln\Shat$ term in this expression is multi-valued so
rotating $\Shat\mapsto e^{2\pi i}\Shat$ gives
$\wvy\mapsto\wvy-2\pi iC\nleft(\mathrm{adj}\right)\Shat$. But
from~\eqref{dvn113} we see that this is exactly cancelled by
rotating $\vartheta\mapsto\vartheta+2\pi
C\nleft(\mathrm{adj}\right)$ in perfect agreement with
$\Shat\mapsto e^{2\pi i}\Shat$ corresponding to a chiral
transformation of the gaugino as $\la\mapsto e^{\pi i}\la$ and the
anomaly equation~\eqref{dvn101}.

Determination of the correct constant $a$ is non-trivial, but can
be found for instance using instanton
calculus,\footnote{Naturally, this does not mean that the gaugino
condensation is a semi-classical phenomenon like the instantons.
Rather, the normalisation factor $a$ can be determined by
calculating $\expect{S^{3h}}$ which is saturated by a
one-instanton and then use the cluster decomposition principle.
Here $h$ is the dual Coxeter number.} monopoles in the theory
compactified to three dimensions or Seiberg-Witten curves (a
review can be found in \cite{0206063} and the constant $a$ for
general classical and exceptional groups is given
in~\cite{0006011}). Naturally, the factor $a$ can be absorbed in
$\La$. However, using the standard normalisation
$C\nleft(\mathrm{fund}\right)=1/2$ we actually find that $a=1$ for
a $\sun{N}$ gauge group and thus the Veneziano-Yankielowicz
superpotential here is ($C\nleft(\mathrm{adj}\right)=N$):
\begin{equation}\label{dvn121}
    W_{\mathrm{VY,}\,\sun{N}}=N\Shat\bigg(1-\ln\frac{\Shat}{\La^3}\bigg).
\end{equation}
Extending this solution to semi-simple groups suggests the claimed
form of the Veneziano-Yankielowicz superpotential in~\eqref{dvn7}.
However, as we discuss in the next subsection we can actually not
use the integrating in technique in the case considered in
section~\ref{dvsecdvun}.

In the derivation of the Veneziano-Yankielowicz superpotential we
have actually ignored the fact that for small $N$ in the classical
groups we have no non-abelian dynamics. Choosing the normalisation
such that $C\nleft(\mathrm{adj}\right)$ is the dual Coxeter number
given in table~\ref{dvtablen3} and absorbing the normalisation $a$
into the scale $\La$ we can then write (following~\cite{0304271};
``sgt'' stands for standard gauge theory):
\begin{equation}\label{dvn121.5}
    \wvy=h_{\mathrm{sgt}}\Shat\bigg(1-\ln\frac{\Shat}{\La^3}\bigg),\qquad
    h_{\mathrm{sgt}}=\left\{ \begin{array}{ll}
    N-\de_{N,1}&\textrm{for }\sun{N}\\
    N+1-\de_{N,0}&\textrm{for }\textrm{Sp}(N)\\
    N-2+\de_{N,1}+2\de_{N,0}&\textrm{for }\textrm{SO}(N)
    \end{array}\right.,
\end{equation}

Let us emphasise that our derivation relied on the assumption of
gaugino condensation. However, in~\cite{0307176} the
Veneziano-Yankielowicz superpotential has been derived in the
Dijkgraaf-Vafa context -- but all things considered, the proof
relies on the supersymmetric Ward identities in the same way as in
the original proof
in~\cite{venezianoyankielowicz}.\footnote{Thanks to J. Wheater for
a discussion on this subject.} Finally, we will later see that the
Veneziano-Yankielowicz superpotential can be seen as a
contribution from the free energy of the matrix model in the
Dijkgraaf-Vafa setup.

\subsection{The Glueball Superpotential -- Our Case}\label{dvsecglueball}

We will end this section by summing up what we have learnt about
the Wilsonian low energy effective superpotential for the
$\Nscr=1$ supersymmetric theory with gauge group $\un{N}$ and
adjoint chiral matter which we used in the Dijkgraaf-Vafa
conjecture in section~\ref{dvsecdvun}. The Lagrangian is given by
$\lagrun$ from~\eqref{dvn20.5}.\footnote{Actually, $\lagrun$ is
the Lagrangian for $\Nscr=2$ supersymmetry broken to $\Nscr=1$ by
a tree-level superpotential, but as we discussed in
section~\ref{dvsecdvunsublagr} this only affected the
normalisation of the \kahler{} terms and is not important in the
proof of the conjecture.} As we saw in
section~\ref{dvsecvacsublagrun} we have to choose around which
classical supersymmetric vacuum to expand, and that the classical
expectation value of the chiral field breaks the gauge group as
$\un{N}\mapsto\un{N_1}\times\ldots\times\un{N_n}$. The massive
gauge multiplets corresponding to this breaking we simply
integrate out. We also proved that if the critical points of the
tree-level superpotential are isolated then the chiral multiplets
in these vacua are massive. Integrating out these massive fields
leaves us with an $\Nscr=1$ supersymmetric pure gauge theory. It
is, however, important to realise that the holomorphic scale,
$\La_{u}$, before integrating out is not the same as the scale,
$\La$, in the low energy theory (here assuming an unbroken gauge
group). Using the simple threshold relation~\eqref{dvn111} we can
match the scales as:
\begin{equation}\label{dvn122}
    \La^3=\La_u^2m,
\end{equation}
where $m$ is the mass of the chiral field. We have here used that
$b_d=3N$, $b_u=2N$ and $C\nleft(\mathrm{adj}\right)=N$ since we
assume $C\nleft(\mathrm{fund}\right)=1/2$ (see~\eqref{dvn111} for
notation). With a broken gauge group we should be more careful.
The gauge fields are in this case not independently coupled since
at the UV energy $\mu_0$ for the tree-level Lagrangian the complex
gauge couplings are all equal to the coupling for the unbroken
$\sun{N}$. So the holomorphic scales $\La_{i,u}$ corresponding to
the factors $\sun{N_i}$ are related as (using~\eqref{dvn88}):
\begin{equation}\label{dvn123}
    \nleft(\frac{\La_{1,u}}{\mu_0}\right)^{2N_1}=\cdots=\nleft(\frac{\La_{n,u}}{\mu_0}\right)^{2N_n}.
\end{equation}
The relation to the low energy scales $\La_i$ is as
in~\eqref{dvn122}, but there can be different masses, $m_i$, for
each factor: $\La_i^3=\La_{i,u}^2m_i$.

We saw in section~\ref{dvsecweffsubnon} that the abelian $\un{1}$
part of $\un{N_i}$ described by the $w_{i\al}$
from~\eqref{dvn14.5} is (perturbatively) weakly coupled at low
energy (IR free). On the other hand, for the non-abelian
$\sun{N_i}$ subgroups we have strong coupling in the IR. Here we
expect, as described in section~\ref{dvsecweffsublore},
confinement and gaugino condensation breaking the gaugino number
as $\Z_{2N_i}\mapsto\Z_2$ and giving $N_i$ inequivalent vacua each
with a mass gap. Since we have confinement, we know that the low
energy theory should be described by singlet gauge fields. When
focusing on the superpotential part of the effective Lagrangian
the relevant, elementary fields are believed to be the traceless
glueball superfields $\Shat_i$ defined in equation~\eqref{dvn4}.
The IR dynamics is described by the \emph{glueball superpotential}
for $\Shat_i$ and $w_{i\al}$,
$\weff\nleft(S_i,w_{i\al},g_k\right)$. As noted in
section~\ref{dvsecweffsubilsintein} the glueball superfield is
generally massive (mass of order $\abs{\La}$) so we should see the
glueball superpotential as obtained by integrating out the chiral
matter fields, but with the glueball superfields integrated in.
Naturally, the result is not just the Veneziano-Yankielowicz
superpotential obtained in the last section, but depends
non-trivially on the bare superpotential couplings, $g_k$.
Actually, as mentioned in section~\ref{dvsecdvun}, the full
superpotential, obtained by taking into account the full path
integral, is conjectured to be the sum of the
Veneziano-Yankielowicz superpotential, accounting for the gauge
dynamics, and $\weffpert$ obtained by integrating out the chiral
fields. The way we obtain $\weffpert$ is to perform the path
integral over the chiral fields while treating $\W_\al$ as a
(constant) background field thus allowing us to get the $\Shat$
dependence. We should also note that the Veneziano-Yankielowicz
superpotential in the case of a broken gauge group can not be
determined using the integrating in technique from the last
subsection since the $\Shat_i$ fields are not independently
coupled as explained above.

The full glueball superpotential then determines the IR dynamics
of $\Shat_i$ and we can e.g. get the expectation values of the
glueball superfields by the equation of motion~\eqref{dvn6.5}. A
non-zero expectation value gives gaugino condensation, chiral
symmetry breaking and the corresponding inequivalent vacua. Also
the tension in the domain walls connecting these vacua can be
determined from the glueball superpotential. We can obtain
$\weffpert$ to a certain order in $\Shat$ by going to the
corresponding loop order for the matrix model Feynman diagrams
using the Dijkgraaf-Vafa conjecture. This gives us the
non-perturbative corrections to a corresponding fractional order
in $\La$ (also called fractional instantons). This is what
Dijkgraaf and Vafa refer to as ``a perturbative window into
non-perturbative physics''~\cite{0208048}. However, we note that
this interplay between perturbative and non-perturbative physics
happens through the Veneziano-Yankielowicz term. But all that we
will prove in the Dijkgraaf-Vafa conjecture is simply the form of
$\weffpert$. We do not prove confinement, mass gaps or $\Shat$
being the elementary field, and we have to add the
Veneziano-Yankielowicz superpotential by hand.

In~\cite{0211069} F. Ferrari has investigated the above theory in
the case of a cubic tree-level superpotential
$\wtree=\tr\nleft(\half m\Phi^2+\frac{1}{3}g\Phi^3\right)$. Using
the Dijkgraaf-Vafa conjecture or rather, which amounts to the
same, using the setup that he used to prove the Dijkgraaf-Vafa
conjecture in~\cite{0210135}, the low-energy effective
superpotential can be found (see also section~\ref{dvsecexample})
and investigated. Assuming that the gauge group is unbroken it is
found that there exist some critical points in the quantum
parameter space given by $8g^2\La^3/m^3=e^{-2\pi ik/N}$ with
$k=0,\ldots, N-1$. In these points the glueball superfield
actually is massless if one assumes the \kahler{} potential to be
well-behaved. We note, however, that from the Dijkgraaf-Vafa
conjecture we know nothing about the \kahler{} potential so this
result should be seen with some reservation. Further, using the
Seiberg-Witten theory it is shown that in these points, for $N$
odd, there is monopole condensation and hence confinement. Thus it
is claimed that we here have a phase with confinement without a
mass gap in contradiction with the lore. For $N$ even we do not
have total confinement and for $N=2$ there is no confinement at
all in these points. Other interesting phenomena are tensionless
domain walls and the $\vartheta$ angle not being periodic. The
investigation of the phases and parameter spaces of the
supersymmetric theories has been carried on in many articles
e.g.~\cite{0301006},~\cite{0301157},~\cite{0303207}, and~\cite{0305225}. For more
references see~\cite{0311066}.

With this example of the interesting results of the Dijkgraaf-Vafa
conjecture we will end this section. To prove the Dijkgraaf-Vafa
conjecture and understand the planar limit in the matrix model we
will in the next section introduce the double line notation
allowing a geometric interpretation of the Feynman diagrams.

%---------New section-----------

\section{Double Line Notation}\label{dvsecdouble}

In this section we will introduce the double line notation for
theories with fields in the adjoint representation. This will lead
us to a topological classification of the Feynman diagrams and the
t' Hooft large $N$ limit. We will use this for the matrix model in
the next section and in the diagrammatic proof of the
Dijkgraaf-Vafa conjecture in the next chapter.

\subsection{Double Line Propagators}

Let us consider a theory in which we have a real field $\Phi^a$
where $a$ is an index in the adjoint representation of $\un{N}$.
$\Phi$ could e.g. be the gauge potential which was the case G. 't
Hooft considered in his article introducing the double line
notation and the large $N$ limit~\cite{'thooftaplanar} (a more
general introduction can be found in~\cite{9905111}). However, we
naturally think of $\Phi$ as a real version of the chiral adjoint
field in the Dijkgraaf-Vafa conjecture. As we see below we can
even use all of this for the hermitian matrix model also
introduced in the conjecture (the double line notation for the
matrix model is introduced in e.g.~\cite{9306153}).

The whole idea of the double line notation is based on the
observation that we can use the generators of $\un{N}$ in the
fundamental representation -- as we have done numerous times above
-- to write the adjoint field as a hermitian matrix:
\begin{equation}\label{dvn124}
    \Phi^a,\quad a=1,\ldots,
    N^2\quad\longleftrightarrow\quad\Phi^{i}_{\phantom{i}j}=\Phi^a(T_{a}^{(\mathrm{fund})})^{i}_{\phantom{i}j},\quad
    i,j=1,\ldots, N.
\end{equation}
We here note that if $\Phi^a$ had been complex then
$\Phi^{i}_{\phantom{i}j}$ could be any complex matrix.
In~\eqref{dvn124} the upper index transforms in the fundamental
representation and the lower index in the anti-fundamental
representation. To be precise: If
$\Phi^a\mapsto(\exp\nleft(-i\al^cT_{c}^{(\mathrm{adj})}\right))^a_{\ph{a}b}\Phi^b$
then
\begin{equation}\label{dvn125}
    \Phi^{i}_{\phantom{i}j}\mapsto
    (e^{-i\al^aT_{a}^{(\mathrm{fund})}})^{i}_{\phantom{i}k}\Phi^{k}_{\phantom{k}l}(e^{i\al^aT_{a}^{(\mathrm{fund})}})^{l}_{\phantom{l}j}
    =(e^{-i\al^aT_{a}^{(\mathrm{fund})}})^{i}_{\phantom{i}k}(e^{-i\al^aT_{a}^{(\textrm{anti-fund})}})^{j}_{\phantom{j}l}\Phi^{k}_{\phantom{k}l},
\end{equation}
where we have used the definition of the anti-fundamental, i.e.
conjugate, representation
$T_{a}^{(\textrm{anti-fund})}=-\big(T_{a}^{(\mathrm{fund})}\big)^T$.~\eqref{dvn125}
is proven using the definition of the adjoint representation
$(T_{a}^{(\mathrm{adj})})^b_{\ph{b}c}\Phi^c=[T_a,\Phi^cT_c]^b$
which also gives
$(T_{a}^{(\mathrm{adj})})^b_{\ph{b}c}=if_{ac}^{\ph{ac}b}$. We can
state~\eqref{dvn125} as: The adjoint representation of $\un{N}$ is
$\boldsymbol{N}\otimes\overline{\boldsymbol{N}}$.

As we have done before we can use the matrix notation to write the
invariant Lagrangian using traces, e.g.:
\begin{equation}\label{dvn126}
    \lagr=\frac{1}{\gs}\left(\frac{1}{2}m\tr\nleft(\Phi^2\right)+\sum_{k}
    \frac{g_k}{k}\tr\nleft(\Phi^k\right)\right).
\end{equation}
Naturally, we could also have a kinetic term, however, we are only
interested in the index structure of the Feynman diagrams, not the
space-time dependence. Actually, the theory could just as well be
the matrix model used in the Dijkgraaf-Vafa conjecture with a
global $\un{N}$ symmetry. We have put a coupling $\gs$ in front of
the Lagrangian (for the gauge theory $\gs$ is the Yang-Mills
coupling squared). This coupling is first needed in the large $N$
expansion and is not necessary for the double line notation. We
could also have multi-trace terms, but we will assume that we do
not have such terms.

\begin{figure}\caption{}\label{dvfign2}
\begin{center}
\includegraphics{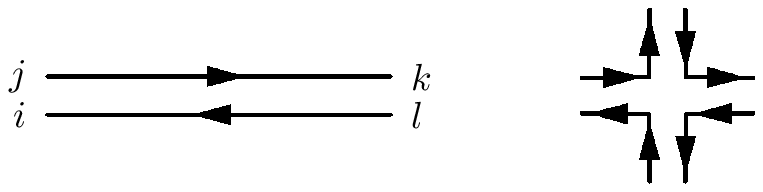}
\end{center}
    \begin{center}
        On the left the double line propagator and on the right the quartic vertex.
    \end{center}
\end{figure}

We will now develop the Feynman rules for the
Lagrangian~\eqref{dvn126}. To this end we can think of
$\Phi^{i}_{\phantom{i}j}$ as a complex particle if $i>j$ and a
real particle if $i=j$. In the usual way we treat
$\Phi^{j}_{\phantom{j}i}=\Phi^{*i}_{\phantom{*i}j}$ and
$\Phi^{i}_{\phantom{i}j}$ as independent. Thus we can see the path
integral as having an integration over each entry of the matrix
$\Phi$ and all entries are independent. The propagator is
determined from the quadratic term in the usual way. We quickly
get (here in Euclidean notation and disregarding the space-time
dependence):
\begin{equation}\label{dvn127}
    \int\DPhi\,
    e^{-\frac{m}{2\gs}\tr(\Phi^2)}\Phi^{i_1}_{\phantom{i_1}j_1}\cdots\Phi^{i_n}_{\phantom{i_n}j_n}=\frac{\partial}{\partial
    J^{j_1}_{\phantom{j_1}i_1}}\cdots\frac{\partial}{\partial
    J^{j_n}_{\phantom{j_n}i_n}}\left.e^{\frac{\gs}{2m}\tr(J^2)}\right\lvert_{J=0},
\end{equation}
where $J$ is the external current which here is a hermitian
matrix. We can directly read off the propagator from
$\gs\!\tr\nleft(J^2\right)/2m=\gs
J^{i}_{\phantom{i}j}\de^l_i\de^j_kJ^{k}_{\phantom{k}l}/2m$:\footnote{\label{footnotesun}In
the case of $\sun{N}$ we have to subtract $\de^i_j\de^k_l/N$ on
the right hand side since in this case $\Phi$ is traceless.}
\begin{equation}\label{dvn128}
    \expect{\Phi^{i}_{\phantom{i}j}\Phi^{k}_{\phantom{k}l}}_0=\frac{\gs}{m}\de^i_l\de^k_j.
\end{equation}
If $\Phi$ is the gauge field then we naturally have the usual
propagator $1/k^2$ instead of $1/m$. The propagator is consistent
with our view of $\Phi^{i}_{\phantom{i}j}$ as complex particles.
If $i>j$ we have
$\expect{\Phi^{i}_{\phantom{i}j}\Phi^{k}_{\phantom{k}l}}_0=\expect{\Phi^{i}_{\phantom{i}j}\Phi^{*l}_{\phantom{l}k}}_0$
which immediately gives us~\eqref{dvn128}. As usual with complex
particles we assign arrows to the propagators. Here we must use
double lines for the propagators since we have two indices. The
upper index, which transformed in the fundamental representation,
is then associated with an incoming arrow and the anti-fundamental
lower index with an outgoing arrow. The double line propagator
corresponding to~\eqref{dvn128} is shown in
figure~\ref{dvfign2}.
The single lines in the double line propagator are called
\emph{index lines} since they are indexed by $i=1,\ldots,N$. The
interaction vertices from~\eqref{dvn126} all have the same index
structure consistent with the double line propagators and can be
derived from~\eqref{dvn127}. They are just proportional to
Kronecker delta functions that connect the index lines and
preserve the directions of these. As an example the quartic vertex
is also shown in figure~\ref{dvfign2}. Let us finally note that if
$\Phi^a$ was complex, and hence $\Phi^{i}_{\phantom{i}j}$ a
general complex matrix, then we would have to assign an extra
overall arrow to the whole double line propagator.

\subsection{Topological Classification of Double Line
Diagrams}\label{dvsecdoublesubtopo}

Let us now look at the connected vacuum diagrams in the double
line notation. These ``fat'' graphs have enough structure to
associate each of them with a Riemann surface or rather a
topological class of these. To do this we view each index line as
a perimeter of a face in a simplicial decomposition of a surface.
To this end we compactify the space by adding a point at infinity
so we also can see the outer index loop as a face on a compact
surface. In our $\un{N}$ case the surface is further oriented
since we have directions on the index lines. The compact oriented
surfaces in $\R^3$ are topologically classified by the Euler
characteristic, $\chi$, which is a topologically invariant
integer. The surfaces are simply given by the sphere, $S^2$, with
$g$ handles added. $g$ is called the genus of the surface. The
Euler characteristic is then given by:
\begin{equation}\label{dvn129}
    \chi=2-2g.
\end{equation}
We can determine the Euler characteristic of the surface
corresponding to a given diagram by simply counting the number of
vertices, V, edges i.e. double line propagators, E, and faces i.e.
index loops, F. Then:
\begin{equation}\label{dvn130}
    \chi=V-E+F.
\end{equation}

\begin{figure}\caption{}\label{dvfign3}
\begin{center}
\includegraphics{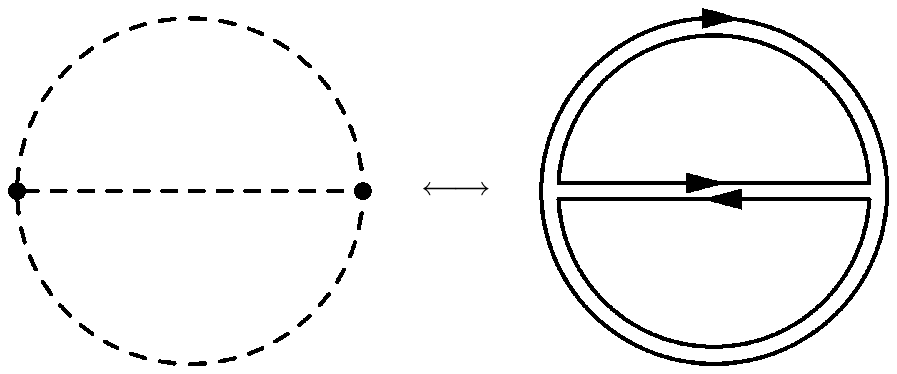}
\end{center}
    \begin{center}
        A planar diagram. On the left in single line adjoint
        representation with interaction vertices denoted by dots. On
        the right the same diagram in double line notation.
    \end{center}
\end{figure}

\begin{figure}[t]\caption{}\label{dvfign4}
\begin{center}
\includegraphics{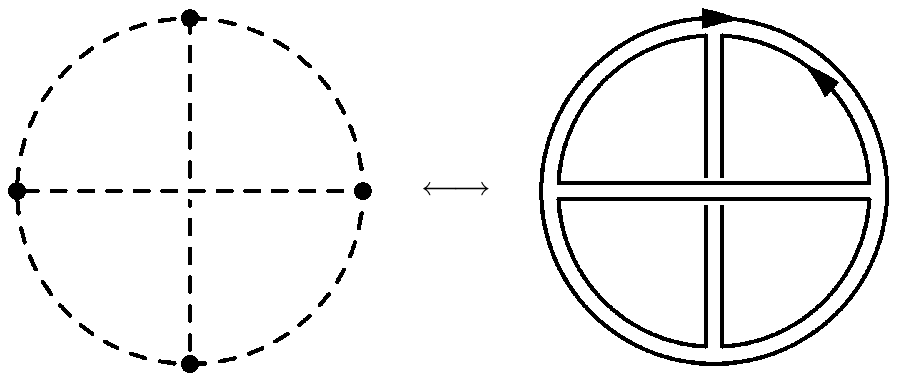}
\end{center}
    \begin{center}
        A diagram with the topology of a torus. On the left in single line adjoint
        representation with interaction vertices denoted by dots. On
        the right the same diagram in double line notation.
    \end{center}
\end{figure}

\begin{figure}\caption{}\label{dvfign5}
\begin{center}
\includegraphics{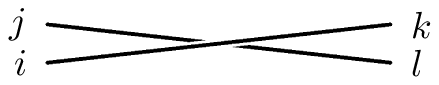}

        The cross-over part of the $\son{N}$ propagator.
    \end{center}
\end{figure}

If the topology of the diagram is that of a sphere, $g=0$, the
diagram is called planar. An example of a planar diagram with
cubic vertices is given in figure~\ref{dvfign3}.
We can simply count $V=2$, $E=3$ and $F=3$ so $\chi=2-3+3=2$. An
example of a diagram with the topology of a torus, $g=1$, is given
in figure~\ref{dvfign4}. Here we count $V=4$, $E=6$ and $F=2$ so
$\chi=4-6+2=0$.
We can obtain surfaces with boundaries if we add fundamental
matter to our Lagrangian. Such matter only has one index so it
will only correspond to a single index line which we can interpret
as a boundary. If we look at $\son{N}$ or $\textrm{USp(N)}$
instead of $\un{N}$ we must also include non-orientable surfaces.
E.g. in the case of $\son{N}$ we can also use~\eqref{dvn124},
however, we must here remember that the fundamental representation
is real and $\Phi$ in this case is antisymmetric. We thus see that
the adjoint of $\son{N}$ is
$\boldsymbol{N}\otimes_{as}\boldsymbol{N}$. Since we can not
distinguish the upper from the lower indices we can not orient the
index lines. Further we have to constrain the path integral to
only include antisymmetric matrices. We could do this by replacing
the antisymmetric matrices with $(\Phi-\Phi^T)/2$ where $\Phi$ is
a hermitian matrix and then integrate over the unconstrained
$\Phi$. Expanding the quadratic terms shows that the propagator
must contain a cross-over term as shown in figure~\ref{dvfign5}
which indeed gives rise to non-orientable surfaces.
In these more general cases the surfaces are characterised by
starting from the sphere $S^2$ and adding $g$ handles, $b$
boundaries and $c$ cross-caps. The Euler characteristic is then
given by~\cite{polchinski1}:
\begin{equation}\label{dvn131}
    \chi=2-2g-b-c.
\end{equation}
We obtain the boundaries by adding fundamental matter single index
lines and the cross-caps from the cross-over part of the $\son{N}$
or $\textrm{USp(N)}$ propagator. We note that $\chi\leq2$ and only
odd with fundamental matter or another group than $\un{N}$.

\subsection['t Hooft Large $N$ Limit]{'t Hooft Large $\boldsymbol{N}$
Limit}\label{dvsecdoublesublargen}

Using the topological classification obtained in the last
subsection we can now explain the 't Hooft large $N$ limit as
introduced in~\cite{'thooftaplanar}. We will assume a $\un{N}$
group and no fundamental matter. For a given connected vacuum
diagram we count the dependence on $N$ and the coupling $\gs$
from~\eqref{dvn126}. For each vertex we have a factor $1/\gs$, for
each double line propagator we have a factor $\gs$ as seen
from~\eqref{dvn128}, and finally we have a factor $N$ for each
index loop since we here sum $\sum_i\de^i_i=N$. Thus we get an
overall factor of:
\begin{equation}\label{dvn132}
    N^F
    \gs^{E-V}=N^{V-E+F}\left(\gs N\right)^{E-V}=N^{\chi}\left(\gs N\right)^{E-V}=N^{2-2g}\left(\gs N\right)^{E-V},
\end{equation}
where we have used~\eqref{dvn129} and~\eqref{dvn130}. We thus see
that by taking $N\rightarrow\infty$ while keeping $\gs N$ fixed
the planar diagrams with $g=0$ are the dominant ones. This is the
't Hooft large $N$ limit also called the planar limit. To connect
with the Dijkgraaf-Vafa conjecture we rewrite the
factor~\eqref{dvn132} and include the dependence on the couplings
$g_k$ from~\eqref{dvn126} (with a minus for Euclidean space) and
the mass $m$ for completeness:
\begin{equation}\label{dvn133}
    \gs^{2g-2}\nleft(\gs N\right)^Fm^{-E}\prod_k(-g_k)^{V_k},
\end{equation}
where $V_k$ is the number of vertices of order $k$ so that
$V=\sum_k V_k$. What we want to calculate is the free energy,
$W_{\mathrm{free}}$, which exactly is the sum over (minus) the
connected vacuum diagrams. We can now group the diagrams
topologically and write (using~\eqref{dvn133}):
\begin{equation}\label{dvn134}
    W_{\mathrm{free}}=\sum_{g\geq0}\gs^{2g-2}\cF_g\nleft(\gs
    N\right),
\end{equation}
where $\cF_g$ is the contribution to the free energy (modulo
factors of $\gs^{2g-2}$) from diagrams with genus $g$ and whose
dependence on $\gs$ and $N$ only is through $\gs N$. In the 't
Hooft large $N$ limit we see that the planar contribution is
dominant since $\gs\rightarrow0$ and thus
$W_{\mathrm{free}}\approx\gs^{-2}\cF_{g=0}$. This
explains~\eqref{dvn10} and the planar limit taken on the matrix
side in the Dijkgraaf-Vafa conjecture in the case of an unbroken
gauge group.

%---------New section-----------

\section{The Matrix Model}\label{dvsecmatrix}
In this section we will investigate the matrix model side of the
Dijkgraaf-Vafa conjecture.

\subsection{The Matrix Model and the Dijkgraaf-Vafa
Conjecture}\label{dvsecmatrixmm}

In the Dijkgraaf-Vafa conjecture for a $\un{N}$ gauge group we are
instructed to calculate the free energy in the planar limit for a
bosonic matrix model with partition function~\eqref{dvn8}:
\begin{equation}\nonumber
    \Zm=\int\DM e^{-\frac{1}{\gs}\wtree(M)},
\end{equation}
where $M$ are $N'\times N'$ hermitian matrices and $\wtree=\tr
    \wpoly\nleft(M\right)$. The
constraint to hermitian matrices requires the couplings in
$\wtree$ to be real contrary to the gauge theory side where $\Phi$
is a general complex matrix. Thus the matrix model is real (real
eigenvalues, real couplings) and we should perform an analytic
continuation after obtaining the wanted free energy to compare
with the gauge theory side in the Dijkgraaf-Vafa conjecture. There
should be no ambiguity in this continuation since we restrict to
the planar limit~\cite{0210135}. Naturally, this is of no concern
in calculations where one simply formally uses $g_k$ as couplings
with no reference to whether they are real or complex. From the
partition function we also see that $M$ must have mass dimension
one and $\gs$ mass dimension three to fit the dimensions of the
couplings. Of course, one can scale these dimensions away using
the holomorphic scale $\La$ which has mass dimension one.

The matrix model does not have any supersymmetry as on the gauge
theory side, but it is gauged in the sense that the potential is
(globally) $\un{N}$ invariant under $M\mapsto UMU^{-1}$ for
$U\in\un{N}$. We will treat this symmetry as an equivalence like
the gauge symmetry. We can e.g. use this to diagonalise the
matrices. Using the same manipulations that led to~\eqref{dvn49}
the classical equation of motion for $M$ can be shown to be
$\wpoly'\nleft(M\right)=0$ . Choosing $M$ to be diagonal we
conclude that the ``vacua'' for the matrix model are obtained
analogously to the supersymmetric vacua by distributing the $N'$
eigenvalues of $M$ over the critical points $a_1,\ldots,a_n$ of
$\wpoly$. This corresponds to the partition of $N'$
in~\eqref{dvn9}. As with the supersymmetric vacua we should think
modulo permutation of the eigenvalues and the vacuum then breaks
the gauge symmetry as:
\begin{equation}\label{dvn135}
    \un{N'}\mapsto\un{N'_1}\times\cdots\times\un{N'_n}.
\end{equation}
In the Dijkgraaf-Vafa conjecture we should, in the terminology of
section~\ref{dvsecdvunsubdv}, here choose $N'_i=0$ if $N_i=0$ to
obtain the same symmetry breaking pattern as on the gauge theory
side. However, due to the realness of the matrix model we here
have a problem since the real form of the polynomial $\wpoly'$
does not necessarily have $n$ real roots and the eigenvalues of
the hermitian matrix $M$ must be real. The solution is probably to
constrain the couplings $g_k$ to certain real intervals and then
analytically continue to all real numbers. E.g. for the quartic
potential $\wpoly=\half mM^2+\tfrac{1}{4}gM^4$ we have the set of
critical points $\big\{0,\pm\sqrt{-m/g}\big\}$. We can thus choose
$m$ positive and $g$ negative to have the same set of critical
points to distribute the eigenvalues over as in the gauge
theory.\footnote{Unfortunately, this would in turn mean that
$\wtree\nleft(M\right)$ is not bounded from below.}

What we should do now is to formulate a perturbation theory for
the fluctuations around the chosen vacuum. Assuming, as in
section~\ref{dvsecdvunsubdv}, that the critical points are
distinct, the vacua will be massive in the sense that the parts
corresponding to unbroken gauge group factors have non-zero
masses. This can be proven in the same way as done for the gauge
theory side in section~\ref{dvsecvacsublagrun}. In the case of an
unbroken gauge group, i.e. if all eigenvalues are equal, the
perturbative expansion is simple. As an example consider a cubic
interaction which has two critical points $a_1$ and $a_2$.
Expanding around $a_1\idmatr_{N'\times N'}$ gives (disregarding a
constant term):
\begin{equation}\label{dvn136}
    \wtree=\half\De\tr\nleft(M^2\right)+\tfrac{g}{3}\tr\nleft(M^3\right),
\end{equation}
where $\De=g(a_1-a_2)$ which we note is non-zero for distinct
critical points. We can then simply expand the partition function
as:
\begin{equation}\label{dvn137}
    \int\DM e^{-\frac{1}{\gs}\wtree}=\int\DM
    e^{-\frac{1}{2\gs}\De\tr\left(M^2\right)}\sum_{n=0}^{\infty}\frac{(-1)^n}{n!}\left(\frac{1}{\gs}\frac{g}{3}\tr\nleft(M^3\right)\right)^n.
\end{equation}

In the case of broken gauge symmetry we have to be more careful
and take into account Faddeev-Popov ghosts in the matrix
model~\cite{0210238}. These ghosts are described by two
Grassmannian matrices $B$ and $C$ which contribute to the
potential as:
\begin{equation}\label{dvn138}
    \wghost=\tr\nleft(B[M,C]\right).
\end{equation}
This ghost term can be derived from the Vandermonde determinant
that, as we will see in the next section, arises from choosing a
gauge such that $M$ is diagonal. From this term we immediately see
that the ghosts are only propagating if we expand around a vacuum
with broken symmetry since a matrix proportional to the identity
will disappear from the commutator. On the gauge theory side a
ghost term of precisely the same form arises with $B$ and $C$
being anticommuting chiral superfields. This is in line with the
Dijkgraaf-Vafa conjecture assuming that the superpotential in the
gauge theory and the potential in the matrix model should be the
same.

\begin{figure}\caption{}\label{dvfign6}
\begin{center}
\includegraphics{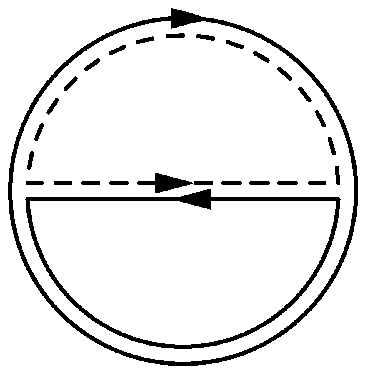}
\end{center}
    \begin{center}
        Example of a planar diagram for the matrix model with
        ghosts. The solid index lines carry indices running
        over $1,\ldots,N'_1$ while the dashed index lines carry
        indices running over $1,\ldots,N'_2$. We have two ghost
        double line propagators and one double line propagator
        associated with $M_{11}$.
    \end{center}
\end{figure}

To be specific, let us consider the cubic model as we did above
(following~\cite{0210238}). We now consider the broken case where
the vacuum expectation value $M_0$ has $N'_1$ eigenvalues equal to
$a_1$ and $N'_2$ eigenvalues equal to $a_2$. We should expand
around this vacuum:
\begin{equation}\label{dvn139}
    M\mapsto M_0+M=\begin{pmatrix}
      a_1\idmatr_{N'_1\times N'_1} & 0 \\
      0 & a_2\idmatr_{N'_2\times N'_2} \\
    \end{pmatrix}+\begin{pmatrix}
      M_{11} & M_{12} \\
      M_{21} & M_{22} \\
    \end{pmatrix},
\end{equation}
where we have written the matrix in block notation ($M_{ij}$ is a
$N'_i\times N'_j$ matrix). Using this block notation, also for the
ghost term, one finds the quadratic terms in the potential to be:
\begin{equation}\label{dvn140}
    \frac{1}{2}\De\tr\nleft(M^2_{11}\right)-\frac{1}{2}\De\tr\nleft(M^2_{22}\right)+\De\tr\nleft(B_{21}C_{12}\right)-\De\tr\nleft(B_{12}C_{21}\right),
\end{equation}
where $\De=(a_1-a_2)$ and we have set $g=1$. We thus see that the
off-diagonal blocks for $M$ do not propagate which is clear since
they are pure gauge terms. On the other hand, only the
off-diagonal blocks for the ghost fields propagate.
Using~\eqref{dvn140} it is trivial to write down the double line
propagators using the same method as in last section. We see that
we have two types of propagators. Firstly the usual double line
propagators with the same type of index running in both index
lines. We have two propagators of this type corresponding to
respectively $M_{11}$ and $M_{22}$ and with indices running over
respectively $1,\ldots,N'_1$ and $1,\ldots,N'_2$. The second type
of double line propagators stems from the ghost terms. In these
propagators the index in one index line runs from $1,\ldots,N'_1$
while the index in the other runs from $1,\ldots,N'_2$. We have
two propagators of this type corresponding to the two ghost terms
in~\eqref{dvn140}. As can be derived from~\eqref{dvn138} we also
get vertices connecting a standard double line propagator with two
ghost propagators. An example of a diagram with ghosts can be seen
in figure~\ref{dvfign6}.

Generalising these results we see that in the case of broken
symmetry we have diagrams in the double line notation in which the
index loops, i.e. the faces, are indexed according to which broken
part they are associated with. With partition $N'=\sum_i N'_i$ the
index loops are indexed by $i$ and the index loop gives a
contribution $N'_i$. As an example the index loops in the diagram
in figure~\ref{dvfign6} give a contribution of $N'^2_1N'_2$. As in
section~\ref{dvsecdoublesubtopo} the connected vacuum diagrams can
be topologically classified. Let $F_i$ be the number of faces
indexed with $i$ then the total number of faces is $F=\sum_i F_i$.
We simply perform the topological classification ``blind'' to the
index $i$ i.e. only depending on $F$ and also we do not
distinguish between the types of double line propagators when
counting edges. Finally, to connect with the Dijkgraaf-Vafa
conjecture we should take the large $N'$ limit, or more precisely,
we should let $\gs\ll1$ while keeping $\gs N'_i$ fixed and
finite.\footnote{We have not been careful with the normalisation
of the ghost term. To take the 't Hooft limit we should also give
this term a $1/\gs$ normalisation.} We can count the $\gs$
dependence in a given diagram and we find, in analogy
with~\eqref{dvn133}, a factor:
\begin{equation}\label{dvn141}
    \gs^{2g-2}\prod_i\nleft(\gs N'_i\right)^{F_i}.
\end{equation}
We can make a topological expansion of the free energy as
in~\eqref{dvn134}. The leading contribution is again from genus
$g=0$ and we get the planar limit of the free energy
$\gs^{-2}\cF_{g=0}\nleft(\gs N_i\right)$. Identifying $S_i=\gs
N'_i$ we see from~\eqref{dvn141} that we get one factor of $S_i$
for every index loop indexed by $i$ as stated at the end of
section~\ref{dvsecdvunsubdvun}. In section~\ref{dvsecexample} we
will give an explicit example of this (in the unbroken case).

\subsection{The Measure}\label{dvsecmatrixsubmeasure}

So far we have not discussed the measure of the matrix model.
However, when we obtained the Feynman rules for double line
propagators, we implicitly assumed in~\eqref{dvn127} that the
normalisation of the measure was such that for the free theory
$\expect{1}_0=1$. This means that in the measure we have to divide
with the volume of the gauge group $\vol\nleft(G\right)$.
In~\cite{0205297} it is found that in the planar limit (here for a
broken gauge group):
\begin{equation}\label{dvn142}
    \frac{1}{\vol\nleft(\un{N'_1}\times\cdots\times\un{N'_n}\right)}\sim e^{\frac{1}{2}\sum_i N'^2_i\ln N'_i+\ldots},
\end{equation}
where we have only kept the part of the planar contribution that
involves $\ln N'_i$. We can here substitute $N'_i=\Shat_i/\gs$ and
compare with the planar free energy
\mbox{$e^{-\gs^{-2}\cF_{g=0}(\Shat_i)}$}. This gives us a
contribution to $\cF_{g=0}$ of the form $-\frac{1}{2}\sum_i
\Shat^2_i\ln \Shat_i$. As noted by Dijkgraaf and Vafa, if we here
extend the connection~\eqref{dvn13} between $\weff$ and
$\cF_{g=0}$ in the Dijkgraaf-Vafa conjecture to also apply for
this term, we see that the essential
$-\sum_iN_i\Shat_i\ln\big(\Shat_i\big)$ part of the
Veneziano-Yankielowicz superpotential is reproduced. We emphasise
that there is no field theoretic proof for this, but it suggests a
tighter relation between the matrix model and the gauge theory
than we can actually prove.

Let us also note that in the case of unbroken symmetry
(following~\cite{0211170}) we can obtain this coupling independent
part of the planar free energy simply by setting the couplings in
the potential to zero and use Gaussian integration:
\begin{equation}\label{dvn143}
    \mu^{-N'^2}\int\DM
    e^{-\frac{1}{2\gs}m\tr\left(M^2\right)}=\left(\frac{2\pi
    \gs}{m\mu^2}\right)^{N'^2/2},
\end{equation}
where $\mu$ is a scale of mass-dimension one introduced to make
the measure dimensionless. We thus get a contribution to
$\cF_{g=0}$ given by:
\begin{equation}\label{dvn144}
    \De\cF_{g=0}=-\frac{1}{2}\Shat^2\ln\nleft(\frac{2\pi\Shat}{N'm\mu^2}\right)=-\frac{1}{2}\Shat^2\ln\nleft(\frac{\Shat}{e^{3/2}m\La_u^2}\right),
\end{equation}
where we in the last line identify $N'\mu^2/2\pi$ with
$e^{3/2}\La_u$ where $\La_u$ is the holomorphic scale before
integrating out the chiral field in the gauge field theory.
Extending the Dijkgraaf-Vafa conjecture as above this exactly
gives the Veneziano-Yankielowicz superpotential if we use the
matching of scales~\eqref{dvn122}: $\La^3=m\La_u^2$.

\subsection{Exact Solution of the Matrix Model}\label{dvsecmatrixexact}
Above we have solved the matrix model perturbatively using
diagrams. However, as we now show, it is also possible to use
non-perturbative techniques to obtain the exact solution of the
matrix model in the planar limit. This was first done
in~\cite{brezinplanar}. Here we will also use~\cite{9306153}.

It is customary to redefine the matrix model coupling to:
\begin{equation}\label{dvn144.5}
    \gm\defi\gs N',
\end{equation}
such that the potential takes the form
$\frac{N'}{\gm}\wtree\nleft(M\right)$. Noting that this potential
only depends on the eigenvalues of $M$ due to the cyclicity of the
trace, we can diagonalise the hermitian matrix as $M=U^\dagger \La
U$ where $U$ is a unitary matrix and $\La$ is a diagonal matrix
consisting of the eigenvalues $\la_i,\,i=1,\ldots,N'$, of $M$. The
integration over $M$ then splits into an integration over the
eigenvalues and a trivial integration over the unitary matrices.
Assuming unit measure for the unitary matrices we get:
\begin{equation}\label{dvn145}
    \Zm=\int\DM
    e^{-\frac{N'}{\gm}\wtree(M)}=\int\prod_{i=1}^{N'}\inted\la_i\,\De^2\nleft(\la\right)e^{-\frac{N'}{\gm}\sum_i\wpoly\left(\la_i\right)},
\end{equation}
where $\De\nleft(\la\right)=\prod_{i<j}\left(\la_j-\la_i\right)$
is the Vandermonde determinant which can be derived using the
Faddeev-Popov method.\footnote{Simply write one as
$1=\int\prod_i\inted\la_i\inted U'
\deltafunk^{(N'^2)}\nleft(U'MU'^\dagger-\mathrm{diag}\nleft(\la_i\right)\right)\De^2\nleft(\la\right)$.
The $\la_i$ integrations constrain $\la_i$ to be the eigenvalues
of $M$ and $\mathrm{diag}\nleft(\la_i\right)=\La$. The
$\delta$-function constrain $U'$ to be in a neighbourhood of the
matrix $U$ that diagonalised $M$: $U'=(1+T)U$ where $T$ is
infinitesimal. Thus
$\left(U'MU'^\dagger-\La\right)_{ij}\simeq[T,\La]_{ij}=T_{ij}\left(\la_j-\la_i\right)$.
The result then follows from performing the integration over the
real and complex part of $T_{ij},\,i<j$.} Exponentiating the
Vandermonde determinants then gives us an effective potential:
\begin{equation}\label{dvn146}
    \Zm=\int\prod_{i}\inted\la_i\,e^{-\frac{N'}{\gm}\sum_i\wpoly\left(\la_i\right)+\sum_{i\neq j}\ln\abs{\la_j-\la_i}}.
\end{equation}

In the large $N'$ limit the exact solution for the free energy
$\gs^{-2}\cF_{g=0}=\frac{N'^2}{\gm^2}\cF_{g=0}$ is found using the
steepest descent method: We simply evaluate the effective
potential in its critical point (the saddle-point), i.e.:
\begin{eqnarray}
  \cF_{g=0} &=& \frac{\gm}{N'}\sum_i\wpoly\nleft(\la_i\right)-\frac{\gm^2}{N'^2}\sum_{i\neq j}\ln\abs{\la_j-\la_i}, \label{dvn147}\\
  0 &=& -\frac{N'}{\gm}\wpoly'\nleft(\la_i\right)+2\sum_{j\neq i}\frac{1}{\la_i-\la_j},\quad i=1,\ldots, N'. \label{dvn148}
\end{eqnarray}
We can describe the distribution of the eigenvalues with the
density of eigenvalues:
\begin{equation}\label{dvn148.5}
    \rho\nleft(\la\right)=\frac{1}{N'}\sum_i\deltafunk\nleft(\la-\la_i\right),\quad\int\dla\,\rho\nleft(\la\right)=1.
\end{equation}
Due to the term from the Vandermonde determinants we see that
there is a Coulomb repulsion between the eigenvalues. This means
that the eigenvalues will spread out evenly and in the large $N'$
limit we get a continuous distribution of eigenvalues and a smooth
density, $\rho$. In general (as we will show below) the support of
$\rho$ is disconnected consisting of maximally $n$ intervals
(cuts), $\supp \nleft(\rho\right)=\bigcup_k C_k$. There is one cut
$C_k$ for each critical point, $a_k$, of $\wpoly$ and $a_k\in
C_k$. From~\eqref{dvn148} we see that in the classical limit
$\gm\rightarrow0$ the eigenvalues are all in the critical points
and the cuts $C_k$ simply shrink to the points $a_k$. Using $\rho$
we can rewrite the sums over $\la$ as integrals,
and~\eqref{dvn147} and~\eqref{dvn148} become:
\begin{eqnarray}
  \cF_{g=0} &=& \gm\int\dla\,\rho\nleft(\la\right)\wpoly\nleft(\la\right)-\gm^2\dashint\dla\,\dla'\,\rho\nleft(\la\right)\rho(\la')\ln\abs{\la-\la'}, \label{dvn149}\\
  0 &=& -\wpoly'\nleft(\la\right)+2\gm\dashint\dla'\frac{\rho(\la')}{\la-\la'},\quad\la\in\supp(\rho). \label{dvn150}
\end{eqnarray}
Here we have principal value diagrams since $j\neq i$ in the sums
over $\la$.

To solve the matrix model it is convenient to introduce the
resolvent:
\begin{equation}\label{dvn151}
    R(z)\defi\frac{1}{N'}\tr\nleft(\frac{1}{M-z}\right)=\int_{-\infty}^\infty\dla\,\frac{\rho(\la)}{\la-z},\quad
    z\in\C\setminus\supp(\rho).
\end{equation}
The resolvent is analytic with branch cuts at $\supp(\rho)$. The
point is that we can determine $R(z)$ which in turn determines
$\rho$ since by redrawing contours of integration,~\eqref{dvn151}
gives:
\begin{equation}\label{dvn152}
    \rho(\la)=\frac{1}{2\pi
    i}\big(R\nleft(\la+i\ep\right)-R\nleft(\la-i\ep\right)\big),\quad\la\in\supp(\rho),
\end{equation}
where $\ep$ is infinitesimal. Indeed we can determine $R(z)$ using
the saddle-point equation which from~\eqref{dvn150} takes the
form:
\begin{equation}\label{dvn153}
    R\nleft(\la+i\ep\right)+R\nleft(\la-i\ep\right)=-\frac{1}{\gm}\wpoly'\nleft(\la\right),\quad\la\in\supp(\rho).
\end{equation}

However, we can also multiply~\eqref{dvn150} with
$\frac{1}{\gm}\frac{\rho(\la)}{\la-z}$ and integrate over $\la$ to
obtain (after a couple of rewritings):\footnote{This equation is
also found in~\cite{0211170} with $R$ being the expectation value
of $\frac{1}{N'}\tr\nleft(\frac{1}{M-z}\right)$. The result is
obtained using the loop equations and that the correlation
functions factor in the large $N'$ limit. This tells us, as we
have already noted, that the saddle-point approximation is exact
in the large $N'$ limit.}
\begin{equation}\label{dvn154}
    R^2(z)-\frac{1}{N'}R'(z)+\frac{1}{\gm}\wpoly'(z)R(z)+f(z)=0,
\end{equation}
where $f(z)$ is a polynomial of degree $n-1$ given by:
\begin{equation}\label{dvn155}
    f(z)=\frac{1}{\gm}\int_{-\infty}^\infty\dla\,\rho(\la)\frac{\wpoly'\nleft(z\right)-\wpoly'\nleft(\la\right)}{z-\la}.
\end{equation}
Since we are in the large $N'$ limit, we can disregard the $R'/N'$
term in~\eqref{dvn154} and thus the equation is purely algebraic.
We solve this most easily by splitting $R$ in its regular and
singular part $R(z)=R_{\mathrm{reg}}(z)+R_{\mathrm{sing}}(z)$. The
regular part is $R_{\mathrm{reg}}(z)=-\wpoly'\nleft(z\right)/2\gm$
which is also directly seen from~\eqref{dvn153}. The singular part
is then determined from~\eqref{dvn154} as:
\begin{equation}\label{dvn156}
    R_{\mathrm{sing}}(z)=\sqrt{\tfrac{1}{4\gm^2}\wpoly'^2(z)-f(z)}.
\end{equation}
Since $\wpoly'^2$ is of order $2n$, we see that $R$ has at most
$2n$ different branch points and hence at most $n$ branch cuts
$C_k$ giving the promised support of $\rho$. However, we have not
really solved the problem yet since $f$ depends on $\rho$. But the
$n$ coefficients in $f$ are actually determined by the $n$ filling
fractions~\cite{0206255}:\footnote{Actually, if we sum all these
conditions we simply get the normalisation
condition~(\ref{dvn148.5}) for $\rho$ which in turn is equivalent
to require $R(z)$ to behave like $-1/z$ for $\abs{z}$ large.}
\begin{equation}\label{dvn157}
    \frac{N'_i}{N'}=-\frac{1}{2\pi i}\oint_{C_i}\inted z\,R(z)=\int_{C_i}\dla\,\rho(\la).
\end{equation}
We see that the filling fraction $N'_i/N'$ is the relative number
of eigenvalues in the $i^{\mathrm{th}}$ cut. Finally we have
solved for $R$ and hence $\rho$ as a function of $\gm$, $N'_i/N'$
and the couplings $g_k$. We can then obtain the free energy
using~\eqref{dvn149}.

In this exact case the relation between the gauge theory side and
the matrix model side in the Dijkgraaf-Vafa conjecture is through
the filling fractions:\footnote{From~(\ref{dvn149})
and~(\ref{dvn157}) we see that the dependence on $\gm$ in the free
energy is always through $\gm N'_i/N'=\gs N'_i$ as it should be.}
\begin{equation}\label{dvn158}
    \Shat_i\defi\gm\frac{N'_i}{N'}=\gs N'_i.
\end{equation}
This definition, naturally, gives the same result as the one we
have in the diagrammatic case~\eqref{dvn11}. We can see this from
the classical limit $\gm\rightarrow0$ where the filling fraction
$N'_i/N'$ simply becomes the relative number of eigenvalues in the
$i^{\mathrm{th}}$ critical point. The form of the exact solution
is used to prove the Dijkgraaf-Vafa conjecture in the original
superstring proof, in the proof using the generalised Konishi
anomaly, and in the proof using the Seiberg-Witten curves.

\subsection{One-Cut Solution}\label{dvsecmatrixsubonecut}

Let us consider the exact solution of the matrix model in the case
of a single cut. This corresponds to the case of unbroken gauge
symmetry where one $N'_i$ equals $N'$ and the rest are zero. We
write the cut as $C_i=[a,b]$. To determine $R$ we use
equation~\eqref{dvn153}. We note, following~\cite{9306153}, that
the homogeneous version of this equation (i.e. setting the right
hand side to zero) has the solution
$g(z)=\sqrt{(z-b)(z-a)}=\exp\nleft(\half\ln(z-b)+\half\ln(z-a)\right)$,
where the logarithms are defined with the usual branch cut along
the negative real axis. Thus $g(\la\pm i\ep)=\pm
i\sqrt{(b-\la)(\la-a)}$ for $\la\in[a,b]$. We can then
rewrite~\eqref{dvn153} as:
\begin{equation}\label{dvn159}
    r(\la+i\ep)-r(\la-i\ep)=-\frac{1}{\gm}\frac{\wpoly'\nleft(\la\right)}{i\sqrt{(b-\la)(\la-a)}},\quad\la\in[a,b],
\end{equation}
where we have defined $r(z)=R(z)/g(z)$. In analogy with the
relation between equation~\eqref{dvn151} and~\eqref{dvn152} we
get:\footnote{We can also prove this using
$1/(\la-\la'-i\ep)-1/(\la-\la'+i\ep)=2i\ep/((\la-\la')^2-\ep^2)=2\pi
i\deltafunk(\la-\la')$.}
\begin{equation}\label{dvn160}
    R(z)=g(z)r(z)=\frac{\sqrt{(z-b)(z-a)}}{2\pi
    \gm}\int_a^b\dla\,\frac{1}{\la-z}\frac{\wpoly'\nleft(\la\right)}{\sqrt{(b-\la)(\la-a)}},\quad z\in\C\setminus[a,b].
\end{equation}
To obtain the solution we should determine $a$ and $b$ using the
constraint~\eqref{dvn157} which simply tells us that $R(z)$
behaves as $-1/z$ for $\abs{z}\rightarrow\infty$. This condition
also rules out the possibility of a regular term
in~\eqref{dvn160}. Using that
${\sqrt{(z-b)(z-a)}/(\la-z)}=-1{-(\la-(b+a)/2)/z}+\cO\nleft(1/z^2\right)$
for large $\abs{z}$, we get the two constraints:
\begin{eqnarray}
  \int_a^b\dla\,\frac{\wpoly'\nleft(\la\right)}{\sqrt{(b-\la)(\la-a)}} &=& 0, \label{dvn161}\\
  \int_a^b\dla\,\frac{\la\wpoly'\nleft(\la\right)}{\sqrt{(b-\la)(\la-a)}} &=&
  2\pi\gm.\label{dvn162}
\end{eqnarray}
These two equations determine $a$ and $b$ as functions of $\gm$
and the couplings $g_k$. Naturally, we should here choose the
solution such that for the critical point $a_i$ with $N'_i=N'$ we
have $a_i\in[a,b]$.

In principle we have solved the problem and we can obtain the free
energy~\eqref{dvn149} using $\rho$ determined from $R$
using~\eqref{dvn152}. However, what we really need in the
Dijkgraaf-Vafa conjecture~\eqref{dvn13} is the derivative of the
free energy $\partial\cF_{g=0}/\partial\gm$ (where
$\Shat\defi\gm$). To obtain this we first note that (as we show in
appendix~\ref{appmatrix}):
\begin{equation}\label{dvn163}
    \frac{\partial}{\partial\gm}\big(\gm\rho\nleft(\la,\gm\right)\big)=\frac{1}{\pi\sqrt{(b-\la)(\la-a)}}.
\end{equation}
Following~\cite{0210135} we can use this and (as we also show in
appendix~\ref{appmatrix})
\begin{equation}\label{dvn164}
    \dashint_a^b\dla\,\frac{\ln\abs{\la-\la'}}{\sqrt{(b-\la)(\la-a)}}=\pi\ln\nleft(\frac{b-a}{4}\right),\quad\forall\la'\in[a,b],
\end{equation}
to rewrite~\eqref{dvn149} as:
\begin{equation}\label{dvn165}
    \frac{\partial}{\partial\gm}\cF_{g=0}=\int_a^b\dla\,\frac{\wpoly\nleft(\la\right)}{\pi\sqrt{(b-\la)(\la-a)}}-2\gm\ln\nleft(\frac{b-a}{4}\right).
\end{equation}
Thus~\eqref{dvn161},~\eqref{dvn162} and~\eqref{dvn165} determines
the effective superpotential $\weffpert$ using the Dijkgraaf-Vafa
conjecture~\eqref{dvsun} for an unbroken gauge group. But here we
should be careful. In the solution of the matrix model we have
actually been working with dimensionless variables i.e. where the
dimensions have been scaled away using the only mass-scale at hand
namely $\La_u$ -- the scale for the gauge theory before
integrating out the chiral fields. At the point where we wish to
use the Dijkgraaf-Vafa conjecture we should restore the dimensions
such that $\gm$ and $\wpoly(\la)$ have dimensions $3$, and thus
$a$ and $b$ dimension $1$, and $\cF_{g=0}$ dimension $6$. The only
place where we will be able to see this, is in the logarithm in
the last term of~\eqref{dvn165} which will take the
form~$\ln\big((b-a)/4\La_u\big)$. We can then identify
$\Shat\defi\gm$ and $\weffpert=N\partial\cF_{g=0}/\partial\gm$.

However, as we show in appendix~\ref{appmatrix}, we can go one
step further and actually perform the integrations if we expand
$\wpoly$ as in~\eqref{dvn1.5}. We then get the following algebraic
equations for the solution (with $m=g_2$):
\begin{subequations}\label{dvonecutsol}
\begin{eqnarray}
  0 &=& \sum_{p=2}^{n+1}g_p\sum_{q=0}^{\left[(p-1)/2\right]}\frac{p-2q}{p}\binom{p}{2q}\binom{2q}{q}\left(\frac{a+b}{2}\right)^{p-2q-1}\left(\frac{b-a}{4}\right)^{2q},\quad \label{dvn166}\\
  \Shat &=& \sum_{p=2}^{n+1}g_p\sum_{q=1}^{\left[p/2\right]}\frac{q}{p}\binom{p}{2q}\binom{2q}{q}\left(\frac{a+b}{2}\right)^{p-2q}\left(\frac{b-a}{4}\right)^{2q}, \label{dvn167}\\
  \weffpert &=&
  N\sum_{p=2}^{n+1}g_p\sum_{q=0}^{\left[p/2\right]}\frac{1}{p}\binom{p}{2q}\binom{2q}{q}\left(\frac{a+b}{2}\right)^{p-2q}\left(\frac{b-a}{4}\right)^{2q}\nonumber\\
  &&-2N\Shat\ln\nleft(\frac{b-a}{4\La_u}\right).\label{dvn168}
\end{eqnarray}
\end{subequations}
These are the same equations as obtained on the gauge theory side
in~\cite{0210135} using factorisation of Seiberg-Witten curves and
the ILS linearity principle (for other examples of Seiberg-Witten theory in the matrix model framework inspired by the Dijkgraaf-Vafa conjecture see e.g.~\cite{0211123},~\cite{0211245},~\cite{0211254},~\cite{0212212},~\cite{0212253},~\cite{0301203},~\cite{0302083}, and~\cite{0305263}). Here we should identify $(a+b)/2$
with $z=\tr(\Phi)/N$ and $(b-a)/4$ with
$\La\big(z,\Shat,g_{p\neq2}\big)$. In this way~\eqref{dvn166}
corresponds to integrating out $z$ (using~\eqref{dvn108})
and~\eqref{dvn167} corresponds to integrating in $\Shat$
(i.e.~\eqref{dvn114}). This gives the relation between the matrix
model and the gauge theory in this proof.

%---------New section-----------

\section{Exact Superpotentials}\label{dvsecexample}

In this section we will use the Dijkgraaf-Vafa conjecture and the
exact solution of the matrix model obtained in the last section to
find the exact effective glueball superpotential in the case of a
cubic tree-level superpotential. We will also briefly discuss the
even tree-level superpotential.

\subsection{Cubic Tree-Level Superpotential}
We consider the case of a cubic tree-level superpotential:
\begin{equation}\label{dvn169}
    \wtree=\tr\nleft(\frac{1}{2}m\Phi^2+\frac{g}{3}\Phi^3\right).
\end{equation}
The planar free energy in the matrix model was first obtained
in~\cite{brezinplanar}. The effective superpotential in the gauge
theory has been obtained already in~\cite{0103067}, and later in
e.g.~\cite{0210238} and~\cite{ferretti,0311066} using the
Dijkgraaf-Vafa conjecture.

The critical points for the cubic potential are $0$ and $-m/g$.
Let us choose the vacuum where all eigenvalues are $0$ and the
gauge symmetry thus is unbroken. Plugging into the single-cut
solution~\eqref{dvonecutsol} gives:
\begin{subequations}\label{dvonecutsolcubic1}
\begin{eqnarray}
    0&=&mz+gz^2+2g\De,\label{dvn170}\\
    \Shat&=&m\De+2gz\De,\label{dvn171}\\
    \weffpert&=&N\left(m\left(\frac{1}{2}z^2+\De\right)+g\left(\frac{1}{3}z^3+2z\De\right)\right)-N\Shat\ln\nleft(\frac{\De}{\La_u^2}\right)\label{dvn172}.
\end{eqnarray}
\end{subequations}
Here we have defined $z=(a+b)/2$ and $\De=((b-a)/4)^2$ as
in~\cite{0211069}. We can use~\eqref{dvn170} to solve for $\De$.
After a couple of rewritings the result is (as also obtained
in~\cite{0211069} using Seiberg-Witten curves):
\begin{subequations}\label{dvonecutsolcubic2}
\begin{eqnarray}
    \Shat&=&-\frac{m^3}{2g^2}\frac{gz}{m}\left(1+\frac{gz}{m}\right)\left(1+2\frac{gz}{m}\right),\label{dvn173}\\
    \weff&=&N\frac{m^3}{g^2}\left(\frac{1}{2}\left(\frac{gz}{m}\right)^2+\frac{1}{3}\left(\frac{gz}{m}\right)^3\right)+N\Shat\ln\nleft(1+2\frac{gz}{m}\right)\nonumber\\
    &&+N\Shat\left(1-\ln\nleft(\frac{\Shat}{m\La_u^2}\right)\right)\label{dvn174}.
\end{eqnarray}
\end{subequations}

In~\eqref{dvonecutsolcubic2} we have actually found the
Veneziano-Yankielowicz superpotential by solving the matrix model
exactly! This result has been pointed out in~\cite{0401101}. It is
in accordance with the result we obtained in
section~\ref{dvsecmatrixsubmeasure}. That we get the exact right
form also tells us that our normalisation of the measure
(essential in equation~\eqref{dvn147}) is well-chosen. Naturally,
in this exact solution of the matrix model there is now no need to
add the Veneziano-Yankielowicz superpotential by hand as
in~\eqref{dvn12} and hence we have exchanged $\weffpert$ with the
more suitable~$\weff$ in~\eqref{dvn174}. We emphasise that the
Dijkgraaf-Vafa conjecture for the relation between the matrix
model and the gauge theory has not been proven for the
$-N\Shat\ln\Shat$ term. In the diagrammatic proof that we will
give in chapter~\ref{chpproof} only the perturbative behaviour of
$\Shat$ is captured. And in the generalised Konishi anomaly
proof~\cite{0211170}, which in fact relates the exact
superpotential to the exact free energy in the matrix model, the
relation is only proven for terms depending on the couplings $g_k$
in the tree-level superpotential.\footnote{Naturally, it is
captured by the non-perturbative methods in the proof using
Seiberg-Witten curves~\cite{0210135}.} Thus the Dijkgraaf-Vafa
conjecture does not give a derivation of the
Veneziano-Yankielowicz superpotential as also discussed above.

Let us briefly look at the case where we also consider the abelian
part of the supersymmetric gauge field strength
(section~\ref{dvsecdvunsubdvun}). We should here remember the
extra double derivative term in~\eqref{dvn17}. Let us assume
that~\eqref{dvn17} can be used for the exact free energy of the
matrix model found above, and let us consider the non-perturbative
term. Due to the double derivative term in~\eqref{dvn17} we should
then have an extra term which is the derivative of the
Veneziano-Yankielowicz superpotential (since~\eqref{dvn174} is the
matrix model free energy differentiated once). Further, we should
remember to replace $\Shat$ with the full glueball superfield $S$.
However, when expanding $S$ in $\Shat$ and $w_\al$ as
in~\eqref{dvn15} we will have cancellations such that the two
non-perturbative terms simply gives the Veneziano-Yankielowicz
superpotential expressed in the traceless glueball superfield
$\Shat$ as in~\eqref{dvn174}. That is, we have no dependence on
the abelian part $w_\al$. This is as expected since this part is
decoupled, as discussed in section~\ref{dvsecdvunsubdvun}, and IR
free. The cancellation is just the same as we saw for the gauge
coupling in~\eqref{dvn18} which is zero in the unbroken case. As
we noted in section~\ref{dvsecdvunsubdvun} it is not clear how to
add the non-perturbative part by hand in the unbroken case when we
consider the abelian parts. However, in the light of the above we
could guess that the full superpotential is obtained by the
extension of the Dijkgraaf-Vafa conjecture where we
use~\eqref{dvn17} to give the full effective superpotential using
the exact (non-perturbative) solution of the matrix model.

Since~\eqref{dvn173} is a cubic equation we can solve it exactly
as:
\begin{equation}\label{dvn175}
    \frac{gz}{m}=A+\frac{1}{12A}-\frac{1}{2},\qquad
    A\defi\sqrt[3]{-\frac{1}{2}\frac{g^2}{m^3}\Shat+\sqrt{-\frac{1}{12^3}+\frac{1}{4}\left(\frac{g^2}{m^3}\Shat\right)^2}}.
\end{equation}
Here we have chosen the solution that satisfies\footnote{Since
$\Shat\defi\gm$, the matrix model classical limit,
$\gm\rightarrow0$, corresponds to $\Shat\rightarrow0$.}
$\lim_{\Shat\rightarrow0}z=0$ thus ensuring $0\in[a,b]$ i.e. we
have chosen the cut around the critical point $0$. Naturally,
there is another solution corresponding to
$\lim_{\Shat\rightarrow0}z=-m/g$. Inserting~\eqref{dvn175}
into~\eqref{dvn174} thus explicitly gives us the exact effective
superpotential.

\begin{figure}\caption{}\label{dvfign7}
\begin{center}
\includegraphics{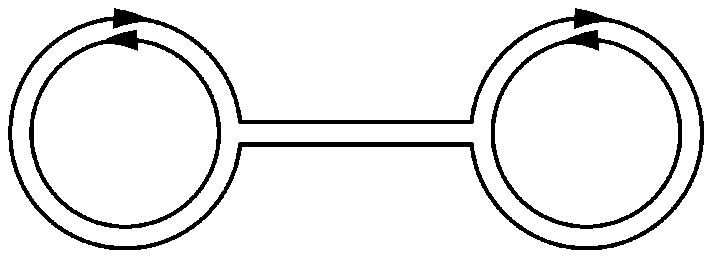}
\end{center}
    \begin{center}
        The dumbbell diagram contributing to the second order term
        in $\weffpert\big(\Shat\big)$.
    \end{center}
\end{figure}

Using~\eqref{dvn175} we can expand the perturbative part of
$\weff$ in a power series in $\Shat$. To third order the result
is:
\begin{equation}\label{dvn176}
    \weffpert=-N\frac{m^3}{g^2}\left(2\left(\frac{g^2}{m^3}\Shat\right)^2+\frac{32}{3}\left(\frac{g^2}{m^3}\Shat\right)^3+\cO\nleft(\Shat^4\right)\right).
\end{equation}
We can compare this with the results that we get from the matrix
model diagrams. The contribution from a given diagram is given
by~\eqref{dvn133} and a combinatorial factor. Since there is no
diagram with less than three index loops the first non-zero term
must be of order $3$ in the matrix model free energy and hence
order 2 in $\weffpert$. There are two types of diagrams with three
index loops. Firstly we find the diagram given in
figure~\ref{dvfign3} which with our choice of tree-level potential
has a combinatorial weight of $1/6$.\footnote{When counting the
number of diagrams one should remember that only the planar
diagrams contribute. If the counting instead is done in the
adjoint single-line notation one should remember that the vertices
are proportional to $\tr\nleft(T_aT_bT_c\right)$. Thus the legs in
the vertices can not be permuted arbitrarily, but only cyclically,
and we should only count the subgroup of diagrams that corresponds
to planar diagrams in the double line notation.} Secondly, we have
the diagram shown in figure~\ref{dvfign7}
which has the combinatorial weight $1/2$. The total contribution
from the two diagrams is then $-\frac{2}{3}m^{-3}g^2(\gs N')^3$
where the minus sign stems from the definition of the free energy.
This gives the contribution $-2Nm^{-3}g^2\Shat^2$ to $\weffpert$
using the Dijkgraaf-Vafa conjecture~\eqref{dvn13} -- in agreement
with~\eqref{dvn176}.

We note that the form of the expansion of
$\weffpert\big(\Shat\big)$ is completely determined by the
analysis we made in section~\ref{dvsecweffsubnonpert}. Remembering
the results obtained there we must demand that $\weff$ has a power
series expansion in $g$. Further we can use the symmetries in
table~\ref{dvtablen2}. This tells us that we have a global
$\un{1}$ symmetry under which $\weff$ and $\Shat$ are invariant,
$m$ has charge $-2$, and $g$ has charge $-3$. Secondly we have a
$\un{1}_R$ symmetry under which $m$, $g$, $\Shat$ and $\weff$ has
charge $2$. This constrains the form of $\weffpert$ to be a power
series expansion of the form:
\begin{equation}\label{dvn177}
    \weffpert=\frac{m^3}{g^2}\sum_k
    c_k\left(\frac{g^2}{m^3}\Shat\right)^k.
\end{equation}
This is exactly the form we get as we can see from~\eqref{dvn174}
and~\eqref{dvn175}. We note that $\weffpert$ is singular in
$m\rightarrow0$ which also is expected since the chiral field has
been integrated out. The possibility of inverse powers of $m$ is
also the reason that we do not have to restrict to linear terms in
the R-charged variables as in~\eqref{dvn103}. We thus see that the
essence of the Dijkgraaf-Vafa conjecture for $\weffpert$ is that
it determines the coefficients in~\eqref{dvn177}: The coefficients
are obtained by combinatorial counting of planar diagrams in the
related matrix model.

\subsection{Even Tree-Level Superpotential}
Let us briefly examine the case of an even tree-level
superpotential. We consider the unbroken case where we expand
around the zero critical point. Here we have a $\la\mapsto-\la$
symmetry in the eigenvalues. Thus $a=-b$ in~\eqref{dvonecutsol}
and the solution reduces to:
\begin{subequations}\label{dvonecutsoleven}
\begin{eqnarray}
  \Shat &=& \frac{1}{2}\sum_{p\geq1}g_{2p}\binom{2p}{p}\left(\frac{b-a}{4}\right)^{2p}, \label{dvn178}\\
  \weffpert &=&
  N\sum_{p\geq1}\frac{g_{2p}}{2p}\binom{2p}{p}\left(\frac{b-a}{4}\right)^{2p}-2N\Shat\ln\nleft(\frac{b-a}{4\La_u}\right).\label{dvn179}
\end{eqnarray}
\end{subequations}
This is actually the general solution one finds for a $\sun{N}$
gauge group as also obtained already in~\cite{0103067}.

%---------New section-----------

\section{Nilpotency of the Glueball Superfield}\label{dvsecnil}

As noted in section~\ref{dvsecdvunsubdvun} the glueball superfield
defined in~\eqref{dvn4} or~\eqref{dvn14} is a sum of terms
proportional to $\W^{\al a}\W^a_\al$ for
$a=1,\ldots,\dim\nleft(G\right)$. Here $G$ is $\sun{N}$ for
$\Shat$, $\un{N}$ for $S$ or we can think of the glueball
superfield for a general gauge group $G$. Since $\W^{\al a}$ is
Grassmannian, we see that the glueball superfield classically is
nilpotent:
\begin{equation}\label{dvn180}
    S^{\dim(G)+1}=0.
\end{equation}
This even holds true in perturbation theory as can be seen using
R-symmetries and dimensional analysis.

However, we can say something even more powerful if we consider
the \emph{chiral ring}. First we define chiral operators as gauge
invariant operators annihilated by the supercharge
$\Qbar_{\aldot}$. One can show that the products of chiral
operators are again chiral operators. By considering chiral
operators modulo terms like $\{\Qbar_{\aldot},\ldots\}$ (where the
dots indicate some gauge invariant operator) the equivalence
classes form a ring which, per definition, is the chiral ring. The
point is that one can show that in a supersymmetric vacuum, which
we will assume in the following, the expectation value of a
product of chiral operators does not depend on the chosen
representatives. In fact, the expectation values factorise and are
space-time independent. There is a one-to-one correspondence to
chiral superfields by noting that the lowest component of a chiral
superfield is a chiral operator. The chiral ring is then obtained
by considering the chiral fields modulo chirally exact terms of
the form $\Dbar\Dbar F$ where $F$ is a superfield for which
$\Dbar^{\aldot}F$ is gauge invariant. Now, in the chiral ring we
have classically and in perturbation theory~\cite{0211170}
(conjectured for all groups, but certainly true for the classical
groups):
\begin{equation}\label{dvn181}
    S^h=0\qquad\textrm{(Chiral ring, perturbation theory)},
\end{equation}
where $h$ is the dual Coxeter number given in
table~\ref{dvtablen3}. Actually, we also have $S^{h-1}\neq0$
(anyway for $\sun{N}$) in the chiral ring. The
relation~\eqref{dvn181} is changed by non-perturbative effects (as
we have seen by instantons) into:
\begin{equation}\label{dvn182}
    S^h=a(G)\La^{3h}\qquad\textrm{(Chiral ring, exact)},
\end{equation}
where $a(G)$ is a normalisation depending on the group. As
mentioned in section~\ref{dvsecweffsubvy} we have $a=1$ for
$\sun{N}$ . Naturally, also the relation~\eqref{dvn180} is changed
non-perturbatively, see also~\cite{0307130}.

We thus see that we should be very careful when employing the
Dijkgraaf-Vafa conjecture for the terms of order $h$ and higher in
$S$. Actually, in~\cite{0303104} (and using the generalised
Konishi anomaly method in~\cite{0304119} and~\cite{0304138}) in
the case of a $\mathrm{Sp}(k)$ gauge group with matter in the form
of a two-index antisymmetric tensor chiral superfield the
superpotential was obtained from the matrix model using the
general Dijkgraaf-Vafa conjecture (as we will introduce in the
next section). The results were compared to superpotentials
obtained earlier in gauge theory using the power of holomorphy. It
was found that the results agreed up to order $k$ and
discrepancies were found at order $h=k+1$ and higher orders. An
even more basic example of discrepancies is found by considering a
$\un{1}$ gauge group with adjoint chiral matter. Here we can think
of the dual Coxeter number as being zero and indeed we have
immediate discrepancy. Using the Dijkgraaf-Vafa conjecture we
really can define a superpotential in $S$ via the matrix model
whereas there should be no glueball superpotential in ordinary
gauge theory (the theory is IR free).

The point here is~\cite{0304271} that the terms in the
superpotential of order $h$ or higher depend on the UV completion
of the theory. To see this let us in the UV turn on the term:
\begin{equation}\label{dvn183}
    \int\dlor\dtotha\sum_{k\geq h}a_k S^k,
\end{equation}
where the glueball superfields here should be considered quantum
mechanically smeared. Due to~\eqref{dvn181} this will not change
the action classically (nor perturbatively), however, in the IR
the term is relevant due to the non-perturbatively corrected
chiral ring relation~\eqref{dvn182}. Thus two theories which agree
classically can differ quantum mechanically. Thus we can split the
superpotential as
$\weff\nleft(S\right)=W_R\nleft(S\right)+W_A\nleft(S\right)$ where
$W_R$ consists of the terms of order less than $h$, and $W_A$ of
the terms of order $h$ and greater. $W_R$ is determined
unambiguously from the tree-level superpotential by integrating
out the matter fields, and the coefficients are given by the
Dijkgraaf-Vafa conjecture. $W_A$, on the other hand, is ambiguous
and should really be seen as a part of the definition of the
quantum theory -- i.e. it depends on the choice of F-term
completion for the theory.

In the case of a $\un{N}$ gauge group with adjoint matter as
considered in section~\ref{dvsecdvun} there is a natural way of
determining the F-term completion. Simply consider $\un{Nk}$ where
$k$ is a (large) positive integer. Here the coefficients of $S^n$
are determined unambiguously up to order $Nk$ by the
Dijkgraaf-Vafa conjecture. And actually we see from~\eqref{dvsun}
that the dependence on $k$ is a simple multiplicative factor. The
F-term completion for a $\un{N}$ theory is then simply defined by
noting that the coefficient of $S^n$ is determined by the
coefficient in the $\un{Nk}$ theory with $Nk>n$ divided by $k$.
Thus the effective superpotential to any order is obtained from
the matrix model using the Dijkgraaf-Vafa conjecture. This can
also be done for the other classical groups, but the dependence on
$N$ is generally not multiplicative.

There is, however, another approach to F-term completion
applicable for any classical group, denoted $\mathrm{G}(N)$, and
any matter representation. Instead of $\mathrm{G}(N)$ we consider
the larger supergroup $\mathrm{G}(N+k|k)$ (which corresponds to
adding $k$ brane/anti-brane pairs). With this group the terms in
the superpotential are unambiguous up to order $N+2k$ and the
coefficients are actually independent of $k$.\footnote{This is due
to the supertrace in an index loop giving $(N+k)-k=N$.} We can
thus take $k\rightarrow\infty$ to determine the full
superpotential -- the F-term completion. In the end we then use
that all of the $\mathrm{G}(N+k|k)$ theories have a Higgs branch
where in the IR the supergroup is Higgsed down to $\mathrm{G}(N)$.
The completion obtained in this way corresponds, in essence, to
treating the $S$'s in the different index loops (in the gauge
theory) as being distinct.

The $G(N+k|k)$-completion also tells us when we have discrepancies
compared to the standard gauge completion. This happens when there
are residual instanton effects in the broken part of the group in
the Higgsing $\mathrm{G}(N+k|k)\mapsto \mathrm{G}(N)$. By the
knowledge of section~\ref{dvsecweffsubinstchiral} a necessary
condition for this is that the third homotopy group
${\pi_3(\mathrm{G}(N+k|k)/\mathrm{G}(N))}\neq0$. In the case of a
$\un{N}$ gauge group with adjoint matter we only have residual
instanton effects for $\un{1+1|1}\mapsto\un{1}$ and this explains
the discrepancy we found for the $\un{1}$ theory. The residual
instanton effects also explain the discrepancy in the
$\mathrm{Sp}(k)$ case considered above. The ambiguity is also
investigated in
e.g.~\cite{0307063},~\cite{0307285},~\cite{0310111},~\cite{0311181},~\cite{0311238},~\cite{0405101}
and~\cite{0406253}.

Thus, in conclusion, we can trust the glueball superpotential
obtained by the Dijkgraaf-Vafa conjecture up to order $h$
unambiguously, whereas the terms of order $h$ and greater
correspond to a choice of F-term completion. In the case of a
$\un{N}$ gauge group ($N> 1$) with adjoint matter we can use the
Dijkgraaf-Vafa conjecture to any order since the choice of
completion here is the natural one. We should then think of $S$ as
an unconstrained elementary field and the relation~\eqref{dvn182}
as obtained on-shell using the equations of motion.

%---------New section-----------

\section{The General Dijkgraaf-Vafa Conjecture}\label{dvsecgeneral}

Now that we have understood the Dijkgraaf-Vafa conjecture in
details in the case of a $\un{N}$ gauge group with a single
adjoint chiral field, let us end this chapter by presenting the
Dijkgraaf-Vafa conjecture in the case of more general gauge groups
and matter representations following~\cite{0208048}. We assume an
$\Nscr=1$ supersymmetric theory with a classical gauge group $G$
i.e. a product with factors of $\un{N}$, $\son{N}$ and
$\mathrm{Sp}(k)$. We further assume that the matter content of the
theory in the form of chiral fields, $\Phi^a$, allows a double
line notation as in section~\ref{dvsecdouble}. Also, we demand
that it is possible to add mass terms to the matter fields. We
will here think of a single adjoint chiral field and flavours,
i.e. chiral fields in the fundamental/anti-fundamental
representation, with Yukawa couplings. One could consider more
exotic matter, but in general the comparison between the gauge
theory and the matrix model should be done diagram by diagram and
we do not have a nice relation to the total free energy of the
matrix model as below~\cite{0303104}. The matter is described by a
tree-level superpotential
$\wtree\nleft(\Phi^a\right)$.\footnote{This should only include
single traces as we discussed at the end of
section~\ref{dvsecdvunsublagr}.}

The classical supersymmetric vacua for our system are
by~\eqref{dvn42} again determined by the critical points of
$\wtree$. We will assume a massive supersymmetric vacuum where the
gauge group is broken to $\prod_iG_i$ where each of the $G_i$'s is
one of the classical groups $\un{N_i}$, $\son{N_i}$ or
$\mathrm{Sp}(k_i)=\mathrm{USp}(N_i=2k_i)$. All fields should be
massive in the vacuum except the $\Nscr=1$ super Yang-Mills part
for the gauge group $\prod_iG_i$. Corresponding to each of the
$G_i$'s we have a glueball superfield $S_i$ defined analogously
to~\eqref{dvn4}. We will here ignore the abelian part as in
section~\ref{dvsecdvunsubdv} -- the result when considering this
is as in section~\ref{dvsecdvunsubdvun}.

The Dijkgraaf-Vafa conjecture once again tells us that the
effective glueball superpotential is determined as a sum of a
superpotential $\weffpert(S_i)$, perturbative in $S_i$, and the
Veneziano-Yankielowicz superpotential (extending~\eqref{dvn120}):
\begin{equation}\label{dvn184}
    \wvy=\sum_i
    C\nleft(\mathrm{adj}(G_i)\right)S_i\bigg(1-\ln\frac{S_i}{a_i\La_i^3}\bigg).
\end{equation}
Here $a_i$ is the normalisation from~\eqref{dvn118} depending on
$G_i$, and $C\nleft(\mathrm{adj}(G_i)\right)$ is the quadratic
invariant for the adjoint representation of $G_i$ which, in a
proper normalisation, equals the dual Coxeter number for $G_i$
given in table~\ref{dvtablen3}.\footnote{We do not include the
small $N$ exceptions as in~(\ref{dvn121.5}) since our
$\mathrm{G}(N+k|k)$-completion discussed in section~\ref{dvsecnil}
has no small $N$ exceptions because we take $k$ large.}

Furthermore, $\weffpert$ is determined by the related matrix model
with potential given by $\frac{1}{\gs}\wtree$. Here the matrices
can now correspond to all of the classical groups $\un{N'}$,
$\son{N'}$ and $\mathrm{USp}(N')$, so e.g. for $\son{N'}$ we have
real antisymmetric matrices. If we have fundamental matter, we
should, naturally, also include integrations for this. In the
matrix model we expand around the saddle-point corresponding to
the breaking of $G\mapsto\prod_iG_i$ where each $G_i$ has a
corresponding $N'_i$. As explained in
section~\ref{dvsecdoublesubtopo} we can classify the matrix model
diagrams topologically. Analogously to
section~\ref{dvsecdoublesublargen} we can make an expansion of the
free energy in powers of the matrix model coupling $\gs$ using the
Euler characteristic $\chi$:
\begin{equation}\label{dvn185}
    W_{\mathrm{free}}=\sum_{\chi}\gs^{-\chi}\cF_\chi\nleft(\gs
    N'_i\right)=\frac{1}{\gs^2}\cF_{\chi=2}\nleft(\gs
    N'_i\right)+\frac{1}{\gs}\cF_{\chi=1}\nleft(\gs
    N'_i\right)+\cO\nleft(\gs^0\right).
\end{equation}
As we will see in the proof of the Dijkgraaf-Vafa conjecture, it
is only diagrams with Euler characteristic $\chi \geq1$ that
contribute. In the case of a $\un{N}$ gauge group with a single
adjoint chiral field this means that only the $\chi=2$ diagrams
contribute corresponding to genus $g=0$. However, in the general
case we can also have a $\chi=1$ contribution. The contribution to
$\cF_{\chi=1}$ comes from diagrams with two different topologies
as we see from~\eqref{dvn131}. We have the diagrams with one
boundary (stemming from the fundamental matter) which have the
topology of the disk $D^2$ and whose contribution to
$\cF_{\chi=1}$ we denote $\cF_{D^2}$. Secondly, we have the
diagrams with one cross-cap with topology of the projective plane
$\mathbb{R}\mathbb{P}^2$ stemming from the cross-over in the
double line propagators of $\son{N}$ and $\mathrm{USp}(N)$, see
figure~\ref{dvfign5}. The contribution from these diagrams to
$\cF_{\chi=1}$ is denoted $\cF_{\mathbb{R}\mathbb{P}^2}$. We thus
have:
\begin{equation}\label{dvn186}
    \cF_{\chi=1}\defi\cF_{D^2}+\cF_{\mathbb{R}\mathbb{P}^2}.
\end{equation}
We have to distinguish between these two contributions since they
have different weights in the Dijkgraaf-Vafa conjecture, even
though the Euler characteristic is the same. To be consistent with
the notation depending on the topology we also define
$\cF_{S^2}\defi\cF_{\chi=2}=\cF_{g=0}$.

The relation to the gauge theory is similarly to~\eqref{dvn11}
through $2C(\mathrm{fund})S_i\defi\gs N'_i$ and we can state the
Dijkgraaf-Vafa conjecture as (for the traceless glueball
superfields):
\begin{thmdvgen}
\begin{subequations}\label{dvgeneral}
\begin{eqnarray}
    \weff(S_i)&=&\weffpert(S_i)+\wvy(S_i),\label{dvn187}\\
    \weffpert(S_i)&=&\sum_i N_i\frac{\partial\cF_{S^2}(S'_i)}{\partial
    S'_i}+4\cF_{\mathbb{R}\mathbb{P}^2}(S'_i)+\cF_{D^2}(S'_i),\label{dvn188}\\
    S'_i&\defi& \gs N'_i,\quad S'_i=2C(\mathrm{fund})S_i.\label{dvn188.1}
\end{eqnarray}
\end{subequations}
\end{thmdvgen}
As we will see in section~\ref{proofsecdiasubreduc} the reason for
including $C(\mathrm{fund})$ is that it appears in the definition
of $S$ in~\eqref{dvn112.5}.\footnote{We note that $S'_i$ is
invariant under the scaling of the gauge group generators
discussed at the end of section~\ref{susysecchiralsubgauge}. This
means that $\weffpert$ is invariant under such scalings as it
should be.} For a $\un{N}$ gauge with adjoint matter and
$C(\mathrm{fund})=1/2$ this, naturally, reduces to~\eqref{dvsun}.

This form of the conjecture is corrected from the original
proposal in~\cite{0208048} in the case of $\son{N}$ and
$\mathrm{USp}(N)$ gauge groups as was first found
in~\cite{0211261} and~\cite{0211291} (also investigated
in~\cite{0210148} and~\cite{0212069}). There it is also shown
(using the loop equations) that
$\cF_{\mathbb{R}\mathbb{P}^2}=\mp\half\partial\cF_{S^2}/\partial
S$ with minus for $\son{N}$ and plus for $\mathrm{USp}(N)$. The
addition of fundamental matter was first done in~\cite{0210291}
(see also~\cite{0211254}, \cite{0212212}, \cite{0305263}, \cite{0211009}, \cite{0211052}, \cite{0211075},
\cite{0211082}, \cite{0211189}, \cite{0211249},
\cite{0211271}, \cite{0212083}, \cite{0212095}, \cite{0212121},
\cite{0212274}, \cite{0301217}, \cite{0303115}, \cite{0305096},
\cite{0306007}, \cite{0307082}, \cite{0307115},
and~\cite{0309084} for the $\un{N}$-case
and~\cite{0211287},~\cite{0212231},~\cite{0301011},~\cite{0301226},~\cite{0306068},
and~\cite{0307190} for the $\son{N}$ and $\mathrm{USp}(N)$-case).
Let us finally mention that the reduction to zero-momentum modes
holds true for any gauge group and any (massive) representation as
shown in~\cite{0304271}.

%-----New Chapter-----------------------------------------------

\chapter[Diagrammatic Proof of the Dijkgraaf-Vafa\ldots]{Diagrammatic Proof of the Dijkgraaf-Vafa
Conjecture}\label{chpproof}

In this chapter we will present the diagrammatic proof for the
form of $\weffpert$ in the Dijkgraaf-Vafa conjecture.
Following~\cite{0211017} the strategy will be as follows: Using
holomorphy we will choose the anti-chiral part of the action
suitably and then integrate it out. This gives us an effective
chiral action for which we can develop perturbative methods. In
the case of a classical gauge group and a single adjoint chiral
field we then show that the path integral reduces to zero-momentum
modes and that we can relate the diagrams to matrix model diagrams
in the conjectured way.

At the end of the chapter we will show the localisation to
zero-momentum modes for general gauge groups and matter
representations.

\section{Setup for Perturbation Theory}

\subsection{Gauge Covariant Notation}\label{proofsecdiasubgauge}
In the following it will be convenient to use a gauge covariant
notation for the superspace derivatives and fields. We will use
the so-called gauge chiral representation (described
in~\cite{0108200}). The point is to define derivatives $\nabla_A$
transforming under gauge transformations with the chiral
superfield $\La\nleft(x,\tha,\thabar\right)$
(section~\ref{susysecchiralsubgauge}) as:
\begin{equation}\label{pron1}
    \nabla_A\mapsto e^{-i\La}\nabla_A e^{i\La}.
\end{equation}
Using the transformation property~\eqref{susyn52} of the vector
superfield $e^{2V}\mapsto e^{-i\La^\dagger}e^{2V}e^{i\La}$, $\La$
being chiral, and $\La^\dagger$ being anti-chiral we get the
covariant derivatives:
\begin{equation}\label{pron2}
    \nabla_\al\defi e^{-2V}D_\al e^{2V},\quad
\nablabar_{\aldot}\defi\Dbar_{\aldot},\quad
\nabla_{\al\aldot}\defi-i\{\nabla_\al,\nablabar_{\aldot}\}.
\end{equation}
We note that $\Dbar_{\aldot}$ is the conjugate of $D_\al$ whereas
the same is not true for $\nablabar_{\aldot}$ and $\nabla_\al$.
Furthermore, we see that:
\begin{equation}\label{pron2.5}
    \nabla_\al=D_\al+e^{-2V}\left(D_\al e^{2V}\right),
\end{equation}
where the last term gives the gauge connections.

Analogous to the chiral fields we define the \emph{covariantly}
chiral fields as being annihilated by the covariant derivative
$\nablabar_{\aldot}$. The covariantly chiral fields are,
naturally, simply the chiral fields. Correspondingly we define the
covariantly anti-chiral fields $\Phicov$ as being annihilated by
$\nablabar_{\aldot}$, however, acting from the right. Given a
chiral field $\Phi$ these take the form:
\begin{equation}\label{pron3}
    \Phicov\defi\Phi^\dagger
e^{2V},\quad\Phicov\overleftarrow{\nabla}_\al=\Phi^\dagger
\overleftarrow{D}_\al e^{2V}=0.
\end{equation}
This also means that $\Phicov$ transforms covariantly as
$\Phicov\mapsto\Phicov e^{i\La}$.

As an analogue of~\eqref{dvn61} we define for a chiral field
$\Phi$:
\begin{equation}\label{pron4}
    \square_+\Phi\defi\frac{1}{16}\nablabar\nablabar\nabla\nabla\Phi.
\end{equation}
Using that (by simple calculation)
\begin{equation}\label{pron5}
    [\nablabar_{\aldot},\nabla_\al^{\ph{\al}\aldot}]=i8\W_\al,
\end{equation}
we easily get:
\begin{equation}\label{pron6}
    \square_+\Phi=\frac{1}{8}\nabla^{\al}_{\ph{\al}\aldot}\nabla_\al^{\ph{\al}\aldot}\Phi-\frac{1}{2}\W^{\al}\nabla_\al\Phi-\frac{1}{2}\nabla^\al\W_\al\Phi.
\end{equation}
Defining
\begin{equation}\label{pron6.5}
    \sqcov=\frac{1}{8}\nabla^{\al}_{\ph{\al}\aldot}\nabla_\al^{\ph{\al}\aldot},
\end{equation}
which for $V=0$ simply is $\square$, and rewriting the last
differentiation in~\eqref{pron6}, where $\nabla^\al$ works on both
$\W_\al$ and $\Phi$, we get:
\begin{equation}\label{pron7}
    \square_+\Phi=\Big(\sqcov-\W^\al
\nabla_\al-\frac{1}{2}\left(\nabla^\al\W_\al\right)\Big)\Phi.
\end{equation}
Here we have used that $\nabla^\al$ obeys the Leibnitz rule, and
we should remember that when $\nabla^\al=e^{-2V}D^\al e^{2V}$
works on $\W_\al$, then $V$ should be in the adjoint
representation.

For a real representation, e.g. the adjoint representation which
is the case of our interest, we have $V^T=-V$ and hence
$(\Phicov)^T=e^{-2V}\Phibar$. $\Phicov$ is then annihilated from
the left: $\nabla_\al(\Phicov)^T=0$. Analogously to~\eqref{pron4}
we then define in a real representation:
\begin{equation}\label{pron8}
    \square_-\big(\Phicov\big)^T\defi\frac{1}{16}\nabla\nabla\nablabar\nablabar\big(\Phicov\big)^T.
\end{equation}
On covariantly chiral fields we then see:\footnote{There is a typo
in this equation in~\cite{0211017}.}
\begin{equation}\label{pron9}
    \nabla\nabla\square_+\Phi=\square_-\nabla\nabla\Phi,
\end{equation}
since $\nabla\nabla\Phi$ is covariantly anti-chiral. We note that,
by definition, $\square_+$ maps chiral fields to chiral fields and
$\square_-$ maps covariantly anti-chiral field´s to covariantly
anti-chiral fields.

\subsection{Anti-Chiral Part}
Let us consider an $\Nscr=1$ supersymmetric gauge theory with
arbitrary chiral matter given by the chiral field $\Phi$. We will
assume that the Lagrangian is renormalisable with the exception
that we allow arbitrary tree-level superpotentials, $\wtree$. We
want to determine the perturbative part of the effective glueball
superpotential, $\weffpert$, (discussed in
section~\ref{dvsecglueball}) obtained by integrating out the
chiral superfield while treating the vector superfield as an
external background field. Thus the relevant part of the action is
given by:
\begin{equation}\label{pron10}
    S\nleft(\Phi,\Phibar\right)=A\int\dlor\dsuper\Phi^\dagger
e^{2V}\Phi+\left(\int\dlor\dtotha\wtree\left(\Phi\right)+\cc\right).
\end{equation}
Here we have allowed for a normalisation $A$ of the \kahler{} term
as we e.g. have in the case of $\lagrun$ in~\eqref{dvn20.5}. We
will demand that the supersymmetric vacuum is massive and for now
consider the case of an unbroken gauge group. For convenience, we
will assume that $\Phi$ transforms in a real representation thus
allowing the mass term $\half m\Phi^T\Phi$. However, our results
until section~\ref{proofsecpert} are valid in any representation.

The main point in this section is that $\weffpert$ can only depend
holomorphically on the couplings in $\wtree$ as explained in
section~\ref{dvsecweffsubholo}. In the notation of~\eqref{dvn13.1}
what we want to determine is then $\Zholo$. We can thus treat both
$\Phi$ and $\Phibar$, and $\wtree(\Phi)$ and
$\overline{\wtree}(\Phibar)$ independently and we can choose the
form of $\overline{\wtree}(\Phibar)$ freely. We
choose:\footnote{One might think that it would easier to take
$\overline{\wtree}\nleft(\Phibar\right)=0$ and then calculate the
partition function. However, the diagrammatic method that we will
use turns out to be easiest~\cite{0211017}.}
\begin{equation}\label{pron11}
    \overline{\wtree}\nleft(\Phibar\right)=\frac{1}{2}\mbar\Phi^\dagger\Phibar=\frac{1}{2}\mbar\Phibar^T\Phibar.
\end{equation}
Thus the action is quadratic in $\Phibar$ and we can simply
integrate it out by completing the square. However, before we do
this let us remark that we can scale the normalisation $A$
in~\eqref{pron10} away by taking $\Phibar\mapsto\Phibar/A$ and
$\mbar\mapsto A^2\mbar$. Since the final result does not depend on
$\mbar$, we can choose $\mbar$ freely and thus absorb the
normalisation $A$ in this way. This was used in
section~\ref{dvsecdvunsublagr} to note that there is no need for
the $\Nscr=2$ supersymmetry of the Lagrangian without the
tree-level superpotential in section~\ref{dvsecdvunsubdv}
and~\ref{dvsecdvunsubdvun}.

Switching to the gauge covariant notation introduced in the last
subsection, the part of the action depending on $\Phibar$ can then
be written:
\begin{equation}\label{pron12}
    \De S = \int\dlor\dsuper\Phicov\Phi+\int\dlor\dtothabar\frac{\mbar}{2}\Phicov\big(\Phicov\big)^T,
\end{equation}
where we have used that $\Phicov\big(\Phicov\big)^T=\Phi^\dagger
e^{2V}e^{2V^T}\Phibar=\Phi^\dagger\Phibar$ in our real
representation. We could change the functional integration over
$\Phibar$ to an integration over $\Phicov$. In the next section we
will change variables from $\Phi$ to $e^{2V}\Phi$ so the Jacobian
under
$\mathcal{D}\Phi\DPhibar\mapsto\mathcal{D}\big(e^{2V}\Phi\big)\mathcal{D}\big(e^{-2V}\Phibar\big)$
should be independent of $V$ in analogy with~\eqref{dvn98} since
there should be no anomalies in a real representation.

We can rewrite the last term of~\eqref{pron12} as:
\begin{equation}\label{pron13}
    \int\dlor\dtothabar\frac{\mbar}{2}\Phicov\big(\Phicov\big)^T=-\frac{1}{4}\int\dlor\dsuper\frac{\mbar}{2}\Phicov\frac{1}{\square_+}\nablabar\nablabar\big(\Phicov\big)^T,
\end{equation}
where $\square^{-1}_+$, naturally, should be defined modulo its
kernel. To prove~\eqref{pron13} we  use the conjugated analogue
of~\eqref{susyn38} to replace $\int\dtotha$ with $-\frac{1}{4}DD$.
Then by the Leibnitz rule for $D_\al$:
\begin{equation}\label{pron14}
    \frac{\mbar}{2\cdot16}DD\Phicov\frac{1}{\square_+}\nablabar\nablabar\big(\Phicov\big)^T=\frac{\mbar}{2\cdot16}\Phi^\dagger
DD
e^{2V}\frac{1}{\square_+}\nablabar\nablabar\big(\Phicov\big)^T=\frac{\mbar}{2\cdot16}\Phicov
\nabla\nabla\frac{1}{\square_+}\nablabar\nablabar\big(\Phicov\big)^T.
\end{equation}
Finally using~\eqref{pron9} to write
$(\square_+)^{-1}=\left(\nabla\nabla\right)^{-1}(\square_-)^{-1}\nabla\nabla$
and using~\eqref{pron8} to note that
$(\square_-)^{-1}\nabla\nabla\nablabar\nablabar\big(\Phicov\big)^T=16\big(\Phicov\big)^T$,
we get the wanted result.

We can then complete the square as:
\begin{multline}\label{pron15}
    \De S
=\int\dlor\dsuper\bigg\{-\frac{\mbar}{8}\left(\Phicov-\frac{1}{4\mbar}(\nabla\nabla\Phi)^T\right)\frac{1}{\square_+}\nablabar\nablabar\left(\Phicov-\frac{1}{4\mbar}(\nabla\nabla\Phi)^T\right)^T\\
+\frac{1}{8\cdot16\mbar}(\nabla\nabla\Phi)^T\frac{1}{\square_+}\nablabar\nablabar\nabla\nabla\Phi\bigg\}.
\end{multline}
The only non-trivial part in completing the square is the term:
\begin{multline}\label{pron16}
    \frac{1}{4\cdot8}(\nabla\nabla\Phi)^T\frac{1}{\square_+}\nablabar\nablabar\big(\Phicov\big)^T=\frac{1}{32}\big(DD(e^{2V}\Phi)^T\big)
e^{2V}\frac{1}{\square_+}\nablabar\nablabar\big(\Phicov\big)^T\\=\frac{1}{32}\Phi^T\nabla\nabla\frac{1}{\square_+}\nablabar\nablabar\big(\Phicov\big)^T,
\end{multline}
where we in the last line have integrated $DD$ by parts which is
allowed since $\dtotha D\sim DDD=0$. As above,
$\nabla\nabla(\square_+)^{-1}\nablabar\nablabar$ can be replaced
with $16$ to show that the term reduces to the wanted
$\half\Phicov\Phi$.

Using~\eqref{pron15} we can integrate $\Phicov$ out by translating
the integration variable to
$\Phicov-\frac{1}{4\mbar}(\nabla\nabla\Phi)^T$. The Gaussian
integration then gives a factor depending on the determinant of
$(\square_+)^{-1}\nablabar\nablabar$ which seems to depend on $V$.
However, we do note that according to the method of unconstrained
superfields introduced in section~\ref{dvsecquantsubsuper} we
should really integrate over $\bar{\Pi}$ defined by (the complex
conjugate of) equation~\eqref{dvn65} i.e. $\Phibar=D D \bar{\Pi}$.
Analogous to the derivations in equations~\eqref{pron14}
and~\eqref{pron16} we can write:
\begin{equation}\label{pron17}
    -\frac{\mbar}{8}\Phicov\frac{1}{\square_+}\nablabar\nablabar\big(\Phicov\big)^T=-2\mbar\Pi^\dagger
D D \bar{\Pi}.
\end{equation}
Here the Gaussian integration just contributes with a constant
independent of the background field $V$ which we can disregard.

The integrating out procedure leaves us with the last term
in~\eqref{pron15} as a contribution to the action for $\Phi$. In
this term we can immediately use the definition~\eqref{pron4} of
$\square_+$ to see that it cancels out. Using~\eqref{susyn38} we
can then rewrite the term to a $\int\dtotha$-term:
\begin{equation}\label{pron18}
    \frac{1}{8\mbar}\int\dlor\dtotha\left(-\frac{1}{4}\Dbar\Dbar\right)\Phi^T\nabla\nabla\Phi=-\frac{1}{2\mbar}\int\dlor\dtotha\Phi^T\square_+\Phi,
\end{equation}
where we again used the definition of $\square_+$. Thus the end
result of this subsection is (using~\eqref{pron7}):
\begin{equation}\label{pron19}
  \Zholo = \int\DPhi e^{iS(\Phi)},
\end{equation}
with
\begin{equation}\label{pron20}
  S(\Phi)
=\int\dlor\dtotha\bigg\{\frac{-1}{2\mbar}\Phi^T\left(\sqcov-\W^\al
\nabla_\al-\frac{1}{2}\left(\nabla^\al\W_\al\right)\right)\Phi+\wtree\nleft(\Phi\right)\bigg\}.
\end{equation}

\subsection{Simplifications}

In this section we will simplify the result~\eqref{pron20} by
taking into consideration the form of the background and which
terms that can contribute.

In order to obtain the glueball superpotential (which has no
derivatives) we can think of the background field $\W_\al$ as
being constant. Following~\cite{0211017} we will further make the
following simplifications:
\begin{itemize}
    \item We choose the background such that $\W_\al$ is also
covariantly constant i.e.:
\begin{equation}\label{pron21}
    \nabla_{\al\aldot}\W_\be=0.
\end{equation}
By the definition of $\sqcov$ in~\eqref{pron6.5} we then conclude
that $\W_\al$ commutes with $\sqcov$. From~\eqref{pron20} we see
that we can choose the propagator in perturbation theory to be
like $\sqcov^{-1}$ and we can treat the $\W^\al \nabla_\al$ term
as an interaction giving $\W_\al$ insertions. With the
assumption~\eqref{pron21} we can thus move such insertions around
in the loops when only considering the space-time part.
    \item We drop the term $\nabla^\al\W_\al$ in~\eqref{pron20}
since we do not consider such contributions. This naturally calls
for a redefinition of $\Zholo$ not to include these terms,
however, we will see below that in a very simple background this
term is not present at all.
    \item Analogous to the gauge chiral representation in
section~\ref{proofsecdiasubgauge} we have a gauge anti-chiral
representation transforming covariantly with respect to
anti-chiral $\Labar$ gauge transformations. Here the covariantly
chiral field is $\Phi'=e^{2V}\Phi$ (and thus $\Phi'^T=\Phi^T
e^{-2V}$) and it is annihilated by
$\nablabar'_{\aldot}=e^{2V}\Dbar_{\aldot}e^{-2V}$. Switching to
this basis we can rewrite~\eqref{pron20} as:
\begin{equation}\label{pron22}
      S=\int\dlor\dtotha\bigg\{\frac{-1}{2\mbar}\Phi'^T\left(\sqcov'-\W'^\al
D_\al\right)\Phi'+\wtree\nleft(\Phi'\right)\bigg\},
\end{equation}
where the point is that we have $D_\al$ instead of $\nabla_\al$.
Here $\sqcov'=e^{2V}\sqcov e^{-2V}$, $\W'_\al=e^{2V}\W_\al
e^{-2V}$ and we have used the gauge invariance of the tree-level
superpotential to write $\wtree(\Phi)=\wtree(\Phi')$. We can then
change the functional integral to $\Phi'$ and drop the primes
since the glueball superfield depends on
$\tr\nleft(\W^\al\W_\al\right)=\tr\nleft(\W'^\al\W'_\al\right)$.
However, let us note that we also can obtain this reduction to
$D_\al$ simply using~\eqref{pron2.5} and dropping the term $\W^\al
e^{-2V}\left(D_\al e^{2V}\right)$ since by~\eqref{susyn53}
and~\eqref{susyn58.5} we see that (in Wess-Zumino gauge) this term
always contains at least one $\thabar$ and hence does not
contribute to the $\int\!\dtotha$-integral.

    \item Finally we can replace $\sqcov$ with the usual
$\square=\partial_\mu\partial^\mu$. This is because the connection
terms in $\sqcov$ generally only appear in the effective action in
order to covariantise derivative terms -- but we have none of
these. However, to see that we can drop the connection terms in
general requires a detailed covariant supergraph analysis or the
assumption of a very simple background as below.
\end{itemize}

Thus the action we will use for our calculation of $\Zholo$ is:
\begin{equation}\label{pron23}
    S(\Phi)=\int\dlor\dtotha\bigg\{\frac{-1}{2\mbar}\Phi^T\left(\square-\W^\al
D_\al\right)\Phi+\wtree\nleft(\Phi\right)\bigg\}.
\end{equation}
Here we actually can replace $D_\al$ with
$\partial/\partial\tha^\al$ using the definition~\eqref{susyn30}
since we again can drop terms depending on $\thabar$.

The above reduction can also be seen by choosing the simplest
possible non-trivial background as done
in~\cite{ferretti,0311066}. Simply set $A_\mu=0$, take the gaugino
field $\la_\al$ to be constant, choose $\labar_{\aldot}=0$, and
disregard the auxiliary field $D$. Then by~\eqref{susyn53}
and~\eqref{susyn59} we have (in Wess-Zumino gauge):
\begin{equation}\label{pron24}
    V=-i\thabar\thabar\tha\la,\quad \W_\al=-i\la_\al.
\end{equation}
By~\eqref{pron2.5} $\nabla_\al=D_\al+2\thabar\thabar\W_\al$ and
after a short calculation~\eqref{pron6.5} then gives:
\begin{equation}\label{pron25}
    \sqcov=\square+2i\W\si^\mu\thabar\partial_\mu+2\thabar\thabar\W\W.
\end{equation}
Inserting into~\eqref{pron7} yields:
\begin{equation}\label{pron26}
    \square_+=\square-\W^\al
D_\al+2i\W\si^\mu\thabar\partial_\mu.
\end{equation}
If we drop the last term since it depends on $\thabar$, we
get~\eqref{pron23} as wanted.

\subsection{Perturbation Theory Setup}\label{proofsecdiasubpert}

In this subsection we will develop the perturbation theory based
on the action~\eqref{pron23}. The crucial point is that we have
been able to integrate $\Phibar$ out since we are only interested
in the F-term. In this way we are only left with
$\Phi\Phi$-propagators contrary to the usual supergraphs developed
in section~\ref{dvsecquantsubsuper}. The new Feynman rules are
also the reason that we escape the perturbative
non-renormalisation theorems from
section~\ref{dvsecquantsubnonrenorm}
and~\ref{dvsecweffsubnon}.\footnote{This should not be
misunderstood as the perturbative non-renormalisation theorems
being wrong. Rather, the perturbation theory for the field $\Phi$
after integrating out $\Phibar$ captures the form of the
non-perturbative corrections from
section~\ref{dvsecweffsubnonpert}.} Since $\Phibar$ has been
integrated out and there is no longer any dependence on $\thabar$,
we can also restrict the superspace to the half-superspace
generated by $\tha$ and think of $\Phi$ as a general superfield in
this space. Thus we do not have to worry about imposing chirality
by e.g. introducing unconstrained superfields as in~\eqref{dvn65}.

The effective action $\weffpert$ is obtained from $\Zholo$ as in
equation~\eqref{dvn13.2}. However, we will here Wick rotate into
Euclidean space:
\begin{equation}\label{pron26.5}
    \Zholo = \int\DPhi e^{-S^{\eucl}(\Phi)}=
e^{-\int\dlor\dtotha\weffpert^{\eucl}},
\end{equation}
where the label E denotes Euclidian space. We note that there is a
simple sign change between the superpotential in Euclidean and
Minkowski space-time:
\begin{equation}\label{pron26.6}
    \weffpert^{\eucl}\nleft(S\right)=-\weffpert\nleft(S\right).
\end{equation}
Equation~\eqref{pron26.5} tells us that we should consider (minus)
the sum of connected Feynman diagrams. In developing the Feynman
rules for these diagrams it will be convenient to use the momentum
space formulation not only for the space-time part, but also for
the fermionic parameters $\tha_\al$. To obtain the fermionic
Fourier transformation we note that by~\eqref{appspinorn34}:
\begin{equation}\label{pron27}
    -4\int\!\dtopi
e^{\tha\pi}=\tha\tha=\deltafunkto\nleft(\tha\right),
\end{equation}
where $\pi_\al$ is the fermionic momentum and we have defined
$\int\!\dtopi\,\pi\pi=1$. Thus the Fourier transformation is given
by:
\begin{eqnarray}
  f\nleft(\tha\right) &=& -4\int\!\dtopi
e^{\tha\pi}\tilde{f}\nleft(\pi\right),\label{pron27.1} \\
  \tilde{f}\nleft(\pi\right) &\defi& \int\!\dtotha
e^{-\tha\pi}f\nleft(\tha\right).\label{pron27.2}
\end{eqnarray}
In this way the spinorial derivative is replaced by a fermionic
momentum:
\begin{equation}\label{pron28}
    \frac{\partial}{\partial\tha^\al}\mapsto\pi_\al,
\end{equation}
which is consistent with the hermitian adjoint of the spinorial
derivative given in~\eqref{appspinorn31.2}.

When we construct the propagator we have two choices: Either to
see the $\W^\al \partial/\partial\tha^\al$ term as giving a vertex
or to include it as a part of the propagator. We will use the
latter and assume that $\W_\al$ is not only constant in
space-time, but also independent of $\tha$ like in the case of the
simple background~\eqref{pron24}. This is sufficient to determine
the form of $\weffpert$. Now $\Zholo$ takes the form:
\begin{equation}\label{pron29}
    \Zholo=e^{-\int\dlor\dtotha\left\{\frac{-1}{2\mbar}\Phi^T\left(\square-\W^\al
\partial/\partial\tha^\al-\meucl\mbar\right)\Phi+\mathrm{interactions}\right\}},
\end{equation}
where we have included the mass term in the inverse propagator
term. Here $\meucl=-m$ is the mass in the Euclidean superpotential
where there is a sign change as in~\eqref{pron26.6}. The sign
change in the two other terms in the inverse propagator has been
absorbed into $\mbar$. The interaction terms are then the
remaining terms in $\wtree^{\eucl}$. The propagator
$\De\nleft(x,\tha;x',\tha'\right)$ is then determined by:
\begin{equation}\label{pron30}
   \frac{-1}{\mbar}\left(\square-\W^\al
\frac{\partial}{\partial\tha^\al}-\meucl\mbar\right)\Delta\nleft(x,\tha;x',\tha'\right)=\deltafunkfire\nleft(x-x'\right)\deltafunkto\nleft(\tha-\tha'\right).
\end{equation}
Using~\eqref{pron27} and expanding
$\deltafunkfire\nleft(x-x'\right)$ similarly we get the
propagator:
\begin{equation}\label{pron31}
    \Delta\nleft(x,\tha;x',\tha'\right)=\int\dmom(-4)\!\int\!\dtopi\frac{\mbar}{p^2+\W^\al\pi_\al+\meucl\mbar}e^{i(x-x')p}e^{(\tha-\tha')\pi},
\end{equation}
where we have repressed the gauge group representation indices.
From this propagator we see that the $\mbar$ dependence explicitly
cancels out since we can rescale $p^2\mapsto\mbar p^2$ and
$\pi_\al\mapsto\mbar\pi_\al$. $\mbar$ then cancels in the fraction
and since the fermionic Jacobian is the inverse of the bosonic
Jacobian, the dependence also cancels here. In the following we
can therefore set $\mbar=1$.

The Feynman rule for the vertices is to multiply with (minus) a
coupling times a gauge group invariant tensor and then integrate
over $x$ and $\tha$. Since $\W_\al$ is merely a (fermionic)
constant, such an integration is over the exponentials from the
propagators~\eqref{pron31} and as usual this gives a delta
function
$(2\pi)^4\deltafunkfire\nleft(\sum_ip_i\right)(-\frac{1}{4})\deltafunkto\nleft(\sum_i\pi_i\right)$
giving bosonic and fermionic momentum conservation at the
vertices. However, since we consider connected diagrams one of
these delta functions simply is
$(2\pi)^4\deltafunkfire\nleft(0\right)(-\frac{1}{4})\deltafunkto\nleft(0\right)$
which expresses the fact that the total incoming momentum in the
diagram is zero. This delta function thus equals an integration
$\int\dlor\dtotha$ and this is our overall integration in the
effective action. We should not worry about the integrand being
$x$ and $\tha$ independent since what we want is the form of the
effective Lagrangian and here it suffices to take $\W^\al$
constant. We can now use the remaining delta functions to remove
some of the momentum integrations leaving us with $L$ integrations
and thus $4L$ independent bosonic loop momenta and $2L$ fermionic
loop momenta where:
\begin{equation}\label{pron32}
    L=E-V+1.
\end{equation}
Here $E$ is the number of propagators and $V$ is the number of
vertices. In conclusion, we are left with the momentum space
Feynman rules:
\begin{eqnarray}
     \vphantom{\bigg(}\includegraphics{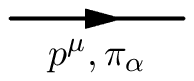} &=& \frac{1}{p^2+\W^\al\pi_\al+\meucl}, \label{pron33}\\
\includegraphics{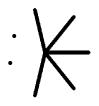} &=& -\geuclk f_{a_1\cdots a_k},\label{pron34}\\
  \weffpert^{\eucl} &=& -\sum\textrm{connected diagrams}.\label{pron35}
\end{eqnarray}
Here $\geuclk=-g_k$ is the Euclidean coupling (with a sign change
as above) for the interaction term which has the form $\geuclk
f_{a_1\cdots a_k}\Phi^{a_1}\ldots\Phi^{a_k}$.

The trick is now to write the propagators, indexed by $i=1,\ldots,
E$, as:
\begin{equation}\label{pron36}
    \int_0^\infty\dsi
e^{-s_i\left(p_i^2+\W^\al\pi_{i\al}+\meucl\right)}=\frac{1}{p_i^2+\W^\al\pi_{i\al}+\meucl},
\end{equation}
where $s_i$ is the so-called Schwinger time variable, and $p_i$
and $\pi_i$ are the momenta flowing through the $i^{\mathrm{th}}$
propagator.\footnote{This is inspired by string theory where we
have a cancellation between the factors obtained from integration
over the space-time momenta and the integration over fermionic
momenta. The Schwinger time variables can be seen as the length of
an edge and are thus the field theory limit of the world-sheet
moduli from string theory.} The point is that we can now factorise
the propagator in the contribution coming from respectively
$p_i^2$, $\W^\al\pi_{i\al}$ and $\meucl$. This gives us a
corresponding factorisation of the amplitude, $A$, of a given
diagram:
\begin{equation}\label{pron37}
    A=\nsymm\int_0^\infty\prod_i\dsi A_{\textrm{bosonic}}(s_i)A_{\textrm{fermionic}}(s_i)e^{-\sum_i s_i
\meucl}\prod_k(- \geuclk)^{V_k},
\end{equation}
where $\nsymm$ is a symmetry factor, $A_{\textrm{bosonic}}$ is the
contribution from the space-time momentum integrations,
$A_{\textrm{fermionic}}$ is the contribution from the fermionic
momenta which also holds all the gauge group index contractions,
and $V_k$ is the number of vertices of order $k$.

It is easy to calculate $A_{\textrm{bosonic}}$. To this end we
introduce the $L$ independent loop momenta $p'_a$ which are
related by a matrix $L_{ia}$ to the momenta flowing through the
propagators, $p_i$, as:
\begin{equation}\label{pron38}
    p^\mu_i=\sum_{a=1}^LL_{ia}p'^\mu_a,\quad i=1,\ldots, E.
\end{equation}
We can then write $A_{\textrm{bosonic}}$ as:
\begin{eqnarray}
    A_{\textrm{bosonic}}&=&\int\prod_{a=1}^L\dmoma \exp
\Big(-\sumlimits_i s_ip^2_i\Big)=\int\prod_{a=1}^L\dmoma
\exp\Big(-\sumlimits_i
s_i\Big(\sumlimits_{a}L_{ia}p'_a\Big)^2\Big)\nonumber\\
&=&\int\prod_{a=1}^L\dmoma \exp\Big(-\sumlimits_{a,b}p'^\mu_a
M_{ab}(s_i)p'_{b\mu}\Big)=\frac{1}{(4\pi)^{2L}}\frac{1}{(\det
M(s_i))^2},\label{pron39}
\end{eqnarray}
where we in the last line have used Gaussian integration and we
have introduced the real symmetric matrix:
\begin{equation}\label{pron40}
    M_{ab}(s_i)=\sum_is_iL_{ia}L_{ib}.
\end{equation}

What we want to prove is that the dependence on the Schwinger time
variables $s_i$ in $A_{\textrm{fermionic}}(s_i)$ cancels the $s_i$
dependence in $A_{\textrm{bosonic}}$ thus giving us a localisation
to zero-momentum modes. Modulo gauge group factors
$A_{\textrm{fermionic}}$ takes the form:
\begin{equation}\label{pron41}
    A_{\textrm{fermionic}}(s_i)\sim(-4)^L\int\prod_{a=1}^L\dtopi'_a
e^{-\sum_i\sum_as_i\W_{(\reprn_i)}^\al L_{ia}\pi'_{a\al}},
\end{equation}
where we have introduced the fermionic loop momenta $\pi'_{a}$
which are related to the propagator momenta like
in~\eqref{pron38}:
\begin{equation}\label{pron42}
    \pi_{i\al}=\sum_{a=1}^LL_{ia}\pi'_{a\al},\quad i=1,\ldots,E.
\end{equation}
In~\eqref{pron41} we have explicitly shown that $\W^\al$ depends
on the representation. Here $\reprn_i$ is the representation that
flows through the $i^{\mathrm{th}}$ propagator, where we allow the
representation to depend on the propagator as we saw was the case
for a broken gauge group. We see that the cancellation of the
dependence on $s_i$ in $A_{\textrm{fermionic}}$ and
$A_{\textrm{bosonic}}$ depends non-trivially on the gauge group
representation. We will show this cancellation for general gauge
groups and matter representations in
section~\ref{proofsecdiasubgeneral}. However, we will start by
considering the case of classical gauge group with an adjoint
chiral field for which we will prove the direct relation to the
matrix model given by the Dijkgraaf-Vafa conjecture.

\section{Reduction to the Matrix Model}\label{proofsecpert}

\subsection{Double Line Notation and $\W^\al$ Insertions}\label{proofsecdiasubdouble}

\begin{figure}\caption{}\label{profign1}
\begin{center}
\includegraphics{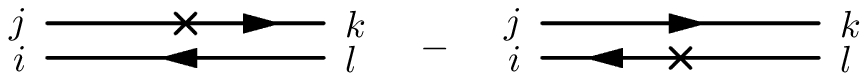}
\end{center}
    \begin{center}
        The inverse double line propagator from the $\W_{(\mathrm{adj})}^\al\pi_{\al}$ term.
        Crosses denote $\W^\al$ insertions.
    \end{center}
\end{figure}

\begin{figure}\caption{}\label{profign2}
\begin{center}
\includegraphics{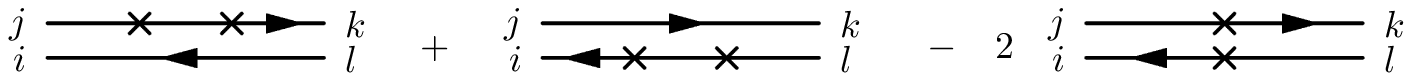}
\end{center}
    \begin{center}
        Double line notation for the second order term of the exponential in the Schwinger representation of the propagator.
        Crosses denote $\W^\al$ insertions.
    \end{center}
\end{figure}

In this section we will continue the evaluation of
equation~\eqref{pron37} for the amplitude $A$ of a given diagram.
But here we restrict the gauge group to be one of the classical
gauge groups $\un{N}$, $\son{N}$ or
$\mathrm{Sp}(k)={\mathrm{USp}(N=2k)}$, and the matter is in the
form of a single adjoint chiral superfield, $\Phi$. Since we have
integrated $\Phibar$ out, we can think of $\Phi$ as a real field
and use the double line notation for diagrams introduced in
section~\ref{dvsecdouble}. Here the difference is that we have a
$\Phi^T\W_{(\mathrm{adj})}^\al \pi_{i\al}\Phi$ term. To determine
the corresponding inverse double line propagator we write the
adjoint field as a hermitian matrix
$\Phi^aT_{a}^{(\mathrm{fund})}$ as in~\eqref{dvn124}. The term now
takes the form $\tr\nleft(\Phi[\W^\al \pi_{i\al},\Phi]\right)$
where $\W^\al=\W^{a\al}T_{a}^{(\mathrm{fund})}$.\footnote{Please
note that when going to the trace formulation we actually get a
factor of $C\nleft(\mathrm{fund}\right)$ as in
$\Phi^a\Phi^a=\tr\nleft(\Phi\Phi\right)/C\nleft(\mathrm{fund}\right)$.
For the $p^2$ and $\W^\al\pi_\al$ terms this factor can be
absorbed in $\mbar$. For the mass term we think of the factor as
redefining the mass. E.g. to have a mass term $\half
m\tr\nleft(\Phi^2\right)$ in $\wtree$ as in~(\ref{dvn1.5}) the
mass term in~(\ref{pron29}) should really have been multiplied
with $C\nleft(\mathrm{fund}\right)$ -- in this way when we go to
the trace formulation we will get the wanted mass term.} The
contribution to the inverse propagator
$\de\De^{-1}\defi\Ga^\al\pi_{i\al}$ is then determined by:
\begin{equation}\label{pron43}
\Phi^{i}_{\phantom{i}j}\left(\Ga^\al\right)^{lj}_{ik}\Phi^{k}_{\phantom{k}l}\defi\tr\nleft(\Phi[\W^\al,\Phi]\right)=\Phi^{i}_{\ph{i}j}\left(\left(\W^\al\right)^{j}_{\ph{j}k}\Phi^{k}_{\ph{k}i}-\Phi\opned{j}{k}\left(\W^\al\right)\opned{k}{i}\right),
\end{equation}
giving us the result:
\begin{equation}\label{pron44}
    \left(\Ga^\al\right)^{lj}_{ik}=\de^l_i\left(\W^\al\right)\opned{j}{k}-\left(\W^\al\right)\opned{l}{i}\de^j_k.
\end{equation}
This inverse propagator corresponds in the double line diagrams to
insertions of $\W^\al$ in the single index lines
(figure~\ref{profign1}) with a sign which is correlated with the
single index lines having opposite directions (and hence being
parallel or anti-parallel with the direction of the fermionic
momentum).
By expanding $\tr\nleft(\Phi[\W^\al \pi_{i\al},[\W^\al
\pi_{i\al},\Phi]]\right)$ we see that two of these inverse double
line propagators multiply as:
\begin{equation}\label{pron45}
    \left(\Ga^\al\pi_{i\al}\Ga^\al\pi_{i\al}\right)^{lj}_{ik}=\left(\Ga^\al\pi_{i\al}\right)^{mj}_{in}\left(\Ga^\al\pi_{i\al}\right)^{ln}_{mk}.
\end{equation}
This simply tells us to join the double lines in the obvious way
and defines the exponentials in the Schwinger representation of
the propagators in $A_{\textrm{fermionic}}$
(equation~\eqref{pron41}). Thus the second order term, shown in
figure~\ref{profign2}, involves double line propagators with two
$\W^\al$ insertions. Since the exponential involves the fermionic
propagator momentum $\pi_{i\al}$, the expansion terminates at
second order and thus figure~\ref{profign1} and~\ref{profign2}
show all the possible double line propagators with insertions;
especially we note that there can be maximally two insertions per
double line propagator.

With the double line notation we see that we should define Feynman
rules giving diagrams where we can have $\W^\al$ insertions in the
index lines. However, the important point is that we should have
exactly $2L$ insertions since the integrations
$\int\prod_{a=1}^L\dtopi_a$ over the $2L$ fermionic loop momenta
in~\eqref{pron41} bring down exactly $2L$ factors of $\W^\al$. On
the other hand, there can be at most two $W^\al$ insertions in an
index loop. This is because
$\tr\nleft(\W^{\al_1}\ldots\W^{\al_n}\right)=0$ for $n\geq3$ as
shown in~\cite{0211170}. We can prove this
following~\cite{ferretti,0311066} by noting that by the
relation~\eqref{pron5}, the definition~\eqref{pron2} and the fact
that $\W_\be$ is chiral we get:
\begin{equation}\label{pron46}
    \{\W_\al,\W_\be\}=\frac{1}{8i}[\nablabar_{\aldot},\nabla_\al^{\ph{\al}\aldot}]\W_\be=-\frac{1}{8}\nablabar_{\aldot}\nablabar^{\aldot}\nabla_\al\W_\be.
\end{equation}
where the vector field $V$ in the definition of $\nabla^\al$ is in
the adjoint. Thus
\begin{multline}\label{pron47}
    \tr\nleft(\W^{\al_1}\ldots\W^{\al_{i}}\W^{\al_{i+1}}\ldots\W^{\al_{n}}\right)=-\tr\nleft(\W^{\al_1}\ldots\W^{\al_{i+1}}\W^{\al_{i}}\ldots\W^{\al_{n}}\right)\\-\frac{1}{8}\Dbar\Dbar\tr\nleft(\W^{\al_1}\ldots\nabla^{\al_{i}}\W^{\al_{i+1}}\ldots\W^{\al_{n}}\right).
\end{multline}
Due to the transformation of the gauge covariant derivative the
last trace is gauge invariant and we conclude that we can
anticommute the $\W^{\al_i}$'s under a trace in the chiral ring
(section~\ref{dvsecnil}). Since the Weyl index $\al$ can only take
two values, traces of more than two $\W^{\al}$'s are zero
(classically) in the chiral ring and can thus be disregarded.

Suppose the diagram under consideration has $F$ index loops (and
$E$ propagators, $L$ momentum loops, and $V$ vertices). We can
thus have maximally $2F$ $\W^\al$ insertions and since we know
that we have $2L$ insertions, we conclude that $F\geq L$. Since
the Euler characteristic $\chi=V-E+F$~\eqref{dvn130} and
$L=E-V+1$~\eqref{pron32} we conclude:
\begin{equation}\label{pron48}
    \chi=F-L+1\geq1.
\end{equation}
Thus for a $\un{N}$ gauge group and an adjoint chiral superfield
only planar diagrams contribute and we conclude that the planar
limit is exact. We note that this result came about simply by
considering the number of fermionic integrations. For the
$\son{N}$ and $\mathrm{USp}(N)$ gauge groups we can also have
diagrams with $\chi=1$ and topology $\mathbb{R}\mathbb{P}^2$ if we
include a single cross-cap using the cross-over double line
propagator from figure~\ref{dvfign5}. If we take into
consideration fundamental matter, we further have the possibility
of diagrams with one boundary (and the topology of a disk) and
hence $\chi=1$.

\subsection{Reduction to the Matrix Model}\label{proofsecdiasubreduc}

We now want to show the reduction to the matrix model using the
results from the last subsection.

Let us first note that an index loop with a single insertion
contributes with a factor $\tr\nleft(\W^\al\right)$. An index loop
with two insertions gives a factor proportional to the glueball
superfield from~\eqref{dvn112.5}:
\begin{equation}\label{pron49}
    \tr\nleft(\W^\al\pi_{i\al}\W^\be\pi_{j\be}\right)=-\frac{1}{2}\tr\nleft(WW\right)\pi_{i}\pi_{j}=16\pi^2C(\mathrm{fund}) S\pi_i\pi_j,
\end{equation}
where we have used the fact that the matrices $\W^\al$ and
$\W^\be$ anticommute under the trace and hence behave as spinors
allowing us to use~\eqref{appspinorn34}.

\begin{figure}\caption{}\label{profign3}
\begin{center}
\includegraphics{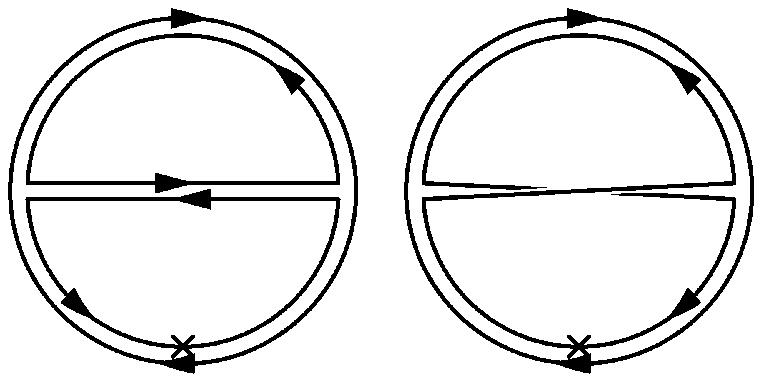}
\end{center}
    \begin{center}
        On the left a planar diagram with an insertion. On the right a non-orientable diagram obtained
        from the diagram on the left by replacing a double line
        propagator with a cross-over.
    \end{center}
\end{figure}

However, we have to care about the sign of an insertion since, as
we found above (figure~\ref{profign1}), the sign depends on
whether the direction of the single index line with the insertion
is parallel or anti-parallel to the direction of the momentum.
This actually also holds true in the case of non-orientable
diagrams. Take the case of a $\son{N}$ gauge group. If we replace
a double line with a cross-cap\footnote{To obtain a cross-cap we
replace a normal double line propagator with the cross-over part
of the $\son{N}$ propagator from figure~\ref{dvfign5}. But the
double line propagator should not have the same face on both
sides. E.g. if we draw the middle propagator in the dumbbell
diagram in figure~\ref{dvfign7} as a cross-over, we essentially
get the same diagram.} as in figure~\ref{profign3}, we will join
two index loops to one and flip the arrows in one of the index
loops i.e. change the order of how to contract the indices. But
for $\son{N}$ $\left(\W_\al\right)\opned{i}{j}$ is antisymmetric
and hence this change in the direction of arrows on the single
lines exactly gives a sign change.
Since $\W_\al$ does not depend on where in the index loop it is
inserted, this means that we can keep track of the signs and the
insertions by introducing an auxiliary set of fermionic variables
$\W'^\al_m,\,m=1,\ldots, F$, corresponding to the index loop
insertions such that $\W_{(\reprn_i)}^\al$ in~\eqref{pron41} is
given by:
\begin{equation}\label{pron50}
    \W_{(\reprn_i)}^\al=\sum_{m=1}^F K_{im}\W'^\al_m.
\end{equation}
Here $K_{im}$ is an $E\times F$ matrix defined to be $+1$ if the
direction of the single index line on the side of the
$m^{\mathrm{th}}$ index loop in the $i^{\mathrm{th}}$ double line
propagator is parallel to the momentum $\pi_i$, $-1$ if it is
anti-parallel, and zero if the $i^{\mathrm{th}}$ double line
propagator is not part of the $m^{\mathrm{th}}$ index loop at all.
If we have the same face on both sides of the propagator, we
should sum the contributions. Thus in this case if the index lines
have opposite directions $K_{im}$ is zero and if they have the
same direction it is $\pm2$. If $K_{im}$ is zero~\eqref{pron50}
implies that we will not have any insertions in that propagator.
For $\chi\geq1$ this is consistent with the fact that in such a
propagator the momentum must be zero and we really do not have any
insertion since $\W^\al\pi_{i\al}=0$. This happens e.g. in the
dumbbell diagram in figure~\ref{dvfign7}. However, if $\chi<1$ we
can have propagators with $K_{im}=0$ and non-zero momentum, see
figure~\ref{dvfign4}. Thus this method only works for $\chi\geq1$,
but, actually, this is all we need by~\eqref{pron48}. The middle
propagator in the non-orientable diagram in figure~\ref{profign3}
gives an example where $K_{im}=2$. This is also consistent since
we only have one variable $\W'^\al_m$ which can give insertions,
but we can have insertions in both sides with the same sign --
hence we need the factor 2. In conclusion
$K_{im}\in\{0,\pm1,\pm2\}$ and we only get $\pm2$ for $\chi=1$.
The advantage is further, as we will see below, that the
contraction of the gauge group indices reduces to thinking of
$\W'^\al_m$ as a simple fermionic variable (not a matrix) and
performing fermionic integrations over these variables. Using
these auxiliary variables we can now prove the reduction to the
matrix model.

\begin{figure}\caption{}\label{profign4}
\begin{center}
\includegraphics{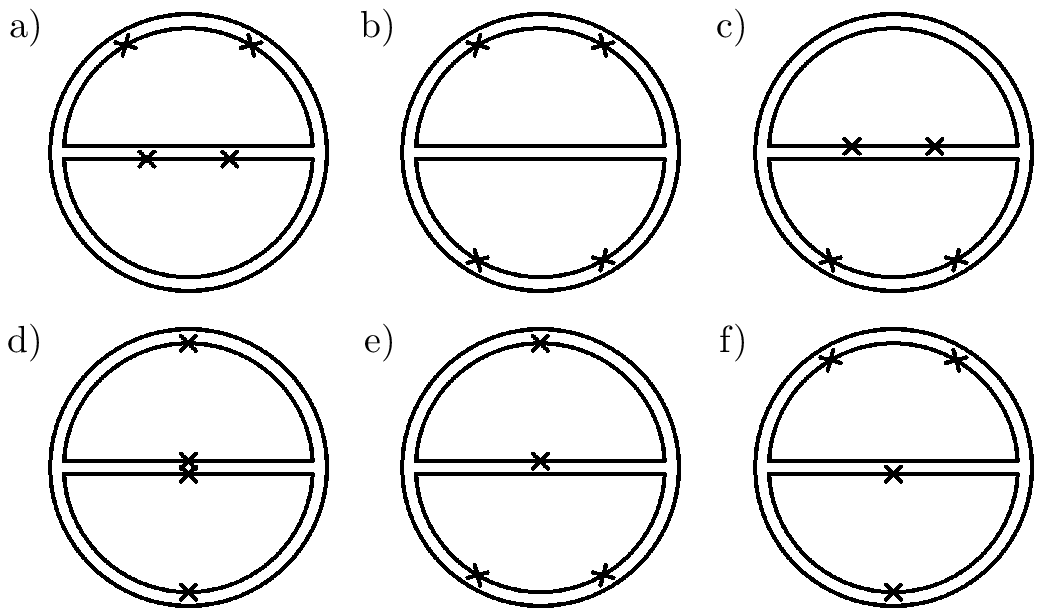}
\end{center}
    \begin{center}
        The stop sign diagram with all the different possible
        distributions of the four insertions. The outer loop has been
        chosen to be without insertions.
    \end{center}
\end{figure}

Let us start by considering a $\un{N}$ gauge group and the
traceless glueball superfield. In this case we have seen that only
planar diagrams contribute to $\weffpert$. Consider such a planar
diagram with amplitude $A$ as in~\eqref{pron37}. Using the new
double line Feynman rules this amplitude splits into amplitudes
from diagrams with the different possible insertions. For a planar
diagram we have $F=L+1$ using~\eqref{pron48}. Thus we have two
possibilities of distributing the $2L$ insertions: Either we have
$2$ index loops with just one insertion and the remaining $L-1$
index loops have two insertions or we have $1$ index loop without
any insertions and the remaining $L$ index loops with two
insertions. In the traceless case we should disregard the first
option because it gives contributions depending on $\tr{\W^\al}$,
which is zero since we choose a background such that the abelian
part of $\W^\al$ is zero. So let us consider the last case. Let us
fix the index loop without any insertions, e.g. as the outer loop.
The remaining index loops have two insertions thus giving a factor
of $S^{F-1}$. However, to calculate the precise value we should
consider all the different ways to distribute the insertions. In
figure~\ref{profign4} all the possible insertions have been shown
in the case of a stop sign diagram.
We can now calculate the dependence on the Schwinger parameters
$s_i$ since an insertion in the $i^{\mathrm{th}}$ propagator gives
a factor $s_i$ as seen from~\eqref{pron41}. Remembering that a
propagator with a double insertion comes with a factor $1/2$ from
expanding the exponential, we then see that the contribution from
the diagrams is proportional to:
\begin{equation}\label{pron50.5}
    \frac{1}{4}\left(s_1^2s_2^2+s_1^2s_3^2+s_2^2s_3^2\right)+\frac{1}{2}\left(s_1s_2^2s_3+s_1s_2s_3^2+s_1^2s_2s_3\right),
\end{equation}
One can check that this actually is $(\det M(s_i))^2/4$ thus
cancelling the $s_i$ dependence in $A_{\textrm{bosonic}}$
from~\eqref{pron39} and proving the reduction to zero-modes for
this diagram. Actually, one has to be a little more careful
in~\eqref{pron50.5}. When calculating diagram d in
figure~\ref{profign4} we get an extra factor $\half$ compared to
the other diagrams which is, however, cancelled by the factor of
$2$ coming with the double line propagators that have insertions
in both sides -- see figure~\ref{profign2}.

\begin{figure}\caption{}\label{profign5}
\begin{center}
\includegraphics{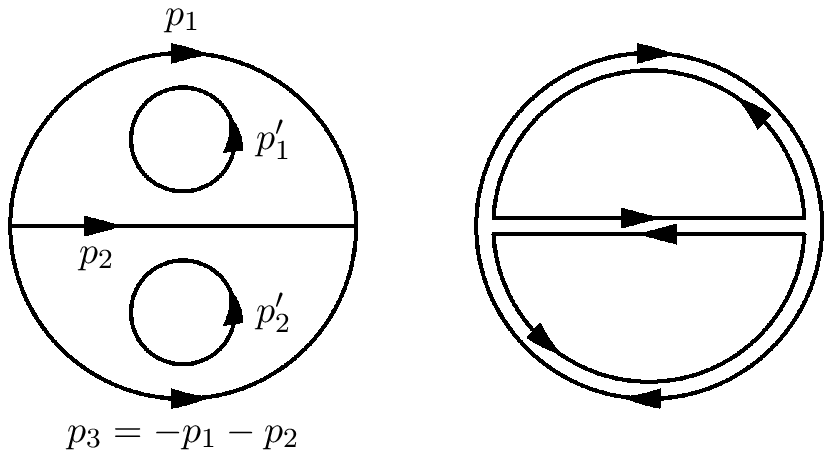}
\end{center}
    \begin{center}
        On the left the stop sign diagram in single line notation
        with the loop momenta $p'_1$ and $p'_2$ chosen to run
        like the inner index loops in the corresponding double line diagram
        (displayed to the right).
    \end{center}
\end{figure}

However, using the auxiliary fermionic variables there is no need
to explicitly do this elaborate expansion into diagrams with
insertions. We simply constrain the matrix $K_{im}$ by removing
the column corresponding to the index loop without insertions thus
ensuring no insertions in this index loop. But the resulting
matrix is, by definition, exactly the matrix $L_{ia}$
from~\eqref{pron38} if we choose the $L$ loop momenta to run as
the $L$ index loops with insertions and, especially, with the same
orientation, see figure~\ref{profign5}.
This is possible since we have a planar, oriented diagram. It is
here a point that the Jacobian for the transformation from the
independent propagator momenta to these loop momenta simply is
one.\footnote{We can sketch a proof of this as follows: Consider
the matrix relating the loop momenta $p'_a$ to the independent
propagator momenta $p_i$. Take a double line propagator at the
boundary. Here the loop momentum, denoted $p'_1$, and the
propagator momentum, denoted $p_1$, is the same up to a sign --
see e.g. figure~\ref{profign5}. Thus the top row in the
transformation matrix is $\pm1$ in the first entry and zero in the
rest. Adjacent to the inner face of this double line propagator we
have another propagator whose propagator momentum, denoted $p_2$,
thus depends on $p'_1$ and some other loop momenta $p'_2$. Thus
the second row in the matrix has something in the two first
entries and zeroes in the rest. Continuing in this way we get a
triangle matrix whose determinant simply is the multiple of the
entries in the diagonal. Since these are $\pm1$, the Jacobian is
1.} We can then rewrite~\eqref{pron50} as
$\W_{(\reprn_i)}^\al=\sum_{a} L_{ia}\W'^\al_a$ where we have used
the same label for the (remaining) auxiliary fermionic variables
as for the momentum loops. The point is that the auxiliary
variables also captures the index contraction structure of the
insertions: By~\eqref{pron49} we can identify $\W'^\al\W'_\al$
with $-32\pi^2C\nleft(\mathrm{fund}\right) S$ where $\W'^\al$ is a
simple fermionic variable, not a matrix. Thus, to obtain
$A_{\textrm{fermionic}}(s_i)$ in~\eqref{pron41} we simply have to
include the $2L$ integrations
$\prod_a(-32)\pi^2C\nleft(\mathrm{fund}\right)S\int\dtow_a$ with
$\int\dtow\,\W'^\al\W'_\al=1$. In this way we get exactly two
insertions in the index loops with insertions and no insertion in
the remaining index loop. Furthermore, this takes care of all
gauge index contractions and we will get the right signs both from
the insertions and from the ordering of the fermionic variables.
Also, we are ensured that we have the right Weyl index
contractions. We then get:
\begin{eqnarray}\label{pron51}
A_{\textrm{fermionic}}(s_i)&=&\left(\prod_{a=1}^L(-32)\pi^2C\nleft(\mathrm{fund}\right)S\int\dtow_a\,(-4)\!\int\dtopi'_a\right)
e^{-\sum_i\sum_{a,b}s_iL_{ia}\W'^\al_a L_{ib}\pi'_{b\al}}\nonumber\\
&=&
\left(-2C\nleft(\mathrm{fund}\right)S\right)^{F-1}(4\pi)^{2L}\int\!\Big(\prod_a\dtow_a(-4)\dtopi'_a\Big)e^{-\sum_{a,b}\W'^\al_a M_{ab}(s_i)\pi'_{b\al}}\nonumber\\
&=&
(-1)^{F-1}\left(2C\nleft(\mathrm{fund}\right)S\right)^{F-1}(4\pi)^{2L}(\det
M(s_i))^2,
\end{eqnarray}
where we have used the definition of $M_{ab}(s_i)$
from~\eqref{pron40} and in the last line we have made a fermionic
Gaussian integration.\footnote{Naturally, the sign and
normalisation from the fermionic Gaussian integration is important
for us. Going through the proof of Gaussian integration, one finds
that the fermionic integrations gives the determinant squared and
$L$ factors of
$\int\dtow\,(-4)\!\int\dtopi'\W'^1\pi'_1\W'^2\pi'_2$ which simply
is 1 by~(\ref{appspinorn32}).} We see that the $s_i$ dependence in
$A_{\textrm{fermionic}}(s_i)$ cancels that of
$A_{\textrm{bosonic}}$ in~\eqref{pron39} thus giving us the wanted
localisation to zero-modes.

The total amplitude~\eqref{pron37} now reduces to:
\begin{equation}\label{pron52}
    A=NF(-1)^{F-1}\left(2C\nleft(\mathrm{fund}\right)S\right)^{F-1}\nsymm\int_0^\infty\prod_{i=1}^E\dsi e^{-\sum_i s_i
\meucl}\prod_k(-\geuclk)^{V_k},
\end{equation}
where we have multiplied with $NF$ since we have $F$ possibilities
in choosing the loop without insertions and this loop contributes
with a factor $N$. The integrations over the Schwinger variables
$s_i$ are now trivial and gives $\meucl^{-E}$. To compare with the
matrix model we should use $\meucl=-m$ and $\geuclk=-g_k$ where
$m$ is the mass and $g_k$ the couplings in $\wtree$ in Minkowski
space. We thus get a factor $(-1)^{V-E}$ which by~\eqref{dvn130}
can be written as $(-1)^{\chi-F}$. With the definition
\begin{equation}\label{pron52.5}
    S'=2C\nleft(\mathrm{fund}\right)S
\end{equation}
we can thus write
\begin{equation}\label{pron53}
    A=(-1)^{\chi-F}(-1)^{F-1}N\frac{\partial}{\partial S'}S'^F
    \nsymm m^{-E}\prod_k(-g_k)^{V_k}.
\end{equation}
From~\eqref{dvn133} we see that the amplitude of the corresponding
planar double line diagram in the matrix model is given by:
\begin{equation}\label{pron53.5}
    A_{\mathrm{matrix}}\nleft(\gs N'\right)=\nsymm\gs^{-2}\nleft(\gs
N'\right)^Fm^{-E}\prod_k(-g_k)^{V_k}.
\end{equation}
Thus, if we identify $S'$ with $\gs N'$, we can write:
\begin{equation}\label{pron54}
    A=-N\frac{\partial}{\partial S'}\gs^2A_{\mathrm{matrix}}\nleft(\gs N'\defi
    S'\right),
\end{equation}
where we have used that $\chi=2$ so $(-1)^{\chi-F}(-1)^{F-1}=-1$.
From~\eqref{pron35} we know that the Euclidian effective
superpotential is obtained as minus the sum of such amplitudes.
Thus by~\eqref{pron26.6} the effective glueball superpotential in
Minkowski space is (plus) the sum of these amplitudes of connected
planar diagrams. On the other hand, by~\eqref{dvn10} such a sum
over $\gs^2A_{\mathrm{matrix}}$ gives $-\cF_{g=0}$. Thus we
finally get:
\begin{equation}\label{pron55}
    \weffpert=N\frac{\partial}{\partial S'}\cF_{g=0}\nleft(\gs N'\defi
    S'\right),
\end{equation}
which simply is~\eqref{dvn13} in the unbroken case with
$C\nleft(\mathrm{fund}\right)=1/2$ and $\Shat=S$. For this case we
have thus proven the Dijkgraaf-Vafa conjecture for the relation
between the perturbative part of the effective glueball
superpotential and the planar limit of the matrix model.

Now, in much the same way we can prove the conjecture for other
gauge groups and matter representations that allow a double line
notation:
\begin{description}
    \item[$D^2$ diagrams:] If we have a classical group
with an adjoint chiral field and include fundamental matter then,
as mentioned above, we should add amplitudes from diagrams with a
single boundary and hence topology $D^2$. Here $\chi=1$ and hence
$F=L$ by~\eqref{pron48} which, naturally, corresponds to one of
the index loops being replaced by a boundary. Since we only have
$L$ index loops, all of these must have two insertions. This gives
us an overall factor of $(-1)^F S'^F$. So in this case we should
keep all the auxiliary variables $\W'^\al_m$ and integrate over
them. By choosing the loop momenta to run as the $L$ index loops,
the matrix $L_{ia}$ is again equal to $K_{im}$. Thus the fermionic
integration is the same as in the case above and we get a factor
that cancels $A_{\textrm{bosonic}}$. In this case we have no
overall $NF$ factor so the amplitude, in analogy
with~\eqref{pron53}, becomes:
\begin{equation}\label{pron56}
    A=(-1)^{\chi-F}(-1)^{F} S'^F
    \nsymm m^{-E}\prod_k(-g_k)^{V_k},
\end{equation}
where $(-1)^{\chi-F}$ as before comes from changing $\meucl$ and
$\geuclk$ to $m$ and $g_k$. Together with $(-1)^F$ this gives an
overall minus sign since $\chi=1$. Using~\eqref{dvn133} (with
$2-2g$ replaced by $\chi=1$) we can identify the last part
of~\eqref{pron56} with $\gs$ times the amplitude from the
corresponding diagram in the matrix model under the identification
$S'\defi\gs N'$. The contribution to $\weffpert$ from these disk
diagrams are again the sum of the amplitudes. The sum over the
corresponding amplitudes in the matrix model times $-\gs$ gives
$\cF_{D^2}$ from~\eqref{dvn186}. Thus we get:
\begin{equation}\label{pron57}
    \De\weffpert(S)=\cF_{D^2}\nleft(S'\defi\gs N'\right),
\end{equation}
as claimed in~\eqref{dvgeneral}.

    \item[$\mathbb{R}\mathbb{P}^2$ diagrams:] In the case of a $\son{N}$ or $\textrm{USp}(N)$
gauge group with adjoint matter we should also consider the
contribution from diagrams with the topology of the projective
plane $\mathbb{R}\mathbb{P}^2$ and hence $\chi=1$. Since the Euler
characteristic is the same as for the $D^2$ diagrams above, this
case is very similar, especially with two insertions in all of the
$F=L$ loops. The only difference is when we choose the matrix
$L_{ia}$. We can once again choose the loop momenta to run as the
index loops and in this way $L_{ia}$ equals $K_{im}$. But we
should here pay special attention to the cross-cap in the form of
a cross-over propagator that joins two oriented loops to one. For
this cross-over propagator, indexed by $i$, and the corresponding
index loop, indexed by $a$, we have $L_{ia}=K_{ia}=\pm2$
corresponding to $p_i=\pm2p'_a$ (see figure~\ref{profign3}). This
means that the Jacobian for the change from the independent
propagator momenta to the loop momenta is not $1$ as above, but
$2$. Since we have $4$ sets of the $L$ bosonic momenta ($p^\mu$)
we get a factor of $2^4$ from these. However, we also have 2 sets
of the fermionic momenta $(\pi^\al)$ and since the Grassmannian
Jacobian is the inverse of the bosonic, this gives a factor of
$(\pm2)^{-2}$. Thus in this case we get a factor $4$ compared to
the contribution from the disk diagrams above:
\begin{equation}\label{pron58}
    \De\weffpert(S)=4\cF_{\mathbb{R}\mathbb{P}^2}\nleft(S'\defi\gs N'\right),
\end{equation}
which finishes the proof of~\eqref{dvgeneral} in the unbroken
case. A proof for this case can also be found in~\cite{0211261}.

    \item[Disconnected diagrams:] Suppose that we have matter which specifically needs a
projection to be traceless. This is e.g. the case for a $\sun{N}$
gauge group with adjoint matter as mentioned in
footnote~\ref{footnotesun} on page~72. In these
cases we should also consider disconnected double line
propagators. One can then prove~\cite{0303104} that the number of
disconnected components is one higher than the number of
disconnected propagators. Thus, in the case of an even tree-level
superpotential these disconnected diagrams will not contribute at
all. However, in general the disconnected diagrams should be
considered say for a $\textrm{USp}(N)$ gauge group with traceless
antisymmetric matter. Since a given amplitude here is a multiple
of factors from the disconnected pieces, we can not give a
relation to the total free energy of the matrix model and the
comparison should be done diagram by diagram.

    \item[Broken gauge group:] We now consider what happens when the vacuum breaks
the gauge group. We will assume that we can make diagrams for the
gauge theory just as we did for the matrix model in the broken
case (section~\ref{dvsecmatrixmm}) which is plausible since the
ghost term has the same form. We should then consider diagrams
where the index loops have an extra label telling to which broken
part, $G_i$, they belong. The $\W^\al$ insertion in a loop with
the label $i$ should then be the gauge field strength,
$\W^\al_{(i)}$, for $G_i$. Two insertions in an index loop then
correspond to the glueball superfield $S_i$ for $G_i$ as
in~\eqref{dvn4} and~\eqref{dvn14}. For the contribution
corresponding to diagrams of topology $D^2$ and
$\mathbb{R}\mathbb{P}^2$ there is no change in the form of the
relation to the matrix model, except that $S$ now has a label $i$
and we should identify $S'_i\defi\gs N'_i$. For the planar
diagrams (and a traceless glueball superfield) we should split the
contribution from a given diagram according to which type of index
loop we choose to be without insertions. Let us assume that we
have chosen the type $i$ and let $F=\sum_jF_j$ where $F_j$ is the
number of index loops of type $j$. Since the loop without
insertion now gives a factor $N_i$ and we have $F_i$ possibilities
in choosing this loop, we get a factor of $N_iF_iS_i^{F_i-1}$
which should be multiplied with factors of $S_j^{F_j}$ from the
remaining types of index loops. We can write this as
$N_i\partial/\partial S_i \prod_i S_i^{F_i}$. Finally, we should
sum over the choice of type of index loop without insertions, $i$.
Thus~\eqref{pron55} becomes:
\begin{equation}\label{pron59}
    \weffpert=\sum_iN_i\frac{\partial}{\partial S'_i}\cF_{g=0}\nleft(S'_i\defi\gs N'_i
    \right),
\end{equation}
as in~\eqref{dvn13}.
\end{description}

In the next section we will finish the proof of the Dijkgraaf-Vafa
conjecture for the form of $\weffpert$ by considering the abelian
part of $\W_\al$. But let us end this section by noting that it is
essential in the localisation to zero-modes that we are in four
space-time dimensions. This is because $A_{\textrm{bosonic}}$
in~\eqref{pron39} is determined by the space-time dimension
whereas $A_{\textrm{fermionic}}$ in~\eqref{pron41} depends on the
dimension of the Weyl spinors. Thus, the cancellation between
$A_{\textrm{bosonic}}$ and $A_{\textrm{fermionic}}$ only takes
place in four space-time dimensions.

\subsection{Abelian Part}\label{proofsecdiasubabelian}

In this section we will finish the proof from the last subsection
by taking into consideration the abelian part of $\W^\al$. We will
think of a $\un{N}$ gauge group with adjoint matter. A sketch of
the proof can be found in~\cite{0211017}, but since we have not
found a detailed diagrammatic proof in the literature, we will be
thorough here.\footnote{In the generalised Konishi anomaly proof
in~\cite{0211170}, the proof for the abelian part is included in a
natural way.}

With a non-zero abelian part of $\W^\al$ we should allow index
loops with a single $\W^\al$ insertion giving a contribution of
$\tr\nleft(\W^\al\right)$. Since we have $2L$ insertions, only
planar diagrams have enough index loops to allow single
insertions. We thus get an extra contribution to $\weffpert$
obtained in~\eqref{pron55} from the planar diagrams in which 2 of
the $F=L+1$ loops have one insertion and the rest have two
insertions. We will consider the case of a broken gauge group
since for adjoint matter the overall $\un{1}$ in $\un{N}$ is
completely decoupled as mentioned in
section~\ref{dvsecdvunsubdvun}.

\begin{figure}\caption{}\label{profign6}
\begin{center}
\includegraphics{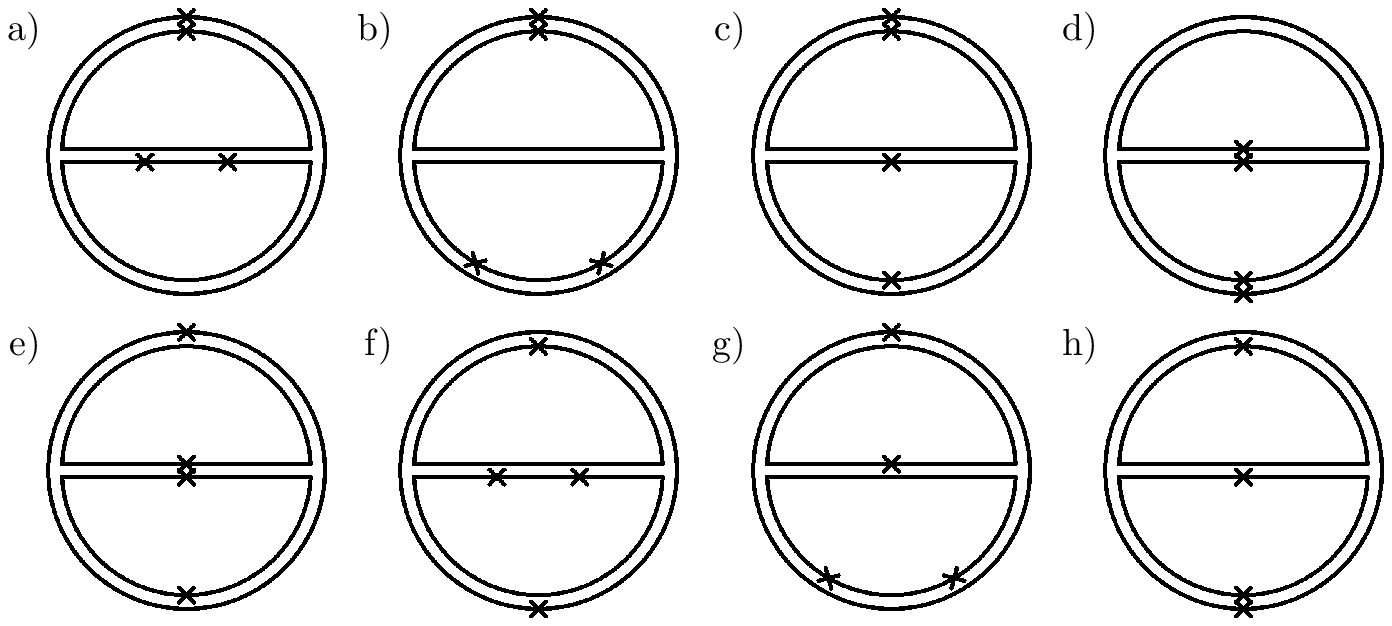}
\end{center}
    \begin{center}
        The stop sign diagram with all the different possible
        distributions of the four insertions in the case where we choose the outer index loop and the upper of the
inner index loops to only have one insertion.
    \end{center}
\end{figure}

Let us first point out that the cancellation between
$A_{\textrm{bosonic}}$ and $A_{\textrm{fermionic}}$ in this case
does not happen in exactly the same way as for the planar diagrams
in the last subsection. Consider as an example the stop sign
diagram. In the last section we considered the insertion pattern
where one of the index loops has no insertions and the rest have
two insertions and we obtained the possible diagrams in
figure~\ref{profign4}. Furthermore, we saw that the dependence on
the Schwinger variables $s_i$ exactly corresponded to the six
terms in the square of the determinant of $M_{ab}$
from~\eqref{pron40} thus cancelling $A_{\textrm{bosonic}}$.
However, in this case we do not have six, but eight possible
insertion distributions. These are shown in figure~\ref{profign6}
where the outer index loop and the upper of the inner index loops
have been chosen to only have one insertion.
We should compare the $s_i$ dependence in these diagrams to the
ones in figure~\ref{profign4}. If we factor out the dependence on
the traces of the $\W_\al$'s which, naturally, is different in the
two set of diagrams, we will actually get that diagrams a, b, c
and d in figure~\ref{profign6} gives $-2$ times respectively the
diagrams a, b, f and c in figure~\ref{profign4}. For diagrams a, b
and c in figure~\ref{profign6} this can be seen by moving the
single insertion in the outer loop to the opposite side of the
double line propagator and thus to the index line in the
corresponding inner loop. This gives us the same $s_i$ dependence,
but we get a minus sign since the insertion change sign and a
factor of 2 from changing a propagator with insertions in both
sides to one with insertions in only one side (remember the factor
2 in figure~\ref{profign2}). For diagram d in
figure~\ref{profign6} one should be more careful and remember
which insertions should be Weyl contracted with each other. For
the remaining diagrams in figure~\ref{profign6} one finds that e
and f are equal and each corresponds to minus diagram d in
figure~\ref{profign4}. In the same way g and h in
figure~\ref{profign6} are equal and each corresponds to minus
diagram e in figure~\ref{profign4}. Thus, all in all, we again get
the determinant of $M_{ab}$ squared and a cancellation between
$A_{\textrm{bosonic}}$ and $A_{\textrm{fermionic}}$, but also an
extra factor $-2$.

For a general diagram we can use the auxiliary variables to see
the reduction to the matrix model. Let $F=\sum_iF_i$ where $F_i$
is the number of index loops of type $i$. Consider the case where
we choose the single insertions to be in the loops of type $i$ and
$j$ respectively. Thus, from these two index loops we get a
contribution of
$\tr\big(\W_{(i)}^\al\pi_{l\al}\big)\tr\big(\W_{(j)}^\be\pi_{k\be}\big)$
and after performing the fermionic momentum integrations we will,
by Lorentz invariance, end up with a factor proportional to:
\begin{equation}\label{pron60}
    \tr\Big(\W_{(i)}^\al\Big)\tr\Big(\W_{(j)\al}\Big)=\frac{(4\pi)^2}{2}w^\al_iw_{j\al},
\end{equation}
where we have introduced $w^\al_i$ from~\eqref{dvn14.5}. Let us
now also introduce the auxiliary variables $\W'^\al_m$. We choose
the labelling such that $m=0$ corresponds to the index loop of
type $i$ with a single insertion and $m=1$ to the index loop of
type $j$ with the other single insertion. We can thus
identify~\eqref{pron60} with $\W'^\al_0\W'_{1\al}$ and to
calculate $A_{\textrm{fermionic}}$ we should include an
integration
\begin{equation}\label{pron61}
    \frac{(4\pi)^2}{2}w^\al_iw_{j\al}\frac{1}{2}\left.\int\denw_{0\al}\denw_1^\al\right\rvert.
\end{equation}
Here
$\frac{1}{2}\left.\int\denw_{0\al}\denw_1^\al\right\rvert\W'^\al_0\W'_{1\al}=1$
and the restriction means that we should set the parts of
$\W'^\al_0$ and $\W'^\al_1$ that are not integrated over to zero,
just as above where we removed the auxiliary variables
corresponding to the index loop without insertions, i.e.:
\begin{equation}\label{pron62}
\left.\denw_{0\al}\denw_1^\al\right\rvert=\left.\denw_{0\al=1}\denw_1^{\al=1}\right\rvert_{\W'_{0\,\al=2}=\W'^{\al=2}_1=0}+\left.\denw_{0\al=2}\denw_1^{\al=2}\right\rvert_{\W'_{0\,\al=1}=\W'^{\al=1}_1=0}
\end{equation}
For the remaining index loops, numbered by $a=2,\ldots, L$, with
two insertions we get glueball superfield contributions and
similar to~\eqref{pron51} we should include the $2(L-1)=2(F-2)$
integrations
\begin{equation}\label{pron63}
    (-1)^{F-2}(4\pi)^{2(F-2)}S_i^{F_i-1}S_j^{F_j-1}
\Big(\prod_{k\neq i,j}S_k^{F_k}\Big)\prod_{a=2}^L\int\dtow_a,
\end{equation}
where we have chosen $C\nleft(\mathrm{fund}\right)=\half$ as in
section~\ref{dvsecdvun}. The fermionic integrations in
$A_{\textrm{fermionic}}$ from~\eqref{pron41} thus take the form:
\begin{equation}\label{pron64}
\frac{1}{2}\left.\int\denw_{0\al}\denw_1^\al\right\rvert\Big(\prod_{a=2}^L\int\dtow_a\Big)\Big(\prod_{a=1}^L(-4)\!\int\dtopi'_a\Big)e^{-\sum_i\sum_{mb}s_iK_{im}\W'^\al_m
L_{ib}\pi'_{b\al}}
\end{equation}
This splits into two contributions by~\eqref{pron62} (which
actually give the same by Lorentz invariance). Let us consider the
last contribution. Using~\eqref{appspinorn17.5}
and~\eqref{appspinorn18} we can write the second integration as
$\int\denw_{0\al=2}\denw_1^{\al=2}=\int\denw_0^{\al=1}\denw_1^{\al=2}$.
We can factorise the exponential in~\eqref{pron64} by writing out
the sum over $\al$ into two contributions (remembering the
restriction in~\eqref{pron62}):
\begin{equation}\label{pron65}
    \left.e^{-\sum_i\sum_{mb}s_iK_{im}\W'^{\al=1}_m
L_{ib}\pi'_{b\,\al=1}}\right\rvert_{\W'^{\al=1}_1=0}
\left.e^{-\sum_i\sum_{mb}s_iK_{im}\W'^{\al=2}_m
L_{ib}\pi'_{b\,\al=2}}\right\rvert_{\W'^{\al=2}_0=0}.
\end{equation}
By choosing the loop momenta to run as the index loops with
$m=1,\ldots,L$ we have, as above, $K_{ia}=L_{ia},\,a=1,\ldots,L$.
The second exponential in~\eqref{pron65} then has the form:
\begin{equation}\label{pron66.5}
    e^{-\sum_i\sum_{a,b=1}^L s_iL_{ia}\W'^{\al=2}_a
L_{ib}\pi'_{b\,\al=2}},
\end{equation}
whereas the first exponential in~\eqref{pron65} is:
\begin{equation}\label{pron66}
    \exp\nleft(-\sum_is_i\bigg(\sum_{a=2}^L L_{ia}\W'^{\al=1}_a+K_{i0}\W'^{\al=1}_0\bigg)
\bigg(\sum_bL_{ib}\pi'_{b\,\al=1}\bigg)\right).
\end{equation}
Now, the trick is that since we have an oriented diagram, the
single index lines in the double line propagators have opposite
directions and by definition of $K_{im}$ we conclude that:
\begin{equation}\label{pron67}
    \sum_m K_{im}=0,
\end{equation}
 for all propagators labelled by $i$. Thus in our case $K_{i0}=-\sum_{a=1}^L L_{ia}$ and we can
rewrite~\eqref{pron66} as:
\begin{equation}\label{pron68}
    \exp\nleft(-\sum_is_i\bigg(\sum_{a=2}^L L_{ia}\big(\W'^{\al=1}_a-\W'^{\al=1}_0\big)-L_{i1}\W'^{\al=1}_0\bigg)
\bigg(\sum_bL_{ib}\pi'_{b\,\al=1}\bigg)\right).
\end{equation}
Since the integrations over $\W'^{\al=1}_a,\,a=2,\ldots,L$, only
work on this exponential, we can safely translate the integrations
over these with $\W'^{\al=1}_0$. The integration over
$\W'^{\al=1}_0$ also only works on this exponential and we can
thus substitute $\W'^{\al=1}_0$ with $-\W'^{\al=1}_1$ and get:
\begin{equation}\label{pron69}
    \exp\nleft(-\sum_is_i\bigg(\sum_{a=1}^L L_{ia}\W'^{\al=1}_a\bigg)
\bigg(\sum_bL_{ib}\pi'_{b\,\al=1}\bigg)\right).
\end{equation}
This is the same form as we got for the second exponential
in~\eqref{pron66.5}. We can now do the same for the first term
in~\eqref{pron62} and thus~\eqref{pron64} becomes:
\begin{equation}\label{pron70}
-2\Big(\prod_{a=1}^L\int\dtow_a\Big)\Big(\prod_{a=1}^L(-4)\!\int\dtopi'_a\Big)e^{-\sum_i\sum_{ab=1}^Ls_iL_{ia}\W'^\al_a
L_{ib}\pi'_{b\al}},
\end{equation}
where the minus sign is from the substitution of $\W'_0$ with
$-\W'_1$ and the factor 2 comes from:
\begin{equation}\label{pron71}
    \frac{1}{2}\int\denw_{1\al}\denw_1^\al=2\int\dtow_1.
\end{equation}
The integration in~\eqref{pron70} is the same as we had
in~\eqref{pron51} which gave $(\det M(s_i))^2$, but now with an
extra factor of $-2$ that we also found when comparing the
diagrams in figure~\ref{profign6} to the diagrams in
figure~\ref{profign4}. We have thus obtained the localisation to
zero-modes.

Collecting the terms from~\eqref{pron61} and~\eqref{pron63} and
using~\eqref{pron39} we get:
\begin{equation}\label{pron71.5}
    A_{\textrm{bosonic}}A_{\textrm{fermionic}}=-2\frac{1}{2}w^\al_iw_{j\al}(-1)^{F-2}S_i^{F_i-1}S_j^{F_j-1}
\Big(\prod_{k\neq i,j}S_k^{F_k}\Big).
\end{equation}
Thus, with this choice of in which type of index loops we should
have only one insertion, the contribution to the total
amplitude~\eqref{pron37} is:
\begin{equation}\label{pron72}
    -a_{i,j}(-1)^{F-2}(-1)^{\chi-F}w^\al_iw_{j\al}S_i^{F_i-1}S_j^{F_j-1}
\Big(\prod_{k\neq i,j}S_k^{F_k}\Big)\nsymm
m^{-E}\prod_k(-g_k)^{V_k},
\end{equation}
where $(-1)^{\chi-F}$ as before comes from the change of $\meucl$
and $\geuclk$ to $m$ and $g_k$. $a_{i,j}$ is a combinatorial
factor telling in how many ways we can choose the index loops with
single insertions. If $i\neq j$ we have $a_{i,j}=F_i F_j$, whereas
for $i=j$ we have $a_{i,i}=\tbinom{F_i}{2}=F_i(F_i-1)/2$. To
obtain the total amplitude we should sum over the choices of $i$
and $j$ i.e. a sum as $\sum_{i\leq j}$. Using that
$\sum_{i<j}=\half\sum_{i\neq j}$ we get the total amplitude for
the planar diagrams having two index loops with single insertions:
\begin{equation}\label{pron72.5}
    A=-\frac{1}{2}\sum_{i,j}w^\al_iw_{j\al}\frac{\partial^2}{\partial S_i\partial
S_j}\Big(\prod_{k}S_k^{F_k}\Big)\nsymm m^{-E}\prod_k(-g_k)^{V_k},
\end{equation}
since $(-1)^{F-2}(-1)^{\chi-F}=1$ for $\chi=2$. The same analysis
that led from~\eqref{pron53} to~\eqref{pron55} gives us the
contribution to $\weffpert$ from the planar diagrams with two
single insertions:
\begin{equation}\label{pron73}
    \De\weffpert=\frac{1}{2}\sum_{i,j}w^\al_iw_{j\al}\frac{\partial^2}{\partial S_i\partial
S_j}\cF_{g=0}\nleft(\gs N'_i\defi
    S_i\right),
\end{equation}
thus finishing the proof of~\eqref{dvn17}.

\begin{figure}\caption{}\label{profign7}
\begin{center}
\includegraphics{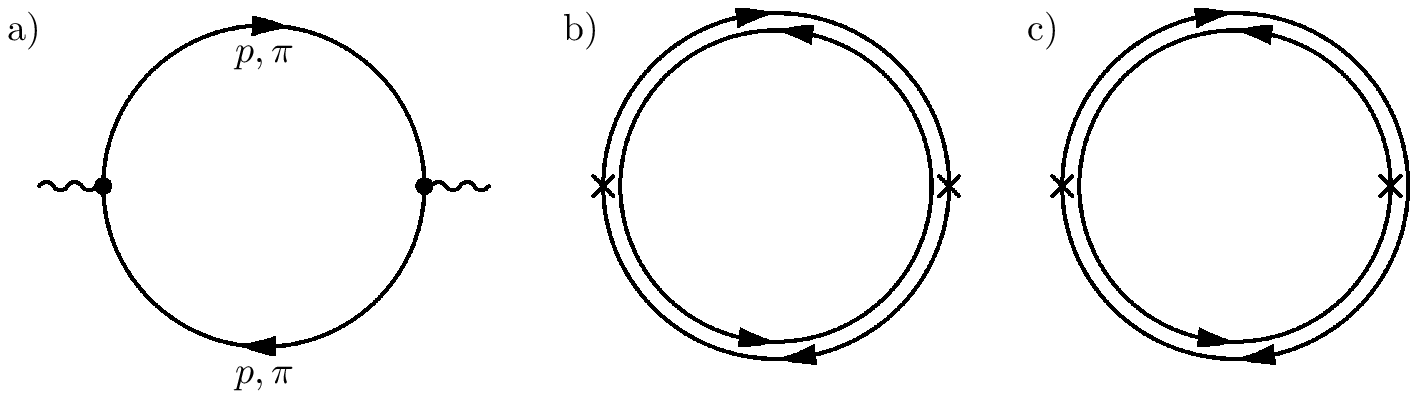}
\end{center}
    \begin{center}
        The one-loop diagram with two $\W^\al$ interaction vertices. a) shows the diagram in single line
        notation while b) and c) show the diagram in double line
        notation with the two possible types of insertion patterns.
    \end{center}
\end{figure}

We have now finished the proof for the form of $\weffpert$ in the
Dijkgraaf-Vafa conjecture. However, we have actually missed one
type of diagram in the proof. If we think of the $\W^\al\pi_\al$
term as an interaction term, the one-loop diagram shown in
figure~\ref{profign7} should also be taken into consideration (as
done in~\cite{0211170}). The diagram has two propagators and two
$\W^\al$ interaction vertices as shown in diagram a) in
figure~\ref{profign7}. Since $\W^\al$ is a constant, there is no
external momentum and the same momentum runs through both
propagators.

Let us use the double line notation for the diagram. The $\W^\al$
interaction vertices takes the same form as in
figure~\ref{profign1}. Since the diagram has one loop, it should
have two $\W^\al$ insertions. The two types of insertion patterns
are shown in b) and c) in figure~\ref{profign7}. In the broken
case the diagrams in which the index loops are of different types
should cancel. Let us therefore consider a diagram in which both
index lines are of the same type $i$. As seen from~\eqref{pron49}
the insertions of type b) give a term proportional to
$-2N_i\cdot8\pi^2 S_i\pi\pi$. Here we have a minus sign since the
insertions sit in different sides of the double lines (but both in
the same index loop), the factor two comes from the choice of
index loop without insertions, and the factor $N_i$ comes from the
trace in that index loop. Correspondingly, the second insertion
pattern in diagram c) in figure~\ref{profign7} gives a
contribution of $-8\pi^2w^\al_iw_{i\al}\pi\pi$. Since one of the
interaction vertices gives an overall chiral superspace
integration, the contribution to the Euclidean effective
superpotential is given by:
\begin{equation}\label{pron74}
    2\frac{1}{2!}\left(-\frac{1}{2}\right)^2\int\dmom(-4)\!\int\!\dtopi\frac{1}{(p^2+\meucl)^2}(-1)(4\pi)^2\left(N_iS_i+\frac{1}{2}w^\al_iw_{i\al}\right)\pi\pi.
\end{equation}
Here the factors in front are a factor $2$ from the symmetry of
the diagram, a factor $1/2!$ from going to second order in the
interaction, and a factor $(-1/2)^2$ from the normalisation of the
interaction term as seen from~\eqref{pron29}. Introducing a
cut-off, $\La_0$, the four-dimensional bosonic momentum
integration gives standardly:
\begin{equation}\label{pron75}
   \int\dmom\frac{1}{(p^2+\meucl)^2}=\frac{1}{(4\pi)^2}\ln\nleft(\frac{\La_0e^{-1}}{\abs{\meucl}}\right).
\end{equation}
Inserting into~\eqref{pron74}, summing over the type of index loop
$i$, and rotating into Minkowski space (giving a sign change)
gives the contribution
 to the glueball superpotential:
\begin{equation}\label{pron76}
    \sum_i
    \left(N_iS_i+\frac{1}{2}w^\al_iw_{i\al}\right)\ln\nleft(\frac{m}{\La_0e^{-1}}\right)=\sum_i
    N_i \Shat_i\ln\nleft(\frac{m}{\La_0e^{-1}}\right),
\end{equation}
where we have used~\eqref{dvn15} to write the result in terms of
the traceless glueball superfield, $\Shat_i$. We see that this is
nothing but the $m$ dependent part of the Veneziano-Yankielowicz
superpotential~\eqref{dvn7} if we use the matching of
scales~\eqref{dvn122}: $\La^3=\La_u^2m$. This is the same
$m$-dependence we obtained in the matrix model in~\eqref{dvn144}.

\section{General Gauge Groups and
Matter Representations}\label{proofsecdiasubgeneral} In the last
section we have proven diagrammatically the localisation to
zero-modes (and the form of $\weffpert$) for the gauge groups and
matter representations that allowed a double line notation.
Following~\cite{0304271} let us end this chapter by briefly
showing the localisation to zero-modes for general gauge groups
and representations.

We will use the same setup as in section~\ref{proofsecdiasubpert},
but we will assume that the constant field strength $\W^\al$ is
abelian, i.e. lies in a Cartan subalgebra. Let us denote the
generators of the Cartan subalgebra by $H_a$ so $\W^\al=\W^{\al
a}H_a$. For the given representation of the chiral field $\reprn$
we can diagonalise the $H^{\repr}_a$'s simultaneously since they
commute:
\begin{equation}\label{pron77}
    \big(H_a^{\repr}\big)\opned{i}{j}\Phi^j=\la_a^i\Phi^i.
\end{equation}
We can thus define the $\dim\repr$ vectors
$\lavect^i={(\la^i_1,\ldots,\la^i_{\mathrm{rank}(G)})}$ where
$\mathrm{rank}(G)$ is the rank of the gauge group. These are
called the \emph{weights} of the representation.\footnote{The
precise mathematical definition of the weights (as given
in~\cite{Cahn}) is that they are the linear functionals, $M^i$, on
the Cartan subalgebra given by $M^i(\sum_a c^aH_a)=\sum_a
c^a\la_a^i$.} Using these we can write:
\begin{equation}\label{pron78}
    \big(\W^\al_{\repr}\big)\opned{i}{j}=\W^{\al a}\la_a^i\de^i_j\defi\Wvect^\al\cdot\lavect^i\de^i_j.
\end{equation}
Since $\W^\al_{\repr}$ is diagonal, we can use a notation where
the propagators just carry a single gauge index sitting on
$\lavect$ instead of two indices.

Let us think of a given diagram. With $\lavect_i$ we denote the
weight propagating through the $i^{\mathrm{th}}$ propagator which
is then equal to $\lavect^j$ where $j$ is the gauge index for the
propagator. The $\lavect_i$'s are like charges for the Cartan
generators and especially we have conservation of charges at the
vertices (due to gauge invariance of the interaction terms). This
means that we can introduce loop Cartan charges,
$\lavect'_a$,\footnote{The loop index $a$ should not be confused
with the adjoint gauge index.} analogous to the bosonic and
fermionic momenta in~\eqref{pron38} and~\eqref{pron42}:
\begin{equation}\label{pron79}
    \lavect_i=\sum_a L_{ia}\lavect'_a.
\end{equation}
For a given choice of $\lavect'_a$ we see that
$A_{\textrm{fermionic}}(s_i)$ from~\eqref{pron41} takes the form
of a factor from the vertices and:
\begin{multline}\label{pron80}
    (-4)^L\int\prod_{a=1}^L\dtopi'_a
e^{-\sum_is_i\Wvect^\al\cdot\lavect_i\pi_{i\al}}=(-4)^L\int\prod_{a=1}^L\dtopi'_a
e^{-\sum_is_i\sum_{ab}L_{ia}\Wvect^\al\cdot\lavect'_aL_{ib}\pi'_{b\al}}\\
=(-4)^L\int\prod_{a=1}^L\dtopi'_a
e^{-\sum_{ab}\Wvect^\al\cdot\lavect'_aM_{ab}\pi'_{b\al}},
\end{multline}
where we have introduced the matrix $M_{ab}$ from~\eqref{pron40}.
Exactly as in~\eqref{pron51} (the integrations over $\W'$ there
was just a bookkeeping device) we see that this gives $(\det
M(s_i))^2$ times a multiple of factors of
$\Wvect^\al\cdot\lavect'_a$. The last factor cancels the $s_i$
dependence in $A_{\textrm{bosonic}}$ from~\eqref{pron39}. We are
then left with the trivial $s_i$ dependence in the mass part of
the propagators (see~\eqref{pron37}) and we have thus obtained the
wanted localisation to zero-modes.

One can pursue the matter and obtain the total amplitude by
remembering the factor from the vertices, which depends on the
choice of $\lavect'_a$, and then sum over the $\lavect'_a$'s (and
there is also an overall sum analogous to the overall superspace
integration). In this way the relation to the matrix model can be
found e.g. for the $\un{N}$ gauge group with adjoint matter as
done in~\cite{0304271}.

%-----Conclusion--------------
\chapter*{Conclusions}
\addcontentsline{toc}{chapter}{Conclusions}

The prime goal of this thesis has been to give a thorough
introduction to the Dijkgraaf-Vafa conjecture, its diagrammatic
proof and the concepts needed to understand the conjecture.

We started in chapter 2 by stating the conjecture and then used
the rest of the chapter to understand the concepts used in the
conjecture. Along the way we got a better understanding of the
conjecture and its context. Using the Seiberg scheme of
holomorphy, symmetries and various limits we proved the
non-renormalisation theorem telling us that the low energy
superpotential, which is what the Dijkgraaf-Vafa conjecture
determines, essentially consists of non-perturbative corrections.
And in the case of a cubic tree-level superpotential we saw that
the Seiberg scheme determined the form of the perturbative part of
the glueball superpotential and thus the essence of the
Dijkgraaf-Vafa conjecture is to determine the coefficients in the
power series for the glueball superfield. Furthermore, we briefly
pointed out that there is an interplay between the ILS linearity
principle and the Dijkgraaf-Vafa conjecture.

We have also seen some of the limitations in the conjecture. We
still need to assume that the glueball superfield is the
fundamental field (confinement). And we can only prove the
relation to the matrix model for the perturbative part of the
glueball superpotential, not that the full glueball superpotential
simply is given by adding the Veneziano-Yankielowicz
superpotential. Furthermore, we have discussed the nilpotency of
the glueball superfield. We have seen that the terms in the
glueball superpotential up to the power of the dual Coxeter number
are determined unambiguously. And we saw how to interpret the full
solution for the glueball superpotential given by the
Dijkgraaf-Vafa conjecture as an F-term completion using
supergroups, but also mentioned the discrepancies that arise here
when comparing to standard gauge theory. We have also pointed out
the difficulties in stating the conjecture in a general form when
considering multi-trace tree-level superpotentials and in the case
of general matter where the relation to the matrix model should be
done diagram by diagram.

In the diagrammatic proof of the conjecture we were able to
integrate out the conjugated chiral field using holomorphy. In
this way we escaped the non-renormalisation theorems and were able
to derive the strong results for the non-perturbative corrections
in the glueball superpotential purely diagrammatically. In the
proof we have been careful and kept track of all the details. In
this way we found the $2C(\mathrm{fund})$ factor in the
identification of the glueball superfield with $\gs N'_i$ in the
matrix model for a general classical gauge group. We have also
seen the cancellation of the intricate sign from rotating to
Euclidean space with the minus sign in the definition of the
glueball superfield. And we have given a thorough diagrammatic
proof in the case where one takes into account the abelian part of
supersymmetric gauge field strength. All in all, we have
understood very basically how the different terms in the
Dijkgraaf-Vafa conjecture arise by considering insertions in
diagrams. We have seen the projection to planar and Euler
characteristic $\chi=1$ diagrams by counting of these insertions.
With the localisation to zero-momentum modes for general gauge
groups and (massive) matter representations we have seen the
generality of the Dijkgraaf-Vafa conjecture.

It is natural to ask if there is a meaning with the diagrams in
the matrix model of higher genera. Indeed one finds that with
gravitational couplings one should take these non-planar diagrams
into account and much of recent research has been done in this
area.

The use of the Dijkgraaf-Vafa conjecture has also been discussed.
We have only briefly mentioned the important results obtained for
the parameter spaces of different theories using the conjecture.
But we have seen how to solve the one-cut case in the matrix model
exactly and with this solution we have used the Dijkgraaf-Vafa
conjecture to obtain the exact effective glueball superpotential
for a cubic tree-level superpotential.

At last we have seen that we can recover the
Veneziano-Yankielowicz superpotential from the matrix model free
energy by considering the measure of the matrix model or by
evaluating the partition function for zero couplings in the
tree-level superpotential. But perhaps most strikingly the term
emerged when we solved the matrix model exactly. Thus there were
no need for adding the Veneziano-Yankielowicz superpotential by
hand. This calls for a proof of this deeper relationship between
the gauge theory and the matrix model.

%-----Acknowledgements--------------
\chapter*{Acknowledgements}
\addcontentsline{toc}{chapter}{Acknowledgements}

I am very grateful to my supervisor Niels Obers for his support
and patience with my never ending questions. I would also like to 
thank Francesco Sannino and Paolo Merlatti for discussions, and
especially Paolo Di Vecchia for his suggestions to the text.

I thank Anja for her patience, suggestions and support.

%-----APPENDIX----------------------------------------------------
\begin{appendix}

%-------New Appendix--------------------------------------------------

\chapter{Notation}\label{appnota}
The topic of supersymmetry suffers under an abundance of
notations. We choose to follow the notation of
\cite{wessandbagger} as strict as possible, however, we shall make
deviations.

We choose to use the ``mostly plus'' metric for the Minkowski
space:
\begin{equation*}
    \eta_{\mu\nu}\sim\begin{pmatrix}
  -1 &  &  &  \\
   & 1 &  &  \\
   &  & 1 &  \\
   &  &  & 1 \\
\end{pmatrix}.
\end{equation*}
This also defines the similarity symbol for use when we do not
have a formal equality (as here for the same matrix/tensor in
index notation and written out with all entries in bracket
notation). Also note that we  use Greek letters $\mu, \nu, \ldots$
for Lorentz indices contrary to \cite{wessandbagger}. Latin
indices $i, j, k$ will be used for spatial coordinates.

This choice of metric then demands that we choose to represent the
four-momentum operator as:
\begin{equation}\label{notan1}
    p_\mu=-i\partial_\mu,
\end{equation}
such that $p^0$ is the Hamiltonian and $p^i$ the usual momentum
operators $-i\partial_i$.

The orientation in Minkowski space is given by the totally
antisymmetric Levi-Civita tensor $\varepsilon^{\mu\nu\rho\sigma}$
defined by:
\begin{equation*}
    \varepsilon^{0123}=-\varepsilon_{0123}\defi+1.
\end{equation*}

$[-,-]$ will be used for commutators and $\{-,-\}$ for
anticommutators.

In order to simplify expressions we choose units such that
$\hslash=c=1$.

Matrices can be in bold font while operators are in normal font --
except abstract Lie algebra operators which can be in calligraphic
font.

The term Lagrangian will also be used for the Lagrangian density
as long as no confusion should be possible.

For notation on spinors please see appendix~\ref{appspinors}.

%------New Appendix----------------------------------------------------

\chapter[Minkowski Space as Cosets\ldots]{Minkowski Space as Cosets in the Poincar\'e Group}\label{apppoincare}
In this appendix we describe how the Minkowski space emerges as
the cosets of the Poincar\'e group modulo the Lorentz group. The
treatment builds on~\cite{0109172}.

The idea is based on the fact that the Poincar\'e algebra is the
semidirect product of the translation group and the Lorentz group
and hence the same for the corresponding algebras. Any point in
Minkowski space can be reached as a translation of the origin --
but not uniquely since any Lorentz transformation keeps the origin
fixed. Hence we must identify Minkowski space with the
translations modulo the Lorentz transformations.

\section{Poincar\'e Algebra}

To be precise let $\Pgen_{\mu}$ be the generators of the
translations and let $\Jgen_{\mu\nu}=-\Jgen_{\nu\mu}$ be the
antisymmetric generators of the Lorentz group such that the total
Poincar\'e algebra is given by (taken from~\cite{weinberg1}):
\begin{eqnarray}\label{apppoincaren1}
[\Pgen_{\mu},\Pgen_{\nu}]&=& 0, \nonumber\\{}
   [\Pgen_{\mu},\Jgen_{\nu\rho}]&=&-i\left(\eta_{\mu\nu}\Pgen_{\rho}-\eta_{\mu\rho}\Pgen_{\nu}\right),  \\{}
   [\Jgen_{\mu\nu},\Jgen_{\rho\sigma}]&=&-i\left(\eta_{\nu\rho}\Jgen_{\mu\sigma}-\eta_{\mu\rho}\Jgen_{\nu\sigma}+\eta_{\mu\sigma}\Jgen_{\nu\rho}-\eta_{\nu\sigma}\Jgen_{\mu\rho}\right). \nonumber
\end{eqnarray}
An element in the Poincar\'e group then has the form:
\begin{equation}\label{apppoincaren1.2}
    e^{-i\tau^{\mu}\Pgen_{\mu}+i\half\omega^{\mu\nu}\Jgen_{\mu\nu}},
\end{equation}
where $\omega_{\mu\nu}=-\omega_{\nu\mu}$. Please note the signs
which have been chosen carefully. In a representation where this
is the (active) coordinate transformation of fields, we want:
\begin{equation}\label{apppoincaren1.5}
    e^{-i\tau^{\mu}P_{\mu}+i\half\omega^{\mu\nu}J_{\mu\nu}}\psi(x)=\psi'(x)=\psi(x_{\textrm{Passive}}),
\end{equation}
where $P$ and $J$ is the representations of $\Pgen$ and $\Jgen$
respectively, and $x_{\textrm{Passive}}$ to first order in $\tau$
and $\omega$ is given by
\begin{equation}\label{apppoincaren1.7}
    x^{\mu}_{\textrm{Passive}}\simeq x^\mu-\omega^{\mu}_{\phantom{\mu}\nu}x^{\nu}-\tau^{\mu}.
\end{equation}
With the signs given in~\eqref{apppoincaren1.2} this is exactly
realised by~\eqref{notan1} i.e. $\Pgen_{\mu}\rightsquigarrow
P_{\mu}=-i\partial_{\mu}$ and
\begin{equation}\label{apppoincare1.8}
    \Jgen_{\mu\nu}\rightsquigarrow J_{\mu\nu}=x_\mu P_\nu-x_\nu
    P_\mu=-i\left(x_\mu\partial_\nu-x_\nu\partial_\mu\right).
\end{equation}
It checks easily that we indeed get the right transformation (to
first order in $\tau$ and $\omega$):
\begin{eqnarray*}
  e^{-i\tau^{\mu}P_{\mu}+i\half\omega^{\mu\nu}J_{\mu\nu}}\psi\nleft(x^\mu\right) &=& \left(1-\tau^{\mu}\partial_{\mu}+\omega^{\mu\nu}x_\mu\partial_\nu\right)\psi\nleft(x^\mu\right) \\
   &=&\psi\nleft(x^\mu-\tau^\mu-\omega^{\mu}_{\phantom{\mu}\nu}x^\nu\right).
\end{eqnarray*}
We also note that $P$ and $J$ satisfy the Poincar\'e
algebra~\eqref{apppoincaren1} as they should.

\section{Minkowski Space as Right Cosets}

Now let us consider the cosets (\cite{0109172} call them right
cosets) given by the Poincar\'e group, $G_{\textrm{Poincar\'e}}$,
modulo the Lorentz group, $G_{\textrm{Lorentz}}$. They have the
form $gG_{\textrm{Lorentz}},\, g\in G_{\textrm{Poincar\'e}}$.
However, we can always use a pure translation as a representative
of the coset and this translation is unique. Hence we have the
unique form of a given coset:
\begin{equation}\label{apppoincaren2}
    \exp(-ix^{\mu}\Pgen_{\mu})G_{\mathrm{Lorentz}}.
\end{equation}
\begin{proof}
Since~\cite{0109172} does not prove this let us do it here. Take
an arbitrary element in the Poincar\'e group
$\exp\nleft(-iy^{\mu}\Pgen_{\mu}+i\half\omega^{\mu\nu}\Jgen_{\mu\nu}\right)$.
We now want to use the Baker-Campbell-Hausdorff formula (here
from~\cite{weinberg3}), which says:
\begin{equation}\label{apppoincaren3}
    e^{\mathcal{A}}e^{\mathcal{B}}=e^{\mathcal{A}+\mathcal{B}+\half[\mathcal{A},\mathcal{B}]+\frac{1}{12}[\mathcal{A},[\mathcal{A},\mathcal{B}]]+\frac{1}{12}[\mathcal{B},[\mathcal{B},\mathcal{A}]]+\cdots},
\end{equation}
where $\mathcal{A}$ and $\mathcal{B}$ are arbitrary elements in a
given Lie algebra. Now let us try to set
$\mathcal{A}=i\half\omega^{\mu\nu}\Jgen_{\mu\nu}$ and
$\mathcal{B}=-ix^{\mu}\Pgen_{\mu}$. From~\eqref{apppoincaren1} we
get that $[\mathcal{B},\mathcal{A}]\sim\Pgen$ and hence
$[\mathcal{B},[\mathcal{B},\mathcal{A}]]=0$. Thus one of the
series in the Baker-Campbell-Hausdorff formula truncates and the
resulting series is linear in $\mathcal{B}$. The formula then
takes the form (from~\cite{0109172}):
\begin{equation}\label{apppoincaren4}
    e^{\mathcal{A}}e^{\mathcal{B}}=e^{\mathcal{C}},\qquad
    C=\mathcal{A}+\left(\frac{-\ad\!\mathcal{A}}{e^{-\ad\!\mathcal{A}}-\mathbf{1}}\right)\mathcal{B},
\end{equation}
where $\mathbf{1}$ is the identity and the adjoint
$\ad\!\mathcal{A}$ is defined, as usual, as the operator
\begin{equation}\label{apppoincaren5}
    \ad\!\mathcal{A}\cdot\mathcal{B}=[\mathcal{A},\mathcal{B}].
\end{equation}
Now let us work out how the adjoint works in our case.
\eqref{apppoincaren1} gives:
\begin{eqnarray}\label{apppoincaren6}
  \ad\!\nleft(i\half\omega^{\mu\nu}\Jgen_{\mu\nu}\right)\cdot \left(-ix^{\mu}\Pgen_{\mu}\right) &=&[\half\omega^{\mu\nu}\Jgen_{\mu\nu},x^{\mu}\Pgen_{\mu}]  \nonumber\\
   &=& ix^\mu\omega_\mu^{\phantom{\mu}\nu}P_\nu.
\end{eqnarray}
By induction:
\begin{equation}\label{apppoincaren7}
\left(\ad\!\nleft(i\half\omega^{\mu\nu}\Jgen_{\mu\nu}\right)\right)^k\cdot\left(-ix^{\mu}\Pgen_{\mu}\right)=-(-1)^ki\mathbf{x}\cdot\boldsymbol{\omega}^k\cdot\mathbf{\Pgen},
\end{equation}
where $\mathbf{x}$ is the row vector made out of $x^\mu$,
$\boldsymbol{\omega}$ is the matrix made from
$\omega_\mu^{\phantom{\mu}\nu}$ and $\mathbf{\Pgen}$ is the column
vector made out of $\Pgen_\nu$. Now~\eqref{apppoincaren4} becomes:
\begin{equation}\label{apppoincaren8}
    e^{i\half\omega^{\mu\nu}\Jgen_{\mu\nu}}e^{-ix^{\mu}\Pgen_{\mu}}=e^{i\half\omega^{\mu\nu}\Jgen_{\mu\nu}-i\mathbf{x}\cdot\left(\frac{\boldsymbol{\omega}}{e^{\boldsymbol{\omega}}-\mathbf{1}}\right)\cdot\mathbf{\Pgen}}.
\end{equation}
Noting that as a real function
$\frac{\boldsymbol{\omega}}{e^{\boldsymbol{\omega}}-\mathbf{1}}$
is non-zero, and that it and its inverse are well defined as
series expansions in $\boldsymbol{\omega}$ we get:
\begin{equation*}
    e^{i\half\omega^{\mu\nu}\Jgen_{\mu\nu}}e^{-i\mathbf{y}\cdot\left(\frac{e^{\boldsymbol{\omega}}-\mathbf{1}}{\boldsymbol{\omega}}\right)\cdot\mathbf{\Pgen}}=e^{-iy^{\mu}\Pgen_{\mu}+i\half\omega^{\mu\nu}\Jgen_{\mu\nu}}.
\end{equation*}
Now taking the inverse and substituting $\mathbf{y}$ and
$\boldsymbol{\omega}$ for $-\mathbf{y}$ and $-\boldsymbol{\omega}$
respectively we get the wanted result:
\begin{equation}\label{apppoincaren9}
    e^{-iy^{\mu}\Pgen_{\mu}+i\half\omega^{\mu\nu}\Jgen_{\mu\nu}}=e^{-i\mathbf{y}\cdot\left(\frac{e^{-\boldsymbol{\omega}}-\mathbf{1}}{-\boldsymbol{\omega}}\right)\cdot\mathbf{\Pgen}}e^{i\half\omega^{\mu\nu}\Jgen_{\mu\nu}}.
\end{equation}
The uniqueness of the translation representative can easily be
seen. Suppose that we had two pure translation  representatives of
the same coset (determined by $x$ and $x'$). Then for suitable
$\omega$ and $\omega'$ we have:
\begin{equation*}
    e^{-ix^\mu\Pgen_\mu}e^{i\half\omega^{\mu\nu}\Jgen_{\mu\nu}}=e^{-ix'^\mu\Pgen_\mu}e^{i\half\omega'^{\mu\nu}\Jgen_{\mu\nu}}.
\end{equation*}
Hence
\begin{equation*}
    e^{-i(x-x')^\mu\Pgen_\mu}=e^{i\half\omega'^{\mu\nu}\Jgen_{\mu\nu}}e^{-i\half\omega^{\mu\nu}\Jgen_{\mu\nu}}\in
    G_{\textrm{Lorentz}},
\end{equation*}
which is only possible for $x=x'$
\end{proof}
Now we see that the well defined relation (bijection) between
Minkowski space and the cosets in the Poincar\'e group modulo the
Lorentz group is given by:
\begin{equation}\label{apppoincaren10}
    x^\mu\in \textrm{Minkowski
    space}\longleftrightarrow\exp(-ix^{\mu}\Pgen_{\mu})G_{\textrm{Lorentz}}\in G_{\textrm{Poincar\'e}}/G_{\textrm{Lorentz}}.
\end{equation}

\section{Action on Minkowski Space}\label{apppoincaresecaction}

Now arises the natural question of how the action of the
Poincar\'e group is expressed in the coset space. The answer is
the simple action of left multiplication of the Poincar\'e group
on the cosets. This is easily checked. Left multiplication with a
pure translation determined by $\tau$ gives:
\begin{equation}\label{apppoincaren10.5}
    e^{-i\tau^{\mu}\Pgen_{\mu}}e^{-ix^{\mu}\Pgen_{\mu}}G_{\textrm{Lorentz}}=e^{-i(x+\tau)^{\mu}\Pgen_{\mu}}G_{\textrm{Lorentz}},
\end{equation}
hence sending $x^\mu\mapsto x^\mu+\tau^\mu$ which is the active
translation, that we want. Now let us check how a Lorentz
transformation determined by $\omega$ works. By the use
of~\eqref{apppoincaren8} and~\eqref{apppoincaren9} we get
\begin{eqnarray}\label{apppoincaren11}
  e^{i\half\omega^{\mu\nu}\Jgen_{\mu\nu}}e^{-ix^{\mu}\Pgen_{\mu}}G_{\textrm{Lorentz}} &=& e^{i\half\omega^{\mu\nu}\Jgen_{\mu\nu}-i\mathbf{x}\cdot\left(\frac{\boldsymbol{\omega}}{e^{\boldsymbol{\omega}}-\mathbf{1}}\right)\cdot\mathbf{\Pgen} }G_{\textrm{Lorentz}} \nonumber\\
   &=&e^{-i\mathbf{x}\cdot\left(\frac{\boldsymbol{\omega}}{e^{\boldsymbol{\omega}}-\mathbf{1}}\right)\cdot\left(\frac{e^{-\boldsymbol{\omega}}-\mathbf{1}}{-\boldsymbol{\omega}}\right)\cdot\mathbf{\Pgen}}e^{i\half\omega^{\mu\nu}\Jgen_{\mu\nu}}G_{\textrm{Lorentz}}\nonumber\\
   &=&e^{-i\mathbf{x}\cdot e^{-\boldsymbol{\omega}}\cdot\mathbf{\Pgen}}G_{\textrm{Lorentz}}\nonumber\\
   &=&e^{-i\left(e^{\boldsymbol{\omega}}\right)^\mu_{\phantom{\mu}\nu}x^\nu\Pgen_\mu}G_{\textrm{Lorentz}},
\end{eqnarray}
where we in the last line have used that $\omega$ is
antisymmetric. Hence this generates the transformation
$x^\mu\mapsto\Lambda^\mu_{\phantom{\mu}\nu}x^\nu$ with
\begin{equation}\label{apppoincaren12}
    \Lambda^\mu_{\phantom{\mu}\nu}=\left(e^{\boldsymbol{\omega}}\right)^\mu_{\phantom{\mu}\nu},
\end{equation}
as expected for an active Lorentz transformation.

Please note that the fact that this is a representation of the
Lorentz group, is tightly bound to the generator $\Pgen^\mu$ being
multiplied with $\La^{-1}$ and not $\La$ as one might expect.
$\Pgen^\mu$ is also a representation of the Lorentz group, but the
transformation matrices are not just multiplied on from the right
like on $x^\mu$, but are multiplied on right next to $\Pgen^\mu$.
I.e. first transforming with $\La_1$ and then with $\La_2$ yields
$\La_1^{-1}\La^{-1}_2\Pgen$, but because of the inverses this is a
representation.

In conclusion, we have seen that the Minkowski space can be
identified with the Poincar\'e group modulo the Lorentz group and
the Poincar\'e transformations are given by left multiplication
under this identification.

%------New Appendix----------------------------------------------------
\chapter{Spinors}\label{appspinors}

In this chapter we will give an introduction to spinors with
emphasis on what we will need in this thesis namely Majorana
spinors, Weyl spinors and the conventions and notation that
follows.\footnote{This appendix is based on~\cite{wessandbagger},
\cite{0109172} and~\cite{9912271}, but with all notation and
definitions altered to coincide with~\cite{wessandbagger}.
However, where~\cite{wessandbagger} has not defined objects we
will try to define everything logically and consistent with
\cite{wessandbagger}.}

\section{Spinorial Representations}

In relativistic quantum mechanics Lorentz invariance is the first
principle. Hence our objects must be representations of the
Lorentz group. The spinorial representations are a special case
since they rather are representations of a double cover of the
Lorentz group -- the spin group.

Let us start by looking at the proper orthochronous Lorentz group
with the six generators $\Jgen_{\mu\nu}=-\Jgen_{\nu\mu}$ with
commutation relations given by~\eqref{apppoincaren1}. There is a
very nice and easy way to list all finite-dimensional
representations of the Lorentz group (here inspired
by~\cite{peskinandschroeder}). First we define the usual
generators of rotations and boosts as:
\begin{equation}\label{appspinorn1}
    L^i=\half\vep^{ijk}\Jgen^{jk},\qquad K^i=\Jgen^{0i}.
\end{equation}
In this basis we get from~\eqref{apppoincaren1.2} that a Lorentz
transformation takes the form:
\begin{equation}\label{appspinorn2}
e^{-i\thetabold\cdot\mathbf{L}-i\betabold\cdot\mathbf{K}},
\end{equation}
with $\theta^i=-\half\vep^{ijk}\omega^{jk}$ and
$\beta_i=\omega_{i0}$ i.e. the turning angle and the rapidity
respectively. The signs of $\theta$ and $\beta$ are chosen to
comply with~\eqref{apppoincaren12}. However, we now want to make a
second change of basis to the six generators defined by:
\begin{equation}\label{appspinorn3}
    \mathbf{J}_\pm=\half\left(\mathbf{L}\pm i\mathbf{K}\right).
\end{equation}
One can show that $\mathbf{J}_+$ and $\mathbf{J}_-$ commute and
both satisfy the commutation relations of angular momentum (i.e.
the SU(2) algebra):
\begin{equation*}
    [J_\pm^i,J_\pm^j]=i\vep^{ijk}J_\pm^k.
\end{equation*}
Consequently, we can write~\eqref{appspinorn2} as:
\begin{equation}\label{appspinorn3.1}
    e^{i(-\thetabold+i\betabold)\cdot\mathbf{J}_+}e^{i(-\thetabold-i\betabold)\cdot\mathbf{J}_-}.
\end{equation}
This shows that the representations of the Lorentz group can be
seen as the complexified representations of
$\textrm{SU(2)}\times\textrm{SU(2)}$ where the last part is the
conjugate of the first. Since the irreducible finite-dimensional
representations of SU(2) are characterised by an integer or half
integer $j$ (corresponding to the dimension of the representation
being $2j+1$), we see that the finite-dimensional representations
of the Lorentz group correspond to pairs of integers or half
integers $(j_+,j_-)$.

Let us investigate the representations with the lowest dimensions.
(0,0) is of course the trivial representation. The two dimensional
representation $(\half,0)$ is the complexified SU(2) spin $\half$
representation. Let us use that
$\mathbf{J}_+=\frac{\boldsymbol{\sigma}}{2}$ and $\mathbf{J}_-=0$
where the Pauli matrices, i.e. the generators of SU(2), as usual
are given by:
\begin{equation}\label{appspinorn4}
    \sigma^1=\begin{pmatrix}
      0 & 1 \\
      1 & 0 \\
    \end{pmatrix},\qquad
    \sigma^2=\begin{pmatrix}
      0 & -i \\
      i & 0 \\
    \end{pmatrix},\qquad
    \sigma^3=\begin{pmatrix}
      1 & 0 \\
      0 & -1 \\
    \end{pmatrix}.
\end{equation}
Hence~\eqref{appspinorn3.1} becomes:
\begin{equation}\label{appspinorn5}
    e^{i(-\thetabold+i\betabold)\cdot\frac{\sigmabold}{2}}=e^{\left(-i\thetabold-\betabold\right)\cdot\frac{\sigmabold}{2}}.
\end{equation}
Since $-i\thetabold-\betabold$ can be any complex number
$\left(-i\thetabold-\betabold\right)\cdot\frac{\sigmabold}{2}$ can
by any complex traceless matrix and hence the represented group is
$\sltoc$. If we look at the representation $(0,\half)$ we get the
same result with a simple change in the sign of $\betabold$ i.e.
the group elements are:
\begin{equation}\label{appspinorn6}
    e^{\left(-i\thetabold+\betabold\right)\cdot\frac{\sigmabold}{2}}.
\end{equation}
Now in order to investigate the representation $(\half,\half)$ we
note that if~\eqref{appspinorn6} works on $\psi$, the action on
$\tilde{\psi}=\psi^T\sigma^2$ is (by the use of
$\sigmabold=-\sigma^2\sigmabold^T\sigma^2$ and
$\left(\sigma^2\right)^2=\idmatr$):
\begin{equation}\label{appspinorn7}
    \tilde{\psi}\mapsto\tilde{\psi}e^{\left(i\thetabold-\betabold\right)\cdot\frac{\sigmabold}{2}}.
\end{equation}
This is the hermitian adjoint of the $(\half,0)$ representation
working from the right. Thus we can see a $(\half,\half)$ object
as a matrix, $\mathbf{A}$, transforming like:
\begin{equation}\label{appspinorn8}
    \mathbf{A}\mapsto \mathbf{M}\mathbf{A}\mathbf{M}^\dagger,
\end{equation}
for $\mathbf{M}\in\sltoc$. Let us now assume that $\mathbf{A}$ is
hermitian. Then by defining
\begin{equation}\label{appspinorn9}
    \sigma^0=-\idmatr,\qquad\sigma^\mu\sim(-\idmatr,\sigmabold),
\end{equation}
we get a one to one correspondence between hermitian matrices and
real 4-vectors by $V_\mu\leftrightarrow V_\mu\sigma^\mu$.
Since~\eqref{appspinorn8} preserves hermiticity we can define a
transformation of a 4-vector like
\begin{equation}\label{appspinorn10}
V_\mu\sigma^\mu\mapsto\mathbf{M}V_\mu\sigma^\mu\mathbf{M}^\dagger\defi
V'_\mu\sigma^\mu.
\end{equation}
Now by calculation one gets that
$\det\nleft(V_\mu\sigma^\mu\right)=-V_\mu V^\mu$. Using this and
the fact that \eqref{appspinorn10} preserves the determinant since
$\det\nleft(\mathbf{M}\right)=1$, we see that this transformation
is a (proper orthochronous by further calculations) Lorentz
transformation. Hence the real part of the representation
$(\half,\half)$ is the same as the defining vector representation
of the Lorentz group. Or actually almost. Since both $\mathbf{M}$
and $-\mathbf{M}$ correspond to the same Lorentz transformation,
we are actually dealing with a double cover of the Lorentz group.
The group that we really are representing, $\sltoc$, is called the
spin group. Actually, it is the simply connected extension of the
Lorentz group.\footnote{Following~\cite{weinberg1} a matrix in
$\sltoc$ can be written as a matrix from SU(2) times the
exponential of a traceless hermitian matrix by the polar
decomposition theorem (like in~(\ref{appspinorn5})). The topology
of SU(2) is the same as the three dimensional ball $S^3$ and the
topology of traceless hermitian matrices is the same as $\R^3$.
Hence $\sltoc$ is topologically equivalent to $S^3\times\R^3$. The
Lorentz group, which now can be seen as $\sltoc/\Z_2$, is
topologically equivalent to $S^3\times\R^3/\Z_2$ and is not simply
connected -- actually, the first homotopy group is $\Z_2$.} The
spin representations are, of course, representations of the spin
group $\sltoc$.

\section{The Clifford Algebra}

The spinorial representations can be investigated using the Dirac
$\gamma$-matrices obeying the Clifford algebra:
\begin{equation}\label{appspinorn11}
    \{\gamma_\mu,\gamma_\nu\}=-2\eta_{\mu\nu}\idmatr.
\end{equation}
The connection to the spin group is through
\begin{equation}\label{appspinorn12}
    \Sigma_{\mu\nu}\defi\frac{i}{4}[\gamma_\mu,\gamma_\nu]
\end{equation}
which one can check obeys the Lorentz
algebra~\eqref{apppoincaren1} and hence generates a spinorial
representation.

Until now we have not said anything about the number of space-time
dimensions. In most of this thesis, however, we shall use four
space-time dimensions. In this case the dimension of the Dirac
matrices must be (at least) four and all representations are
unitarily equivalent. We will use the Weyl basis:
\begin{equation}\label{appspinorn12.5}
    \gamma^\mu=\begin{pmatrix}
      0 & \sigma^\mu \\
      \sigmabar^\mu & 0 \\
    \end{pmatrix},
\end{equation}
where $\sigmabar$ is defined as
\begin{equation}\label{appspinorn13}
    \sigmabar^\mu\sim\left(-\idmatr,-\sigmabold\right).
\end{equation}
Please note that $\gamma^0$ is hermitian while $\gamma^i$ is
antihermitian. In the index notation we give the $\ga$-matrices
indices $(\ga^\mu)_a^{\ph{a}b}$ and spinors index down, $\psi_a$.

Taking the full Clifford algebra as a real algebra it is
isomorphic to the algebra of real $4\times 4$ matrices and thus
has a natural real four-dimensional irreducible representation --
the Majorana spinor. However, looking at the complexified Clifford
algebra we get a complex four-dimensional irreducible
representation -- the Dirac spinors. Both are of course
transforming with $\Sigma_{\mu\nu}$ as generators under Lorentz
transformations. Connecting back to the
\mbox{$(j_+,j_-)$-notation} the Dirac spinors correspond to
$(\half,0)\oplus(0,\half)$ and the Majorana spinors are the
subrepresentation fixed by complex conjugation.

Another way to obtain the Majorana spinors is to look at how to
define the conjugate of a spinor, $\psi$. There are two
definitions -- the Dirac conjugate, $\psibar_\textrm{D}$, and the
Majorana conjugate, $\psibar_\textrm{M}$, defined as:
\begin{eqnarray}\label{appspinorn14}
  \psibar_\textrm{D} &=& \psi^\dagger \gamma_0 \\
  \psibar_\textrm{M} &=& \psi^T C,\label{appspinorn14.1}
\end{eqnarray}
where $C$ is the charge conjugation matrix defined by $C\gamma_\mu
C^{-1}=-\gamma_\mu^T$. In our case $C=i\gamma^2\gamma^0$. Now the
Majorana spinors are defined as those Dirac spinors that obey
$\psibar_\textrm{D}=\psibar_\textrm{M}$. This can be rewritten as
the Majorana reality condition:
\begin{equation}\label{appspinorn14.5}
    \psi^*=-i\gamma_2\psi.
\end{equation}

Actually, $C$ is not a matrix like the $\gamma$-matrices, but
rather it transforms like a bilinear form. We define it to have
indices up, $C^{ab}$. Hence by the use of~\eqref{appspinorn14.1}
it can be used to raise indices on Majorana spinors, but one
should be careful since it is antisymmetric. We define
 raising by
$\psi^a=C^{ab}\psi_b$ and hence lowering by $\psi_a=C_{ab}\psi^a$
where $C_{ab}=\left(C^{-1}\right)^{ab}$.

We will not proceed any further in this direction, but rather note
that the Dirac and Majorana representations are not irreducible as
representations of the spin group. To see this define
\begin{equation}\label{appspinorn15}
    \gamma^5=\gamma^0\gamma^1\gamma^2\gamma^3.
\end{equation}
This matrix has the property that it anticommutes with all other
$\gamma$-matrices. Hence it commutes with $\Sigma_{\mu\nu}$. By
Schur's lemma the spinorial representation is reducible and
reduces corresponding to the eigenvalues of $\gamma^5$. Actually:
\begin{equation}\label{appspinorn16}
    \gamma^5=\begin{pmatrix}
      -i\idmatr & 0 \\
      0 & i\idmatr \\
    \end{pmatrix},
\end{equation}
showing that the four dimensional representations split into two
(irreducible) representations -- the Weyl spinors. The two
representations are conjugate and are exactly the representations
$(\half,0)$ and $(0,\half)$ that we have already encountered.

\section{Weyl Spinors}
The two-dimensional representations we uncovered in last section
are of course linked with the fact that $\sltoc$ has a natural
two-dimensional complex representation acting as a matrix. A two
dimensional spinor hence transforms as $\psi_\alpha\mapsto
M_\alpha^{\phantom{\alpha}\beta}\psi_\beta$ where $\mathbf{M}$
belongs to $\sltoc$. We will use Greek indices (running from 1 to
2) for Weyl spinors while Dirac and Majorana spinors have Latin
indices. However, we also get a representation (i.e. we keep the
group multiplication structure) if we work with the conjugated,
the transposed inverse or the hermitian inverse matrix. This is
summarised in table~\ref{tableappspinorn1} where we also see the
notation for the representations (we are here
following~\cite{wessandbagger}). Please note that we use dots on
indices that transform in the conjugate representation.
\begin{table}
\caption{}\label{tableappspinorn1}\centering
\begin{tabular}{|l|l|l|l|}
  \hline Name & Matrix & Notation & Transformation \\\hline
   Fundamental & $\mathbf{M}$ & $\psi_{\alpha}$ & $\psi'_{\alpha}=M_{\alpha}^{\phantom{\alpha}\beta}\psi_\beta$ \\\hline
   Conjugate & $\mathbf{M}^*$ & $\psibar_{\aldot}$ & $\psibar'_{\aldot}=\left(M^*\right)_{\aldot}^{\phantom{\aldot}\bedot}\psibar_{\bedot}$ \\\hline
   Dual & $\left(\mathbf{M}^T\right)^{-1}$ & $\psi^{\alpha}$ & $\psi'^{\alpha}=M^{-1\phantom{\be}\al}_{\phantom{-1}\be}\psi^\be$ \\\hline
   Conjugate dual & $\left(\mathbf{M}^\dagger\right)^{-1}$ & $\psibar^{\aldot}$ & $\psibar'^{\aldot}=\left(M^*\right)^{-1\phantom{\bedot}\aldot}_{\phantom{-1}\bedot}\psibar^{\bedot}$ \\\hline
\end{tabular}
\newline\newline Representations of $\sltoc$ and their notation.
\end{table}

But not all of these representations are inequivalent. Since we
are dealing with $2\times 2$ matrices with unit determinant, we
have the following relations:
\begin{eqnarray}\label{appspinorn17}
    \vep_{\al\be}=\Mnedop{\al}{\ga}\Mnedop{\be}{\de}\vep_{\ga\de},\nonumber\\
    \vep^{\al\be}=\vep^{\ga\de}\Mnedop{\ga}{\al}\Mnedop{\de}{\be},
\end{eqnarray}
where the antisymmetric tensors are defined by (as usual the
tensor with indices up is the inverse of the tensor with indices
down)
\begin{equation}\label{appspinorn17.5}
    \vep^{12}=-\vep^{21}=-\vep_{12}=\vep_{21}=1.
\end{equation}
Thus we see that the antisymmetric tensors are invariant.
Consequently, they can be used to raise and lower indices to
define invariant inner products. But one has to be careful since
we are dealing with antisymmetric tensors. We define raising and
lowering as:
\begin{equation}\label{appspinorn18}
    \psi^\alpha\defi\vep^{\al\be}\psi_\be,\qquad\psi_\al\defi\vep_{\al\be}\psi^\be.
\end{equation}
This is a consistent definition in itself since
$\psi^\alpha=\vep^{\al\be}\psi_\be=\vep^{\al\be}\vep_{\be\ga}\psi^\ga\stackrel{!}{=}\psi^\al$
by the use of $\vep^{\al\be}\vep_{\be\ga}=\de^\al_\ga$. But the
definition is also consistent with table~\ref{tableappspinorn1}
since:
\begin{eqnarray*}
    \psi^\al=\vep^{\al\be}\psi_\be &\mapsto&
    \vep^{\al\be}\Mnedop{\be}{\ga}\psi_\ga=\Mnedop{\de}{\ep}M^{-1\phantom{\ep}\al}_{\phantom{-1}\ep}\vep^{\de\be}\Mnedop{\be}{\ga}\psi_\ga \\
    &&=M^{-1\phantom{\ep}\al}_{\phantom{-1}\ep}\vep^{\ep\ga}\psi_\ga=M^{-1\phantom{\be}\al}_{\phantom{-1}\be}\psi^\be.
\end{eqnarray*}
Consequently, the fundamental representation and the dual
representation are equivalent. The same holds true for the
conjugate and the conjugate dual representations -- raising and
lowering are defined in exactly the same way with the same
definition of the $\vep$-tensor with dotted indices. The invariant
products are defined as:
\begin{equation}\label{appspinorn19}
    \psi\chi\defi\psi^\al\chi_\al,\qquad\psibar\chibar\defi\psibar_{\aldot}\chibar^{\aldot}.
\end{equation}
The reason for these contraction conventions will become clear
later. But please note that the ordering is important since e.g.
$\psi^\al\chi_\al=\vep^{\al\be}\psi_\be\chi_\al=-\psi_\al\chi^\al$.
Keeping in mind that half-integer spinors must be represented by
anticommuting Grassmann numbers we actually see that these
contractions commute since e.g.
$\psi\chi=\psi^\al\chi_\al=-\chi_\al\vep^{\al\be}\psi_\be=\chi^\be\psi_\be=\chi\psi$.
Looking back at the transformation~\eqref{appspinorn10} we see
that the index structure of the $\sigma$-matrices is:
\begin{equation}\label{appspinorn20}
    \sigmaop{\mu}{\al}{\be}.
\end{equation}
Hence we get yet another Lorentz scalar by the contraction
$\psi^\al\sigmaop{\mu}{\al}{\be}\partial_\mu\chi^{\bedot}$. We can
also use the $\vep$-matrices to raise and lower the indices on the
$\sigma$-matrices using the definition~\eqref{appspinorn18}. Using
this we actually get:
\begin{equation}\label{appspinorn21}
    \sigmabar^{\aldot\be}=\sigma^{\be\aldot}=\vep^{\aldot\gadot}\vep^{\be\de}\sigma_{\de\gadot}.
\end{equation}
Inserting this index structure in~\eqref{appspinorn12} we get:
\begin{eqnarray}\label{appspinorn22}
    \Sigma_{\mu\nu} &=& \frac{i}{4}\begin{pmatrix}
      \sigmaned{\mu}{\al}{\ga}\sigmabarned{\nu}{\ga}{\be}-\sigmaned{\nu}{\al}{\ga}\sigmabarned{\mu}{\ga}{\be} & 0 \\
      0 & \sigmabarned{\mu}{\al}{\ga}\sigmaned{\nu}{\ga}{\be}-\sigmabarned{\nu}{\al}{\ga}\sigmaned{\mu}{\ga}{\be} \\
    \end{pmatrix} \nonumber\\
    &\defi& i\begin{pmatrix}
      \left(\sigma_{\mu\nu}\right)_\al^{\ph{\al}\be} & 0 \\
      0 & \left(\sigmabar_{\mu\nu}\right)^{\aldot}_{\ph{\aldot}\bedot} \\
    \end{pmatrix},
\end{eqnarray}
where we also defined $\sigma_{\mu\nu}$ and $\sigmabar_{\mu\nu}$.
The diagonal Lorentz generators (the benchmark of the Weyl basis
for the $\gamma$-matrices) directly show us how the Dirac spinor,
$\Psi_{\textrm{D}}$, and the Majorana spinor, $\Psi_{\textrm{M}}$,
split into two Weyl spinors:
\begin{equation}\label{appspinorn23}
    \Psi_{\textrm{D}}\sim\begin{pmatrix}
      \psi_\al \\
      \chibar^{\aldot} \\
    \end{pmatrix},\qquad
    \Psi_{\textrm{M}}\sim\begin{pmatrix}
      \psi_\al \\
      \psibar^{\aldot} \\
    \end{pmatrix}.
\end{equation}
Here we have used the Majorana reality
condition~\eqref{appspinorn14.5} in writing the Majorana spinor.
We have also used how to complex conjugate the Weyl spinor as will
be defined in the next subsection. Rewriting the $\Sigma_{\mu\nu}$
generators to $\mathbf{K}$ and $\mathbf{L}$ we can easily see that
the upper spinor transforms according to~\eqref{appspinorn5}, i.e.
like $(\half,0)$, while the lower spinor transforms according
to~\eqref{appspinorn6}, i.e. like $(0,\half)$.\footnote{We note
that the transformation~(\ref{appspinorn6}) is not exactly the
conjugate of~(\ref{appspinorn5}). In order to get the exact
conjugate we have to do, as we did when going
from~(\ref{appspinorn6}) to the exactly conjugate representation
in~(\ref{appspinorn7}). Since the conjugate representation (that
we used in table~\ref{tableappspinorn1}) and~(\ref{appspinorn6})
are equivalent, it does not matter which one we choose the dotted
indices to transform in -- as long as we consistently choose the
same everywhere.}

\section{Complex Conjugation}

We define complex conjugation on Weyl spinors as the involution:
\begin{equation}\label{appspinorn24}
    \left(\psi_\al\right)^*\defi\psibar_{\aldot},\qquad
    \left(\psi^\al\right)^*=\psibar^{\aldot},
\end{equation}
where the last part simply followed since we chose the
$\vep$-tensor with and without dotted indices to be the same. This
definition fits with table~\ref{tableappspinorn1}. However, when
complex conjugation works on products of anticommuting spinors it
reverses the order (or equivalently it adds an appropriate sign),
e.g.:
\begin{equation}\label{appspinorn25}
    \left(\psi_\al\chi_\be\right)^*=\chibar_{\bedot}\psibar_{\aldot}.
\end{equation}
Now we see the reason for the placement of the indices
in~\eqref{appspinorn19} because with these definitions we get the
nice equation:
\begin{equation}\label{appspinorn26}
    \left(\psi\chi\right)^*=\left(\psi^\al\chi_\al\right)^*=\chibar_{\aldot}\psibar^{\aldot}=\chibar\psibar=\psibar\chibar.
\end{equation}

Using the definition of complex conjugation we can show that the
inner product of two Majorana spinors is real. First we find the
Majorana conjugate in the Weyl spinor formalism:
(using~\eqref{appspinorn14.1}):
\begin{equation}\label{appspinorn27}
    \Psibar_{\textrm{M}}=\left(\psi^\al,\psibar_{\aldot}\right)
\end{equation}
Hence the product of two Majorana spinors ($\Psi$ and $\Phi=\begin{pmatrix}
  \phi_\al \\
  \bar{\phi}^{\aldot} \\
\end{pmatrix}$) becomes:
\begin{equation}\label{appspinorn28}
    \Psibar_{\textrm{M}}\Phi=\psi\phi+\psibar\bar{\phi}.
\end{equation}
This is real by~\eqref{appspinorn26}.

\section{Differentiation of Spinors}

The last thing we have to settle is the differentiation of Weyl
spinors. As usual for Grassmannian differentiation it should
fulfil:
\begin{equation}\label{appspinorn29}
    \{\partialop{\al},\psi^\be\}=\de^\be_\al,
\end{equation}
and the same with indices lowered. But now we have to be careful
when defining raising, lowering and conjugation of the
differential. Using the definition of raising we see:
\begin{equation*}
    \{-\vep^{\al\be}\partialop{\be},\psi_\ga\}=-\vep^{\al\be}\vep_{\ga\de}\{\partialop{\be},\psi^\de\}=-\vep^{\al\be}\vep_{\ga\de}\de^\de_\be=\de^\al_\ga.
\end{equation*}
Consequently we have:
\begin{equation}\label{appspinorn30}
    -\vep^{\al\be}\partialop{\be}=\partialned{\al}.
\end{equation}

Also when complex conjugating spinorial derivatives we have to be
a bit careful. When conjugating~\eqref{appspinorn29} we have to
reverse the order of the products, however, conjugation should not
change on which part the derivative acts. So instead of changing
order we preserve the order and add a compensating sign. For a
general field $A$ we then have:
\begin{equation}\label{appspinorn30.5}
    \{\partialop{\al},A\}^*=(-1)^{|A|}\{\left(\partialop{\al}\right)^*,\left(A\right)^*\},
\end{equation}
where $|A|$ is 0 if $A$ is bosonic, and 1 if it is fermionic.
Using this on~\eqref{appspinorn29} gives
\begin{equation}\label{appspinorn31}
    \left(\partialop{\al}\right)^*=-\partialbarop{\al}.
\end{equation}
So complex conjugation on differentials does not work in the same
way as on spinors.\footnote{This could have been solved if we had
defined the $\vep$-tensor with dotted indices with a minus sign
contrary to the $\vep$-tensor without dotted indices.}

The hermitian adjoint is defined by:
\begin{equation}\label{appspinorn31.1}
    \int\dx\textrm{d}\psi\textrm{d}\psibar B^*\partialop{\al}A\defi\int\dx\textrm{d}\psi\textrm{d}\psibar
    \left(\left(\partialop{\al}\right)^\dagger B\right)^*A,
\end{equation}
where $\int\textrm{d}\psi\textrm{d}\psibar$ is the integration
over the spinorial degrees of freedom while $A$ and $B$ are two
arbitrary fields. Taking care of all signs, also when doing the
integration by parts, this gives
\begin{equation}\label{appspinorn31.2}
    \left(\partialop{\al}\right)^\dagger=\partialbarop{\al},
\end{equation}
contrary to the usual derivative.

\section{Space-time Dimensions}\label{appspinorsecdim}
Even though this thesis deals with four space-time dimensions let
us appreciate that it is only in a few space-time dimensions that
it is possible to construct an $\Nscr=1$ supersymmetric Yang-Mills
theory. Following~\cite{polchinski2} let us look at which
spinorial representations it is possible to make in a general
dimension. We will assume that the signature of the metric is
Lorentzian, i.e. we are looking at the group $\textrm{SO}(D-1,1)$
where $D$ is the space-time dimension.

In every space-time dimension it is possible to make the Dirac
representation which after imposing the Dirac equation has
dimension (i.e. number of real degrees of freedom)
$2^{\left[D/2\right]}$. The brackets means the least whole number
closest to $D/2$.

However, in even dimensions this representation is not irreducible
since we can make a chirality matrix like $\gamma^5$. This cuts
the Dirac representation into two inequivalent Weyl
representations both of dimension $2^{\left[D/2-1\right]}$. The
two representations are conjugate if $D\equiv 0 \pmod{4}$ and they
are self-conjugate if $D\equiv 2 \pmod{4}$.

The Majorana reality condition defines the Majorana representation
and cuts the dimension to the half: $2^{\left[D/2-1\right]}$.
However, if the Majorana condition is to be consistent one must
require $D\equiv 0,1,2,3,4 \pmod{8}$.

A last representation can be obtained if we impose both the
Majorana condition and the Weyl condition. But this requires that
the Weyl representation is self-conjugate. Hence this
Majorana-Weyl representation is only possible if $D\equiv 2
\pmod{8}$. The dimension is then $2^{\left[D/2-2\right]}$.

When we look at supersymmetry we find that the number of bosonic
and fermionic degrees of freedom must be the same. The number of
bosonic degrees of freedom in the Yang-Mills model is $D-2$ (the
transverse directions in the gauge field $A_\mu$). This must fit
with one of the above irreducible representations. Simple counting
now tells us in which space-time dimensions and with which
representations it is possible to make a supersymmetric Yang-Mills
theory. The result can be seen in table~\ref{tableappspinorn2}.
\begin{table}
\caption{}\label{tableappspinorn2}
\begin{flushright}
\begin{tabular}{|l|l|}
  \hline D & Spinor representation \\\hline
  3 & Majorana \\\hline
  4 & Majorana or Weyl \\\hline
  6 & Weyl \\\hline
  10 & Majorana-Weyl \\\hline
\end{tabular}
\newline
\begin{center}
The possible space-time dimensions, D, for which supersymmetric
Yang-Mills theory can be realised and the corresponding
representations. The table is taken from~\protect{\cite{0109172}}.
\end{center}
\end{flushright}
\end{table}
We see that four space-time dimensions is one of the few
dimensions in which it is possible. Please note that due
to~\eqref{appspinorn23} the Majorana and the Weyl representations
are equivalent.

\section{Weyl Spinor Algebra}

Using our definitions we can find a lot of useful identities for
Weyl spinors. Without proof we give the following relations for
$\psi$, $\chi$ and $\tha$ being anticommuting Weyl spinors. The
contractions follow the logic of~\eqref{appspinorn19}, $\sigma$
having indices~\eqref{appspinorn20} and $\sigmabar$ having
indices~\eqref{appspinorn21} such that e.g.
$\chi\sigma^\mu\psibar=\chi^\al\sigmaop{\mu}{\al}{\be}\psibar^{\bedot}$
(the relations are taken from~\cite{wessandbagger}):
\begin{eqnarray}\label{appspinorn32}
  \tha^\al\tha^\be &=&-\half\vep^{\al\be}\tha\tha,  \nonumber\\
  \tha_\al\tha_\be &=&\half\vep_{\al\be}\tha\tha, \nonumber\\
  \thabar^{\aldot}\thabar^{\bedot} &=&\half\vep^{\aldot\bedot}\thabar\thabar,  \nonumber\\
  \thabar_{\aldot}\thabar_{\bedot} &=&-\half\vep_{\aldot\bedot}\thabar\thabar.
\end{eqnarray}
\begin{equation}\label{appspinorn33}
    \tha\sigma^\mu\thabar\tha\sigma^\nu\thabar=-\half\tha\tha\thabar\thabar\eta^{\mu\nu}.
\end{equation}
\begin{equation}\label{appspinorn34}
    \left(\tha\psi\right)\left(\tha\chi\right)=-\half\left(\psi\chi\right)\left(\tha\tha\right).
\end{equation}
\begin{equation}\label{appspinorn35}
    \left(\thabar\psibar\right)\left(\thabar\chibar\right)=-\half\left(\psibar\chibar\right)\left(\thabar\thabar\right).
\end{equation}
\begin{eqnarray}\label{appspinorn36}
  \chi\sigma^\mu\psibar &=& -\psibar\sigmabar^\mu\chi, \nonumber\\
  \left(\chi\sigma^\mu\psibar\right)^\dagger &=&
  \psi\sigma^\mu\chibar.
\end{eqnarray}
\begin{eqnarray}\label{appspinorn37}
  \chi\sigma^\mu\sigmabar^\nu\psi &=& \psi\sigma^\nu\sigmabar^\mu\chi, \nonumber\\
  \left(\chi\sigma^\mu\sigmabar^\nu\psi\right)^\dagger &=&
  \psibar\sigmabar^\nu\sigma^\mu\chibar.
\end{eqnarray}
\begin{equation}\label{appspinorn38}
    \left(\psi\tha\right)\chibar_{\bedot}=-\half\left(\tha\sigma^\mu\chibar\right)\left(\psi\sigma_\mu\right)_{\bedot}.
\end{equation}
\begin{equation}\label{appspinorn38.5}
    \tr\nleft(\si^{\mu}\sigmabar^{\nu}\right)=-2\eta^{\mu\nu}.
\end{equation}
\begin{equation}\label{appspinorn39}
    \tr\nleft(\si^{\mu\nu}\si^{\rho\de}\right)=-\frac{1}{2}\left(\eta^{\mu\rho}\eta^{\nu\de}-\eta^{\mu\de}\eta^{\nu\rho}\right)-\frac{i}{2}\vep^{\mu\nu\rho\de}.
\end{equation}
\begin{equation}\label{appspinorn40}
    \vep^{\al\be}\ddthaop{\al}\ddthaop{\be}\tha\tha=4.
\end{equation}

%------New Appendix----------------------------------------------------

\chapter{$\Nscr=2$ Superspace and The Prepotential}\label{appn2super}
In this appendix we will continue the analysis of the $\Nscr=2$
supersymmetric Yang-Mills theory from
section~\ref{susysecn2subren}. But here we will not assume
renormalisability of the Lagrangian.

In order to get the most general $\Nscr=2$ supersymmetric pure
Yang-Mills Lagrangian with no constriction of renormalisability it
is an ease to use the $\Nscr=2$ superspace formulation
(following~\cite{9912271} and~\cite{9701069}). This is a simple
extension of the $\Nscr=1$ superspace from
section~\ref{susysecN1subsuper} where the $\tha^\al$-coordinates
now carry an extra $\suto_R$ index: $\tha^\al_i$ with $i=1,2$.
Thus we have one $\tha^\al$-coordinate for each supercharge. The
new index can be raised and lowered as the spinor index using the
$\suto$ invariant antisymmetric tensor $\vep_{ij}$.

The supercharges are linearly realised on superfields in the same
way as in the $\Nscr=1$ case -- we just have to put the index $i$
on the $\tha$'s and the $Q$'s. We also get the covariant
derivatives $D^i_\al$ and $\Dbar_{\aldot i}$ where as an example:
\begin{equation}\label{susyn84}
    \Dbar_{\aldot i}=-\ddthabarop{\al i}-i\tha^\be_i\sigmaop{\mu}{\be}{\al}\partial_\mu.
\end{equation}

A chiral superfield is then defined as:
\begin{equation}\label{susyn85}
    \Dbar_{\aldot i}\Phi=0,\qquad\aldot=\dot{1},\dot{2},\ i=1,2.
\end{equation}
In analogy with the $\Nscr=1$ case a differentiable function of
chiral superfields is again a chiral superfield. Also the
variation under supersymmetry of the component with four $\tha$'s
and no $\thabar$'s is a total derivative. Thus a possible
Lagrangian is:
\begin{equation}\label{susyn86}
    \lagr=\int\dfourtha\Phi,
\end{equation}
where the indices on the differentials are the $\suto_R$ indices.

The chiral field is not irreducible as in the $\Nscr=1$ case. This
means that we have to impose further constraints. To do this we
introduce the $\cG$-valued supergauge fields $A_{\al i}$ and
$\bar{A}_{\aldot i}$. With these we define the gauge-covariant
version of the supersymmetric covariant derivatives as:
\begin{equation}\label{susyn87}
    \tilde{D}_{\al i}=D_{\al i}+iA_{\al i}\qquad \bar{\tilde{D}}_{\aldot i}=\Dbar_{\aldot i}+i\bar{A}_{\aldot
    i}.
\end{equation}
The constraint needed to generate an $\Nscr=2$ gauge field out of
a chiral $\cG$-valued field $W$ (not to be confused with the
$\Nscr=1$ superpotential) turns out to be:
\begin{equation}\label{susyn88}
    \tilde{D}^{\al i}\tilde{D}_{\al}^jW=\bar{\tilde{D}}_{\aldot}^i\bar{\tilde{D}}^{\aldot
    j}W^\dagger.
\end{equation}
Expanding this field as a power series in $\tha^2$ gives the
$\Nscr$=1 superfield components:
\begin{equation}\label{susyn89}
    W(\tilde{x}_+,\tha^1,\tha^2)=\Phi\nleft(\tilde{x}_+,\tha^1\right)+\sqrt{2}\tha^{\al2}\W_\al\nleft(\tilde{x}_+,\tha^1\right)+\tha^2\tha^2G\nleft(\tilde{x}_+,\tha^1\right),
\end{equation}
where $\tilde{x}_+^\mu=x^\mu+i\tha_i\si^\mu\thabar^i$ is the
$\Nscr=2$ version of $x_+$. $\Phi$ and $\W$ are the $\Nscr=1$
chiral field and gauge field strength respectively corresponding
to the splitting of the $\Nscr=2$ gauge supermultiplet into the
$\Nscr=1$ chiral and gauge supermultiplet. $\W$ is based on the
vector superfield $V$. $G$ may be expressed as:
\begin{equation}\label{susyn90}
    G\nleft(\tilde{x}_+,\tha^1\right)=-\int\dtothabar^1\Phi\nleft(\tilde{x}_+-i\tha_1\si\thabar^1,\tha^1,\thabar^1\right)^\dagger
    e^{2[-,V(\tilde{x}_+-i\tha_1\si\thabar^1,\tha^1,\thabar^1)]},
\end{equation}
where $[-,V]$ simply means that the adjoint is working to the left
with $V$ to the right in the commutators. We see that the field
$W$ has the component fields of the $\Nscr=2$ gauge multiplet
(after eliminating the auxiliary fields). The most general gauge
invariant supersymmetric Lagrangian involving $W$ is now:
\begin{equation}\label{susyn91}
    \lagr_{\Nscr=2}=-\frac{1}{8\pi i}\int\dfourtha\tr\cF\nleft(W\right)+\cc,
\end{equation}
where $\cF$ is any holomorphic function. $\tr\cF$ is called the
\emph{prepotential}. The trace ensures gauge invariance since the
fields transform in the adjoint representation. We can expand this
in components as we did in~\eqref{susyn68} using~\eqref{susyn36}
and~\eqref{susyn37}:
\begin{equation}\label{susyn92}
    \lagr_{\Nscr=2}=\frac{1}{16\pi i}\left(\int\dtotha^1\cF_{ab}\nleft(\Phi\right)\W^{\al
    a}\W^b_\al+2\int\dtotha^1\dtothabar^1\left(\Phi^\dagger
    e^{2[-,V]}\right)^a\cF_a\nleft(\Phi\right)\right)+\cc
\end{equation}
Here $\cF_a$ and $\cF_{ab}$ are simply the derivatives
$\cF_a=\partial\cF/\partial\Phi^a$ and
$\cF_{ab}=\partial^2\cF/\partial\Phi^a\partial\Phi^b$ respectively
where we have made an abuse of notation and used the name $\cF$
for $\tr\cF$ as a function of $\Phi^a$. Setting
$\cF=\half\tau\Phi^a\Phi^a$ we get the renormalisable
Lagrangian~\eqref{susyn81}.\footnote{To see that $[-,V]$ (working
to the left) is the same as $\ad\!V$ (working to the right) we
have to rewrite the sum over the adjoint indices as a trace and
use the cyclic properties of the trace.} We can also see that the
\kahler{} potential (without $1/4\pi$) is $\im\nleft(\Phi^{\dagger
a}\cF_a\nleft(\Phi\right)\right)$ and thus the \kahler{} metric
according to~\eqref{susyn72} is
$g_{ab}=\im\nleft(\cF_{ab}\right)$.

%------New Appendix----------------------------------------------------

\chapter{Calculation of Integrals in the Matrix Model}\label{appmatrix}

In this appendix we calculate the integrals needed in
section~\ref{dvsecmatrixsubonecut}.

Let us first derive equation~\eqref{dvn163}:
\begin{equation}
    \frac{\partial}{\partial\gm}\big(\gm\rho\nleft(\la,\gm\right)\big)=\frac{1}{\pi\sqrt{(b-\la)(\la-a)}}.\nonumber
\end{equation}
In principle we could obtain this by differentiating the
equation~\eqref{dvn160} we have obtained for $R(z)$ -- remembering
that $a$ and $b$ depend on $\gm$. However, it is more easy to note
that (following~\cite{9306153}):
\begin{equation}\label{appmatn1}
    \Om(z)\defi\frac{\partial\gm
    R(z)}{\partial\gm}=\int_{-\infty}^\infty\dla\,\frac{1}{\la-z}\frac{\partial}{\partial\gm}\big(\gm\rho\nleft(\la,\gm\right)\big),
\end{equation}
is an analytic function with branch cut $[a,b]$ and fulfils: It
has no regular part due to~\eqref{dvn153}, like $R(z)$ it must
behave like $-1/z$ for large $\abs{z}$, and since $R(z)$ behaves
like $\sqrt{z-a}$ close to the branch point $a$, the derivative
$\Om(z)$ can behave at most as $1/\sqrt{z-a}$ (and the same for
$b$). This determines $\Om$ uniquely to be:
\begin{equation}\label{appmatn2}
    \Om(z)=-\frac{1}{\sqrt{(z-b)(z-a)}}.
\end{equation}
Using~\eqref{appmatn1} we then obtain in analogy
with~\eqref{dvn152}:
\begin{equation}\label{appmatn3}
    \frac{\partial}{\partial\gm}\big(\gm\rho\nleft(\la,\gm\right)\big)=\frac{1}{2\pi
    i}\left(\Om(\la+i\ep)-\Om(\la-i\ep)\right)=\frac{1}{\pi\sqrt{(b-\la)(\la-a)}},
\end{equation}
as wanted.

Let us now derive~\eqref{dvn164} following~\cite{0210135}:
\begin{equation*}
    I(\la')\defi\dashint_a^b\dla\,\frac{\ln\abs{\la-\la'}}{\sqrt{(b-\la)(\la-a)}}=\pi\ln\nleft(\frac{b-a}{4}\right),\quad\forall\la'\in[a,b].
\end{equation*}
In analogy with equation~\eqref{dvn153} we have:
\begin{equation}\label{appmatn4}
    \frac{\partial
    I}{\partial\la'}=\frac{1}{2}\nleft(h(\la'+i\ep)+h(\la'-i\ep)\right),\quad\la'\in\R,
\end{equation}
where
\begin{equation}\label{appmatn5}
    h(z)=\int_a^b\dla\,\frac{1}{(z-\la)\sqrt{(b-\la)(\la-a)}}.
\end{equation}
$h$ is analytic in $\C\setminus[a,b]$ and in analogy
with~\eqref{dvn152} we have:
\begin{equation}\label{appmatn6}
    h(\la'+i\ep)-h(\la'-i\ep)=-\frac{2\pi
    i}{\sqrt{(b-\la')(\la'-a)}},\quad\la'\in[a,b].
\end{equation}
From the definition~\eqref{appmatn5} we also see (using
equation~\eqref{appmatn14} below) that $h\sim\pi/z$ for large
$\abs{z}$. This determines $h$ uniquely as:
\begin{equation}\label{appmatn7}
    h(z)=\frac{\pi}{\sqrt{(z-b)(z-a)}},\quad z\in\C\setminus[a,b].
\end{equation}
Thus using~\eqref{appmatn4} we get:
\begin{equation}\label{appmatn8}
    \frac{\partial I}{\partial\la'}(\la')=\begin{cases}0&\textrm{for
    $\la'\in[a,b]$}\\h(\la')&\textrm{for $\la'\in\R\setminus[a,b]$}
    \end{cases}.
\end{equation}
For $\la'>b$ we then get:
\begin{equation}\label{appmatn9}
    I(\la')=\pi\ln\nleft(2\sqrt{(\la'-b)(\la'-a)}+2\la'-a-b\right)-\pi\ln4,\quad\la'>b,
\end{equation}
where the constant of integration has been fixed using that
$I(\la')\sim\pi\ln(\la')$ for large $\la'$. Using the continuity
of $I(\la')$ we then get the wanted result by plugging $\la'=b$
into~\eqref{appmatn9}.

Let us finally show how to derive equations~\eqref{dvonecutsol}.
From equations~\eqref{dvn161},~\eqref{dvn162} and~\eqref{dvn165}
we see that basically all we need is the integral:
\begin{equation}\label{appmatn10}
    I(p)\defi\int_a^b\dla\frac{\la^p}{\sqrt{(b-\la)(\la-a)}}.
\end{equation}
Defining $\la'=\la-(a+b)/2$ we get:
\begin{eqnarray}\label{appmatn11}
    I(p)&=&\int_{-(b-a)/2}^{(b-a)/2}\dla'\frac{\left(\la'+\frac{b+a}{2}\right)^p}{\sqrt{\left(\frac{b-a}{2}\right)^2-\la'^2}}=
    \sum_{k=0}^p\binom{p}{k}\int_{-(b-a)/2}^{(b-a)/2}\dla'\frac{\la'^k\left(\frac{b+a}{2}\right)^{p-k}}{\sqrt{\left(\frac{b-a}{2}\right)^2-\la'^2}}\nonumber\\
    &=&\sum_{k=0}^p\binom{p}{k}\left(\frac{b+a}{2}\right)^{p-k}\left(\frac{b-a}{2}\right)^{k}\int_{-1}^{1}\dla'\frac{\la'^k}{\sqrt{1-\la'^2}}\nonumber\\
    &=&\sum_{q=0}^{\left[p/2\right]}\binom{p}{2q}\left(\frac{b+a}{2}\right)^{p-2q}\left(\frac{b-a}{2}\right)^{2q}2\int_{0}^{1}\dla'\frac{\la'^{2q}}{\sqrt{1-\la'^2}},
\end{eqnarray}
where we have used that the integral of an odd function is zero in
the last line. Substituting $\la'=\sin y$ and using integrating by
parts we see that:
\begin{equation}\label{appmatn12}
    \int_{0}^{1}\dla'\frac{\la'^{k}}{\sqrt{1-\la'^2}}=\int_0^{\pi/2}\inted
    y\,\sin^k y=\frac{k-1}{k}\int_0^{\pi/2}\inted
    y\,\sin^{k-2} y.
\end{equation}
Thus:
\begin{equation}\label{appmatn13}
    \int_{0}^{1}\dla'\frac{1}{\sqrt{1-\la'^2}}=\frac{\pi}{2},
\end{equation}
and hence
\begin{equation}\label{appmatn13.5}
    \int_{0}^{1}\dla'\frac{\la'^{2q}}{\sqrt{1-\la'^2}}=\frac{1\cdot3\cdots(2q-1)}{2\cdot4\cdots(2q)}\frac{\pi}{2}.
\end{equation}
Using that
$\binom{2q}{q}=2^{2q}\frac{1\cdot3\cdots(2q-1)}{2\cdot4\cdots(2q)}$
we finally get:
\begin{equation}\label{appmatn14}
    I(p)=\pi\sum_{q=0}^{\left[p/2\right]}\binom{p}{2q}\binom{2q}{q}\left(\frac{b+a}{2}\right)^{p-2q}\left(\frac{b-a}{4}\right)^{2q}.
\end{equation}
We then obtain~\eqref{dvn166} and~\eqref{dvn168} simply by
expanding $\wpoly(\la)$ in respectively~\eqref{dvn161}
and~\eqref{dvn165}.~\eqref{dvn167} is obtained by expanding
\begin{equation}\label{appmatn15}
    \int_a^b\dla\,\frac{(\la-(b+a)/2)\wpoly'\nleft(\la\right)}{\sqrt{(b-\la)(\la-a)}}=2\pi\gm,
\end{equation}
which is~\eqref{dvn162} with the addition of~\eqref{dvn161} times
$-(b+a)/2$.~\eqref{appmatn15} is the original condition that
$R(z)\sim-1/z$ for large $\abs{z}$ in equation~\eqref{dvn160}. The
reason we write equation~\eqref{dvn167} in this way is for
comparison with the result obtained on the gauge theory side using
Seiberg-Witten theory in~\cite{0210135}.

\end{appendix}

%--------------------------------------------------------------------------------------------

%The following two lines is for BibTeX only:
\addcontentsline{toc}{chapter}{Bibliography}

\bibliographystyle{utphys}
\bibliography{PeterBib}

\end{document}